\documentclass[journal]{IEEEtran}
\ifCLASSINFOpdf
\else
\fi
\usepackage{nicefrac}       % compact symbols for 1/2, etc.
\usepackage{amsfonts,amsthm,amsopn,mathrsfs}
\usepackage{algpseudocode}
\usepackage{algorithm}
\usepackage{enumitem,cite,url,bm,amsmath,amssymb,graphicx,subfigure,enumitem,float,pifont,xfrac,mathtools}
\usepackage{multirow}
\usepackage{newtxmath}

\usepackage{hyperref}
\hypersetup{colorlinks=black, pdfstartview=FitV, linkcolor=blue, citecolor=black, plainpages=false, pdfpagelabels=true, urlcolor=blue}
\usepackage[all]{hypcap}

\usepackage{color}% Include colors for document elements
\usepackage{dcolumn}% Align table columns on decimal point
\definecolor{00BCB4}{RGB}{0,188,180}
\definecolor{C4E86B}{RGB}{196,232,107}
\definecolor{49DEB2}{RGB}{73,222,178}
\definecolor{FF4747}{RGB}{255,71,71}
\definecolor{FF3561}{RGB}{255,53,97}
\definecolor{CD2C6C}{RGB}{205,44,108}
\definecolor{10B48E}{RGB}{16,180,142}
\definecolor{8FDBD8}{RGB}{143,219,216}

\usepackage{etoolbox}
\makeatletter
\patchcmd{\maketitle}
 {\def\@makefnmark}
 {\def\@makefnmark{}\def\useless@macro}
 {}{}
\makeatother

\usepackage{tikz}
\usetikzlibrary{bayesnet}
\usetikzlibrary{arrows,shapes,snakes,automata,backgrounds,petri}

\newcommand{\vo}{{\bm o}}
\newcommand{\vw}{{\bm w}}
\newcommand{\vx}{{\bm x}}
\newcommand{\vy}{{\bm y}}
\newcommand{\vz}{{\bm z}}
\newcommand{\vi}{{\bm i}}

\newcommand{\vq}{{\bm q}}

\newcommand{\mA}{{\bm A}}
\newcommand{\mK}{{\bm K}}

\newcommand{\mH}{{\bm H}}

\newcommand{\mX}{{\bm X}}
\newcommand{\mO}{{\bm O}}

\usepackage{chngcntr}
\usepackage{apptools}
\AtAppendix{\counterwithin{lemma}{section}}

\newtheorem{finding}{Finding}

\newenvironment{sparse_signal_model}[1][\ding{113} Sparse signal model:\\]{\begin{trivlist}\item[\hskip \labelsep {\bfseries #1}]}{\end{trivlist}}
\newenvironment{1_bit_quantization_noise_model}[1][\ding{113} 1-bit quantization noise model:\\]{\begin{trivlist}\item[\hskip \labelsep {\bfseries #1}]}{\end{trivlist}}
\newenvironment{multi_bit_quantization_noise_model}[1][\ding{113} Multi-bit quantization noise model:\\]{\begin{trivlist}\item[\hskip \labelsep {\bfseries #1}]}{\end{trivlist}}

\algnewcommand{\IIf}[1]{\State\algorithmicif\ #1\ \algorithmicthen}
\algnewcommand{\EndIIf}{\unskip\ \algorithmicend\ \algorithmicif}

\algnewcommand\algorithmicinput{\textbf{Input:}}
\algnewcommand\algorithmicoutput{\textbf{Output:}}
\algnewcommand\Input{\item[\algorithmicinput]}%
\algnewcommand\Output{\item[\algorithmicoutput]}%

\allowdisplaybreaks

\begin{document}
\title{Approximate Message Passing with Parameter Estimation for Heavily Quantized Measurements}

\author{Shuai Huang, Deqiang Qiu, and Trac D. Tran, ~\IEEEmembership{Fellow,~IEEE}% <-this % stops a space
\thanks{\copyright 2022 IEEE. Personal use of this material is permitted. Permission from IEEE must be obtained for all other uses, in any current or future media, including reprinting/republishing this material for advertising or promotional purposes, creating new collective works, for resale or redistribution to servers or lists, or reuse of any copyrighted component of this work in other works.}
\thanks{This work was partially supported by the National Science Foundation under grants NSF-CCF-1117545, NSF-CCF-1422995 and NSF-ECCS-1443936. (Corresponding author: Trac D. Tran.)}
\thanks{Shuai Huang and Deqiang Qiu are with Emory university, Atlanta, GA 30322 USA. Trac D. Tran is with Johns Hopkins University, Baltimore, MD 21218 USA. (email: shuai.huang@emory.edu; deqiang.qiu@emory.edu; trac@jhu.edu).}}% <-this % stops a space

\maketitle

\begin{abstract}
Designing efficient sparse recovery algorithms that could handle noisy quantized measurements is important in a variety of applications -- from radar to source localization, spectrum sensing and wireless networking. We take advantage of the approximate message passing (AMP) framework to achieve this goal given its high computational efficiency and state-of-the-art performance. In AMP, the signal of interest is assumed to follow certain prior distribution with unknown parameters. Previous works focused on finding the parameters that maximize the measurement likelihood via expectation maximization -- an increasingly difficult problem to solve in cases involving complicated probability models. In this paper, we treat the parameters as unknown variables and compute their posteriors via AMP. The parameters and signal of interest can then be jointly recovered. Compared to previous methods, the proposed approach leads to a simple and elegant parameter estimation scheme, allowing us to directly work with 1-bit quantization noise model. We then further extend our approach to general multi-bit quantization noise model. Experimental results show that the proposed framework provides significant improvement over state-of-the-art methods across a wide range of sparsity and noise levels. 
\end{abstract}


% \fi
\IEEEpeerreviewmaketitle

\section{Introduction}
Compressive sensing allows us to recover a large signal family with sparse prior information at lower sampling rates \cite{Decode05,RUP06,CS06,SRRP06}. In this paper, we are interested in recovering a sparse signal $\vx\in\mathbb{R}^N$ given the measurement matrix $\mA\in\mathbb{R}^{M\times N}$ and the \emph{heavily quantized} measurements $\vy\in\mathbb{R}^M$. The problem itself is ill-posed, and we have to rely on the sparse prior of the signal for recovery. In the extreme case where $\vy\in\{-1,+1\}^M$, we have the classic \emph{``1-bit compressive sensing (CS)''} problem originally proposed in \cite{Boufounos:1bitCS:2008}. This 1-bit CS set-up actually arises from a few practical applications. One such application involves the channel estimation problem in the massive multiple-input-multiple-output (MIMO) communication system, where the channel matrix is approximately sparse in the angle domain, and quantized measurements are acquired using low-cost low-resolution analog-to-digital converters (ADC) \cite{Larsson:MIMO:2014,Wen:Channel:2015}. 

As the number of quantization bits increases, power consumption of the ADC grows exponentially, along with the drastically increased cost and difficulty in hardware design \cite{Walden:ADC:1999}. Currently, these issues make it either too expensive or impractical to deploy high-resolution ADCs in base stations and portable devices \cite{Murmann:1997-2020}. As a result, there has been a growing interest in low-resolution ADCs that output $1\sim4$ bits in recent years \cite{Wen:LowADC:2016,Jacobsson:LowADC:2017,Mo:LowADC:2018}. Particularly, 1-bit ADC is much preferred in wideband millimeter wave communication systems that require high sampling frequency \cite{Mo:1bitADC:2015,Jacobsson:1bitADC:2015,Choi:1bitADC:2016,L1:1bitADC:2016}.

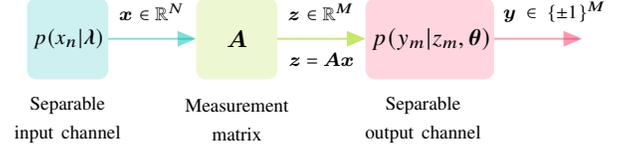
\begin{figure}[tbp]
\begin{tabular}{c}
\scalebox{0.9}{
% See the files LICENSE_LPPL and LICENSE_GPL for more details.
\definecolor{00BCB4}{RGB}{0,188,180}
\definecolor{C4E86B}{RGB}{196,232,107}
\definecolor{FF3561}{RGB}{255,53,97}
%\definecolor{F775A9}{RGB}{247,117,169}

%\beginpgfgraphicnamed{model-pca}
\begin{tikzpicture}

  % Define nodes
  \fill[00BCB4, opacity=0.2, rounded corners] (0,0) rectangle (1.2,1.2);
  \node[text width=1cm, align=center] at (0.6,0.6) {\small$p(x_n|\boldsymbol\lambda)$};
  \node[text width=2cm, align=center] at (0.6, -0.6) {\footnotesize Separable input channel};
  
  \draw[00BCB4, opacity=0.5, thick, ->] (1.2,0.6) -- (2.5,0.6) ;
  \node[text width=1cm, align=center] at (1.85, 1) {\footnotesize $\vx\in\mathbb{R}^N$};
  
  \fill[C4E86B, opacity = 0.3, rounded corners] (2.5,0) rectangle (3.7,1.2);
  \node[text width=1cm, align=center] at (3.1,0.6) {$\mA$};
  %\node[text width=1cm, align=center] at (3.7,0.5) {\footnotesize $M\times N$};
  \node[text width=2cm, align=center] at (3.1,-0.6) {\footnotesize Measurement matrix};
  
  \draw[C4E86B, opacity = 1, thick, ->] (3.7,0.6) -- (5,0.6);
  \node[text width=1cm, align=center] at (4.35, 1) {\footnotesize $\vz\in\mathbb{R}^M$};
  \node[text width=1cm, align=center] at (4.35, 0.3) {\footnotesize $\vz=\mA\vx$};
  
  \fill[FF3561, opacity=0.2, rounded corners] (5,0) rectangle (6.9,1.2);
  \node[text width=1cm, align=center] at (5.6, 0.6) {$p(y_m|z_m,\boldsymbol\theta)$};
  \node[text width=2cm, align=center] at (5.85, -0.6) {\footnotesize Separable output channel};
  
  \draw[FF3561, opacity=0.5, thick, ->] (6.9,0.6) -- (8.2, 0.6);
  \node[text width=2.5cm, align=center] at (7.8, 1) {\footnotesize $\vy\in\{\pm1\}^M$};

\end{tikzpicture}
%\endpgfgraphicnamed 
}
\end{tabular}
\caption{A probabilistic view of the sparse signal recovery \cite{Rangan:GAMP:2011}: the signal $\vx$ follows a prior distribution $p(x_n|\boldsymbol\lambda)$, the noiseless measurements $\vz$ are corrupted by noise $\vw$, producing the noisy measurements $\vy$ that follow the distribution $p(y_m|z_m,\boldsymbol\theta)$. The distribution parameters $\boldsymbol\lambda,\boldsymbol\theta$ are unknown and need to be estimated.}
\label{fig:bayesian_model}
\end{figure}

Under the Bayesian setting illustrated in Fig. \ref{fig:bayesian_model}, approximate message passing (AMP) can be employed to perform probabilistic inference on the factor graph of sparse signal recovery task \cite{Pearl:1988,Kschischang:2001,Koller:2009}. The distribution parameters in AMP are unknown in practice, and hence, need to be estimated. In this paper, we shall focus on developing new recovery methods within the AMP framework \cite{Donoho:AMP:2009,Baron:2010,Rangan:GAMP:2011}. We are especially motivated from the fact that AMP is computationally efficient for large-scale problems and achieves state-of-the-art performances in channel estimation from low-bit measurements \cite{Wen:Channel:2015,Wen:LowADC:2016,Mo:LowADC:2018,Bellili:Lap:2019,Myers:1bit:2019}. Compared to previous AMP approaches that either manually tune the noise parameter or rely on an approximated noise model, our approach works with the true quantization noise model and automatically estimates the parameters on the fly.

Inspired by the long history of treating the distribution parameters as random variables in mathematical statistics \cite{Bickel:2015}, we proposed an extension to AMP with additive-white-Gaussian-noise (AWGN) model in \cite{PE_GAMP17}, where the posteriors of the signal and parameters were computed and used for recovery. There are two main drawbacks of this approach: it was not designed to handle heavily quantized measurements, and it relied on standard gradient descent to find the parameters (which implied incremental searches and was thus computationally intensive). In this paper, we incorporate the quantization noise model and present a computationally efficient approach to perform parameter estimation. Experimental results show that the proposed AMP with built-in parameter estimation (AMP-PE) generally performs much better than other state-of-the-art methods, and matches the performance of the oracle AMP where the true distribution parameters are known.

\subsection{Prior Art}
Depending on how the sparse prior is enforced on $\vx$, various approaches have been proposed to solve the 1-bit CS problem. The $l_1$-norm $\|\vx\|_1$ was minimized in \cite{Boufounos:1bitCS:2008} subject to linear inequality constraints produced by 1-bit measurements and a nonlinear (unit) $l_2$-norm constraint. Linear and convex programming formulations that minimize the $l_1$-norm subject to convex constraints can also be derived \cite{Plan:1bitCS:2013,Plan:1bitCS:2013:2}. With a suitably constructed data-fidelity term for quantized measurements, the $l_1$-norm can also be used as a regularizer to promote sparse solutions \cite{Zymnis:Quantized_CS:2010,Zhang:Robust_CS_l1:2014}. The binary iterative hard thresholding (BIHT) algorithm was proposed in \cite{Jacques:1bitCS:2013}, and it imposed a constraint on the sparsity of the signal, i.e. $\|\vx\|_0\leq E$, where $E$ is the number of nonzero entries in $\vx$. BIHT can be further generalized to quantized iterative hard thresholding (QIHT) for multi-bit measurements \cite{Jacques:QIHT:2013}.

Alternatively, the sparse prior can be interpreted from a probabilistic perspective and then utilized in the signal recovery process via AMP \cite{Donoho:AMP:2009}. Its denoising formulation was first introduced in \cite{AMP10}, then studied extensively in \cite{Guo:SURE:2015,Metzler:Denoising:2016,Ma:AMP_Denoise:2016}, and its Bayesian formulation in the form of belief propagation was later introduced in \cite{Rangan:GAMP:2011}. In this paper, we adopt the Bayesian formulation termed ``generalized approximate message passing'' (GAMP) \cite{Rangan:GAMP:2011}. In order to estimate distribution parameters, expectation maximization (EM) \cite{Dempster:EM:1977} was used to maximize the measurement likelihood in \cite{Vila:BG:2011,Vila:EMGM:2013,Kamilov:PE:2014} or the Bethe free entropy in \cite{Krzakala:2012:1,Marc:IPC:2009}. However, these approaches often involve a high level of computational complexity in cases of complicated probability models such as the quantization noise model. As a result, AWGN model was adopted in \cite{risi2014massive, Wang:Multiuser:2015,Wen:Channel:2015,Bellili:Lap:2019} to approximate the quantization noise model, which leads to sub-optimal performance. The approach in \cite{Mo:LowADC:2018} adopted the true quantization noise model but assumed the noise distribution parameter was already known, i.e. it needed to be manually tuned. In this paper, by treating the parameters as random variables and recovering them jointly with the signal, we design a new GAMP-based algorithm with wider applicability that can directly work with any complicated quantization noise model.

Researchers have proposed and analyzed a few AMP algorithms to solve the 1-bit CS problem with different quantization noise models \cite{Kamilov:1bit:2012,Yang:AMPquan:2013,Musa:GAMP1bit:2016,Movahed:1bit:2016,Kafle:1bit:2019}, where the noise parameters were either pre-specified or needed to be tuned manually. The noise $\vw$ was added before quantization in \cite{Jacques:1bitCS:2013,Yang:AMPquan:2013,Movahed:1bit:2016}, whereas it was added after quantization in \cite{Musa:GAMP1bit:2016}. In this paper, we work with quantized measurements from the ADC whose input-referred noise is added before quantization. In Section \ref{sec:multibitCS}, we shall further extend the 1-bit quantization noise model to a general quantization noise model for multi-bit measurements.

When entries of the measurement matrix $\mA$ are i.i.d. zero-mean Gaussian, the asymptotic behavior of AMP in the large system limit can be characterized by state evolution, which predicts how the variables in AMP evolve through the iterations \cite{Donoho:AMP:2009,Bayati:SE:2011}. The GAMP formulation adopted in this paper also agrees with the state evolution \cite{Rangan:GAMP:2011}, and consistent parameter estimation can be guaranteed \cite{Kamilov:PE:2014}. A new belief propagation formulation termed ``vector approximate message passing'' (VAMP) was proposed in \cite{Rangan:VAMP:2017}, and its state evolution applies to a broader class of random matrices $\mA$ that are right-orthogonally invariant. Although it is still an open question as to how the state evolution analysis can be established for general measurement matrices, AMP has been employed with success in real applications like channel estimation \cite{Wen:Channel:2015,Wen:LowADC:2016,Mo:LowADC:2018} and phase retrieval \cite{Schniter:PR:2015,Metzler:PR:2015}. In practice, operations like damping and mean removal are quite effective in preventing divergence of the algorithm for non-Gaussian measurement matrices \cite{Rangan:DampingCvg:2014,Vila:DampingMR:2015}.

\subsection{Main Contributions and Paper Outline}
Following the practice of treating distribution parameters as variables in mathematical statistics \cite{Bickel:2015}, we perform parameter estimation in a simpler fashion by maximizing their posteriors in AMP. This allows us to consider the true quantization noise model using the proposed AMP-PE framework, where all distribution parameters can be efficiently estimated. This is different from previous AMP-based approaches that either manually tune the noise parameter or use an approximated noise model. Building upon our earlier work with AWGN model in \cite{PE_GAMP17}, we offer the following main contributions:
\begin{itemize}
    \item 1-bit and multi-bit quantization noise models are incorporated to AMP-PE to pave the way for channel estimation in the massive MIMO systems with multi-bit ADCs.
    \item A computationally efficient approach that combines EM and the second-order method is proposed to perform parameter estimation. To the best of our knowledge, this is the first AMP approach that could estimate the noise variance in a true quantization noise model.
    \item State evolution analysis is derived, and it empirically predicts the performance of AMP-PE for the random Gaussian measurement system in the large system limit as $N\rightarrow\infty$ and $\frac{M}{N}$ is fixed.
\end{itemize}

This paper proceeds as follows. In Section \ref{sec:amp_pe_intro}, we introduce the AMP-PE framework. Section \ref{sec:1bitCS} presents the sparse signal model, the 1-bit quantization noise model, and discusses how the distribution parameters can be estimated efficiently. In Section \ref{sec:multibitCS}, we extend our approach to the multi-bit quantization noise model. Next, we derive in Section \ref{sec:se} the state evolution recursions, and empirically verify that they predict the performance of AMP-PE for large random Gaussian matrices. We then compare AMP-PE with other state-of-the-art methods in Section \ref{sec:exp}, and conclude the paper with a discussion in Section \ref{sec:con}. Additional experimental results and discussions are given in the Supplemental Material.

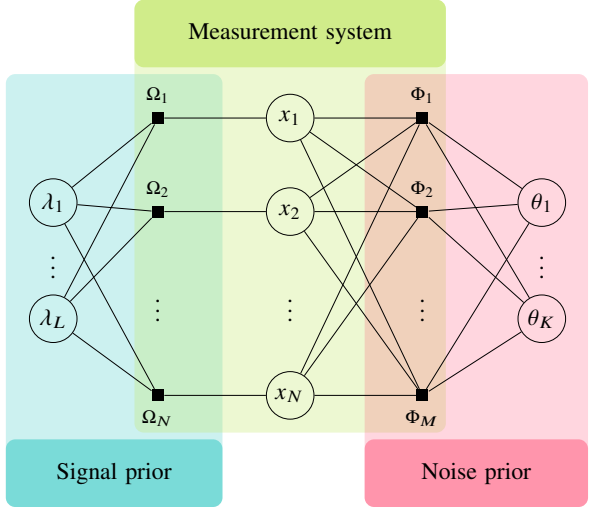
\begin{figure}[tbp]
\begin{center}
\begin{tabular}{c}
\scalebox{0.9}{
% model_pca2.tex
%
% Copyright (C) 2010,2011 Laura Dietz
% Copyright (C) 2012 Jaakko Luttinen
%
% This file may be distributed and/or modified
%
% 1. under the LaTeX Project Public License and/or
% 2. under the GNU General Public License.
%
% See the files LICENSE_LPPL and LICENSE_GPL for more details.
\definecolor{00BCB4}{RGB}{0,188,180}
\definecolor{C4E86B}{RGB}{196,232,107}
\definecolor{FF3561}{RGB}{255,53,97}
%\definecolor{F775A9}{RGB}{247,117,169}

%\beginpgfgraphicnamed{model-pca}
\begin{tikzpicture}

  % Define nodes
  
  \fill[00BCB4, opacity=0.2, rounded corners] (-0.7,-4.5) rectangle (2.5,1.9);
  \fill[C4E86B, opacity=0.3, rounded corners] (1.2,-3.4) rectangle (5.8,3);
  \fill[FF3561, opacity=0.2, rounded corners] (4.6,-4.5) rectangle (7.9,1.9);

  \fill[00BCB4, opacity=0.4, rounded corners] (-0.7,-4.5) rectangle (2.5,-3.5);
  \node[text width=3cm, align=center] at (0.9,-4) {Signal prior};
  \fill[C4E86B, opacity=0.7, rounded corners] (1.2,2) rectangle (5.8,3);
  \node[text width=4.5cm, align=center] at (3.5,2.5) {Measurement system};
  \fill[FF3561, opacity=0.4, rounded corners] (4.6,-4.5) rectangle (7.9,-3.5);
  \node[text width=3cm, align=center] at (6.25,-4) {Noise prior};

  \node[latent,fill=00BCB4!20] (lambda_1) {$\lambda_1$};
  %\factor[left=of lambda_1] {Lambda_1-f} {above:$\Lambda_1$} {lambda_1} {};
  \node[latent,fill=00BCB4!20, below = 1 of lambda_1] (lambda_L) {$\lambda_L$};
  %\factor[left=of lambda_L] {Lambda_L-f} {above:$\Lambda_L$} {lambda_L} {};
  
  \path (lambda_1) -- node[auto=false]{\vdots} (lambda_L);
  %\path (Lambda_1-f) -- node[auto=false]{\vdots} (Lambda_L-f);

  \node[latent,fill=C4E86B!30, above=0.533 of lambda_1, xshift=3.5cm] (x_1) {$x_1$};
  \node[latent,fill=C4E86B!30, below=0.666 of x_1] (x_2) {$x_2$};
  \node[latent,fill=C4E86B!30, below=2 of x_2] (x_N) {$x_N$};
  
  \path (x_2) -- node[auto=false]{\vdots} (x_N);
  
  \factor[left=0 of x_1, xshift=-1.5cm] {Omega_1-f} {above:$\Omega_1$} {x_1, lambda_1, lambda_L}{};
  \factor[left=0 of x_2, xshift=-1.5cm] {Omega_2-f} {above:$\Omega_2$} {x_2, lambda_1, lambda_L}{};
  \factor[left=0 of x_N, xshift=-1.5cm] {Omega_N-f} {below:$\Omega_N$} {x_N, lambda_1, lambda_L}{};
  \path(Omega_2-f) -- node[auto=false]{\vdots} (Omega_N-f);
  
  \node[latent,fill=FF3561!20, right=0 of lambda_1, xshift=6.5cm] (theta_1) {$\theta_1$};
  \node[latent,fill=FF3561!20, right=0 of lambda_L, xshift=6.5cm] (theta_K) {$\theta_K$};
  \path(theta_1) -- node[auto=false]{\vdots} (theta_K);
  
  %\factor[right=of theta_1] {Theta_1-f} {above:$\Theta_1$}{theta_1}{};
  %\factor[right=of theta_K] {Theta_K-f} {above:$\Theta_K$}{theta_K}{};
  %\path(Theta_1-f) -- node[auto=false]{\vdots} (Theta_K-f);

  \factor[right=0 of x_1, xshift=1.5cm] {Phi_1-f} {above:$\Phi_1$} {theta_1, theta_K, x_1, x_2, x_N} {};
  \factor[right=0 of x_2, xshift=1.5cm] {Phi_2-f} {above:$\Phi_2$} {theta_1, theta_K, x_1, x_2, x_N} {};
  \factor[right=0 of x_N, xshift=1.5cm] {Phi_M-f} {below:$\Phi_M$} {theta_1, theta_K, x_1, x_2, x_N} {};
  
  \path(Phi_2-f) -- node[auto=false]{\vdots} (Phi_M-f);
\end{tikzpicture}
%\endpgfgraphicnamed

%%% Local Variables: 
%%% mode: tex-pdf
%%% TeX-master: "example"
%%% End:  
}
\end{tabular}
\end{center}
\caption{The factor graph of the sparse signal recovery task: ``$\bigcirc$'' represents the variable node, and ``$\blacksquare$'' represents the factor node.}
\label{fig:factor_graph_pegamp}
\end{figure}

\section{AMP with Built-in Parameter Estimation}
\label{sec:amp_pe_intro}
This section introduces the extended AMP-PE framework from our earlier work \cite{PE_GAMP17} where the distribution parameters were treated as unknown variables. As shown in Fig. \ref{fig:factor_graph_pegamp}, the factor graph can be divided into three blocks: the signal prior block that contains the signal distribution parameters $\boldsymbol\lambda=\{\lambda_1,\cdots,\lambda_L\}$, the measurement system block that contains the signal of interest $\vx=[x_1\ x_n \cdots\ x_N]^T$, and the noise prior block that contains the noise distribution parameters $\boldsymbol\theta=\{\theta_1,\cdots,\theta_K\}$. Inference tasks that compute the posteriors $p(\boldsymbol\lambda|\vy),\ p(\vx|\vy),\ p(\boldsymbol\theta|\vy)$ rely on the ``messages'' passed among different nodes. Taking the messages between the factor node $\Phi_m$ and the variable node $x_n$ for example, we use the following notations:
\begin{itemize}
\item $\Delta_{\Phi_m\rightarrow x_n}$ denotes the message from $\Phi_m$ to $x_n$.
\item $\Delta_{x_n\rightarrow \Phi_m}$ denotes the message from $x_n$ to $\Phi_m$.
\end{itemize}
Both $\Delta_{\Phi_m\rightarrow x_n}$ and $\Delta_{x_n\rightarrow \Phi_m}$ can be viewed as functions of $x_n$, and they are expressed in the ``$\log$'' domain throughout this paper.

Starting with the \emph{measurement system} block, we can write the messages exchanged among the nodes in the $(t+1)$-th message passing iteration as follows:
\begin{align*}
\begin{split}
\Delta^{(t+1)}_{\Phi_m\rightarrow x_n}=\ &C+\log\int\Big[\vphantom{\sum_{j\neq n}\Delta^{(t)}_{x_j\rightarrow\Phi_m}}\Phi_m\left(y_m, \vx, \boldsymbol\theta\right)\\
&\times\exp\Big(\sum_{j\neq n}\Delta^{(t)}_{x_j\rightarrow\Phi_m}+\sum_v\Delta^{(t)}_{\theta_v\rightarrow\Phi_m}\Big)\Big]\ d\boldsymbol\theta \ d(\vx\backslash x_n),
\end{split}\\
\Delta^{(t+1)}_{x_n\rightarrow \Omega_n}=\ &\sum_i\Delta^{(t+1)}_{\Phi_i\rightarrow x_n},\\
\Delta^{(t+1)}_{\Omega_n\rightarrow x_n}=\ &C+\log\int\Omega_n(x_n,\boldsymbol\lambda)\cdot\exp\Big(\sum_u\Delta^{(t+1)}_{\lambda_u\rightarrow\Omega_n}\Big)\ d\boldsymbol\lambda,\\
\Delta^{(t+1)}_{x_n\rightarrow \Phi_m}=\ &\Delta^{(t+1)}_{\Omega_n\rightarrow x_n}+\sum_{i\neq m}\Delta^{(t+1)}_{\Phi_i\rightarrow x_n}\,,
\end{align*}
where $C$ (by abuse of notation\footnote{Note that the $C$ in $\Delta^{(t+1)}_{\Phi_m\rightarrow x_n}$ and the $C$ in $\Delta^{(t+1)}_{\Omega_n\rightarrow x_n}$ are in fact different, they are both some constants in the $(t+1)$-th iteration.}) denotes a normalization constant that does not depend on the messages in the previous $t$-th iteration, $\vx\backslash x_n$ is the vector $\vx$ with its $n$-th entry $x_n$ removed, the noise prior distribution $\Phi_m(y_m,\vx,\boldsymbol\theta)=p(y_m|\vx,\boldsymbol\theta)$ is encoded at the factor node $\Phi_m$, and the signal prior distribution $\Omega_n(x_n,\boldsymbol\lambda)= p(x_n|\boldsymbol\lambda)$ is encoded at the factor node $\Omega_n$.

In the \emph{signal prior} block, the messages exchanged among the nodes in the $(t+1)$-th iteration are:
\begin{align*}
\begin{split}
\Delta^{(t+1)}_{\Omega_n\rightarrow \lambda_l}=\ &C+\log\int\Big[\Omega_n(x_n,\boldsymbol\lambda)\\
&\times\exp\Big(\Delta^{(t+1)}_{x_n\rightarrow \Omega_n}+\sum_{u\neq l}\Delta^{(t)}_{\lambda_u\rightarrow\Omega_n}\Big)\Big]\ d(\boldsymbol\lambda\backslash\lambda_l)\ dx_n,
\end{split}\\
\Delta^{(t+1)}_{\lambda_l\rightarrow\Omega_n}=\ &C+\sum_{j\neq n}\Delta^{(t+1)}_{\Omega_j\rightarrow\lambda_l}\,,
\end{align*}
where $\boldsymbol\lambda\backslash\lambda_l$ is the set $\boldsymbol\lambda$ with its element $\lambda_l$ removed.

In the \emph{noise prior} block, we have the following messages in the $(t+1)$-th iteration:
\begin{align*}
\begin{split}
\Delta^{(t+1)}_{\Phi_m\rightarrow\theta_k}=\ &C+\log\int\Big[\Phi_m\left(y_m, \vx, \boldsymbol\theta\right)\\
&\times\exp\Big(\sum_j\Delta^{(t+1)}_{x_j\rightarrow\Phi_m}+\sum_{v\neq k}\Delta^{(t)}_{\theta_v\rightarrow\Phi_m}\Big)\Big]\ d(\boldsymbol\theta\backslash\theta_k)\ d\vx,
\end{split}\\
\Delta^{(t+1)}_{\theta_k\rightarrow\Phi_m}=\ &\sum_{i\neq m}\Delta^{(t+1)}_{\Phi_i\rightarrow\theta_k}\,,
\end{align*}
where $\boldsymbol\theta\backslash\theta_k$ is the set $\boldsymbol\theta$ with its element $\theta_k$ removed. 

The posteriors of the signal $\vx$ and the distribution parameters $\boldsymbol\lambda,\boldsymbol\theta$ can then be expressed as 
\begin{subequations}
\label{eq:sm_post_dist}
\begin{align}
\label{eq:pm_x}
\begin{split}
p(x_n|\vy)&\propto\exp\Big(\Delta^{(t+1)}_{\Omega_n\rightarrow x_n}+\sum_m\Delta^{(t+1)}_{\Phi_m\rightarrow x_n}\Big)\,,
\end{split}\\
\label{eq:pm_lambda}
\begin{split}
p(\lambda_l|\vy)&\propto\exp\Big(\sum_n\Delta^{(t+1)}_{\Omega_n\rightarrow\lambda_l}\Big)\,,
\end{split}\\
\label{eq:pm_theta}
\begin{split}
p(\theta_k|\vy)&\propto\exp\Big(\sum_m\Delta^{(t+1)}_{\Phi_m\rightarrow\theta_k}\Big)\,.
\end{split}
\end{align}
\end{subequations}

\subsection{Parameter Estimation}
\label{sec:pe}
The distribution parameters $\boldsymbol\lambda,\boldsymbol\theta$ can be estimated by maximizing the posteriors in \eqref{eq:pm_lambda} and \eqref{eq:pm_theta}:
\begin{subequations}
\begin{align}
    \label{eq:lambda_est}
    \hat{\lambda}_l^{(t+1)} &= \arg\max_{\lambda_l}p(\lambda_l|\vy)=\arg\max_{\lambda_l}\sum_n\Delta^{(t+1)}_{\Omega_n\rightarrow\lambda_l}\,,\\
    \label{eq:theta_est}
    \hat{\theta}_k^{(t+1)} &= \arg\max_{\theta_k}p(\theta_k|\vy)=\arg\max_{\theta_k}\sum_m\Delta^{(t+1)}_{\Phi_m\rightarrow\theta_k}\,.
\end{align}
\end{subequations}
We shall combine EM and the second-order method to find the maximizing parameters in \eqref{eq:lambda_est}, \eqref{eq:theta_est}, which turns out to be a much simpler alternative to previous EM-based approaches that maximize the measurement likelihood \cite{Vila:EMGM:2013,Kamilov:PE:2014}. As discussed later in Section \ref{subsec:comparison_with_pre}, this subtle modification allows us to consider the much more complicated quantization noise models that often arise from applications such as channel estimation in the massive MIMO systems.

Using the estimated parameters $\hat{\boldsymbol\lambda}$ and $\hat{\boldsymbol\theta}$, we can further simplify the messages passed from the factor nodes to the variable nodes as follows:
\begin{subequations}
\label{eq:sp_simplified_messages}
\begin{align}
\label{eq:delta_phi_x_simplifed}
\begin{split}
\Delta^{(t+1)}_{\Phi_m\rightarrow x_n}=\ &C+\log\int\Big[\Phi_m\Big(y_m, \vx, \hat{\boldsymbol\theta}^{(t)}\Big)\\
&\times\exp\Big(\sum_{j\neq n}\Delta^{(t)}_{x_j\rightarrow\Phi_m}\Big)\Big]\ d(\vx\backslash x_n)\,,
\end{split}\\
\label{eq:delta_omega_x_simplified}
\Delta^{(t+1)}_{\Omega_n\rightarrow x_n}=\ &C+\log\Omega_n\Big(x_n,\hat{\boldsymbol\lambda}^{(t+1)}\Big)\,,\\
\label{eq:delta_omega_lambda_simplified}
\begin{split}
\Delta^{(t+1)}_{\Omega_n\rightarrow \lambda_l}=\ &C+\log\int\left[\Omega_n\left(x_n,\lambda_l,\hat{\boldsymbol\lambda}^{(t)}\backslash\hat{\lambda}_l^{(t)}\right)\right.\\
&\times\left.\exp\left(\Delta^{(t+1)}_{x_n\rightarrow \Omega_n}\right)\right]\ dx_n\,,
\end{split}\\
\label{eq:delta_phi_theta_simplified}
\begin{split}
\Delta^{(t+1)}_{\Phi_m\rightarrow\theta_k}=\ &C+\log\int\Big[\Phi_m\Big(y_m, \vx, \theta_k, \hat{\boldsymbol\theta}^{(t)}\backslash\hat{\theta}_k^{(t)}\Big)\\
&\times\exp\Big(\sum_j\Delta^{(t+1)}_{x_j\rightarrow\Phi_m}\Big)\Big]\ d\vx\,,
\end{split}
\end{align}
\end{subequations}
where $\hat{\boldsymbol\lambda}^{(t)}$, $\hat{\boldsymbol\theta}^{(t)}$ are the estimated parameters from the previous $t$-th iteration. 

Exact message passing is generally difficult to compute, so approximated message passing (AMP) is often a good alternative. In the latter case, the distributions are approximated by a family of simpler distributions such as the Gaussians in AMP, where a chosen divergence measure such as the Kullback-Leibler divergence between the true and approximated distributions is minimized \cite{Minka:2001,Minka:Div:2005,Wainwright:Graph:2008}. In this paper, only the messages $\Delta_{\Phi_m\rightarrow x_n}$ and $\Delta_{x_n\rightarrow\Phi_m}$ are computed using the GAMP formulation that adopts the Gaussian approximations \cite{Rangan:GAMP:2011}. The rest of the messages are computed \emph{exactly} according to the above formulas.

\section{1-Bit Compressive Sensing via AMP}
\label{sec:1bitCS}
In this section, we present the sparse signal model and the 1-bit quantization noise model under the Bayesian setting. We then show how both the signal and the distribution parameters can be jointly recovered via AMP. 
\begin{sparse_signal_model}
The entries of the sparse signal $\vx$ are assumed to be i.i.d.
\begin{align}
    p(\vx|\boldsymbol\lambda)=\prod_n p(x_n|\boldsymbol\lambda)\,.
\end{align}

\begin{figure}[tbp]
\centering
\includegraphics[height=0.24\textwidth]{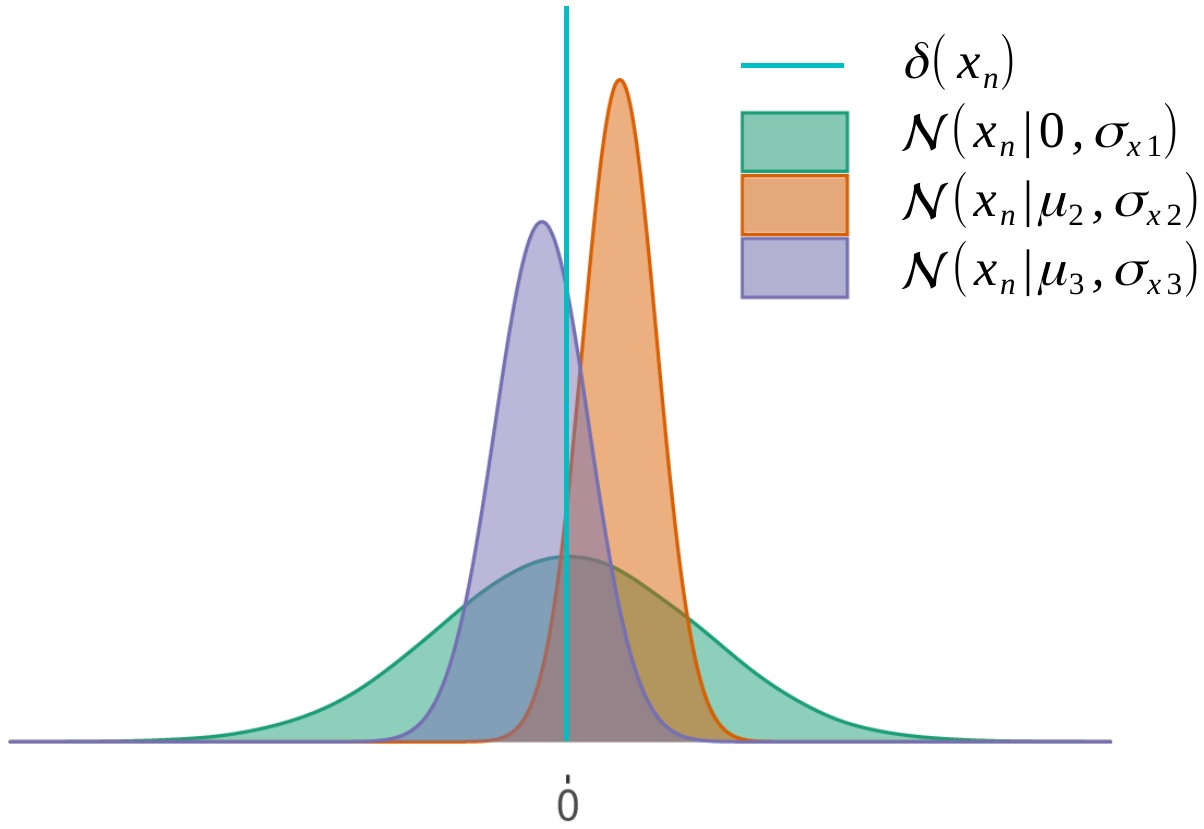}
\caption{The Bernoulli-Gaussian mixture distribution in \eqref{eq:bgm} is used to model the sparse signal prior. In order to model some heavy-tailed distribution like the Cauchy distribution, the first Gaussian component has zero-mean and a large-enough variance to cover the domain of $x_n$.}
\label{fig:bgm}
\end{figure}

We can first use the Bernoulli distribution to model the outcome of determining whether $x_n$ is nonzero, and then use the Gaussian mixture distribution to model the nonzero entries. As shown in Fig. \ref{fig:bgm}, the Bernoulli and Gaussian mixture (BGM) distribution is used to model the sparse signal $x_n$:
\begin{align}
\label{eq:bgm}
\begin{split}
    p(x_n|\boldsymbol\lambda) =& (1-\kappa)\cdot\delta(x_n)+\kappa\cdot\sum_{i=1}^I\xi_i\cdot\mathcal{N}(x_n|\mu_i,{\gamma_x}_i)\,,
\end{split}
\end{align}
where $\delta(x_n)$ is the Dirac delta function, $\kappa$ is the probability that $x_n$ takes a non-zero value, $\xi_i$ is the Gaussian mixture weight, $\mu_i$ and ${\gamma_x}_i$ are the mean and variance of the $i$-th Gaussian component, $\mathcal{N}(x_n|\mu_i,{\gamma_x}_i)$ is the Gaussian probability density function, and $I$ is the number of mixture components. The parameter set $\boldsymbol\lambda$ is then
\begin{align}
\label{eq:parameter_set_lambda}
\boldsymbol\lambda=\left\{\left.\kappa,\xi_i,\mu_i,{\gamma_x}_i\ \right|\ i=1,\cdots,I\right\}\,.
\end{align}

In order to accommodate the case that $x_n$ follows some heavy-tailed distribution like the Cauchy distribution, we choose the first Gaussian component to be zero-mean, i.e. $\mu_1=0$, and ensure its variance ${\gamma_x}_1$ is large enough to cover the domain of $x_n$. In practice, we can initialize ${\gamma_x}_1$ with a reasonably large number like the variance of the least-squares solution. As shown later in the experiments, this zero-mean Gaussian component proves effective in compensating for the mismatch between the BGM distribution and other distributions like the Cauchy or Laplace distribution.
\end{sparse_signal_model}

\begin{figure}[tbp]
\centering
\includegraphics[height=0.24\textwidth]{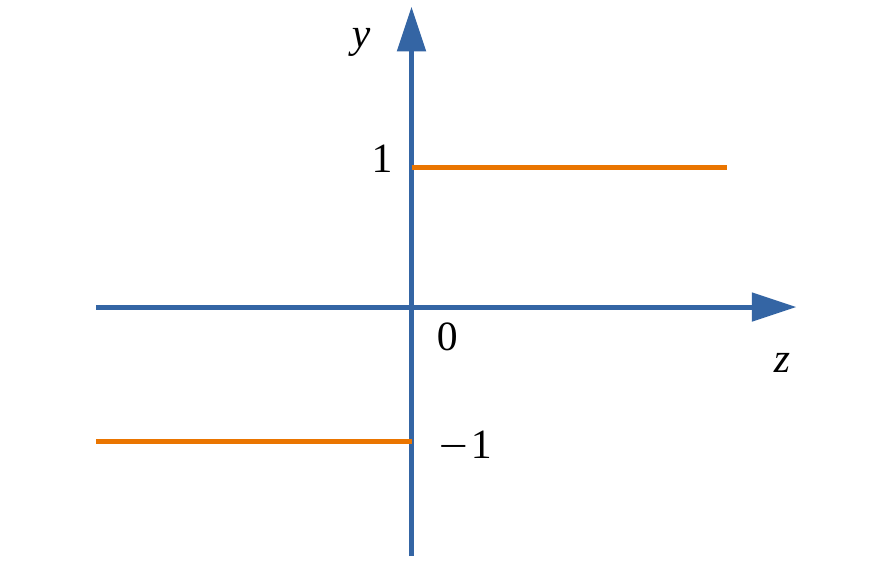}
\caption{The quantizer $\mathcal{Q}$ in \eqref{eq:1bit_quantization} outputs the sign of the input $z$ in 1-bit compressive sensing.}
\label{fig:1bit_quantizer}
\end{figure}

\begin{1_bit_quantization_noise_model}
The noisy 1-bit measurements $\vy$ are given by
\begin{align}
\label{eq:quantized_measurements}
\vy=\mathscr{Q}(\mA\vx+\vw)\,,
\end{align}
where $\mathscr{Q}(\cdot)$ is the quantizer, $\vw$ is the i.i.d. additive white Gaussian noise (AWGN) with $w_m\sim\mathcal{N}(0,\gamma_w)$, and $\gamma_w$ is the noise variance. The noise $\vw$ is added to the noiseless measurements $\vz=\mA\vx$ before quantization. As shown in Fig. \ref{fig:1bit_quantizer}, the quantizer $\mathscr{Q}$ produces 1-bit output by computing the sign of an input $z_m+w_m$:
\begin{align}
\label{eq:1bit_quantization}
    y_m=\mathscr{Q}(z_m+w_m)=\left\{
    \begin{array}{l}
    +1  \\
    -1 
    \end{array}
    \quad
    \begin{array}{l}
    \textrm{if }z_m+w_m>0\,,\\
    \textrm{if }z_m+w_m\leq0\,.
    \end{array}
    \right.
\end{align}
From \eqref{eq:1bit_quantization}, we can see that the magnitude information is lost in 1-bit CS. In this case, the parameter set $\boldsymbol\theta$ is simplified to
\begin{align}
\label{eq:parameter_set_theta}
    \boldsymbol\theta=\left\{\gamma_w\right\}\,.
\end{align}
\end{1_bit_quantization_noise_model}

The proposed AMP with built-in parameter estimation (AMP-PE) algorithm is summarized in Algorithm \ref{alg:amp_pe}. The damping and mean removal operations are often incorporated in AMP algorithms to ensure the convergence for ill-conditioned or non-zero-mean measurement matrices \cite{Rangan:DampingCvg:2014,Vila:DampingMR:2015}. Note that in \eqref{eq:compute_s_m}-\eqref{eq:compute_tau_s_m} and \eqref{eq:compute_x_n}-\eqref{eq:compute_tau_x_n}, we need to compute the posterior means and variances of $\vz$ and $\vx$ respectively. For the sparse signal model and 1-bit quantization noise model introduced earlier, their expressions can be concisely derived as shown in the rest of this section. Certain detailed derivations are deferred to Appendix \ref{app:sec:1_bit_cs}.

\subsection{Nonlinear Updates for Sparse Signal Model}
The BGM model in \eqref{eq:bgm} is chosen as the sparse signal prior. The posterior mean and variance of the signal $x_n$ in \eqref{eq:compute_x_n} and \eqref{eq:compute_tau_x_n} from Algorithm \ref{alg:amp_pe} are derived in this subsection. To simplify the notations, we remove the superscript ``$(t)$'' that denotes AMP iteration in the following derivations. 

The distribution $\exp\left(\sum_m\Delta_{\Phi_m\rightarrow x_n}\right)$ in \eqref{eq:pm_x} is approximated by a Gaussian distribution $\mathcal{N}(x_n|r_n,{\tau_r})$ in GAMP, where $r_n$ is a ``dummy'' variable and can be viewed as a Gaussian-noise corrupted version of $x_n$ with variance ${\tau_r}$. According to the sum-product message passing, the posterior of the signal $\vx$ in \eqref{eq:pm_x} can then be approximated as
\begin{align*}
    p(x_n|\vy,\boldsymbol\lambda)\approx\frac{1}{\Psi(r_n,\boldsymbol\lambda)} p(x_n|\boldsymbol\lambda)\cdot\mathcal{N}(x_n|r_n,{\tau_r})\,,
\end{align*}
where $\Psi(r_n,\boldsymbol\lambda)$ is the normalizing constant
\begin{align}
\label{eq:normalization_psi}
    \Psi(r_n,\boldsymbol\lambda)=\int p(x_n|\boldsymbol\lambda)\cdot\mathcal{N}(x_n|r_n,{\tau_r})\ dx_n\,.
\end{align}

We can compute posterior mean of $x_n$ in \eqref{eq:compute_x_n}
\begin{align}
\label{eq:posterior_x_n_bgm}
\begin{split}
    \hat{x}_n&=\mathbb{E}\left[x_n|r_n,{\tau_r},\boldsymbol\lambda\right]=\int x_n\cdot p(x_n|\vy,\boldsymbol\lambda)\ dx_n\,.
\end{split}
\end{align}
The posterior expectation of $x_n^2$ is
\begin{align}
\label{eq:posterior_x_sq_n_bgm}
\begin{split}
\mathbb{E}\left[x_n^2|r_n,{\tau_r},\boldsymbol\lambda\right] =\int x_n^2\cdot p(x_n|\vy,\boldsymbol\lambda)\ dx_n\,.
\end{split}
\end{align}
The posterior variance of $x_n$ in \eqref{eq:compute_tau_x_n} is then
\begin{align*}
    \tau_x=\mathbb{E}\left[x_n^2|r_n,{\tau_r},\boldsymbol\lambda\right]-\left(\mathbb{E}\left[x_n|r_n,{\tau_r},\boldsymbol\lambda\right]\right)^2\,.
\end{align*}

\begin{algorithm}[tbp]
\caption{The AMP-PE algorithm }\label{alg:amp_pe}
\begin{algorithmic}[1]
\Input $\hat{\vx}^{(0)}, \boldsymbol\tau_x^{(0)}, \vq^{(0)}, \boldsymbol\tau_q^{(0)}, \hat{\boldsymbol\lambda}^{(0)}, \hat{\boldsymbol\theta}^{(0)}$.
\For{$t=\{0,1,\cdots,T\}$}
	\State Output \emph{nonlinear} update: For each $m=1,\cdots,M$
	\begin{subequations}
	\begin{align}
	\label{eq:compute_s_m}
	s_m^{(t)}&=\frac{1}{{\tau_q}^{(t)}}\left(\mathbb{E}\left[z_m\left|q_m^{(t)}, {\tau_q}^{(t)}, y_m, \hat{\boldsymbol\theta}^{(t)}\right.\right]-q_m^{(t)}\right)\\
	\label{eq:compute_tau_s_m}
	{\tau_s}^{(t)}&=\frac{1}{M}\sum_m\frac{1}{{\tau_q}^{(t)}}\left(1-\frac{1}{{\tau_q}^{(t)}}\textrm{Var}\left[z_m\left|q_m^{(t)}, {\tau_q}^{(t)}, y_m, \hat{\boldsymbol\theta}^{(t)}\right.\right]\right).
	\end{align}
	\end{subequations}
	\State Input \emph{linear} update: For each $n=1,\cdots,N$
	\begin{subequations}
	\begin{align}
	{\tau_r}^{(t)}&=\left[\frac{1}{N}\|\mA\|_F^2\cdot {\tau_s}^{(t)}\right]^{-1}\\
	r_n^{(t)}&=x_n^{(t)}+{\tau_r}^{(t)}\sum_mA_{mn}\cdot s_m^{(t)}\,.
	\end{align}
	\end{subequations}
	\State Estimate the input parameters: For each $l=1,\cdots,L$
	\begin{align}
	    \label{eq:amp_pe_est_input}
	    \hat{\lambda}_l^{(t+1)}&=\arg\max_{\lambda_l}\ \sum_n\Delta_{\Omega_n\rightarrow\lambda_l}^{(t+1)}\,.
	\end{align}
	\State Input \emph{nonlinear} update: For each $n=1,\cdots,N$
	\begin{subequations}
	\begin{align}
	\label{eq:compute_x_n}
	\hat{x}_n^{(t+1)} &= \mathbb{E}\left[x_n\left|r_n^{(t)}, {\tau_r}^{(t)}, \hat{\boldsymbol\lambda}^{(t+1)}\right.\right]\\
	\label{eq:compute_tau_x_n}
	{\tau_x}^{(t+1)}&=\frac{1}{N}\sum_n\textrm{Var}\left[x_n\left|r_n^{(t)}, {\tau_r}^{(t)}, \hat{\boldsymbol\lambda}^{(t+1)}\right.\right]\,.
	\end{align}
	\end{subequations}
	\State Output \emph{linear} update: For each $m=1,\cdots,M$
	\begin{subequations}
	\begin{align}
	{\tau_q}^{(t+1)}&=\frac{1}{M}\|\mA\|_F^2\cdot {\tau_x}^{(t+1)}\\
	q_m^{(t+1)}&=\sum_nA_{mn}\cdot \hat{x}_n^{(t+1)}-{\tau_q}^{(t+1)}\cdot s_m^{(t)}\,.
	\end{align}
	\end{subequations}
	\State Estimate the output parameters: For each $k=1,\cdots,K$
	\begin{align}
	    \label{eq:amp_pe_est_output}
	    \hat{\theta}_k^{(t+1)}&=\arg\max_{\theta_k}\ \sum_m\Delta_{\Phi_m\rightarrow\theta_k}^{(t+1)}\,.
	\end{align}
	\If {$\hat{\vx}^{(t+1)}$ reaches convergence}
		\State $\hat{\vx}=\hat{\vx}^{(t+1)}$ and \textbf{break};
	\EndIf
\EndFor
\State\Return The recovered signal $\hat{\vx}$;
\end{algorithmic}
\end{algorithm}

\subsection{Nonlinear Updates for 1-Bit Quantization Noise Model}
Under the 1-bit quantization noise model in \eqref{eq:quantized_measurements}, the posterior mean and variance of the noiseless measurement $z_m$ in \eqref{eq:compute_s_m} and \eqref{eq:compute_tau_s_m} from Algorithm \ref{alg:amp_pe} are derived in this subsection. Since the noisy measurement $y_m$ is binary, we can compute its probability as
\begin{align*}
    \mathrm{Pr}\left(y_m=1|z_m,\boldsymbol\theta\right)&=\int_{-\infty}^0\mathcal{N}(u|z_m,\gamma_w)\ du,\\
    \mathrm{Pr}\left(y_m=-1|z_m,\boldsymbol\theta\right)&=1-\int_{-\infty}^0\mathcal{N}(u|z_m,\gamma_w)\ du\,.
\end{align*}

The prior distribution of $z_m$ can be computed via sum-product message passing, and it is approximated by a Gaussian distribution $\mathcal{N}\left(z_m|q_m,{\tau_q}\right)$ in GAMP, where $q_m$ is another ``dummy'' variable and can be viewed as a Gaussian-noise corrupted version of $z_m$ with variance ${\tau_q}$. Using the Bayes' theorem, we can approximate the posterior of $z_m$ as
\begin{align*}
    p(z_m|y_m,\boldsymbol\theta)\approx\frac{1}{\mathcal{U}_0(q_m,y_m,\boldsymbol\theta)}p(y_m|z_m,\boldsymbol\theta)\cdot\mathcal{N}(z_m|q_m,{\tau_q}),
\end{align*}
where $\mathcal{U}_0(q_m,y_m,\boldsymbol\theta)$ is the normalizing constant
\begin{align}
\label{eq:normalization_U_1bit}
\begin{split}
    \mathcal{U}_0(q_m,y_m,\boldsymbol\theta)&=\int p(y_m|z_m,\boldsymbol\theta)\cdot\mathcal{N}(z_m|q_m,{\tau_q})\ dz_m\,.
\end{split}
\end{align}
We can compute the posterior mean of $z_m$ in \eqref{eq:compute_s_m} as 
\begin{align}
\label{eq:posterior_mean_z_m_1bit}
\begin{split}
    \mathbb{E}\left[z_m\left|q_m, {\tau_q}, y_m, \boldsymbol\theta\right.\right]&=\int z_m\cdot p(z_m|y_m,\boldsymbol\theta)\ dz_m\,.
\end{split}
\end{align}
The posterior expectation of $z_m^2$ is 
\begin{align}
\label{eq:posterior_mean_z_m_sq_1bit}
\begin{split}
    \mathbb{E}\left[z_m^2\left|q_m, {\tau_q}, y_m, \boldsymbol\theta\right.\right]&=\int z_m^2\cdot p(z_m|y_m,\boldsymbol\theta)\ dz_m\,.
\end{split}
\end{align}
The posterior variance of $z_m$ in \eqref{eq:compute_tau_s_m} is then
\begin{align*}
\begin{split}
    \mathbb{E}\left[z_m^2\left|q_m, {\tau_q}, y_m, \boldsymbol\theta\right.\right]-\left(\mathbb{E}\left[z_m\left|q_m, {\tau_q}, y_m, \boldsymbol\theta\right.\right]\right)^2\,.
\end{split}
\end{align*}

\subsection{Parameter Estimation for Sparse Signal Model}
\label{subsec:pe_sparse_signal_model}
We now show how to estimate the signal prior parameters $\boldsymbol\lambda$ in \eqref{eq:lambda_est}. Combining \eqref{eq:lambda_est}, \eqref{eq:delta_omega_lambda_simplified}, \eqref{eq:bgm} and \eqref{eq:parameter_set_lambda} yields
\begin{align}
\label{eq:pe_lambda_signal_prior}
\begin{split}
    \hat{\boldsymbol\lambda} = \arg\max_{\boldsymbol\lambda}\ \sum_n\log\big[&(1-\kappa)\cdot\mathcal{N}(r_n|0,{\tau_r})\\
    &+\sum_i\kappa\xi_i\cdot\mathcal{N}(r_n|\mu_i,{\gamma_x}_i+{\tau_r})\big]\,.
\end{split}
\end{align}
The detailed derivation is given in the Supplemental Material. Standard gradient descent was previously used to solve the above \eqref{eq:pe_lambda_signal_prior} in \cite{PE_GAMP17}. However, incremental searches along the gradient direction significantly slow down the algorithm -- a major disadvantage when the problem size is large. In this paper, we shall rely on EM to estimate the signal prior parameters $\boldsymbol\lambda$, and then switch to the second-order method to estimate the noise prior parameters $\boldsymbol\theta$. The second-order method computes the search step size adaptively based on the current solution, which is more computationally efficient.

Next, we turn our focus to the ``\emph{inner}'' parameter estimation (PE) iteration indexed by ``$e$''. In the $(e+1)$-th PE iteration, we use EM to solve \eqref{eq:pe_lambda_signal_prior}: the dummy variable $r_n$ is treated as the observation, and the latent variable $c(r_n)\in\{0,1,\cdots,I\}$ decides which mixture component $r_n$ is from. Let $f(\boldsymbol\lambda)$ be the objective function from the expectation step. We maximize $f(\boldsymbol\lambda)$ in the $(e+1)$-th PE iteration as follows
\begin{align}
\label{eq:signal_prior_pe_obj_concise}
\begin{split}
    \hat{\boldsymbol\lambda}^{(e+1)}=&\arg\max_{\boldsymbol\lambda} f(\boldsymbol\lambda)\,.
\end{split}
\end{align}
Detail expression of $f(\boldsymbol\lambda)$ is given in \eqref{eq:signal_prior_pe_obj} of Appendix \ref{app:sec:1_bit_cs:C}. The mixture weights $\kappa,\xi_i$, the Gaussian mean $\mu_i$ and variance ${\gamma_x}_i$ that maximize $f(\boldsymbol\lambda)$ all have closed-form update formulas.

\subsection{Parameter Estimation for 1-Bit Quantization Noise Model}
We next present how to estimate the noise prior parameters in \eqref{eq:theta_est} for the 1-bit quantization noise model. Combining \eqref{eq:theta_est}, \eqref{eq:delta_phi_theta_simplified}, \eqref{eq:1bit_quantization} and \eqref{eq:parameter_set_theta}, we have
\begin{align}
\label{eq:g1_theta}
\begin{split}
    \hat{\boldsymbol\theta}=\arg\max_{\boldsymbol\theta}g_1(\boldsymbol\theta)=\arg\max_{\boldsymbol\theta}\ \sum_m\log\left[\mathcal{U}_0(q_m,y_m,\boldsymbol\theta)\right],
\end{split}
\end{align}
where $g_1(\boldsymbol\theta)=\log p(\boldsymbol\theta|\vy)$. The detailed derivation is given in the Supplemental Material. We could not obtain a closed-form update for the noise variance $\gamma_w$ that maximizes $g_1(\boldsymbol\theta)$. Instead, we seek to maximize the second-order approximation of $g_1(\boldsymbol\theta)$ at the estimated $\boldsymbol\theta^{(e)}$ from the previous $e$-th PE iteration
\begin{align}
\label{eq:2nd_order_g1_theta}
    g_1(\boldsymbol\theta)\approx g_1(\boldsymbol\theta^{(e)})+g_1^\prime\cdot\big(\boldsymbol\theta-\boldsymbol\theta^{(e)}\big)+\frac{g_1^{\prime\prime}}{2}\cdot\big(\boldsymbol\theta-\boldsymbol\theta^{(e)}\big)^2\,,
\end{align}
where $g_1^\prime, g_1^{\prime\prime}$ are the first and second order derivatives of $g_1(\boldsymbol\theta)$ with respect to $\gamma_w$. 

Note that the second-order method does not always give us the maximizing solution. We should use gradient descent to solve it occasionally when $g_1^{\prime\prime}\geq0$, since the second-order method would give us the minimizing solution in that case. When $g_1^{\prime\prime}<0$, the update for $\gamma_w$ can then be obtained as
\begin{align}
\label{eq:1bit_awgn_variance_update}
    \gamma_w^{(e+1)}=\gamma_w^{(e)}-\frac{g_1^\prime}{g_1^{\prime\prime}}\,.
\end{align}

\subsection{Comparison with Previous EM-based Approaches}
\label{subsec:comparison_with_pre}
In this subsection, we discuss the differences between the proposed AMP-PE approach and previous approaches that also utilize an EM-style strategy to estimate the distribution parameters. First, AMP-PE only uses EM to estimate the signal prior parameters, it then switches to the second-order method to estimate the noise prior parameter. Whereas previous approaches use EM to estimate \emph{both} the signal prior and noise prior parameters \cite{Vila:BG:2011,Vila:EMGM:2013,Krzakala:2012:1}. Second, AMP-PE maximizes the posteriors of parameters in a different fashion. In the \emph{signal prior} block, the dummy variable $r_n$ is treated as the observation, and its mixture label $c(r_n)$ is treated as the hidden variable. This leads to the objective function $f(\boldsymbol\lambda)$ in \eqref{eq:signal_prior_pe_obj_concise}. In the \emph{noise prior} block, the objective function is $g_1(\boldsymbol\theta)$ in \eqref{eq:g1_theta}. Previous approaches essentially try to maximize the likelihood of measurements. The noisy measurement $y_m$ is treated as the observation, the signal $x_n$ and the noiseless measurement $z_m$ are treated as the hidden variables. This leads to the following drastically different objective functions of previous approaches:
\begin{align}
    \label{eq:previous_em_signal}
    \max_{\boldsymbol\lambda}&\ \sum_n\int p\big(x_n|\vy,\boldsymbol\lambda^{(t)}\big)\log p(\vy,x_n|\boldsymbol\lambda)\ dx_n\,,\\
    \label{eq:previous_em_noise}
    \max_{\boldsymbol\theta}&\ \sum_m\int p\big(z_m|y_m,\boldsymbol\theta^{(t)}\big)\log p(y_m,z_m|\boldsymbol\theta)\ dz_m\,.
\end{align}

For the BGM model in \eqref{eq:bgm}, the optimization problem in \eqref{eq:previous_em_signal} can be solved easily with closed-form solutions. However, when it comes to the quantization noise model in \eqref{eq:1bit_quantization}, the problem in \eqref{eq:previous_em_noise} does not have closed-form solutions. Its objective function is much more complicated and computationally prohibitive compared to the one from AMP-PE in \eqref{eq:g1_theta}. We offer the detailed derivation of \eqref{eq:previous_em_noise} in Appendix \ref{app:sec:1_bit_cs:E}.

\section{Multi-Bit Compressive Sensing via AMP}
\label{sec:multibitCS}
With our new formulation, the 1-bit compressive sensing problem can be generalized to the multi-bit setting in a straightforward fashion when we increase the accuracy level of the quantizer $\mathscr{Q}$. Within the AMP framework, the sparse signal model is kept intact. The only adjustment required is to update the quantization noise model.

\begin{figure}[tbp]
\centering
\includegraphics[height=0.24\textwidth]{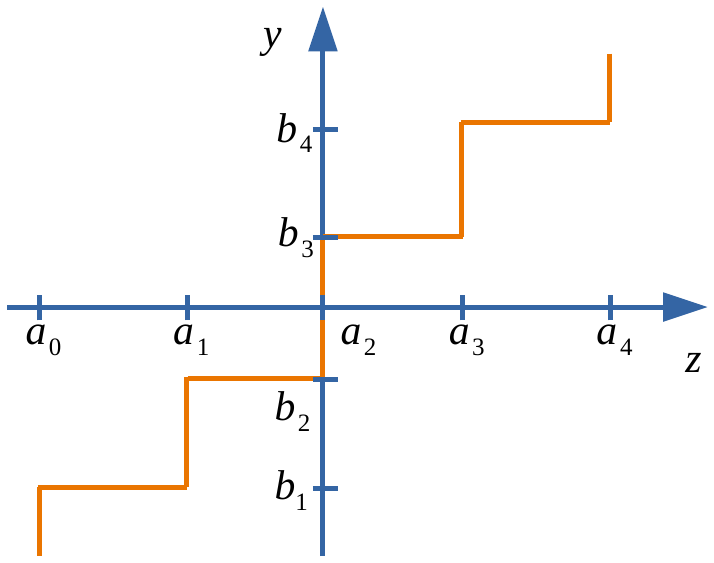}
\caption{The quantizer $\mathcal{Q}$ in \eqref{eq:multi_bit_quantization} outputs multi-bit measurements in multi-bit compressive sensing.}
\label{fig:multi_bit_quantizer}
\end{figure}

\begin{multi_bit_quantization_noise_model}
Let $B$ denote the number of bits of the quantized measurement $y_m\in\{b_i\ |\ i=1,\cdots,2^B\}$, where $b_i$ is the quantization symbol. As shown in Fig. \ref{fig:multi_bit_quantizer}, the quantizer $\mathscr{Q}$ can be written as
\begin{align}
\label{eq:multi_bit_quantization}
    y_m=\mathscr{Q}(z_m+w_m)=b_i,\quad\textrm{if }z_m+w_m\in\left[a_{i-1}, a_i\right)\,,
\end{align}
where $a_{i-1}$ and $a_i$ are the lower and upper bounds for the quantizer $\mathscr{Q}$ to output $b_i$. Under the pre-quantization noise model $w_m\sim\mathcal{N}(0,\gamma_w)$, the parameter set $\boldsymbol\theta$ contains the noise variance $\boldsymbol\theta=\left\{\gamma_w\right\}$.
\end{multi_bit_quantization_noise_model} We only offer here the concise equation updates and leave the detailed derivations to Appendix \ref{app:sec:multi_bit_cs}.

\subsection{Nonlinear Updates for Multi-Bit Quantization Noise Model}
For the multi-bit quantization noise model in \eqref{eq:multi_bit_quantization}, the posterior mean and variance of the noiseless measurement $z_m$ in \eqref{eq:compute_s_m} and \eqref{eq:compute_tau_s_m} from Algorithm \ref{alg:amp_pe} are derived as follows. We can first compute probability of the noisy measurement $y_m$ as
\begin{align*}
\begin{split}
    \mathrm{Pr}\left(y_m=b_i|z_m,\boldsymbol\theta\right)&=\int_{a_{i-1}}^{a_i}\mathcal{N}(u|z_m,\gamma_w)\ du\,.
\end{split}
\end{align*}
We shall assume that $y_m=b_i$ in the following derivations. According to the sum-product message passing, the posterior of $z_m$ can be approximated as 
\begin{align*}
    p(z_m|y_m,\boldsymbol\theta)\approx\frac{1}{\mathcal{V}_0(q_m,y_m,\boldsymbol\theta)}p(y_m|z_m,\boldsymbol\theta)\cdot\mathcal{N}(z_m|q_m,{\tau_q})\,,
\end{align*}
where $q_m$ is a dummy variable in AMP as before and can be viewed as a Gaussian-noise corrupted version of $z_m$ with variance ${\tau_q}$, while $\mathcal{V}_0(q_m,y_m,\boldsymbol\theta)$ is the normalizing constant
\begin{align}
\label{eq:normalization_V_multi_bit}
\begin{split}
    &\mathcal{V}_0(q_m,y_m,\boldsymbol\theta)=\int p(y_m|z_m,\boldsymbol\theta)\cdot\mathcal{N}(z_m|q_m,{\tau_q})\ dz_m\,.
\end{split}
\end{align}
We can compute the posterior mean of $z_m$ in \eqref{eq:compute_s_m} as 
\begin{align}
\label{eq:posterior_mean_z_m_multi_bit}
\begin{split}
    \mathbb{E}\left[z_m\left|q_m, {\tau_q}, y_m, \boldsymbol\theta\right.\right]&=\int z_m\cdot p(z_m|y_m,\boldsymbol\theta)\ dz_m\,,
\end{split}
\end{align}
and the posterior expectation of $z_m^2$ as 
\begin{align}
\label{eq:posterior_mean_z_m_sq_multi_bit}
\begin{split}
    \mathbb{E}\left[z_m^2\left|q_m, {\tau_q}, y_m, \boldsymbol\theta\right.\right]&=\int z_m^2\cdot p(z_m|y_m,\boldsymbol\theta)\ dz_m\,.
\end{split}
\end{align}
The posterior variance of $z_m$ in \eqref{eq:compute_tau_s_m} can be expressed as
\begin{align*}
\begin{split}
    \mathbb{E}\left[z_m^2\left|q_m, {\tau_q}, y_m, \boldsymbol\theta\right.\right]-\left(\mathbb{E}\left[z_m\left|q_m, {\tau_q}, y_m, \boldsymbol\theta\right.\right]\right)^2\,.
\end{split}
\end{align*}

\subsection{Parameter Estimation for Multi-Bit Quantization Noise Model}
Similarly, we can show how to estimate the noise prior parameters in \eqref{eq:theta_est} for the multi-bit quantization noise model. Combining \eqref{eq:theta_est}, \eqref{eq:multi_bit_quantization} and \eqref{eq:parameter_set_theta} yields
\begin{align}
\begin{split}
    \hat{\boldsymbol\theta}=\arg\max_{\boldsymbol\theta}g_2(\boldsymbol\theta)=\arg\max_{\boldsymbol\theta}\ \sum_m\log\left[\mathcal{V}_0(q_m,y_m,\boldsymbol\theta)\right]\,,
\end{split}
\end{align}
where $g_2(\boldsymbol\theta)=\log p(\boldsymbol\theta|\vy)$. Since no closed-form solution could be obtained, we opt for maximizing the second-order approximation of $g_2(\boldsymbol\theta)$ at $\boldsymbol\theta^{(e)}$ from the previous $e$-th PE iteration
\begin{align}
\label{eq:2nd_order_g2_theta}
    g_2(\boldsymbol\theta)\approx g_2(\boldsymbol\theta^{(e)})+g_2^\prime\cdot\big(\boldsymbol\theta-\boldsymbol\theta^{(e)}\big)+\frac{g_2^{\prime\prime}}{2}\cdot\big(\boldsymbol\theta-\boldsymbol\theta^{(e)}\big)^2\,,
\end{align}
where $g_2^\prime, g_2^{\prime\prime}$ are the first and second order derivatives of $g_2(\boldsymbol\theta)$ with respect to $\gamma_w$. When $g_2^{\prime\prime}\geq0$, standard gradient descent can be employed to update $\gamma_w$. When $g_2^{\prime\prime}<0$, the update for $\gamma_w$ is then
\begin{align}
\label{eq:multi_bit_awgn_variance_update}
    \gamma_w^{(e+1)}=\gamma_w^{(e)}-\frac{g_2^\prime}{g_2^{\prime\prime}}\,.
\end{align}

\subsection{Channel Estimation from Low-Resolution ADCs}
\label{subsec:MIMO_CE}
We next introduce the channel estimation problem in a single-user MIMO system with low-resolution ADCs. Although the channel impulse response matrix $\mH$ from transmitters to receivers is dense in the antenna domain, it can be made sparse by transforming it to the angular domain. 

Following the problem setup in \cite{Mo:LowADC:2018}, we let $N_t$ denote the number of antennas at the transmitter, $N_r$ denote the number of antennas at the receiver, and $L$ denote the number of symbol intervals in the delay spread. The quantized measurements $\vy_d\in\mathbb{C}^{N_r}$ from the receivers at time $d$ are:
\begin{align}
\label{eq:channel_quantized_measurements}
    \vy_d=\mathscr{Q}\left(\sum_{l=0}^{L-1}\mH_l\vo_{d-l}+\vw_d\right)\,,
\end{align}
where $\mathscr{Q}$ is the quantizer that operates on the real and complex coefficients separately, $\mH_l\in\mathbb{C}^{N_r\times N_t}$ is the $l$-th channel matrix in the antenna domain, $\vo_d\in\mathbb{C}^{N_t}$ is the transmitted symbol at time $d$, and $\vw_d$ is the i.i.d. AWGN noise. 

As discussed in \cite{Sayeed:MIMOChannel:2002}, the channel matrix $\mH_l$ can be transformed into its sparse form $\mX_l\in\mathbb{C}^{N_r\times N_t}$ in the angular domain:
\begin{align}
    \label{eq:angle_domain_channel}
    \mH_l=\mathcal{S}_{N_r}\mX_l\mathcal{S}_{N_t}^*\,,
\end{align}
where $\mathcal{S}_{N_r}\in\mathbb{C}^{N_r\times N_r}$ and $\mathcal{S}_{N_t}\in\mathbb{C}^{N_t\times N_t}$ are the steering matrices for the receiver and transmitter arrays respectively, and they are constructed from unitary discrete Fourier transform (DFT) matrices. 

Substituting \eqref{eq:angle_domain_channel} into \eqref{eq:channel_quantized_measurements} and letting $\mA_d(\cdot)$ denote the resulting linear measurement operator, we can simplify the notations of the quantized measurement model as follows
\begin{align*}
    \vy_d=\mathscr{Q}(\mA_d(\vx)+\vw_d)\,.
\end{align*}
The channel estimation problem tries to recover the sparse channel coefficient vector $\vx$ from the quantized measurements $\{\vy_d|d=1,\cdots,N_d\}$ that are collected at different times.

\section{State Evolution Analysis}
\label{sec:se}

The asymptotic behavior of the AMP algorithm in the large system limit can be described by the state evolution recursions when $\mA$ is a random Gaussian matrix \cite{Bayati:SE:2011,Rangan:GAMP:2011,Kamilov:PE:2014}. State evolution tracks how the variables evolve through the message passing iterations. In particular, the asymptotic mean squared error (MSE) of the recovered signal $\hat{\vx}$ is a direct indication of AMP's performance: $\textnormal{MSE}(\hat{\vx})=\frac{1}{N}\|\vx-\hat{\vx}\|_2^2$.

Following the state evolution analysis in \cite{Rangan:GAMP:2011,Kamilov:PE:2014}, we demonstrate that the proposed AMP-PE approach for the 1-bit and multi-bit CS problems also obeys the state evolution equations. Specifically, the state evolution recursions occur between the following two sets of variables:
\begin{itemize}
    \item Output channel variables $\{Z,Q,W,Y\}$: The quantized measurement $Y$ depends on the noiseless measurement $Z$, the pre-quantization noise $W$ and the quantizer $\mathscr{Q}(\cdot)$: $Y=\mathscr{Q}(Z+W)$, where $W\sim\mathcal{N}(0,\gamma_w)$. The two variables $(Z,Q)\sim\mathcal{N}({\boldsymbol 0};\mK_q)$, where $Q$ is a dummy variable and $\mK_q\in\mathbb{R}^{2\times 2}$ is the covariance matrix. The distribution parameter set $\Theta=\{\gamma_w\}$.
    \item Input channel variables $\{X,R\}$: The signal $X$ follows the distribution in \eqref{eq:bgm} with mean $\mathbb{E}[X]=0$ and variance $\textnormal{Var}[X]=\nu_x$. The dummy variable $R$ depends on $X$ and follows the distribution $p(R|X)=\mathcal{N}(R|X,\overline{\tau}_r)$. The distribution parameter set $\Lambda=\{\kappa,\xi_i,\mu_i,{\gamma_x}_i\ |\ i=1,\cdots,I\}$.
\end{itemize}
State evolution recursions of the AMP-PE approach are summarized in Algorithm \ref{alg:se_amp_pe} and can be empirically verified by the Monte Carlo simulation. We summarize our finding as follows.

\begin{finding}
Let the initial AMP-PE estimation $\hat{\vx}^{(0)}=\boldsymbol 0$, and the state evolution recursions be initialized with $\hat{X}^{(0)}=0$, $\overline{\tau}_x^{(0)}=\nu_x$ in Algorithm \ref{alg:se_amp_pe}. In the large system limit as $N\rightarrow\infty$ while the sampling ratio $\beta=\frac{M}{N}$ is fixed, the MSE of AMP-PE estimation $\hat{\vx}^{(t+1)}$ is predicted by $\overline{\tau}_x^{(t+1)}$ in \eqref{eq:se_tau_x}:
\begin{align*}
    \textnormal{MSE}(\hat{\vx}^{(t+1)})\approx\overline{\tau}_x^{(t+1)}\,.
\end{align*}
\end{finding}

\begin{algorithm}[tbp]
\caption{The AMP-PE state evolution}\label{alg:se_amp_pe}
\begin{algorithmic}[1]
\Input $\overline{\tau}_x^{(0)},\mK_x^{(0)}$.
\State When $t=0$, $\overline{\tau}_x^{(0)}=\nu_x$, and the covariance matrix $\mK_x^{(0)}$ of the variables $(X,\hat{X}^{(0)})$ is
\begin{align}
    \mK_x^{(0)}=\left[\begin{array}{cc} \nu_x &\nu_x-\overline{\tau}_x^{(0)}\\ \nu_x-\overline{\tau}_x^{(0)} & \nu_x-\overline{\tau}_x^{(0)} \end{array}\right] = \left[\begin{array}{cc} \nu_x &0\\ 0 & 0 \end{array}\right]\,.
\end{align}
\For{$t=\{0,1,\cdots,T\}$}
	\State \emph{Output update}: 
	\begin{subequations}
	\begin{align}
        \overline{\tau}_q^{(t)} &= \beta^{-1}\cdot\overline{\tau}_x^{(t)},\quad\mK_q^{(t)}=\beta^{-1}\cdot\mK_x^{(t)}\\
        \label{eq:se_amp_pe_est_output}
        \hat{\Theta}^{(t)} &= \arg\max_{\Theta}\ p(\Theta|Q,\overline{\tau}_q^{(t)},Y)\\
        \label{eq:se_tau_r}
        \overline{\tau}_r^{(t)} &=
        \mathbb{E}\left[\overline{\tau}_q^{(t)}\left(1-\frac{1}{\overline{\tau}_q^{(t)}}\textnormal{Var}\left[Z|Q,,\overline{\tau}_q^{(t)},Y,\hat{\Theta}^{(t)}\right]\right)^{-1}\right]\,,
    \end{align}
	\end{subequations}
	where the expectation $\mathbb{E}[\cdot]$ in \eqref{eq:se_tau_r} is with respect to the variables $\{Z,Q,W,Y\}$.
	\State \emph{Input update}: 
	\begin{subequations}
	\begin{align}
	    \label{eq:se_amp_pe_est_input}
	    \hat{\Lambda}^{(t)} &= \arg\max_{\Lambda}\ p(\Lambda|R,\overline{\tau}_r^{(t)})\\
	    \label{eq:se_tau_x}
        \overline{\tau}_x^{(t+1)} &= \mathbb{E}\left[\textnormal{Var}\left[X|R,\overline{\tau}_r^{(t)},\hat{\Lambda}^{(t)}\right]\right]\,,
	\end{align}
	where the expectation $\mathbb{E}[\cdot]$ in \eqref{eq:se_tau_x} is with respect to the variables $\{X,R\}$.
	\begin{align}
	    \label{eq:se_x_hat}
	    \hat{X}^{(t+1)} &= \mathbb{E}\left[X|R,\overline{\tau}_r^{(t)},\hat{\Lambda}^{(t)}\right]\\
	    \mK_x^{(t+1)} &= \textnormal{cov}(X,\hat{X}^{(t+1)})\,,
	\end{align}
	where the expectation $\mathbb{E}[\cdot]$ in \eqref{eq:se_x_hat} is with respect to $X$.
	\end{subequations}
\EndFor
\State\Return The predicted MSE values $\left\{\overline{\tau}_x^{(0)}, \overline{\tau}_x^{(1)},\cdots,\overline{\tau}_x^{(T)}\right\}$.
\end{algorithmic}
\end{algorithm}

The expectations in \eqref{eq:se_tau_r}, \eqref{eq:se_tau_x} and \eqref{eq:se_x_hat} are computed using the Monte Carlo simulation. Note that the covariance matrix $\mK_q$ has the following form \cite{Rangan:GAMP:2011}:
\begin{align}
    \mK_q^{(t)}=\left[\begin{array}{cc} \beta^{-1}\cdot\nu_x &\beta^{-1}\cdot\nu_x-\overline{\tau}_q^{(t)}\\ \beta^{-1}\cdot\nu_x-\overline{\tau}_q^{(t)} & \beta^{-1}\cdot\nu_x-\overline{\tau}_q^{(t)} \end{array}\right]\,.
\end{align}
Since $(Z,Q)\sim\mathcal{N}({\boldsymbol 0};\mK_q)$, we can get
\begin{align}
    p(Q)&=\mathcal{N}(0;\beta^{-1}\cdot\nu_x-\overline{\tau}_q^{(t)})\\
    p(Z|Q)&=\mathcal{N}(Q;\overline{\tau}_q^{(t)})\,.
\end{align}
During the simulation, the variables $\{Z,Q,W,Y\}$ can be generated in the order of dependency $Q\rightarrow Z\xrightarrow{W} Y$; and the variables $\{X,R\}$ can be generated in the order $X\rightarrow R$. The parameter estimation problems in \eqref{eq:se_amp_pe_est_output} and \eqref{eq:se_amp_pe_est_input} are solved by the second-order method and EM respectively in the same way as before.

We next compare the MSEs of AMP-PE estimations with the state evolution recursions for random Gaussian measurement matrix. Here, the matrix $\mA$ has i.i.d. entries: $A_{mn}\sim\mathcal{N}(0,\frac{1}{M})$ and the nonzero entries of $\vx$ are independently generated from $\mathcal{N}(0,1)$. We vary the signal length $N\in\{1000,5000,10000\}$ and fix the sampling ratio $\beta=\frac{M}{N}=2$. The sparsity level $\frac{E}{N}$ of the signal $\vx$ is set to $50\%$ and white Gaussian noise $\vw$ is added to $\vz=\mA\vx$ such that the pre-quantization (pre-QNT) SNR=30dB. The quantized measurement $y_m$ is computed as $y_m=\mathcal{Q}(z_m+w_m)$. For each experimental setting, we run $100$ random trials and compute the MSE of the recovered signal $\hat{\vx}$ in every iteration of the AMP-PE algorithm. 

\begin{figure*}[htbp]
\centering
\subfigure[]{
\label{fig:se_50_30_1000}
\includegraphics[height=0.23\textwidth]{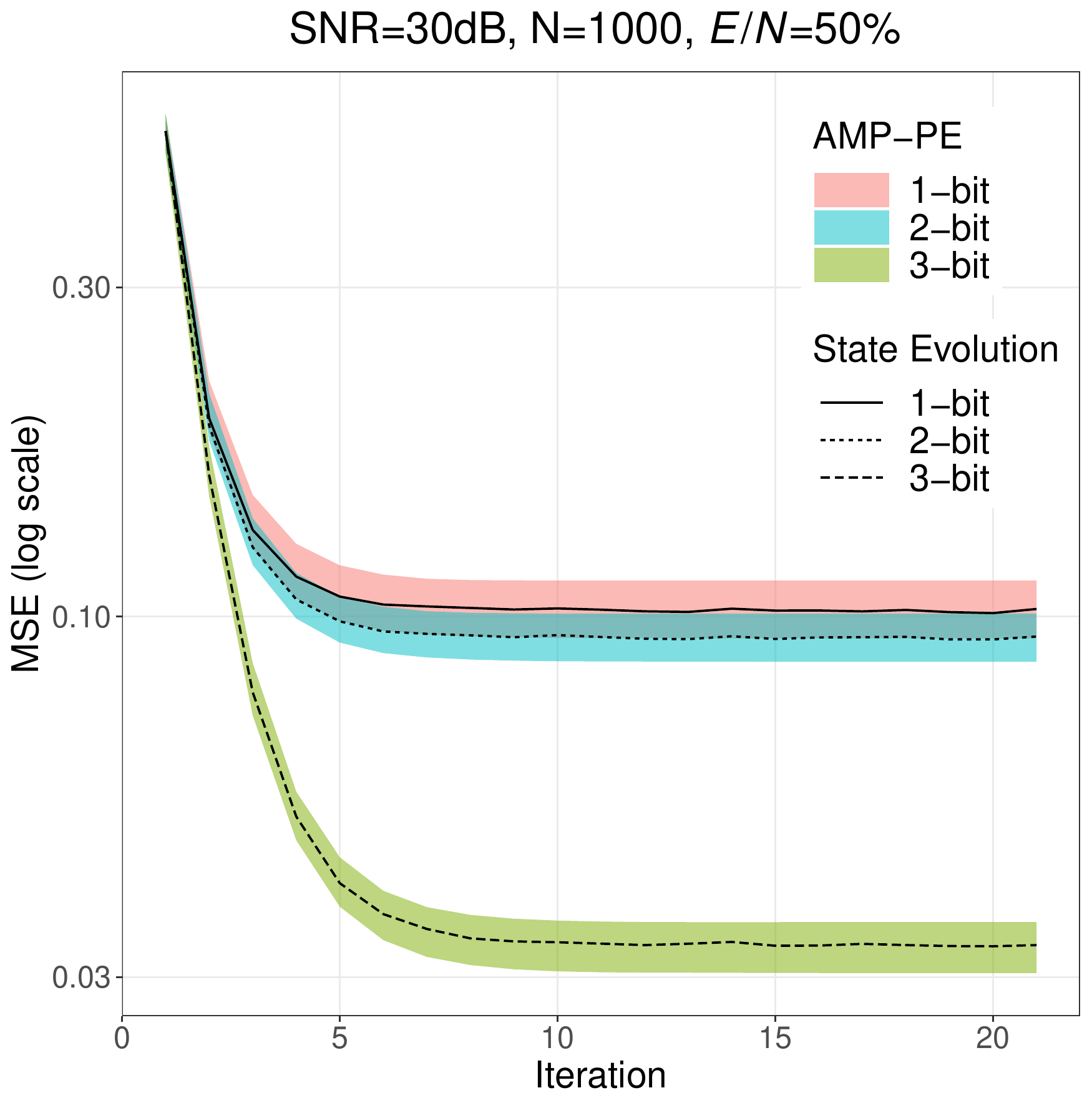}}
\subfigure[]{
\label{fig:se_50_30_5000}
\includegraphics[height=0.23\textwidth]{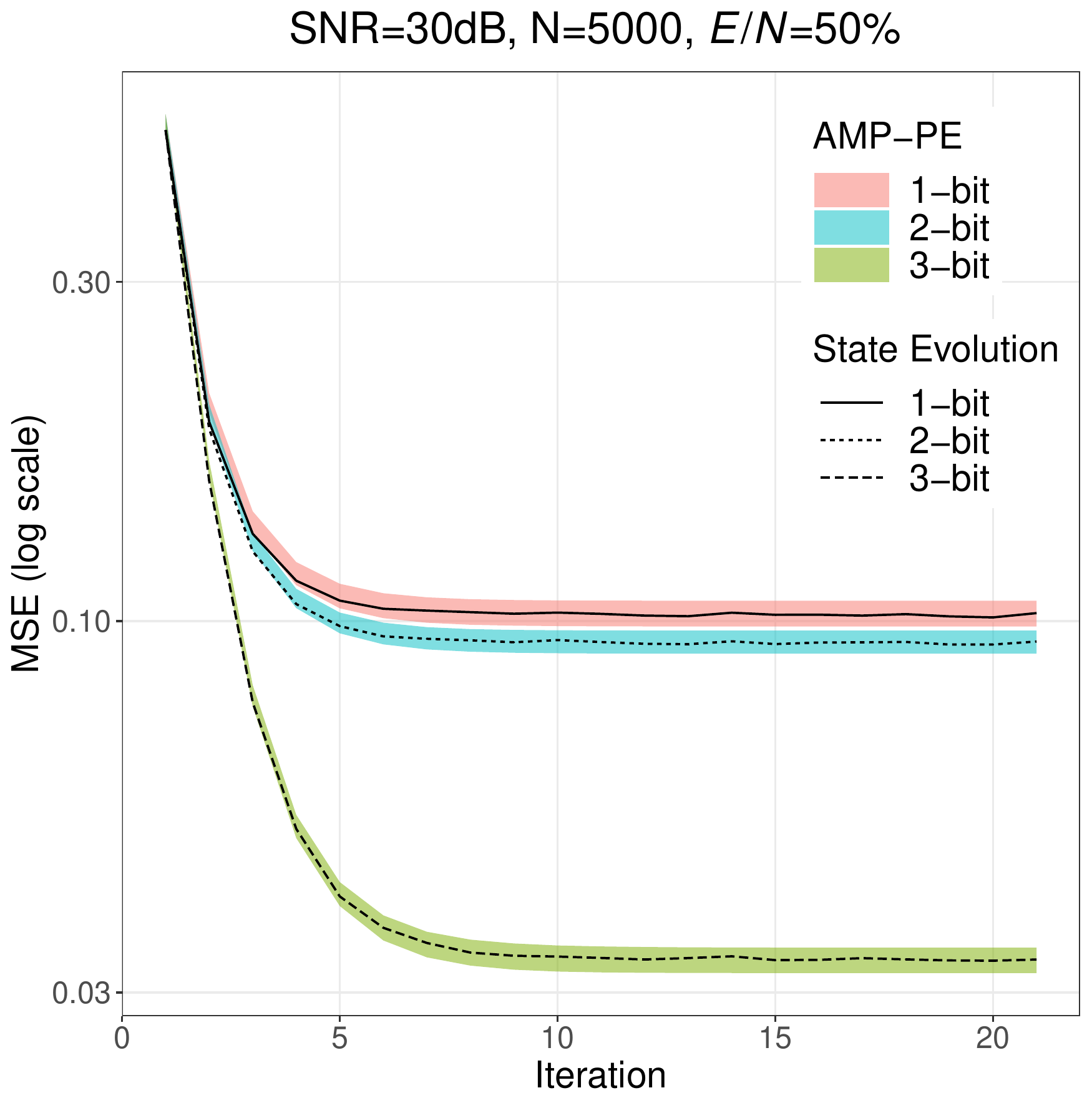}}
\subfigure[]{
\label{fig:se_50_30_10000}
\includegraphics[height=0.23\textwidth]{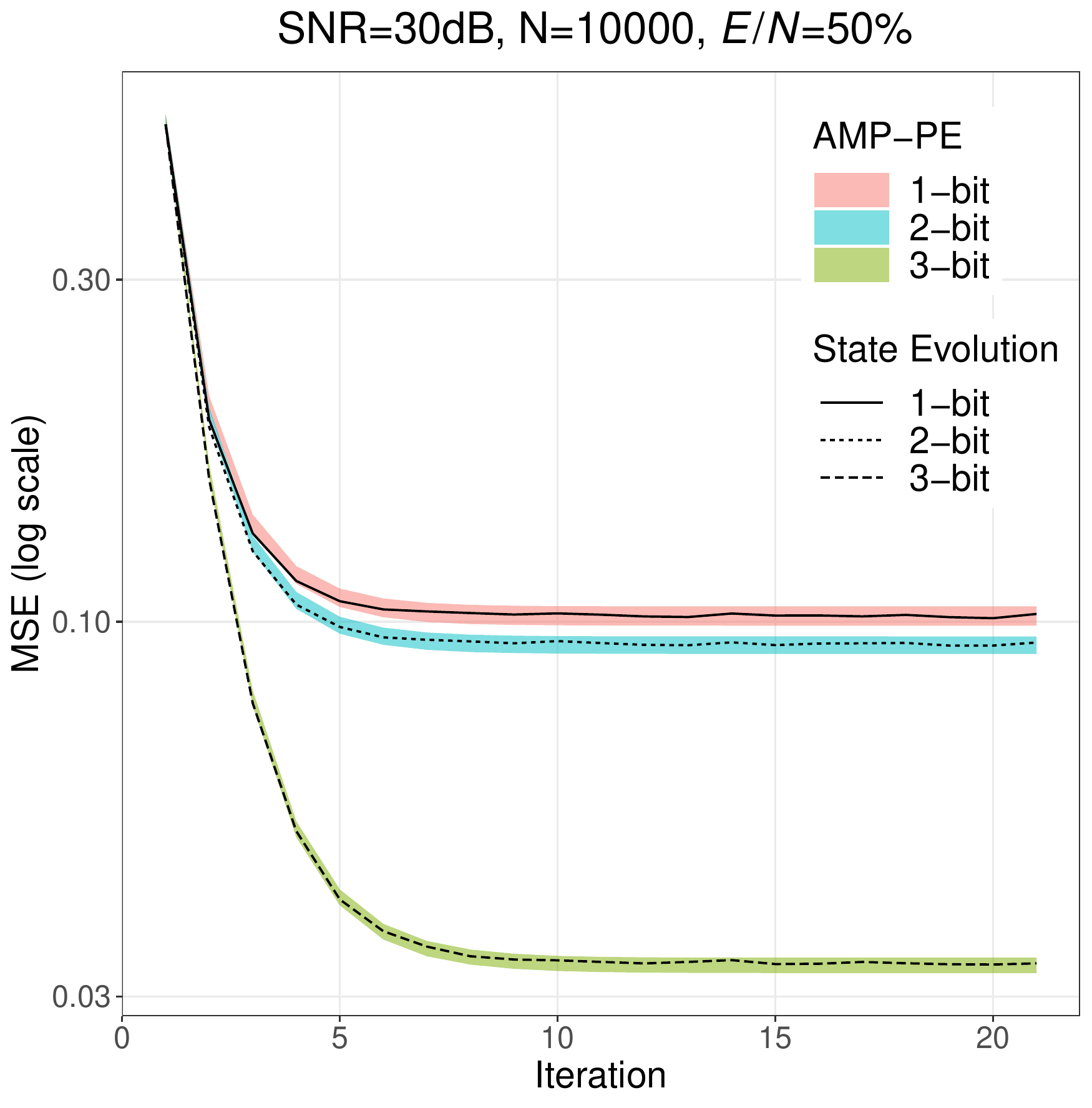}}
\subfigure[]{
\label{fig:se_100_10_10000}
\includegraphics[height=0.23\textwidth]{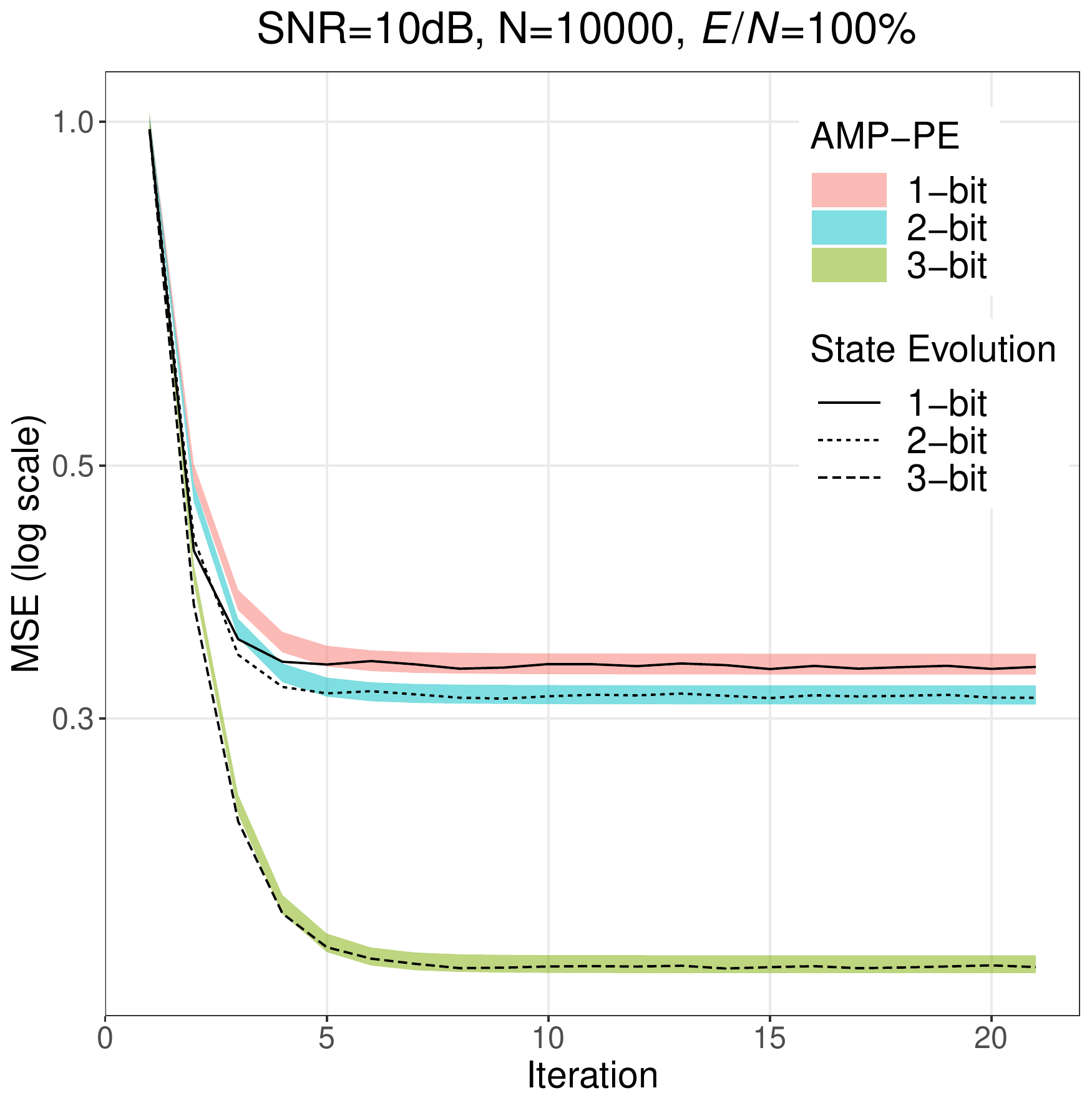}}
\caption{As the signal length $N\rightarrow\infty$, MSEs of AMP-PE estimations are predicted by state evolution recursions for random Gaussian measurement matrix. We set the sampling ratio $\beta=\frac{M}{N}=2$. (a)-(c): Sparsity level $\frac{E}{N}=50\%$, pre-quantization SNR=30dB; (d) $\frac{E}{N}=100\%$, pre-quantization SNR=10dB.}
\label{fig:se_compare}
\end{figure*}

Fig. \ref{fig:se_compare} depicts a comparison of the MSEs from AMP-PE and the state evolution recursions. The MSEs from AMP-PE are illustrated by the shaded area between the two lines corresponding to ``mean-MSE $\pm$ standard deviation'' across 100 random trials. Convergence is reached after around 10 iterations in this case. Since state evolution is derived in the large system limit as $N\rightarrow\infty$, state evolution recursions are more accurate in predicting the MSEs from AMP-PE as $N$ gets larger. As shown in Fig. \ref{fig:se_50_30_1000}-\ref{fig:se_50_30_10000}, we can see that state evolution recursions could accurately predict the MSEs of AMP-PE estimations when $N=10000$.

With $N$ set to $10000$, we perform comparisons under different pre-QNT SNRs $\in\{30\textnormal{dB},20\textnormal{dB},10\textnormal{dB}\}$ and different sparsity levels $\frac{E}{N}\in\{10\%,50\%,100\%\}$. The results are given in the Supplemental Material. In particular, we would like to highlight the results shown in Fig. \ref{fig:se_100_10_10000} where pre-QNT SNR=10dB and $\frac{E}{N}=100\%$. For 1-bit CS and 2-bit CS, we can see that state evolutions do not always fall within the confidence regions in the first few iterations. Since $N$ is already large enough, the finite sample effect no longer plays a major role here. The main reason for this deviation is that the parameter estimation problems in \eqref{eq:amp_pe_est_input},\eqref{eq:amp_pe_est_output} and \eqref{eq:se_amp_pe_est_output},\eqref{eq:se_amp_pe_est_input} are nonconvex. The computed maximizing parameters are only locally optimal; hence, they are slightly different between the AMP-PE algorithm and its state evolution. As shown in the Supplemental Material, such deviation could be corrected when both the AMP-PE and its state evolution use the same ground-truth parameters. Analyzing the nonconvex optimization landscape is an NP-hard problem \cite{Murty1987SomeNP,Sun2016}, the reason why the deviation becomes more pronounced in the low-SNR or high-sparsity-level regimes remains an open question.

\section{Experimental Results}
\label{sec:exp}

\begin{figure*}[htbp]
\centering
\subfigure{
\includegraphics[height=0.23\textwidth]{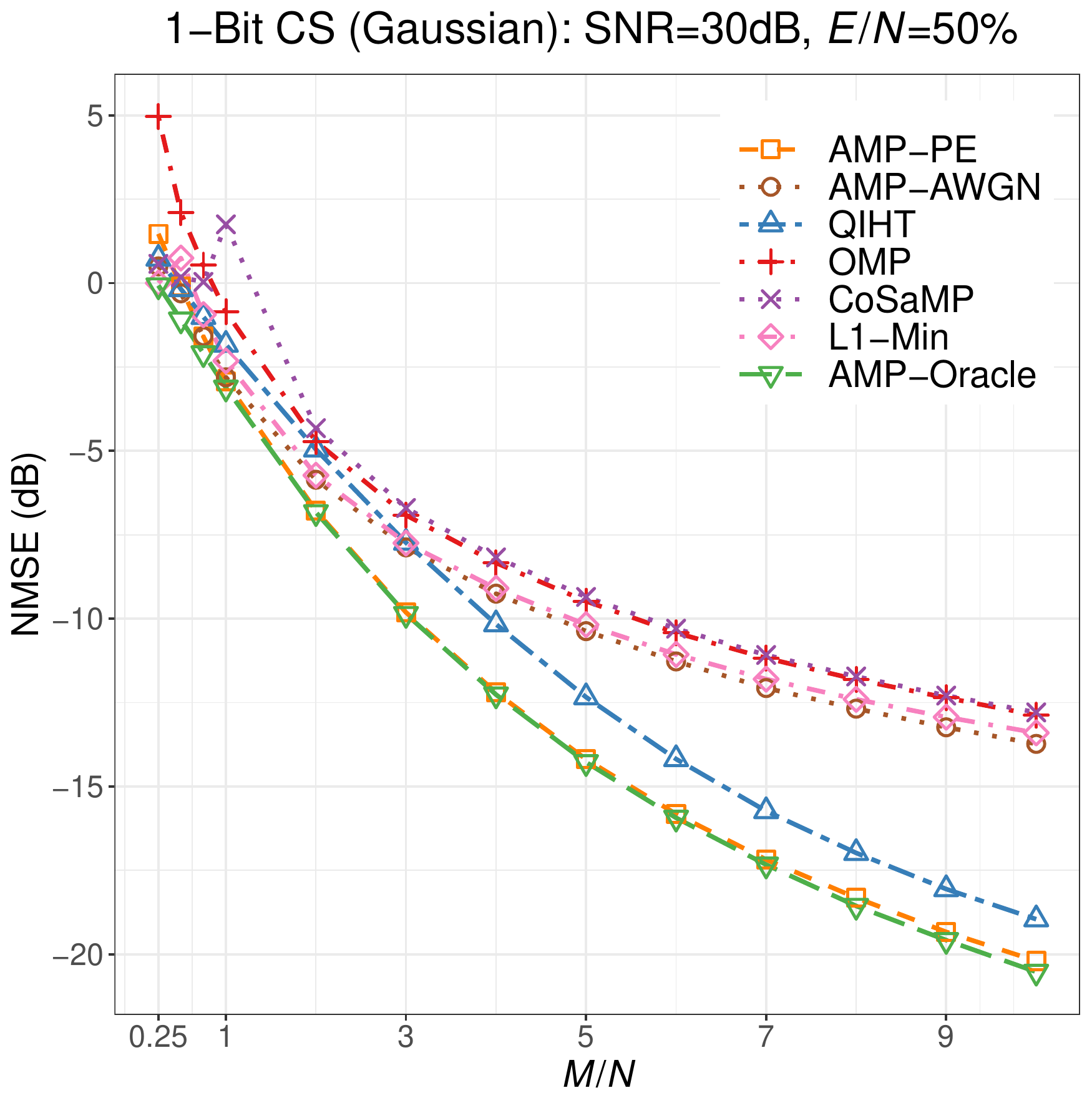}}
\subfigure{
\includegraphics[height=0.23\textwidth]{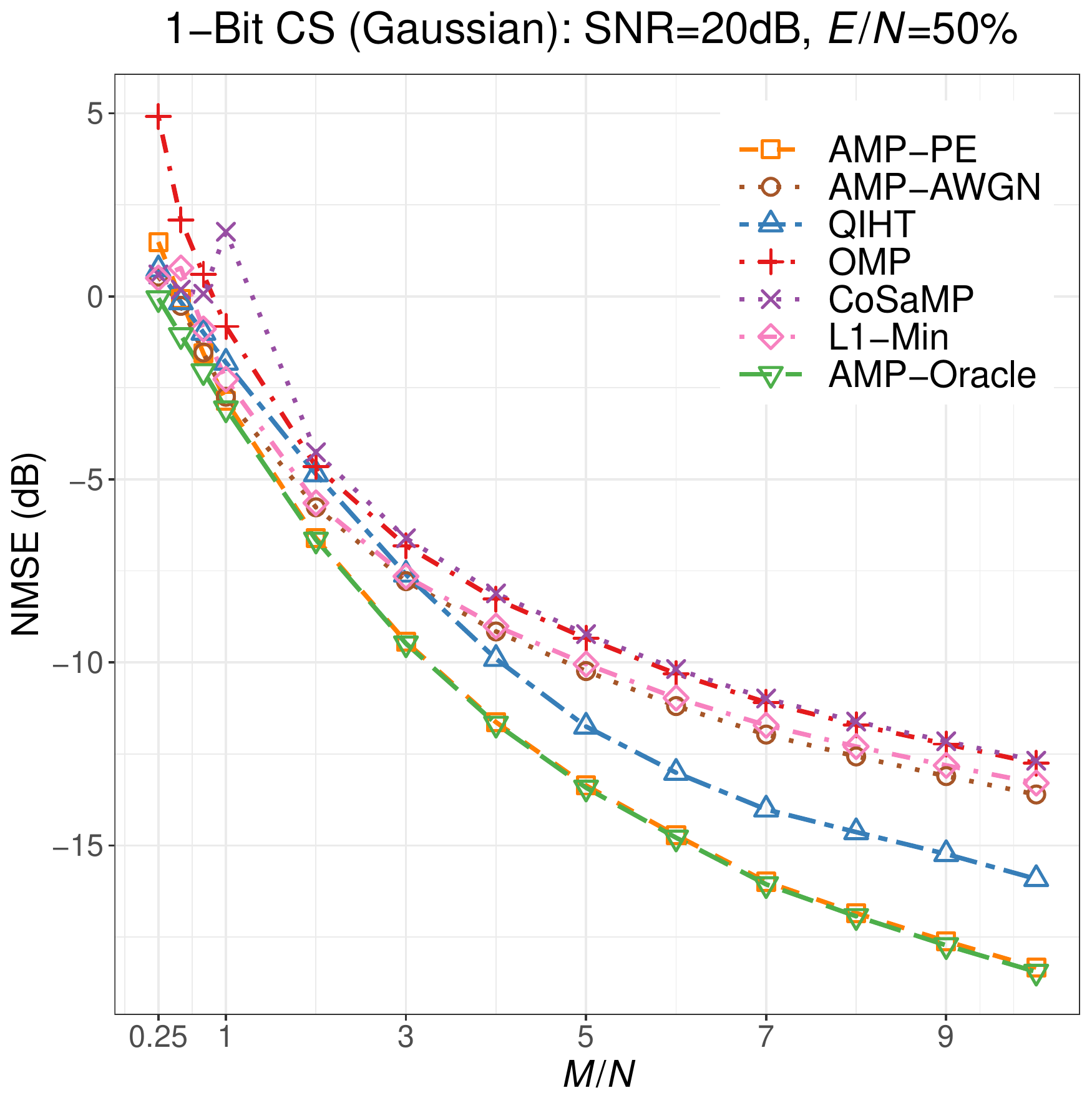}}
\subfigure{
\includegraphics[height=0.23\textwidth]{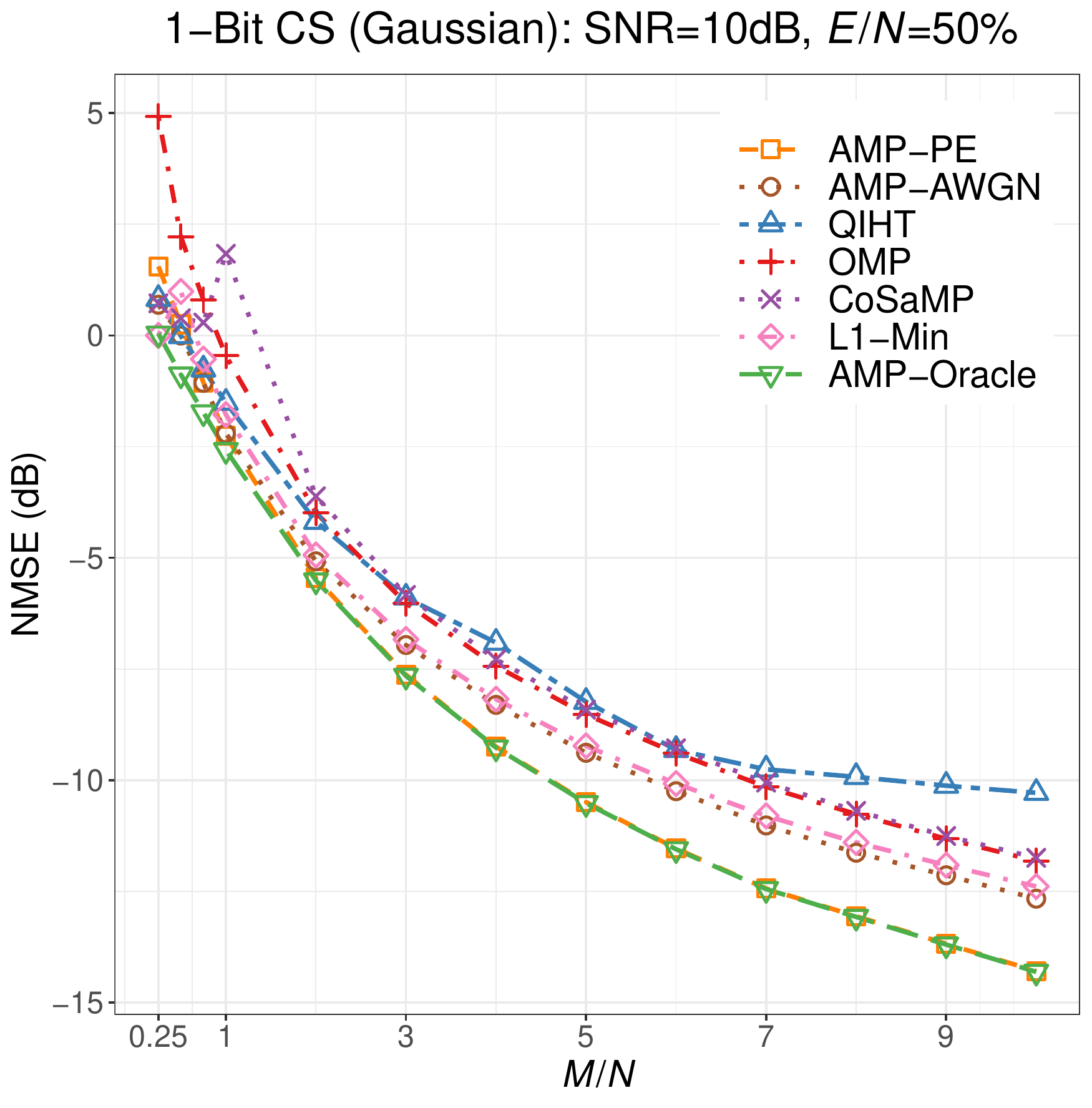}}
\caption{Comparison of different approaches in solving 1-bit CS. Nonzero entries of the signal follow the Gaussian distribution. The sampling ratio $\frac{M}{N}\in\{0.25,\cdots,10\}$, the sparsity level $\frac{E}{N}=50\%$. The pre-quantization SNR varies from $30$dB, $20$dB to $10$dB.}
\label{fig:1bit_experiments_gaussian}
\end{figure*}

\begin{figure*}[htbp]
\centering
\subfigure{
\includegraphics[height=0.23\textwidth]{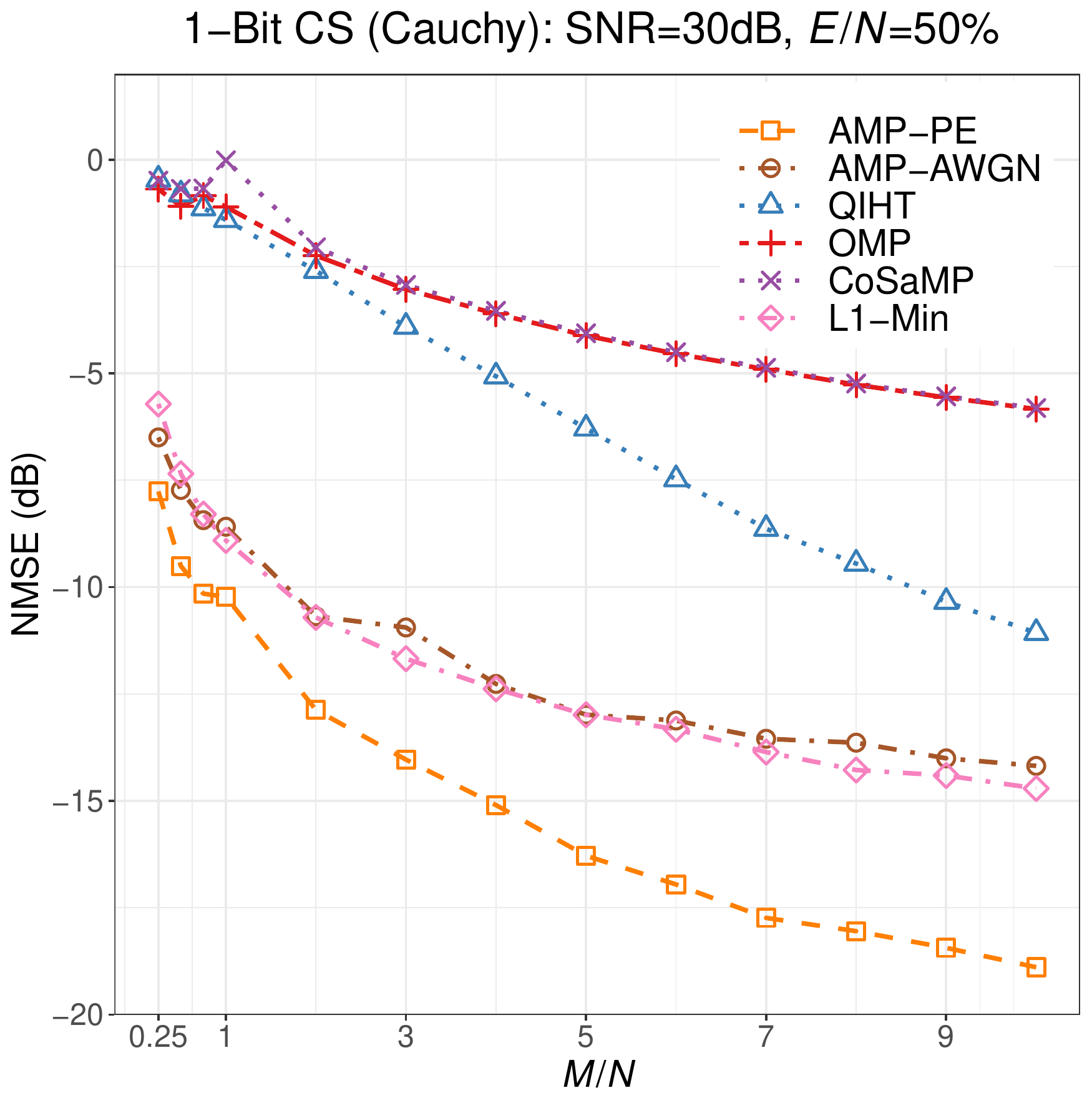}}
\subfigure{
\includegraphics[height=0.23\textwidth]{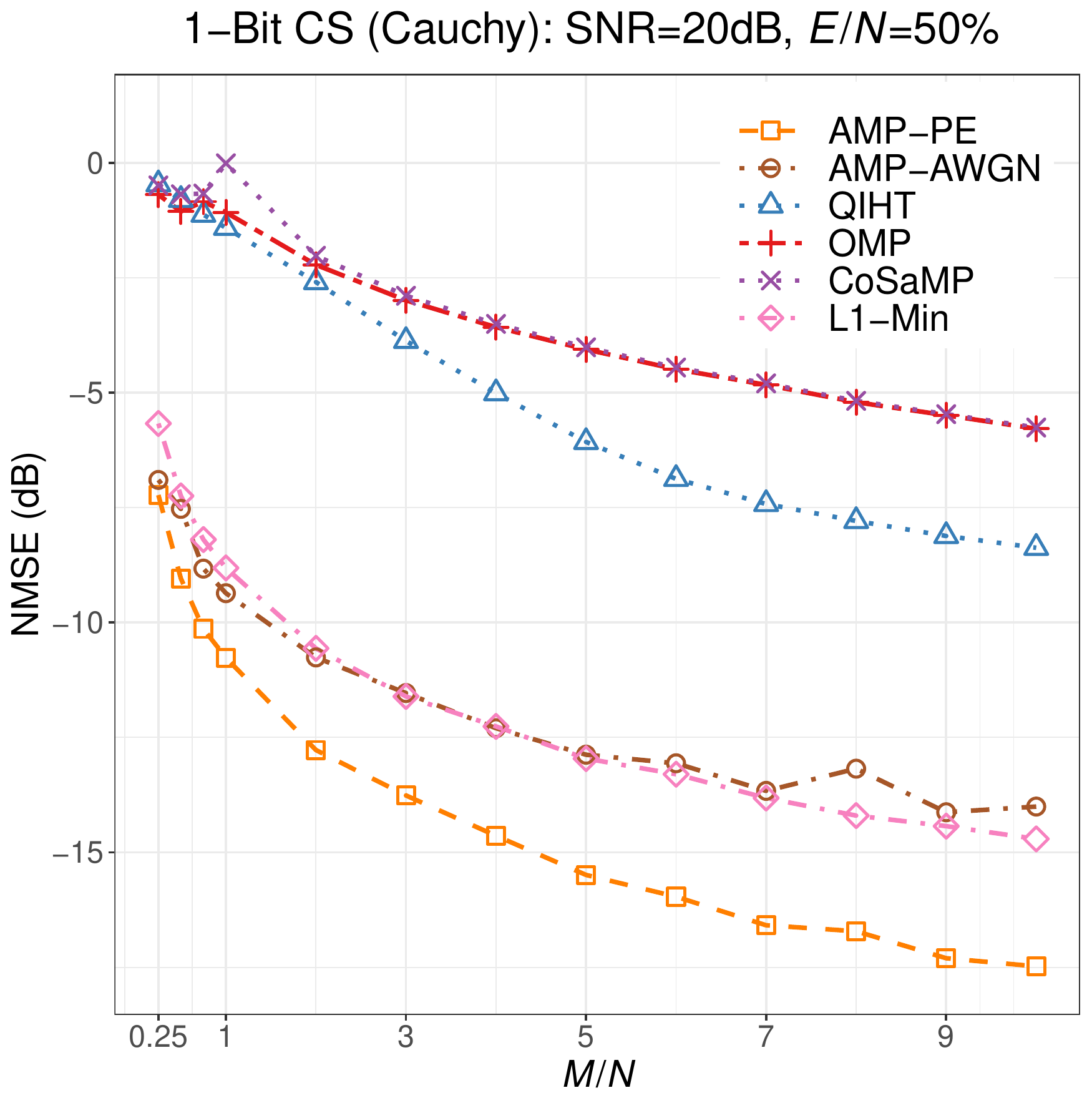}}
\subfigure{
\includegraphics[height=0.23\textwidth]{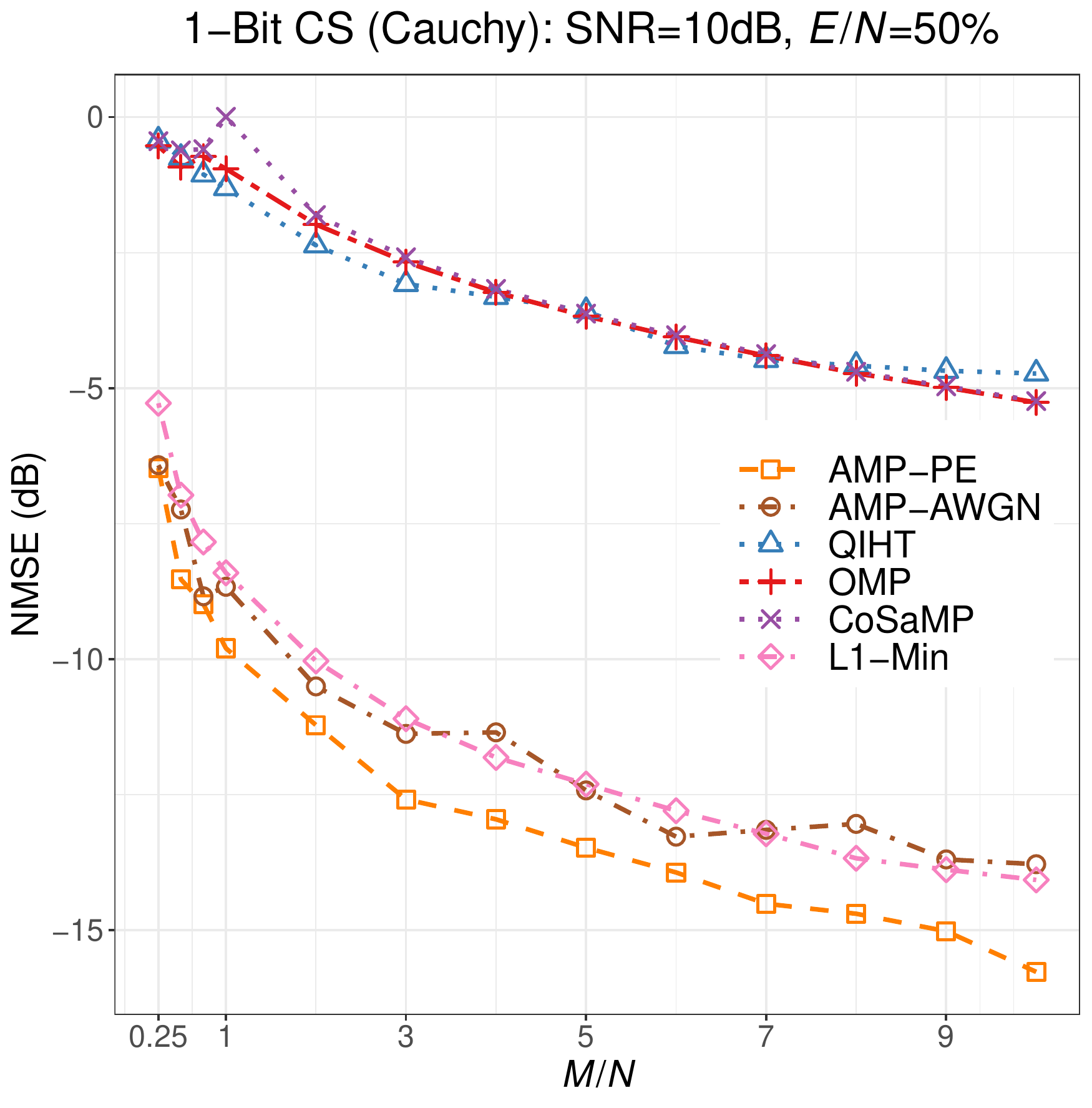}}
\caption{Comparison of different approaches in solving 1-bit CS. Nonzero entries of the signal follow the Cauchy distribution. The sampling ratio $\frac{M}{N}\in\{0.25,\cdots,10\}$, the sparsity level $\frac{E}{N}=50\%$. The pre-quantization SNR varies from $30$dB, $20$dB to $10$dB.}
\label{fig:1bit_experiments_cauchy}
\end{figure*}

This section offers detailed comparisons between the proposed AMP with built-in parameter estimation (AMP-PE) algorithm and other state-of-the-art sparse recovery methods such as QIHT \cite{Jacques:QIHT:2013}, OMP \cite{OMP07,Cai:OMP:2011}, CoSaMP \cite{CoSaMP09}, $l_1$-norm minimization (L1-Min) \cite{WaveletAnalysis94,ScaleBasis99}, and the AMP with an approximated AWGN model (AMP-AWGN) \cite{Vila:BG:2011,Vila:EMGM:2013}. The proposed AMP-PE works with the true quantization noise model, and it recovers the signal and parameters jointly. On the other hand, methods like QIHT, OMP, CoSaMP and L1-Min require extensive parameter tuning processes. In order for OMP, CoSaMP, L1-Min and AMP-AWGN to work with quantization measurements, the quantization symbol $b_i$ is replaced by the value $\frac{a_{i-1}+a_i}{2}$ according to the quantizer defined in \eqref{eq:multi_bit_quantization}. This is the typical approximation scheme that has been adopted in the literature \cite{risi2014massive,Wang:Multiuser:2015,Wen:Channel:2015,Bellili:Lap:2019}. Additionally, results from AMP-PE with the true distribution parameters (AMP-Oracle) are provided as reference when available, they correspond to the best results that could be obtained under the AMP framework. 

In our experiments, signals of varying sparsity levels are recovered from quantized measurements under different noise levels. The normalized mean squared error (NMSE) of the recovered signals are computed to evaluate all the methods. Experimental results show that the proposed AMP-PE generally performs much better than the other methods across various sparsity and noise levels. Reproducible code and data are available at \urlstyle{tt}\url{https://github.com/shuai-huang/1Bit-CS}

\subsection{1-Bit and Multi-Bit Compressive Sensing}
\label{subsec:1bit_cs_exp}
For the sparse signal recovery experiments, the signal length is set to $N=1000$. We first fix the sparsity level $\frac{E}{N}=50\%$ and vary the sampling ratio $\frac{M}{N}\in\{0.25,0.5,0.75,1,2,\cdots,10\}$. We would like to see how AMP-PE holds up against the mismatch between the assumed BGM prior and the actual signal prior. We thus generate nonzero entries of $\vx$ from three different distributions: {\em (i)} Gaussian distribution $\mathcal{N}(0,1)$; {\em (ii)} Cauchy distribution with density function $p(x)=\frac{1}{\pi}\cdot\frac{1}{x^2+1}$; and {\em (iii)} Laplace distribution with density function $p(x)=\frac{1}{2}\exp(-|x|)$.

Entries of the measurement matrix $\mA$ are randomly generated from the distribution $\mathcal{N}(0,1)$. We add white Gaussian noise $\vw\sim\mathcal{N}(0,\gamma_w)$ to the noiseless measurements $\vz=\mA\vx$ before quantization. The noise variance $\gamma_w$ is chosen such that the pre-QNT SNR of $\vz+\vw$ varies from 30dB, 20dB to 10dB. The 1-bit measurement $y_m$ is obtained by applying the element-wise quantizer $\mathcal{Q}$: $y_m=\mathcal{Q}(z_m+w_m)$. For each combination of $\left\{\frac{M}{N}, \textnormal{pre-QNT SNR}\right\}$, the average NMSE across 100 random trials is again computed for each method. 

When the nonzero entries of $\vx$ are generated from the Gaussian distribution, the recovery results are shown in Fig. \ref{fig:1bit_experiments_gaussian}. All methods are initialized with an all-zero vector $\vx^{(0)}=\boldsymbol 0$. For the signal prior in AMP-PE and AMP-AWGN, the initial sparsity ratio is set to $\kappa=0.1$, the number of Gaussian mixtures is set to $I=2$, and the initial Gaussian mixture means as well as variances are estimated from the least-squares solution via k-means clustering. For the noise prior in AMP-PE and AMP-AWGN, the initial noise variance is set to $\gamma_w=10^{-6}$. Damping operations are only applied to the estimated input and output parameters. Take $\hat{\lambda}_l^{(t+1)}$ for example. After it is computed in \eqref{eq:amp_pe_est_input}, it goes through the following damping operation:
\begin{align}
\label{eq:damp_on_lambda}
    \hat{\boldsymbol\lambda}^{(t+1)} = \hat{\boldsymbol\lambda}^{(t)}+\eta\cdot\big(\hat{\boldsymbol\lambda}^{(t+1)}-\hat{\boldsymbol\lambda}^{(t)}\big),
\end{align}
where the damping rate $\eta=0.2$.

From Fig. \ref{fig:1bit_experiments_gaussian}, we can see that the proposed AMP-PE matches the performance of the oracle AMP. Since AMP-PE adopts the true quantization noise model, it performs much better than AMP-AWGN where an approximated AWGN model is used. For QIHT, OMP and CoSaMP, the sparsity level $\frac{E}{N}$ is assumed to be known. For the L1-Min approach, we tune its regularization parameter on a separate training dataset. The proposed AMP-PE does not require any parameter tuning -- it treats the distribution parameters as variables and jointly recovers them along with the signal of interest. When the pre-QNT SNR is 30dB, we can observe that AMP-PE and QIHT perform much better than others as the sampling ratio $\frac{M}{N}$ increases, since they could work with quantized measurements directly. AMP-PE is more robust to noise than QIHT when the pre-QNT SNR level is reduced to 20dB and 10dB. In general, AMP-PE achieves state-of-the-art performances in all cases.

When nonzero entries are generated from the Cauchy distribution, the recovery results are shown in Fig. \ref{fig:1bit_experiments_cauchy}. The number of Gaussian mixtures is set to $I=5$, and the rest of initial parameter settings are kept the same as in the Gaussian case. AMP-PE and AMP-AWGN adopt the BGM distribution in \eqref{eq:bgm} as the sparse signal prior. The first Gaussian component in the BGM prior has zero-mean and a variance that is large enough to cover the domain of $x$. This allows AMP-PE and AMP-AWGN to overcome the mismatch between BGM and Cauchy distribution in the sparse signal model. We can see that AMP-PE, AMP-AWGN and L1-Min generally perform better than QIHT, OMP and CoSaMP. When the pre-QNT SNR is 30dB or 20dB, AMP-PE has much better and robust performances than AMP-AWGN and L1-Min, especially when more measurements become available. When the pre-QNT SNR is 10dB, performances of AMP-PE, AMP-AWGN and L1-Min tend to be similar, with AMP-PE in the slight lead.

When nonzero entries are generated from Laplace distribution, AMP-PE also achieves leading performances. The results and detailed discussions are given in Supplemental Material.

Finally, we vary the sparsity level $\frac{E}{N}\in\{10\%,50\%,100\%\}$, and perform the recovery experiments with 2-bit and 3-bit measurements. The results are also given in the Supplemental Material. When the sparsity level $\frac{E}{N}$ is small, AMP-PE generally performs significantly better than the other methods, especially when the sampling ratio $\frac{M}{N}$ increases. For the multi-bit experiments, similar conclusions can be drawn: AMP-PE achieves leading performances across various sparsity and noise levels.

\subsection{Complexity and Runtime Comparisons}

\begin{figure}[tbp]
\centering
\includegraphics[height=0.27\textwidth]{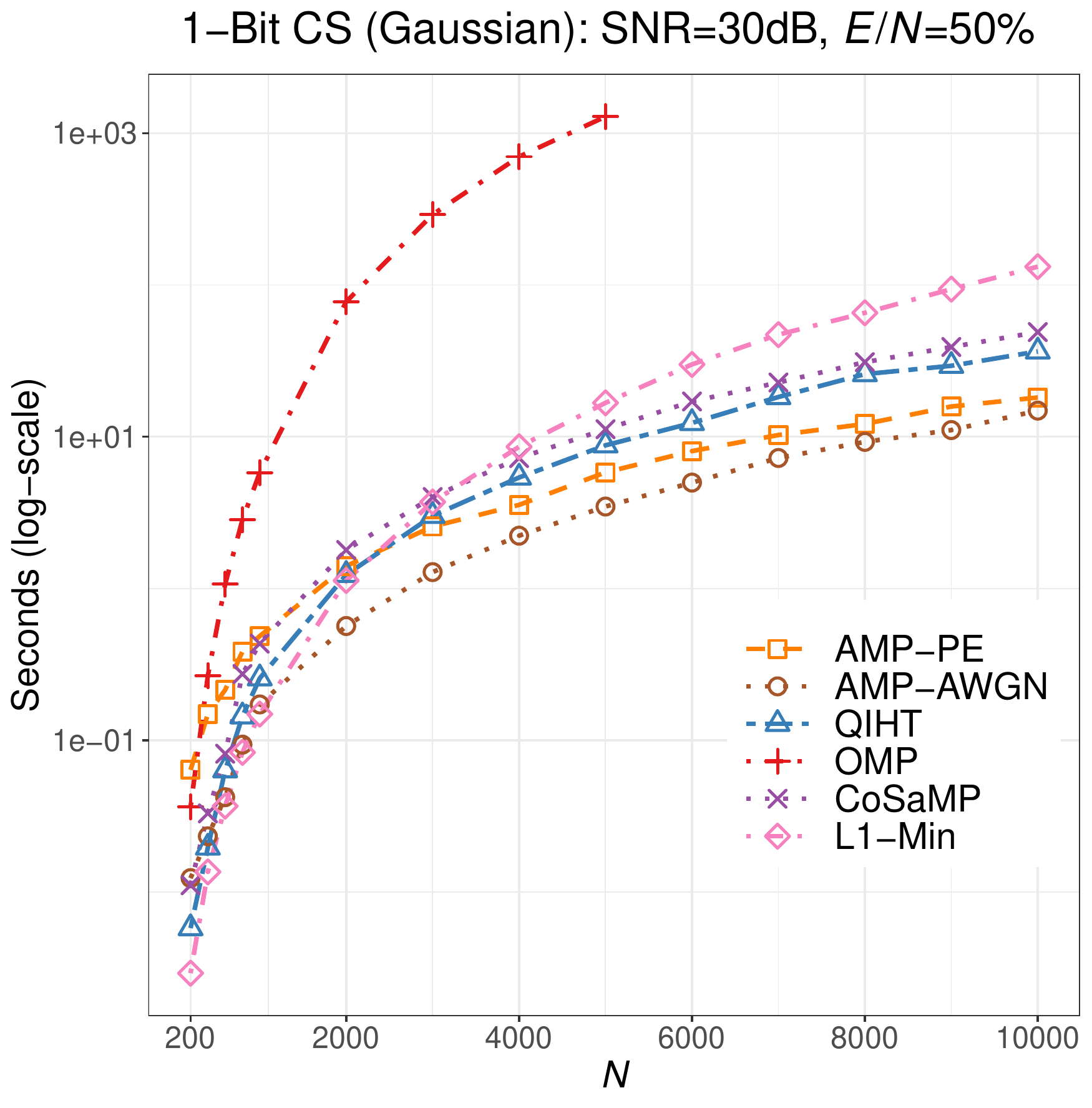}
\caption{Runtime comparison of different approaches with respect to the problem size $N$.}
\label{fig:run_time}
\end{figure}

\begin{figure*}[htbp]
\centering
\subfigure{
\includegraphics[width=0.08\textwidth]{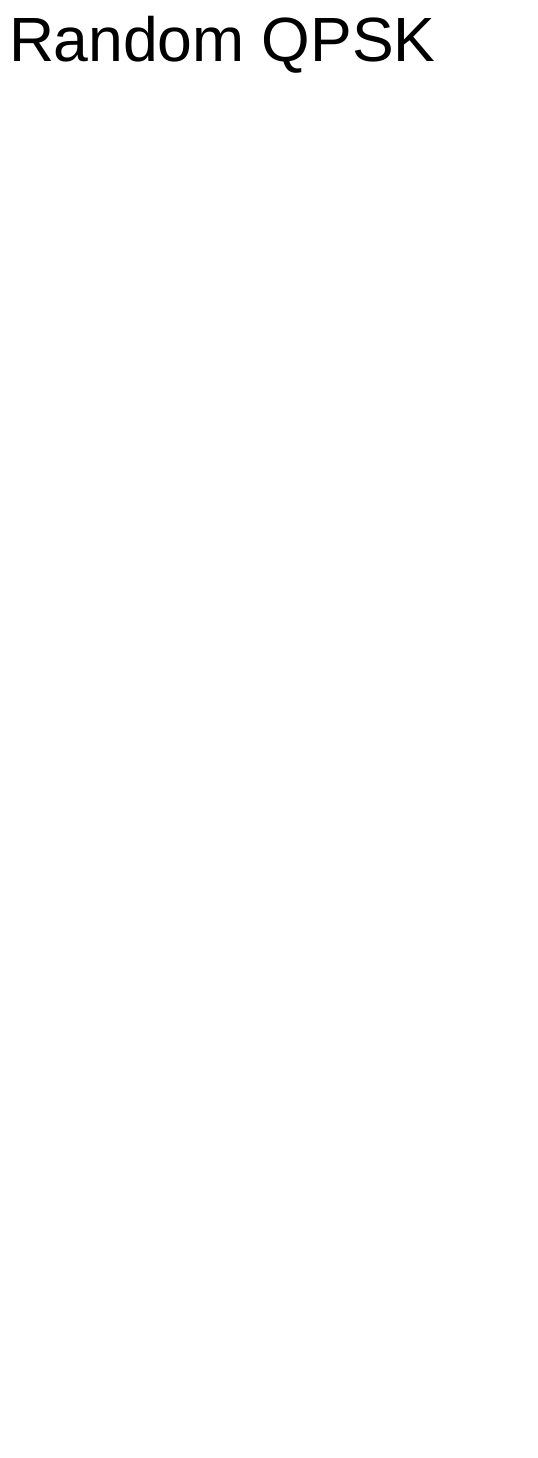}}
\subfigure{
\includegraphics[height=0.23\textwidth]{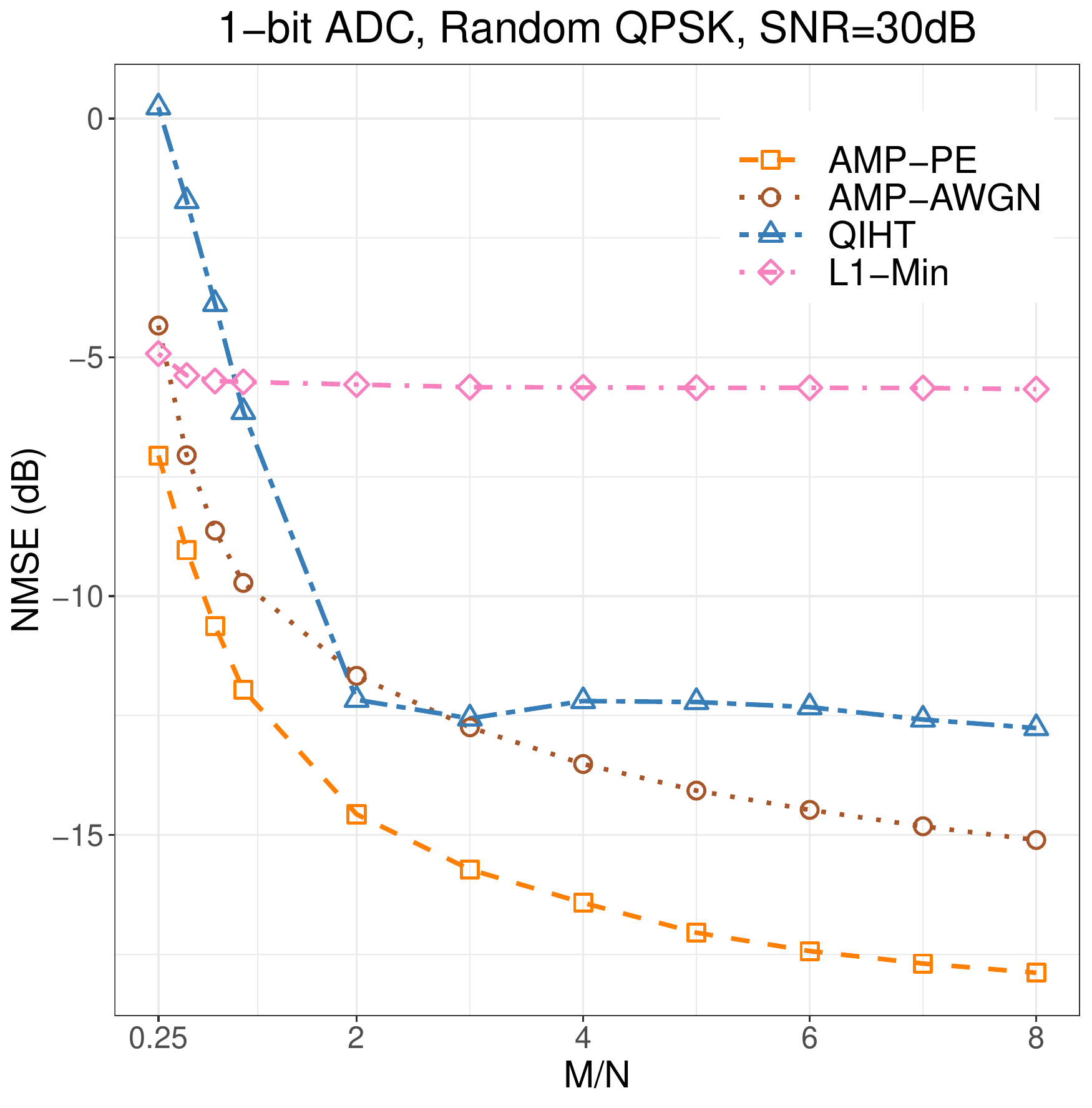}}
\subfigure{
\includegraphics[height=0.23\textwidth]{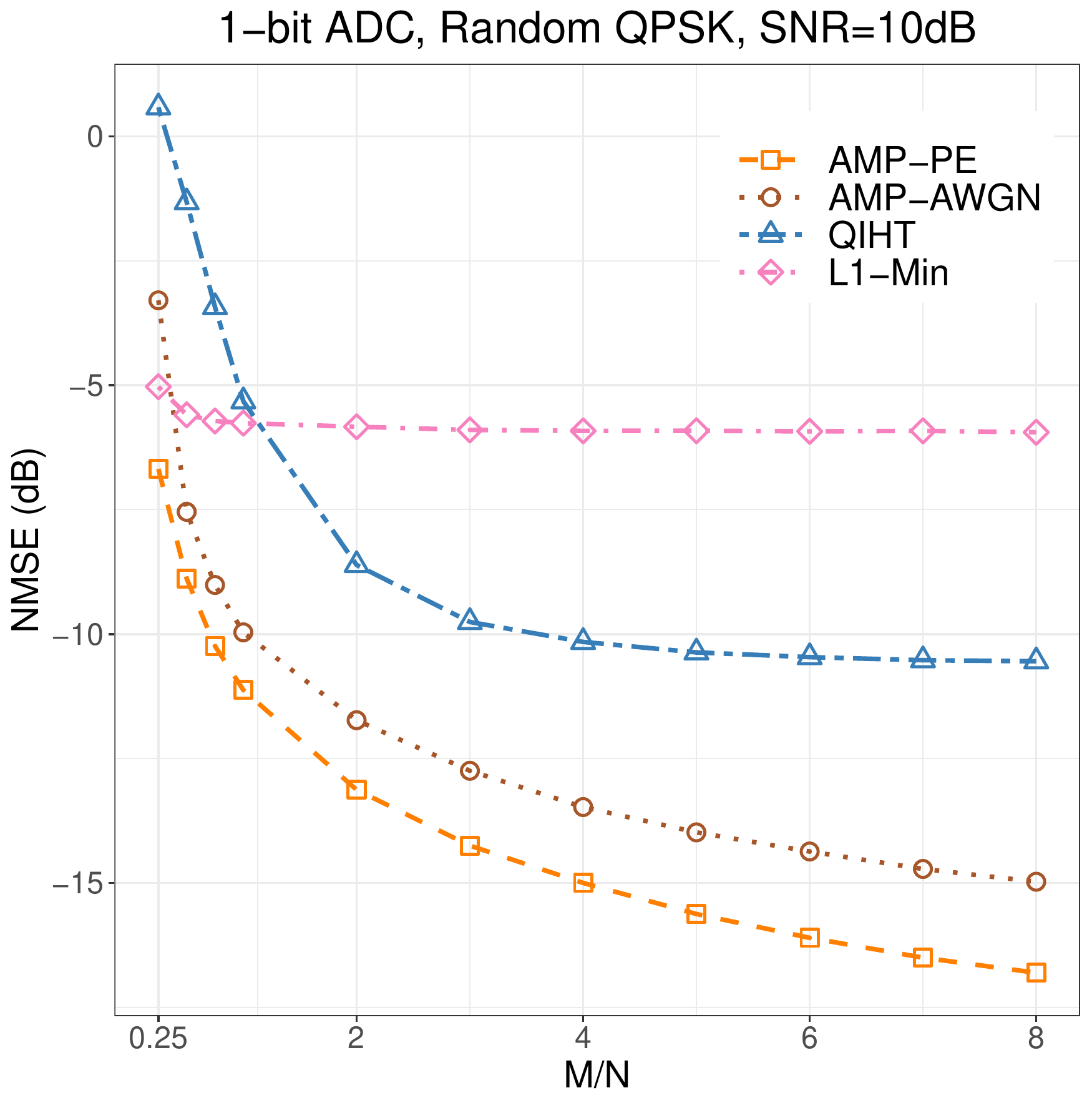}}
\subfigure{
\includegraphics[height=0.23\textwidth]{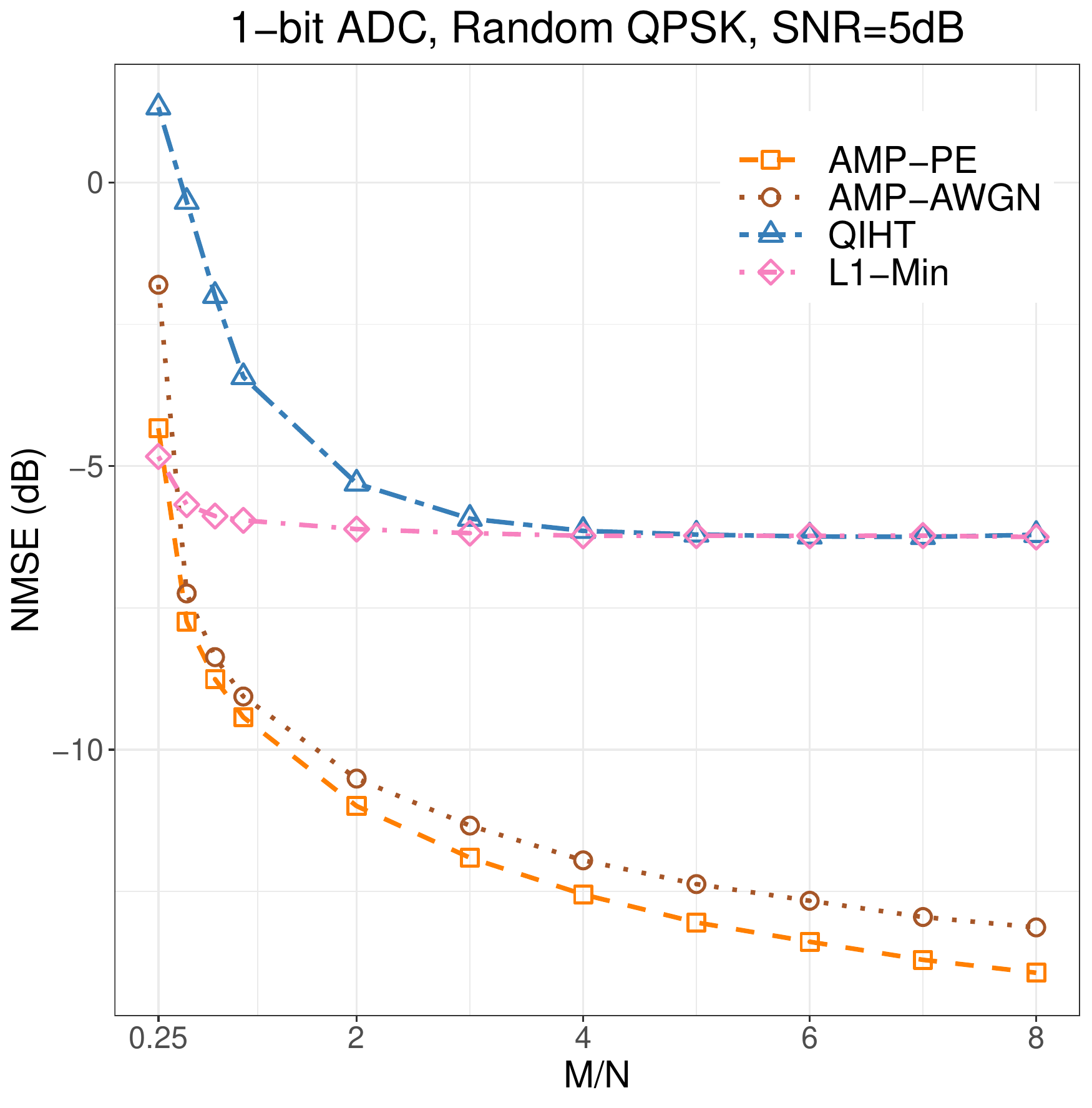}}\\
\subfigure{
\includegraphics[width=0.08\textwidth]{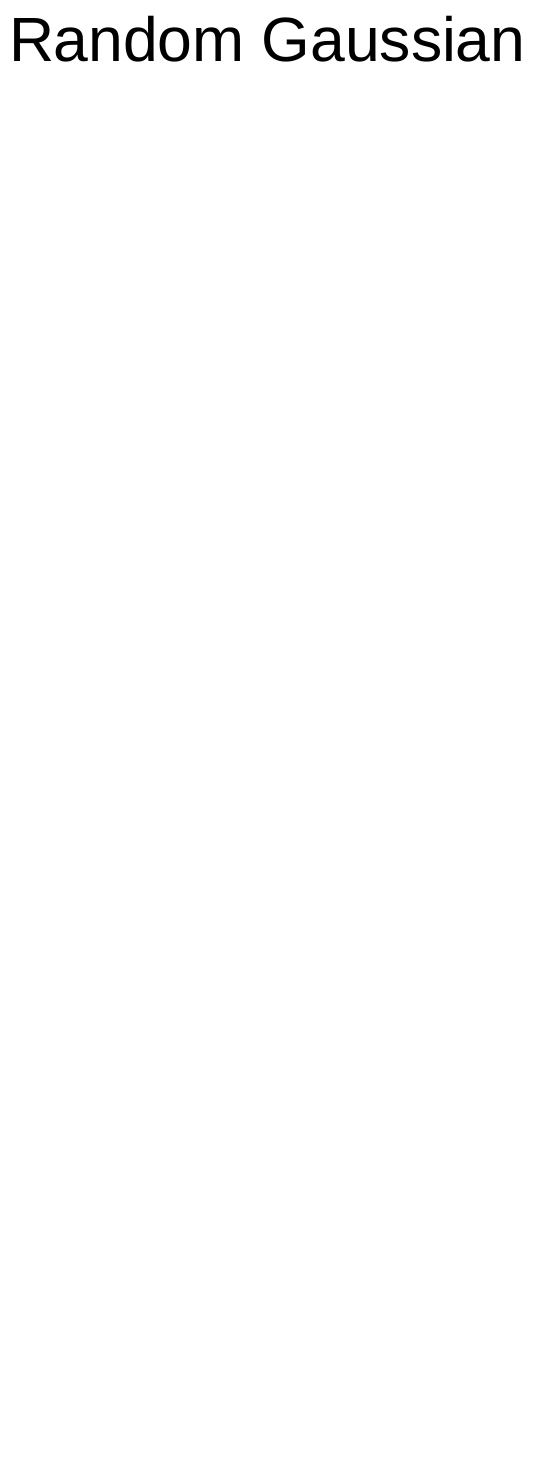}}
\subfigure{
\includegraphics[height=0.23\textwidth]{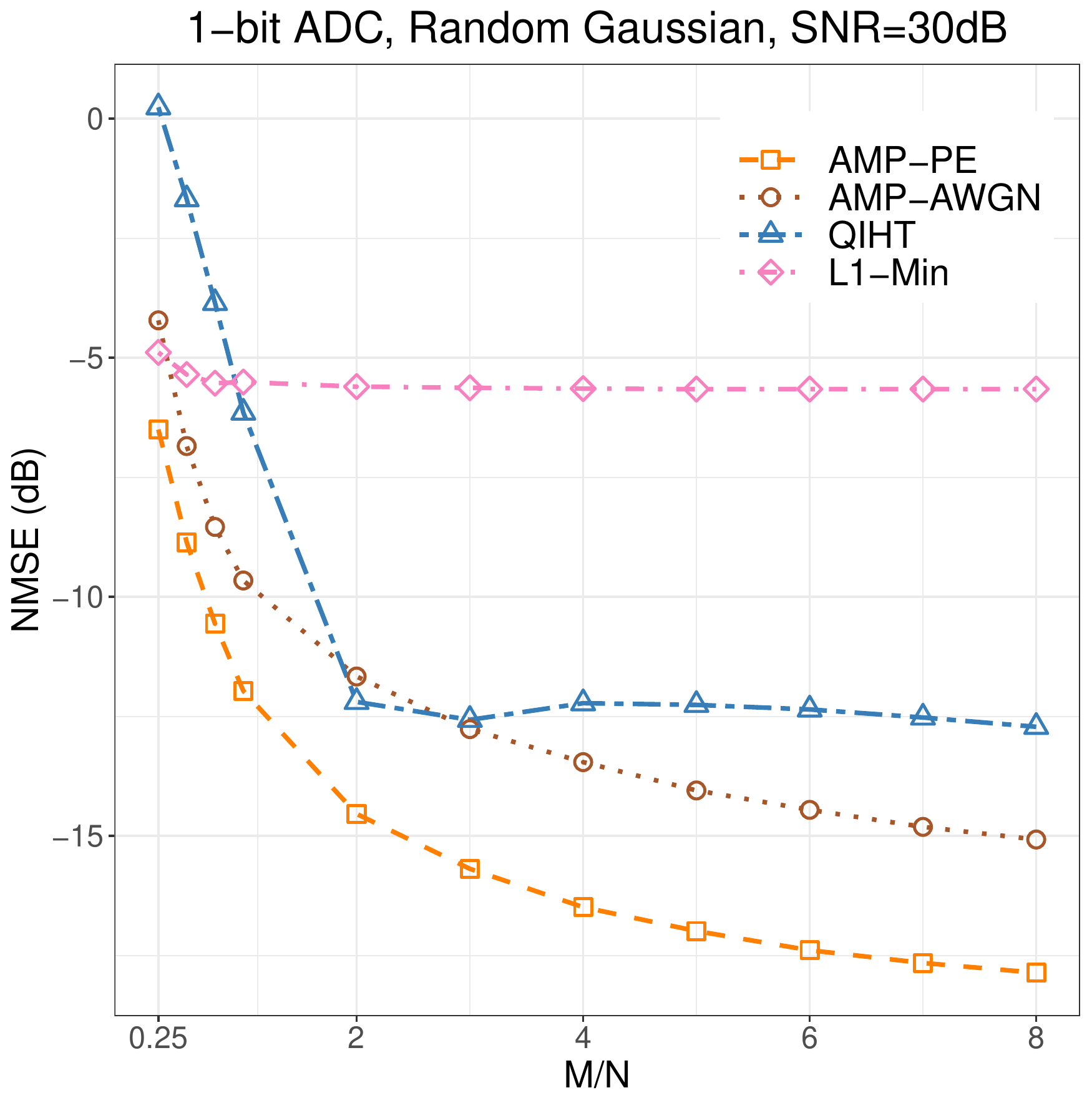}}
\subfigure{
\includegraphics[height=0.23\textwidth]{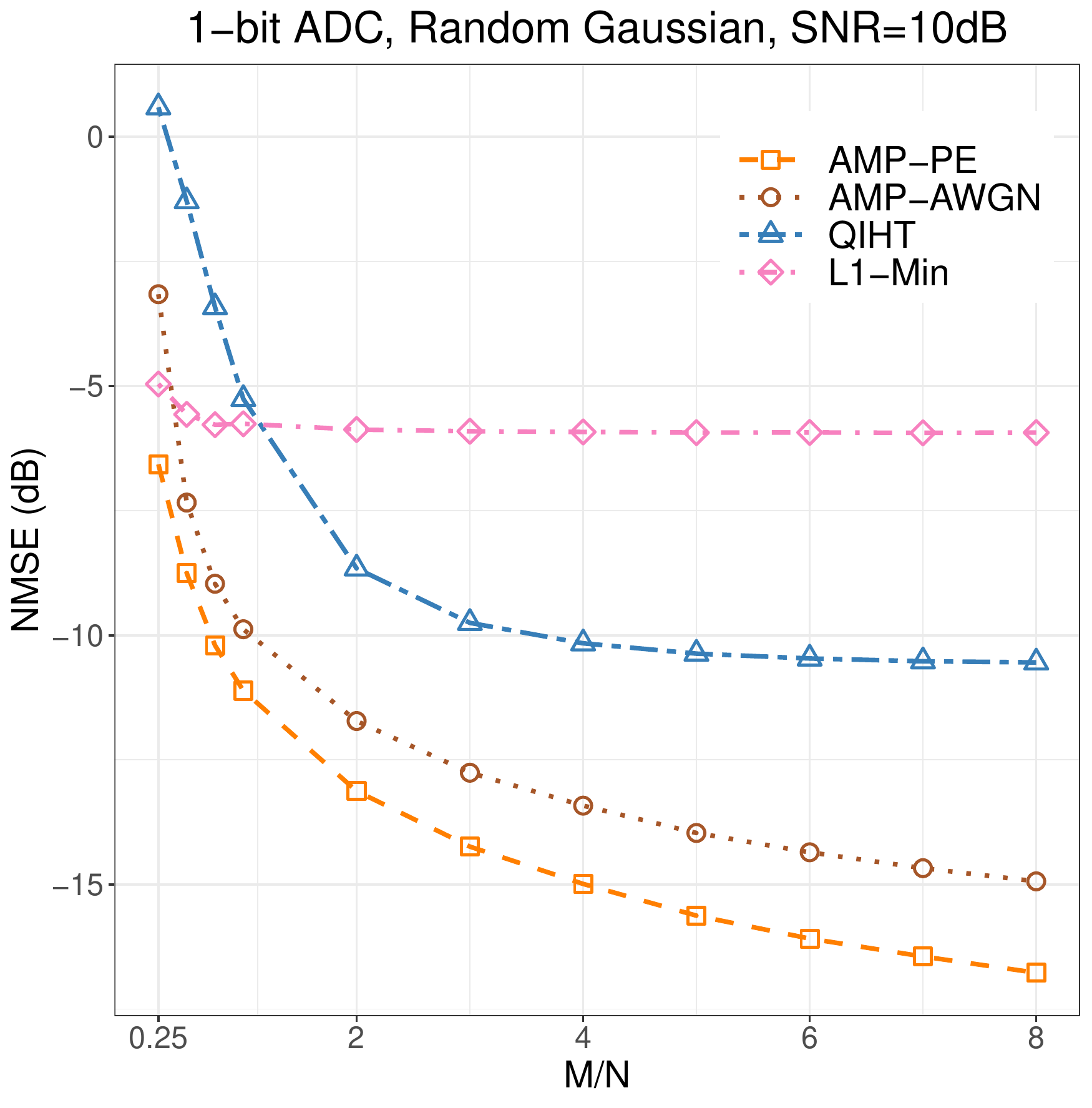}}
\subfigure{
\includegraphics[height=0.23\textwidth]{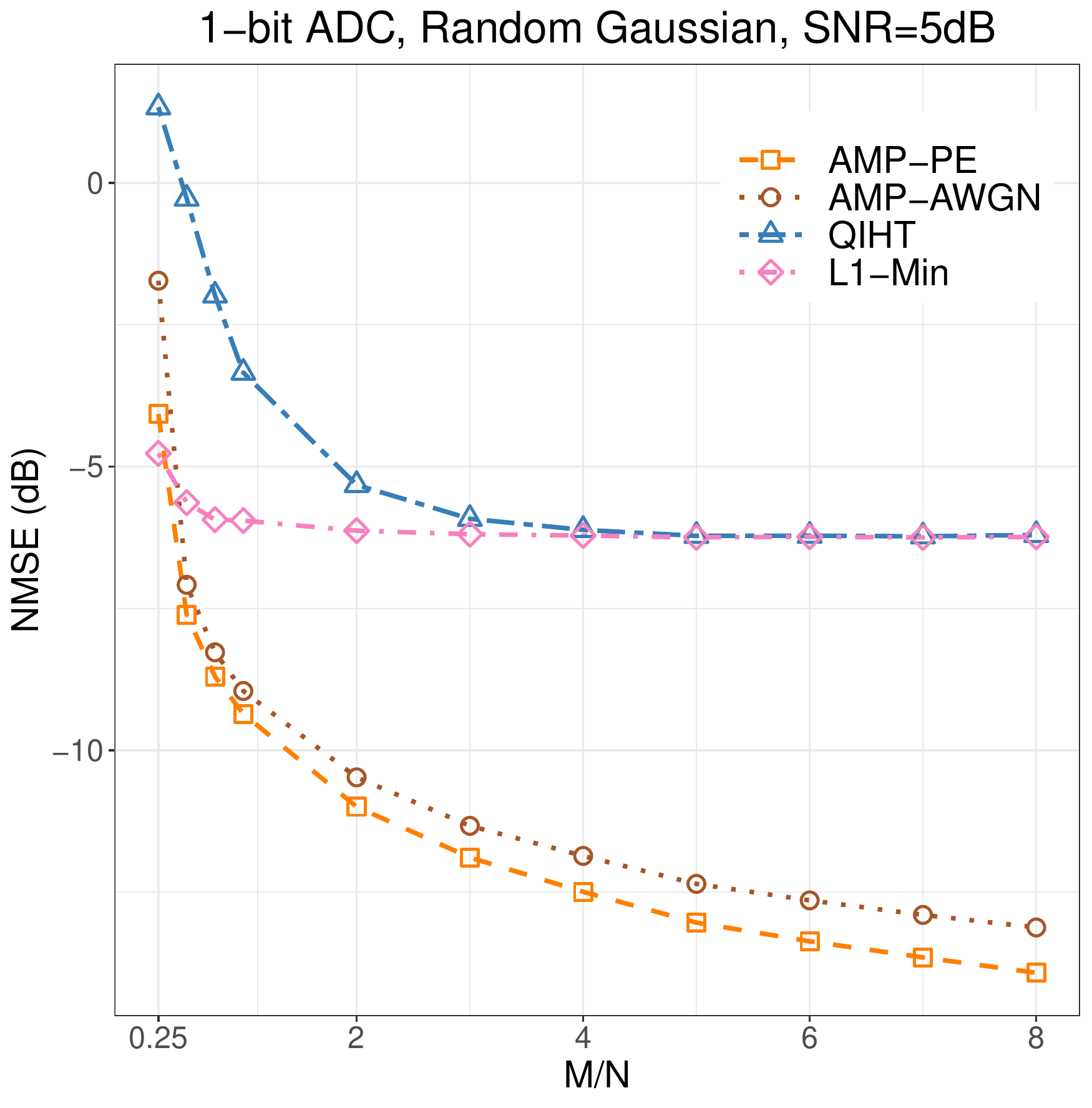}}
\caption{Comparison of different approaches in solving channel estimation from 1-bit measurements. The sampling ratio $\frac{M}{N}\in\{0.25,\cdots,8\}$. Two transmission sequences are used: random QPSK and random Gaussian. The pre-quantization SNR varies from $30$dB, $10$dB to $5$dB.}
\label{fig:1bit_ce_experiments}
\end{figure*}

The computational complexities of AMP-PE, AMP-AWGN, QIHT, CoSaMP and L1-Min are all $\mathcal{O}(MN)$, whereas the computational complexity of OMP is $\mathcal{O}(EMN)$. We compare the runtimes of different algorithms on a workstation equipped with ``Intel Xeon CPU E5-2650 (2.30GHz)'' and 128 GB RAM. One CPU core is reserved for testing. We investigate how the runtime scales with respect to the problem size $N$, aka the signal length. We fix the sparsity level $\frac{E}{N}=50\%$, the sampling ratio $\frac{M}{N}=5$, and set the pre-QNT SNR=30dB. We vary the problem size $N\in\{200,\cdots,10000\}$, and compute the average runtime across 100 random trials for each method. Due to the higher complexity of OMP, we only run the experiments with $N\leq 5000$ for OMP. 

The runtime comparisons are shown in Fig. \ref{fig:run_time}. When $N$ is relatively small ($N\leq 2000$), the complexities of AMP-PE and AMP-AWGN are dominated by the parameter estimation operations. Their runtimes are thus longer compared to the other methods that use pre-specified parameters. However, when $N$ is relatively large ($N\geq 3000$), the complexities of AMP-PE and AMP-AWGN are dominated by the linear and nonlinear updates in Algorithm \ref{alg:amp_pe}. Numerical experiments show that their runtimes are shorter than the other methods. Although the AWGN model adopted by AMP-AWGN leads to sub-optimal performances, it is simpler than the true quantization noise model adopted by AMP-PE. Hence, AMP-AWGN is always faster than AMP-PE.

\subsection{Channel Estimation}
Our final experiment is on estimating the massive MIMO channel introduced in Section \ref{subsec:MIMO_CE} from heavily quantized measurements \cite{MIMO_Bayesian:Huang:2021}. Specifically, there are $4$ clusters in the channel with a delay spread of $L=16$ symbol intervals. The transmitter has $N_t=64$ antennas, and the receiver has $N_r=64$ antennas. The dimensionality of the channel coefficient vector $\vx$ is then $N=N_tN_rL=65536$. Let $\mathscr{T}\in\mathbb{C}^{N_d}$ denote the training sequence used by the transmitter with its $k$-th entry $\mathscr{T}[k]=\frac{1}{\sqrt{2}}\mathscr{t}_{k1}+\frac{1}{\sqrt{2}}\mathscr{t}_{k2}\cdot\vi$. The following two transmission sequences are used
\begin{itemize}
    \item Random quadrature-phase-shift-keying (QPSK) \cite{Haykin:2009}: $\mathscr{t}_{k1}$ and $\mathscr{t}_{k2}$ follow i.i.d. Rademacher distribution.
    \item Random Gaussian: $\mathscr{t}_{k1}$ and $\mathscr{t}_{k_2}$ follow i.i.d. Gaussian distribution $\mathcal{N}(0,1)$.
\end{itemize}
The training symbols $\vo_d$ in \eqref{eq:channel_quantized_measurements} are transmitted at different times: $1\leq d\leq N_d$, where $N_d$ varies from $256$ to $8192$. Grouping the symbols together, we have an $N_t\times N_d$ matrix $\mO=[\vo_1\ \cdots\ \vo_{N_d}]$. The first row of $\mO$ is the training sequence $\mathscr{T}$, the $i$-th row of $\mO$ is obtained by circularly shifting $\mathscr{T}$ with step size $(i-1)L$. At the receiver, this produces different numbers of measurements $M=N_rN_d$, and the sampling ratio $\frac{M}{N}$ thus varies from $0.25$ to $8$. We add white Gaussian noise $\vw$ to the noiseless measurements $\vz$ so that the pre-QNT SNR of $\vz+\vw$ decreases from $30$dB, $10$dB to $5$dB, corresponding to the low-noise (SNR$>$22dB), moderate-noise (22dB$\geq$SNR$>$6dB) and high-noise regimes (SNR$\leq$6dB) as recommended in \cite{Zhang:TR:2011}.

Since the noiseless measurement $\vz$ is obtained through a linear operator $\mA(\cdot)$ in this case: $\vz=\mA(\vx)$, we thus focus on comparing the approaches that could work with linear operators directly, i.e. AMP-PE, AMP-AWGN, QIHT and L1-Min. Since we do not know the true distribution parameters, we could not compare them with AMP-oracle. For the two training sequences (random QPSK and random Gaussian), the resulting measurement matrices have zero-means. However, the entries in both matrices are not independent and identically distributed Gaussian variables. We thus need additional damping operations to achieve the best performances of AMP-PE and AMP-AWGN. Apart from the damping operations in \eqref{eq:damp_on_lambda} on the estimated input and output parameters, the damping operation is also applied on the recovered $\hat{\vx}^{(t+1)}$ after its computation in \eqref{eq:compute_x_n}:
\begin{align*}
    \hat{\vx}^{(t+1)} = \hat{\vx}^{(t)}+\eta\cdot\big(\hat{\vx}^{(t+1)}-\hat{\vx}^{(t)}\big),
\end{align*}
where the damping rate $\eta$ is set to $0.1$. The parameters of QIHT and L1-Min are tuned to achieve their best performances. The average NMSE across $100$ random trials is computed for each method. The recovery results from 1-bit measurements are shown in Fig. \ref{fig:1bit_ce_experiments}. 

For the two transmission sequences, we can see that the performances under random QPSK and random Gaussian sequences are generally similar. For the three noise regimes, we can see that AMP-PE performs significantly better than the other methods in the low-noise (pre-QNT SNR=$30$dB) and moderate-noise (pre-QNT SNR=$10$dB) regimes. The performances of AMP-PE and AMP-AWGN become similar in the high-noise regime (pre-QNT SNR=$5$dB), indicating that the AWGN model is a good approximation of the quantization noise model when the noise level is high. 

For the 2-bit and 3-bit recovery experiments, AMP-PE also achieves similar leading performances. The recovery results are given in the Supplemental Material.

\section{Conclusion}
\label{sec:con}
Taking a probabilistic perspective, we explore the 1-bit and multi-bit compressive sensing problems via the proposed AMP-PE framework where the signal and noise distribution parameters are treated as variables and jointly recovered. This leads to a much simpler method to estimate the parameters by maximizing their posteriors. It allows us to venture into the more complicated quantization noise area that is computationally prohibitive for previous AMP methods. An efficient approach that combines EM and the second-order method is then introduced to compute the maximizing parameters. For random Gaussian measurement matrices, the performance of AMP-PE can be accurately predicted through state evolution analysis. Extensive realistic experimental results show that AMP-PE generally performs much better than other state-of-the-art sparse recovery methods across a wide range of sparsity and noise levels. 

The stability (or reliability) of an algorithm is a key factor in deciding its adoption in practical applications. AMP has long been criticized on its lack of convergence guarantees for general measurement matrices. The damping and mean removal operations have been proposed to alleviate this issue, and are often shown to be quite effective. Initialization also plays an important role in both ensuring the convergence and recovering an accurate signal. In our experiments, we observe that the parameter initialization (rather than the variable initialization) correlates most closely to the algorithm's success. Previous works often overlooked the influence of this initialization issue, and they focused on establishing the convergence conditions for a specific class of random matrices. We believe that it would be worth pursuing how the initialization affects the algorithm's convergence behavior. Additionally, the state evolution of VAMP holds for a broader class of measurement matrices than AMP, and we would like to further investigate VAMP-based algorithm in our future work. Notwithstanding these drawbacks, our proposed framework offers an efficient joint recovery of the signal and parameters from a probabilistic perspective, and pushes forward the state-of-the-art performance level in 1-bit and multi-bit compressive sensing problems.

\begin{appendices}
\counterwithin{assumption}{section}
\counterwithin{theorem}{section}
\renewcommand\thetable{\thesection\arabic{table}}
\renewcommand\thefigure{\thesection\arabic{figure}}

\section{AMP-PE Update Equations for 1-Bit CS}
\label{app:sec:1_bit_cs}
\subsection{Nonlinear Updates for Sparse Signal Model}
The normalizing constant $\Psi(r_n,\boldsymbol\lambda)$ in \eqref{eq:normalization_psi} is
\begin{align*}
    \Psi(r_n,\boldsymbol\lambda)=(1-\kappa)\mathcal{N}(r_n|0,{\tau_r})+\sum_i\kappa\xi_i\cdot\mathcal{N}(r_n|\mu_i,{\gamma_x}_i+{\tau_r})\,.
\end{align*}
The posterior mean of $x_n$ in \eqref{eq:posterior_x_n_bgm} is
\begin{align*}
    \hat{x}_n=\frac{1}{\Psi(r_n,\boldsymbol\lambda)}\sum_i\kappa\xi_i\cdot\mathcal{N}(r_n|\mu_i,{\gamma_x}_i+{\tau_r})\frac{\mu_i{\tau_r}+r_n{\gamma_x}_i}{{\gamma_x}_i+{\tau_r}}.
\end{align*}
The posterior mean of $x_n^2$ in \eqref{eq:posterior_x_sq_n_bgm} is
\begin{align*}
\begin{split}
\mathbb{E}\left[x_n^2|r_n,{\tau_r},\boldsymbol\lambda\right] =& \frac{1}{\Psi(r_n,\boldsymbol\lambda)}\sum_i\kappa\xi_i\cdot\mathcal{N}(r_n|\mu_i,{\gamma_x}_i+{\tau_r})\\
&\times\left(\frac{{\gamma_x}_i{\tau_r}}{{\gamma_x}_i+{\tau_r}}+\left(\frac{\mu_i{\tau_r}+r_n{\gamma_x}_i}{{\gamma_x}_i+{\tau_r}}\right)^2\right)\,.
\end{split}
\end{align*}

\subsection{Nonlinear Updates for 1-Bit Quantization Noise Model}
To simplify the notations, we define the following terms
\begin{align}
\label{eq:q_bar}
&\overline{q}_m\coloneqq\frac{q_m}{\sqrt{{\tau_q}+\gamma_w}}\,,\\
\label{eq:h0_q_bar}
&h_0(q_m)\coloneqq\int_{-\infty}^0\mathcal{N}(u|q_m,{\tau_q}+\gamma_w)\ du=\frac{1}{2}\textrm{erfc}\left(\sqrt{\frac{1}{2}}\cdot\overline{q}_m\right)\,,
\end{align}
where $\textrm{erfc}(\cdot)$ is the complementary error function. The normalizing constant $\mathcal{U}_0(q_m,y_m,\boldsymbol\theta)$ in \eqref{eq:normalization_U_1bit} is then
\begin{align*}
\mathcal{U}_0(q_m,y_m,\boldsymbol\theta)=\big(1-h_0(q_m)\big)\delta(y_m-1)+ h_0(q_m)\delta(y_m+1)\,.
\end{align*}
We further define the following terms
\begin{align}
    \begin{split}
        h_1(q_m)\coloneqq&\int z_m 
        \int_{-\infty}^0\mathcal{N}(u|z_m,\gamma_w)\ du\cdot\mathcal{N}(z_m|q_m,{\tau_q})\ dz_m\\
        =&q_m\cdot h_0(q_m)-{\tau_q}\cdot\mathcal{N}(q_m|0,{\tau_q}+\gamma_w),
    \end{split}
\end{align}

\begin{align}
    \begin{split}
        h_2(q_m)\coloneqq&\int z_m^2\int_{-\infty}^0\mathcal{N}(u|z_m,\gamma_w)\ du\cdot\mathcal{N}(z_m|q_m,{\tau_q})\ dz_m\\
        =&(q_m^2+{\tau_q})\cdot h_0(q_m)\\
        &-q_m\cdot\frac{{\tau_q}^2+2{\tau_q}\gamma_w}{{\tau_q}+\gamma_w}\cdot\mathcal{N}(q_m|0,{\tau_q}+\gamma_w),
    \end{split}
\end{align}

\begin{align}
    \mathcal{U}_1(q_m,y_m,\boldsymbol\theta)\coloneqq&\big(q_m-h_1(q_m)\big)\delta(y_m-1)+ h_1(q_m)\delta(y_m+1),\\
    \begin{split}
    \mathcal{U}_2(q_m,y_m,\boldsymbol\theta)\coloneqq&\big(q_m^2+{\tau_q}-h_2(q_m)\big)\delta(y_m-1)\\
    &+ h_2(q_m)\delta(y_m+1)\,.
    \end{split}
\end{align}
The posterior mean of $z_m$ in \eqref{eq:posterior_mean_z_m_1bit} is 
\begin{align*}
    \mathbb{E}\left[z_m\left|q_m, {\tau_q}, y_m, \boldsymbol\theta\right.\right]=\frac{1}{\mathcal{U}_0}\mathcal{U}_1\,.
\end{align*}
The posterior mean of $z_m^2$ in \eqref{eq:posterior_mean_z_m_sq_1bit} is
\begin{align*}
    \mathbb{E}\left[z_m^2\left|q_m, {\tau_q}, y_m, \boldsymbol\theta\right.\right]=\frac{1}{\mathcal{U}_0}\mathcal{U}_2\,.
\end{align*}

\subsection{Parameter Estimation for Sparse Signal Model}
\label{app:sec:1_bit_cs:C}
The objective function $f(\boldsymbol\lambda)$ from the expectation step is
\begin{align}
\label{eq:signal_prior_pe_obj}
\begin{split}
    &f(\boldsymbol\lambda)=\sum_n\psi_0(r_n)\cdot\log\left[(1-\kappa)\cdot\mathcal{N}(r_n|0,{\tau_r})\right]\\
    &+\sum_n\sum_i\psi_i(r_n)\cdot\log\left[\kappa\xi_i\cdot\mathcal{N}(r_n|\mu_i,{\gamma_x}_i+{\tau_r})\right]\,,
\end{split}
\end{align}
where $\psi_0(r_n),\psi_i(r_n)$ are posteriors of the latent variable $c(r_n)$:
\begin{align*}
    \psi_0(r_n)&=\frac{1}{\Psi_e(r_n)}\big(1-\kappa^{(e)}\big)\cdot\mathcal{N}(r_n|0,{\tau_r}),\\
    \psi_i(r_n)&=\frac{1}{\Psi_e(r_n)}\kappa^{(e)}\xi_i^{(e)}\cdot\mathcal{N}\big(r_n|\mu_i^{(e)},{\gamma_x}_i^{(e)}+{\tau_r}\big),\\
    \begin{split}
    \Psi_e(r_n)&=\big(1-\kappa^{(e)}\big)\cdot\mathcal{N}(r_n|0,{\tau_r})\\
    &\quad+\sum_i\kappa^{(e)}\xi_i^{(e)}\cdot\mathcal{N}\big(r_n|\mu_i^{(e)},{\gamma_x}_i^{(e)}+{\tau_r}\big)\,.
    \end{split}
\end{align*}
Note that the first Gaussian component is zero-mean, i.e. $\mu_1=0$. The mixture weights $\kappa,\xi_i$, the Gaussian mixture mean $\mu_i$ and variance ${\gamma_x}_i$ that maximize $f(\boldsymbol\lambda)$ have the following closed-form update equations:
\begin{align*}
    \kappa^{(e+1)} &= \frac{\sum_n\sum_i\psi_i(r_n)}{\sum_n\psi_0(r_n)+\sum_n\sum_i\psi_i(r_n)},\\
    \xi_i^{(e+1)} &= \frac{\sum_n\psi_i(r_n)}{\sum_n\sum_i\psi_i(r_n)},\\
    \mu_i^{(e+1)} &= \frac{\sum_n\psi_i(r_n)\cdot\frac{r_n}{{\gamma_x}_i^{(e)}+{\tau_r}}}{\sum_n\psi_i(r_n)\cdot\frac{1}{{\gamma_x}_i^{(e)}+{\tau_r}}}\quad \textnormal{for }i\geq 2,\\
    {\gamma_x}_i^{(e+1)} &= \frac{\sum_n\psi_i(r_n)\cdot\left(r_n-\mu_i^{(e)}\right)^2}{\sum_n\psi_i(r_n)}-\tau_r\,.
\end{align*}

\subsection{Parameter Estimation for 1-Bit Quantization Noise Model}
We compute the first and second order derivatives of $h_0(q_m)$ in \eqref{eq:h0_q_bar} with respect to $\gamma_w$ as follows:
\begin{align*}
    \frac{\partial h_0}{\partial\gamma_w}=&\frac{\overline{q}_m}{2\sqrt{2\pi}(\gamma_w+{\tau_q})}\cdot\exp\left(-\frac{\overline{q}_m^2}{2}\right),\\
    \frac{\partial^2h_0}{\partial\gamma_w^2}=&\frac{\overline{q}_m^3-3\overline{q}_m}{4\sqrt{2\pi}(\gamma_w+{\tau_q})^2}\cdot\exp\left(-\frac{\overline{q}_m^2}{2}\right)\,.
\end{align*}

We also maximize the second order approximation of $g_1(\boldsymbol\theta)$ in \eqref{eq:2nd_order_g1_theta}. The first and second order derivatives of $g_1(\boldsymbol\theta)$ with respect to $\gamma_w$ are 
\begin{align*}
    g_1^\prime(\gamma_w) &= \sum_m\frac{1}{\mathcal{U}_0}\cdot\frac{\partial\mathcal{U}_0}{\partial\gamma_w},\\
    \begin{split}
    g_1^{\prime\prime}(\gamma_w) &= \sum_m-\frac{1}{\left(\mathcal{U}_0\right)^2}\left[\frac{\partial\mathcal{U}_0}{\partial\gamma_w}\right]^2+\frac{1}{\mathcal{U}_0}\cdot\frac{\partial^2\mathcal{U}_0}{\partial\gamma_w^2}\,,
    \end{split}
\end{align*}
where the first and second order derivatives of $\mathcal{U}_0$ with respect to $\gamma_w$ are
\begin{align*}
    \frac{\partial\mathcal{U}_0}{\partial\gamma_w}&=-\frac{\partial h_0}{\partial\gamma_w}\cdot\delta(y_m-1)+\frac{\partial h_0}{\partial\gamma_w}\cdot\delta(y_m+1),\\
    \frac{\partial^2\mathcal{U}_0}{\partial\gamma_w^2}&=-\frac{\partial^2 h_0}{\partial\gamma_w^2}\cdot\delta(y_m-1)+\frac{\partial^2 h_0}{\partial\gamma_w^2}\cdot\delta(y_m+1)\,.
\end{align*}

\subsection{Comparison with Previous EM-based Approaches}
\label{app:sec:1_bit_cs:E}
For the 1-bit quantization noise model, we derive detailed expression of the objective function in \eqref{eq:previous_em_noise} used by previous EM-based approaches for parameter estimation. We first define the following terms
\begin{align}
    &\widetilde{z}_m\coloneqq\frac{z_m}{\sqrt{{\gamma_w}}}\,,\\
    &\widetilde{h}_0(z_m)\coloneqq\int_{-\infty}^0\mathcal{N}(u|z_m,{\gamma_w})\ du=\frac{1}{2}\textnormal{erfc}\left(\sqrt{\frac{1}{2}}\cdot\widetilde{z}_m\right)\,,\\
    &\widetilde{\mathcal{U}}_0(z_m,y_m,\boldsymbol\theta)\coloneqq\big(1-\widetilde{h}_0(z_m)\big)\delta(y_m-1)+ \widetilde{h}_0(z_m)\delta(y_m+1)\,.
\end{align}
Since $p(y_m,z_m|\boldsymbol\theta)=p(y_m|z_m,\boldsymbol\theta)\cdot\mathcal{N}(z_m|q_m,\tau_q)$, the objective function in \eqref{eq:previous_em_noise} can also be changed to
\begin{align*}
\label{eq:obj_previous_em_noise}
\begin{split}
    &\sum_m\int p\big(z_m|y_m,\boldsymbol\theta^{(t)}\big)\log p(y_m|z_m,\boldsymbol\theta)\ dz_m\\
    &=\sum_m\frac{1}{\mathcal{U}_0\big(q_m,y_m,\boldsymbol\theta^{(t)}\big)}\\
    &\quad\times\int p\Big(y_m|z_m,\boldsymbol\theta^{(t)}\Big)\cdot \mathcal{N}(z_m|q_m,{\tau_q})\cdot\log\widetilde{\mathcal{U}}_0(z_m,y_m,\boldsymbol\theta)\ dz_m\,.
\end{split}
\end{align*}
It is much more complicated than the objective function of the AMP-PE approach, which is simply $\sum_m\log[\mathcal{U}_0(q_m,y_m,\boldsymbol\theta)]$ in \eqref{eq:g1_theta}.

\section{AMP-PE Update Equations for Multi-Bit CS}
\label{app:sec:multi_bit_cs}

\subsection{Nonlinear Updates for Multi-Bit Quantization Noise Model}
Let $\overline{a}_i$, $\overline{a}_{i-1}$ be defined as follows
\begin{align}
    \overline{a}_{i}&\coloneqq\frac{a_{i}}{\sqrt{\gamma_w+{\tau_q}}}\,,\\
    \overline{a}_{i-1}&\coloneqq\frac{a_{i-1}}{\sqrt{\gamma_w+{\tau_q}}}\,.
\end{align}
The normalizing constant $\mathcal{V}_0(q_m,y_m,\boldsymbol\theta)$ in \eqref{eq:normalization_V_multi_bit} is 
\begin{align*}
    \mathcal{V}_0(q_m,y_m,\boldsymbol\theta)=\frac{1}{2}\left(\textrm{erf}\left(\sqrt{\frac{1}{2}}\left(\overline{a}_i-\overline{q}_m\right)\right)-\textrm{erf}\left(\sqrt{\frac{1}{2}}\left(\overline{a}_{i-1}-\overline{q}_m\right)\right)\right)\,,
\end{align*}
where $\textrm{erf}(\cdot)$ is the error function.

We further define the following terms:
\begin{align}
\begin{split}
    &\mathcal{V}_1(\overline{q}_m,y_m,\boldsymbol\theta)\coloneqq\int z_m\cdot p(y_m|z_m,\boldsymbol\theta)\mathcal{N}(z_m|q_m,{\tau_q})\ dz_m\\
    &=q_m\mathcal{V}_0(q_m,y_m,\boldsymbol\theta)-{\tau_q}\mathcal{N}(a_i|q_m,{\tau_q}+\gamma_w)\\
    &\quad+{\tau_q}\mathcal{N}(a_{i-1}|q_m,{\tau_q}+\gamma_w)\,,
\end{split}\\
\begin{split}
    &\mathcal{V}_2(q_m,y_m,\boldsymbol\theta)\coloneqq\int z_m^2\cdot p(y_m|z_m,\boldsymbol\theta)\mathcal{N}(z_m|q_m,{\tau_q})\ dz_m\\
    &=\big(\tau_q+q_m^2\big)\cdot\mathcal{V}_0(q_m,y_m,\boldsymbol\theta)\\
    &\quad -\left(\frac{{\tau_q}^2(a_i-q_m)}{{\tau_q}+\gamma_w}+2{\tau_q}q_m\right)\mathcal{N}(a_i|q_m,{\tau_q}+\gamma_w)\\
    &\quad +\left(\frac{{\tau_q}^2(a_{i-1}-q_m)}{{\tau_q}+\gamma_w}+2{\tau_q}q_m\right)\mathcal{N}(a_{i-1}|q_m,{\tau_q}+\gamma_w)\,.
\end{split}
\end{align}
The posterior mean of $z_m$ in \eqref{eq:posterior_mean_z_m_multi_bit} is
\begin{align*}
\begin{split}
    \mathbb{E}\left[z_m\left|q_m, {\tau_q}, y_m, \boldsymbol\theta\right.\right]=\frac{1}{\mathcal{V}_0}\mathcal{V}_1\,.
\end{split}
\end{align*}
The posterior expectation of $z_m^2$ in \eqref{eq:posterior_mean_z_m_sq_multi_bit} is 
\begin{align*}
\begin{split}
    \mathbb{E}\left[z_m^2\left|q_m, {\tau_q}, y_m, \boldsymbol\theta\right.\right]=\frac{1}{\mathcal{V}_0}\mathcal{V}_2\,.
\end{split}
\end{align*}

\subsection{Parameter Estimation for Multi-Bit Quantization Noise Model}
We maximize the second order approximation of $g_2(\boldsymbol\theta)$ in \eqref{eq:2nd_order_g2_theta}. The first and second order derivatives of $g_2(\boldsymbol\theta)$ with respect to $\gamma_w$ are 
\begin{align*}
    g_2^\prime(\gamma_w) &= \sum_m\frac{1}{\mathcal{V}_0}\cdot\frac{\partial\mathcal{V}_0}{\partial\gamma_w},\\
    \begin{split}
    g_2^{\prime\prime}(\gamma_w) &= \sum_m-\frac{1}{\left(\mathcal{V}_0\right)^2}\left[\frac{\partial\mathcal{V}_0}{\partial\gamma_w}\right]^2+\frac{1}{\mathcal{V}_0}\cdot\frac{\partial^2\mathcal{V}_0}{\partial\gamma_w^2}\,,
    \end{split}
\end{align*}
where the first and second order derivatives of $\mathcal{V}_0$ with respect to $\gamma_w$ are
\begin{align*}
\begin{split}
    \frac{\partial\mathcal{V}_0}{\partial \gamma_w}&=-\frac{a_i-q_m}{2({\tau_q}+\gamma_w)}\mathcal{N}(a_i|q_m,{\tau_q}+\gamma_w)\\
    &\quad+\frac{a_{i-1}-q_m}{2({\tau_q}+\gamma_w)}\mathcal{N}(a_{i-1}|q_m,{\tau_q}+\gamma_w),
\end{split}\\
\begin{split}
    \frac{\partial^2\mathcal{V}_0}{\partial \gamma_w^2}&=\frac{3(a_i-q_m)}{4({\tau_q}+\gamma_w)^2}\mathcal{N}(a_i|q_m,{\tau_q}+\gamma_w)\\
    &\quad-\frac{(a_i-q_m)^3}{4({\tau_q}+\gamma_w)^3}\mathcal{N}(a_i|q_m,{\tau_q}+\gamma_w)\\
    &\quad-\frac{3(a_{i-1}-q_m)}{4({\tau_q}+\gamma_w)^2}\mathcal{N}(a_{i-1}|q_m,{\tau_q}+\gamma_w)\\
    &\quad+\frac{(a_{i-1}-q_m)^3}{4({\tau_q}+\gamma_w)^3}\mathcal{N}(a_{i-1}|q_m,{\tau_q}+\gamma_w)\,.
\end{split}
\end{align*}

\end{appendices}


\ifCLASSOPTIONcaptionsoff
  \newpage
\fi



%\bibliographystyle{IEEEtran}
\bibliographystyle{IEEEbib}
\bibliography{refs}

\newpage
\onecolumn

\setcounter{section}{0}
\begin{center}
{\LARGE Supplementary Material}
\end{center}

\vspace{5em}

The Supplemental Material contains additional derivations of the formulas and experimental results for the paper ``\emph{Approximate Message Passing with Parameter Estimation for Heavily Quantized Measurements}''.
\newpage

\section{Derivations of the Objective Functions for Parameter Estimation}

\subsection{Parameter Estimation in Signal Prior}
We first derive (22) that is used to find the maximizing parameters of the signal prior.

As discussed earlier, the distribution $\exp\left(\sum_m\Delta^{(t+1)}_{\Phi_m\rightarrow x_n}\right)$ in (1a) is approximated by a Gaussian distribution $\mathcal{N}(x_n|r_n,{\tau_r})$ in GAMP, where $r_n$ is a ``dummy'' variable. We then have
\begin{align}
\label{eq:one}
    \exp\left(\sum_m\Delta^{(t+1)}_{\Phi_m\rightarrow x_n}\right)\approx\mathcal{N}(x_n|r_n,{\tau_r})\,.
\end{align}

We need the following messages to derive the posterior distribution $p(\lambda_l|\vy)$
\begin{align}
    \label{eq:two}
    \Delta^{(t+1)}_{x_n\rightarrow \Omega_n}=\ &\sum_i\Delta^{(t+1)}_{\Phi_i\rightarrow x_n},\\
    \label{eq:three}
    \begin{split}
    \Delta^{(t+1)}_{\Omega_n\rightarrow \lambda_l}=\ &C+\log\int\left[\Omega_n\left(x_n,\lambda_l,\hat{\boldsymbol\lambda}^{(t)}\backslash\hat{\lambda}_l^{(t)}\right)\times\exp\left(\Delta^{(t+1)}_{x_n\rightarrow \Omega_n}\right)\right]\ dx_n\,.
    \end{split}
\end{align}
Combing \eqref{eq:one},\eqref{eq:two},\eqref{eq:three}, we have
\begin{align}
    \label{eq:four}
    \Delta^{(t+1)}_{\Omega_n\rightarrow \lambda_l}=\ &C+\log\int\left[\Omega_n\left(x_n,\lambda_l,\hat{\boldsymbol\lambda}^{(t)}\backslash\hat{\lambda}_l^{(t)}\right)\times\mathcal{N}(x_n|r_n,{\tau_r})\right]\ dx_n\,.
\end{align}
Since the BGM signal prior $\Omega_n(x_n,\boldsymbol\lambda)$ is
\begin{align}
    \label{eq:five}
    \Omega_n(x_n,\boldsymbol\lambda)=&p(x_n|\boldsymbol\lambda)= (1-\kappa)\cdot\delta(x_n)+\kappa\cdot\sum_{i=1}^I\xi_i\cdot\mathcal{N}(x_n|\mu_i,{\gamma_x}_i)\,.
\end{align}
Plugging \eqref{eq:five} into \eqref{eq:four} and dropping the superscript to simplify the notations, we can get 
\begin{align}
    \Delta_{\Omega_n\rightarrow \lambda_l} = C+ \log\left[(1-\kappa)\cdot\mathcal{N}(r_n|0,{\tau_r})+\sum_i\kappa\xi_i\cdot\mathcal{N}(r_n|\mu_i,{\gamma_x}_i+{\tau_r})\right].
\end{align}
The maximizing parameter $\boldsymbol\lambda$ can then be computed as
\begin{align}
\begin{split}
    \hat{\boldsymbol\lambda} &= \arg\max_{\boldsymbol\lambda} p(\boldsymbol\lambda|\vy)=\arg\max_{\boldsymbol\lambda}\sum_n\Delta_{\Omega_n\rightarrow \lambda_l}\\
    &=\arg\max_{\boldsymbol\lambda}\sum_n\log\left[(1-\kappa)\cdot\mathcal{N}(r_n|0,{\tau_r})+\sum_i\kappa\xi_i\cdot\mathcal{N}(r_n|\mu_i,{\gamma_x}_i+{\tau_r})\right]\,.
\end{split}
\end{align}

\subsection{Parameter Estimation in Noise Prior}
We next derive (24) that is used to find the maximizing parameters of the noise prior. 

The following change of variable is performed in GAMP formulation:
\begin{subequations}
\begin{align}
    \label{eq:y_m_x}
    p(y_m|\boldsymbol\theta)&=\int p(y_m|\vx,\boldsymbol\theta)p(\vx)\ d\vx\\
    \label{eq:y_m_z}
    &=\int p(y_m|z_m,\boldsymbol\theta)p(z_m)\ dz_m\,.
\end{align}
\end{subequations}
We first have
\begin{align}
    p(y_m|\vx,\boldsymbol\theta)&=\Phi_m(y_m,\vx,\boldsymbol\theta)\,,\\
    p(x_j)&\propto\exp\Big(\Delta^{(t+1)}_{x_j\rightarrow\Phi_m}\Big)\,.
\end{align}
Using \eqref{eq:y_m_x}, we can get
\begin{align}
\label{eq:phi_theta_1}
    \begin{split}
    \Delta^{(t+1)}_{\Phi_m\rightarrow\theta_k}%&=\log p\Big(\theta_k|y_m,\hat{\boldsymbol\theta}^{(t)}\backslash\hat{\theta}_k^{(t)}\Big)\\
    &=C+\log p\Big(y_m|\theta_k, \hat{\boldsymbol\theta}^{(t)}\backslash\hat{\theta}_k^{(t)}\Big)\\
    &=C+\log \int p\Big(y_m|\vx,\theta_k, \hat{\boldsymbol\theta}^{(t)}\backslash\hat{\theta}_k^{(t)}\Big)p(\vx)\ d\vx\\
    &=C+\log\int\Big[\Phi_m\Big(y_m, \vx, \theta_k, \hat{\boldsymbol\theta}^{(t)}\backslash\hat{\theta}_k^{(t)}\Big)\times\exp\Big(\sum_j\Delta^{(t+1)}_{x_j\rightarrow\Phi_m}\Big)\Big]\ d\vx\,,
    \end{split}
\end{align}
where $C$ is some normalization constant. In the GAMP formulation, the prior distribution of $z_m$ is approximated by a Gaussian
\begin{align}
    p(z_m)\approx \mathcal{N}\left(z_m|q_m,{\tau_q}\right)\,,
\end{align}
where $q_m$ is another ``dummy'' variable with variance $\tau_q$.
Using \eqref{eq:y_m_z}, we can get
\begin{align}
\label{eq:phi_theta_2}
    \begin{split}
    \Delta^{(t+1)}_{\Phi_m\rightarrow\theta_k}&=C+\log p\Big(y_m|\theta_k, \hat{\boldsymbol\theta}^{(t)}\backslash\hat{\theta}_k^{(t)}\Big)\\
    &=C+\log\int p\Big(y_m|z_m,\theta_k, \hat{\boldsymbol\theta}^{(t)}\backslash\hat{\theta}_k^{(t)}\Big)p(z_m)\ dz_m\\
    &=C+\log\int p\Big(y_m|z_m,\theta_k, \hat{\boldsymbol\theta}^{(t)}\backslash\hat{\theta}_k^{(t)}\Big)\cdot\mathcal{N}\left(z_m|q_m,{\tau_q}\right)\ dz_m\\
    &=C+\log\mathcal{U}_0\left(q_m,y_m,\theta_k,\hat{\boldsymbol\theta}^{(t)}\backslash\hat{\theta}_k^{(t)}\right)\,.
    \end{split}
\end{align}
Using \eqref{eq:phi_theta_2} to compute $\Delta^{(t+1)}_{\Phi_m\rightarrow\theta_k}$ is easier than using \eqref{eq:phi_theta_1}. We thus use \eqref{eq:phi_theta_2} to calculate the posterior $p(\boldsymbol\theta|\vy)$. 

Dropping the superscript to simplify the notations, we can compute the maximizing parameters $\boldsymbol\theta$ as follows
\begin{align}
    \hat{\boldsymbol\theta} = \arg\max_{\boldsymbol\theta}\ p(\boldsymbol\theta|\vy)=\arg\max_{\boldsymbol\theta}\ \sum_m\Delta_{\Phi_m\rightarrow\theta_k}=\arg\max_{\boldsymbol\theta}\ \sum_m\log\mathcal{U}_0(q_m,y_m,\boldsymbol\theta)\,.
\end{align}

\newpage

\section{State Evolution Analysis}

\begin{figure*}[htbp]
\centering
\subfigure{
\label{fig:se_10_0}
\includegraphics[width=0.3\textwidth]{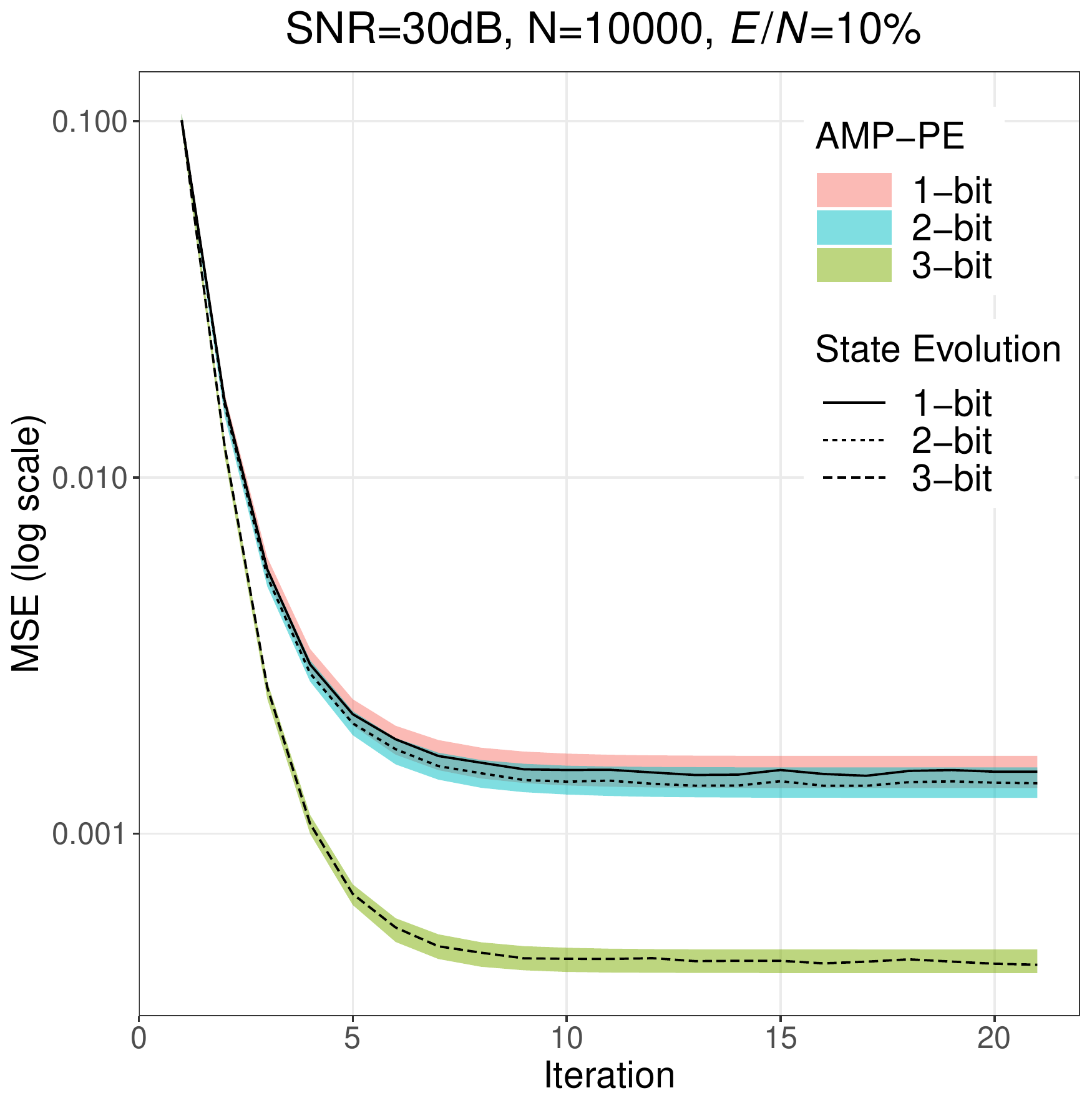}}
\subfigure{
\label{fig:se_10_10}
\includegraphics[width=0.3\textwidth]{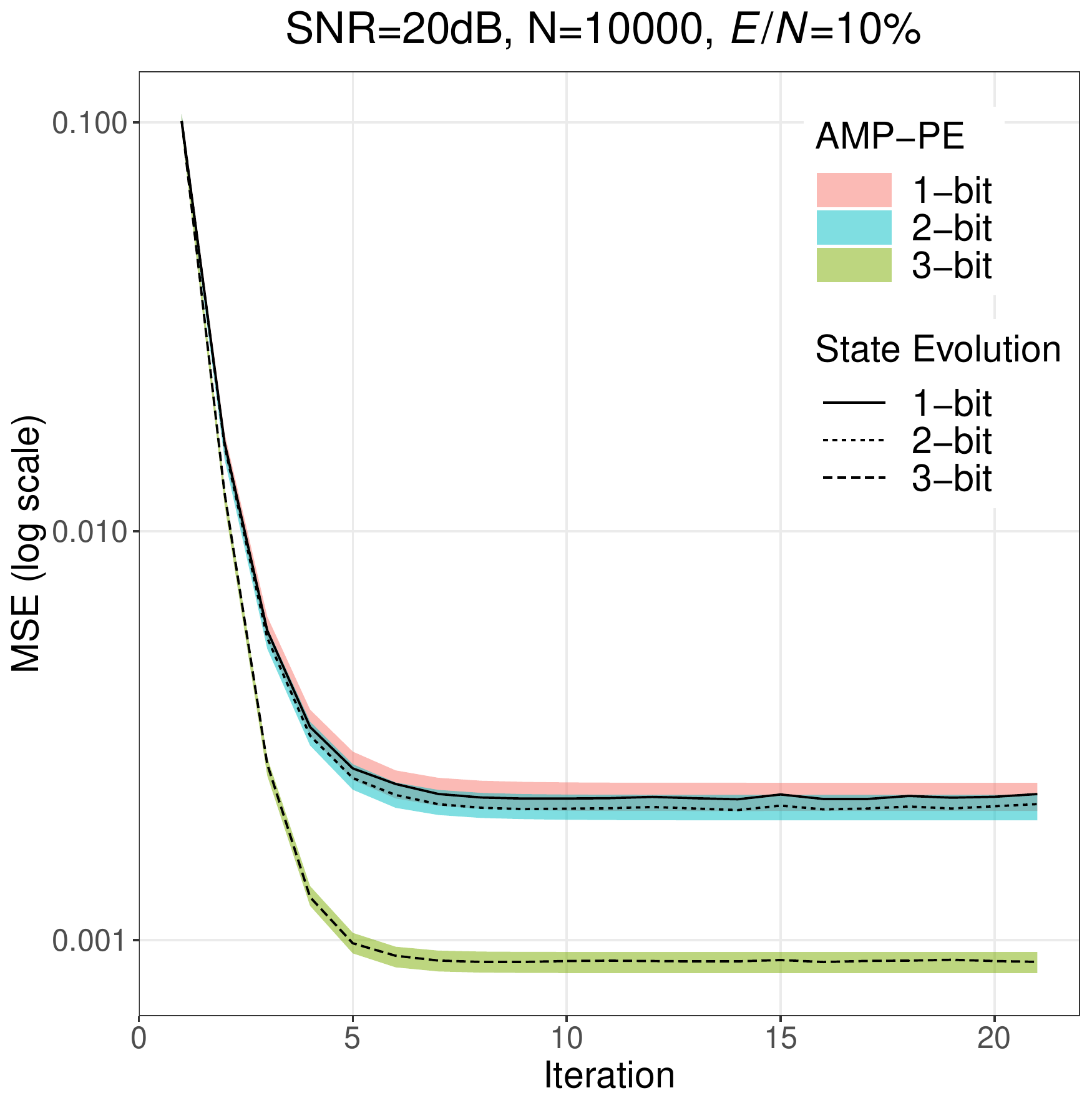}}
\subfigure{
\label{fig:se_10_50}
\includegraphics[width=0.3\textwidth]{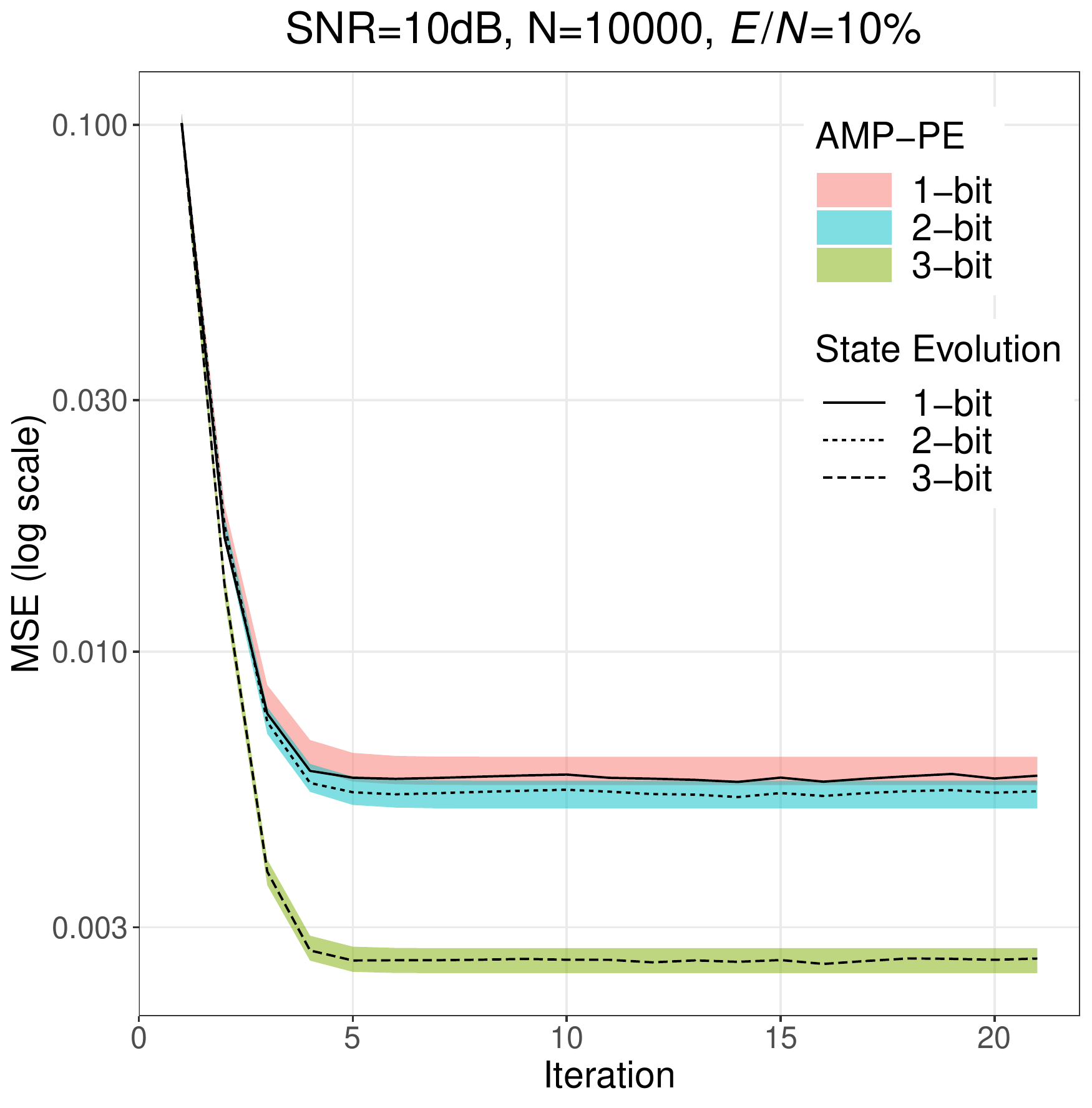}}\\

\subfigure{
\label{fig:se_50_0}
\includegraphics[width=0.3\textwidth]{figures/SE_s50_n30_10000_compare.pdf}}
\subfigure{
\label{fig:se_50_10}
\includegraphics[width=0.3\textwidth]{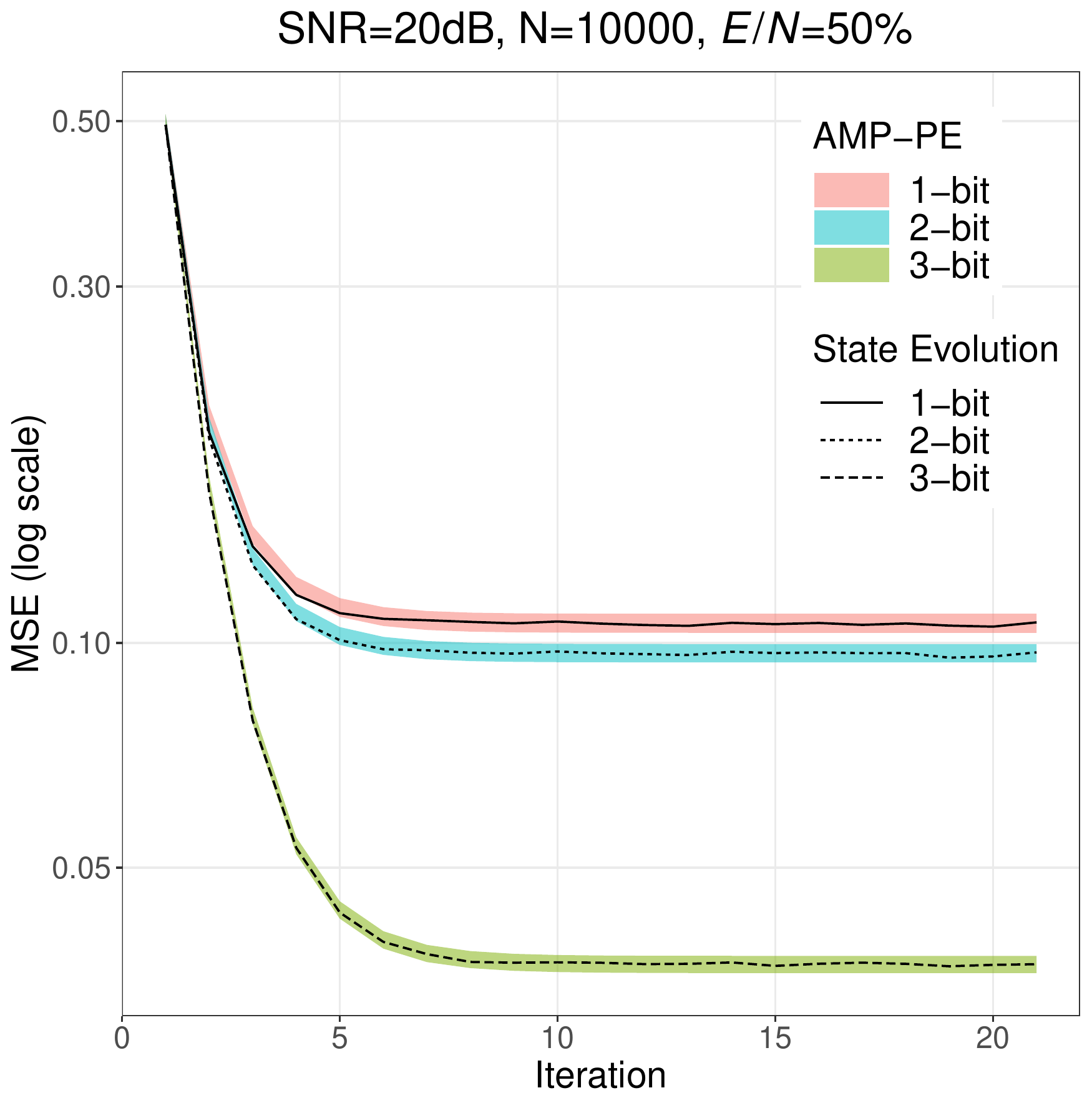}}
\subfigure{
\label{fig:se_50_50}
\includegraphics[width=0.3\textwidth]{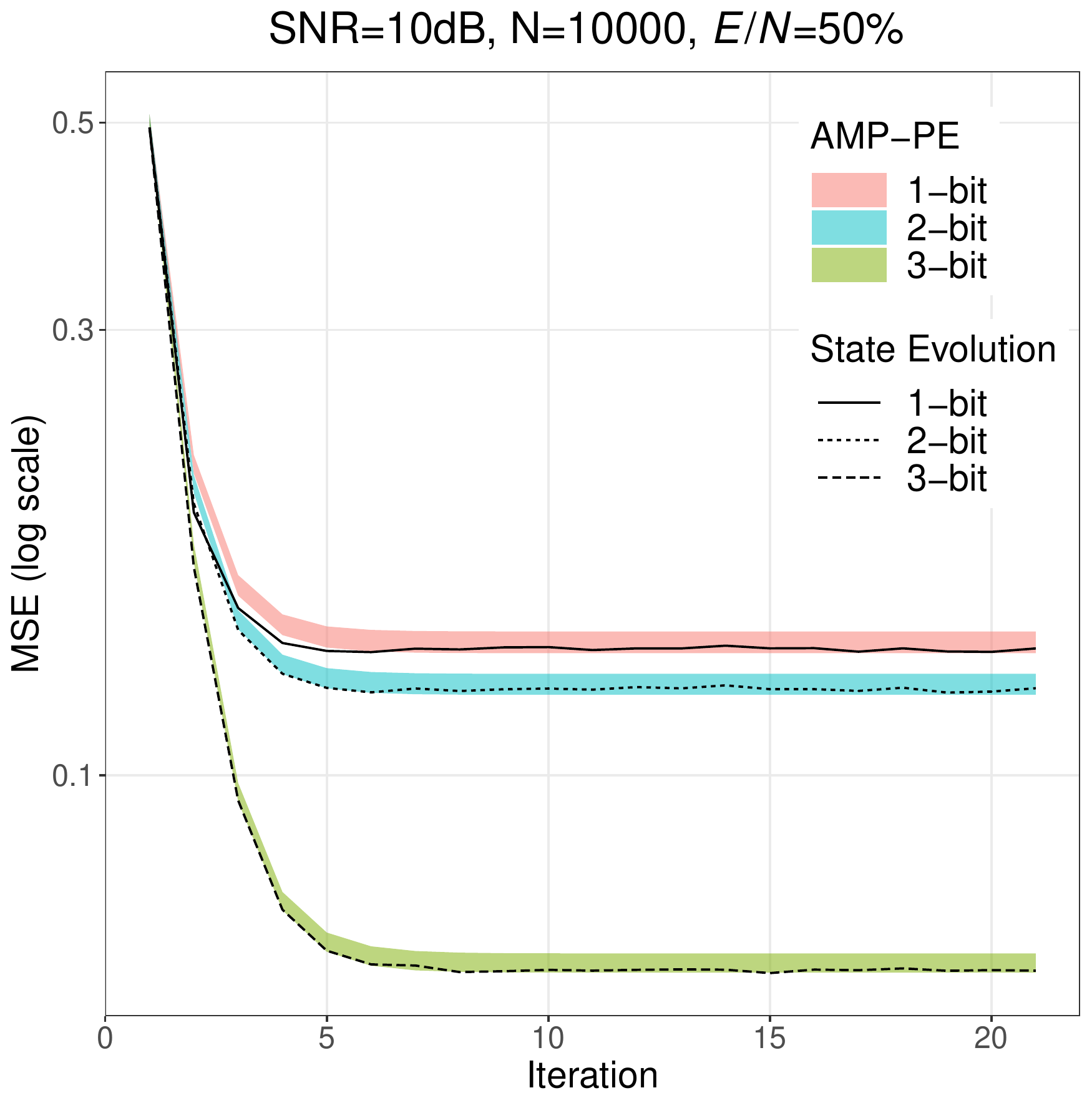}}\\

\subfigure{
\label{fig:se_100_0}
\includegraphics[width=0.3\textwidth]{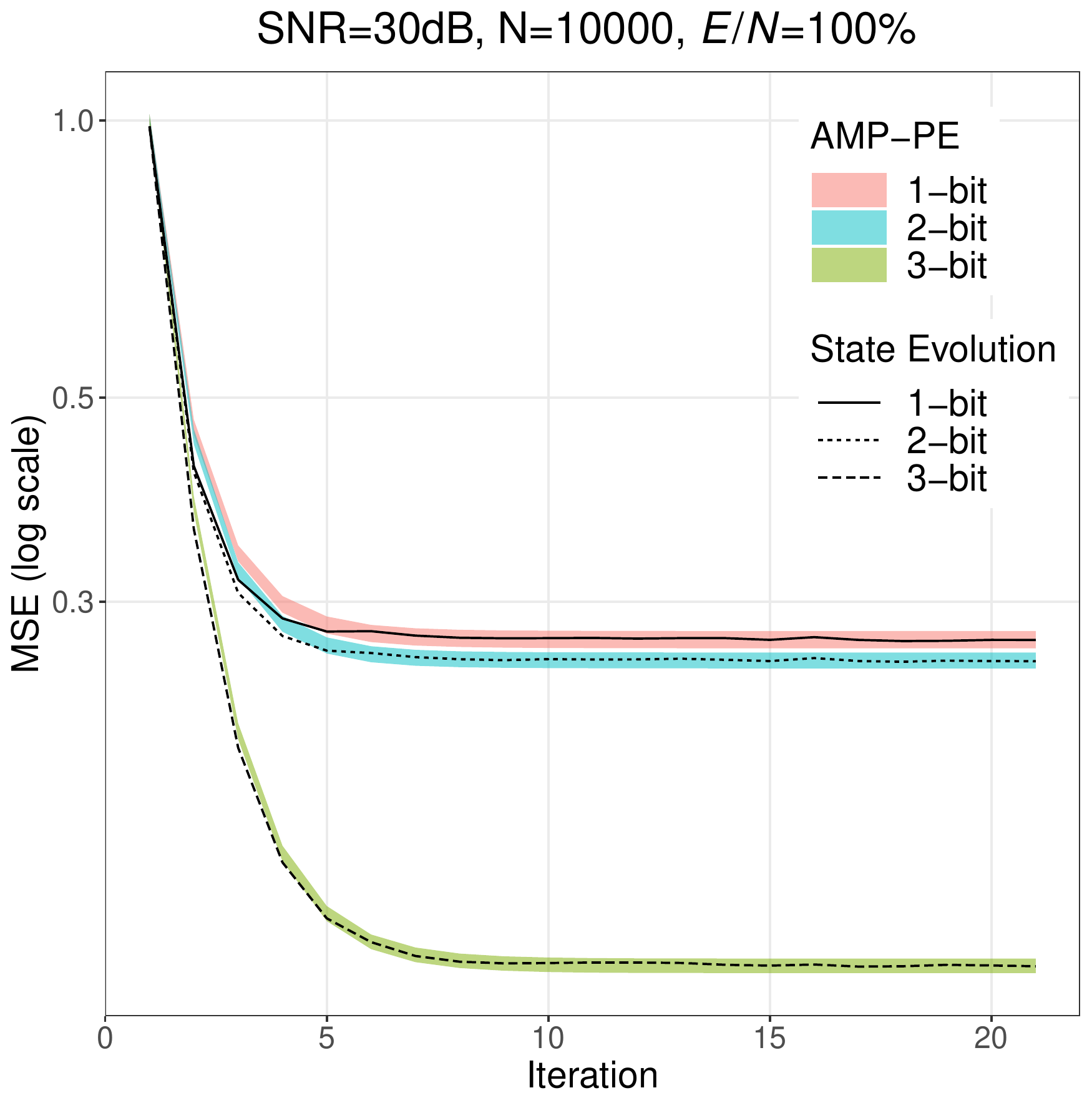}}
\subfigure{
\label{fig:se_100_10}
\includegraphics[width=0.3\textwidth]{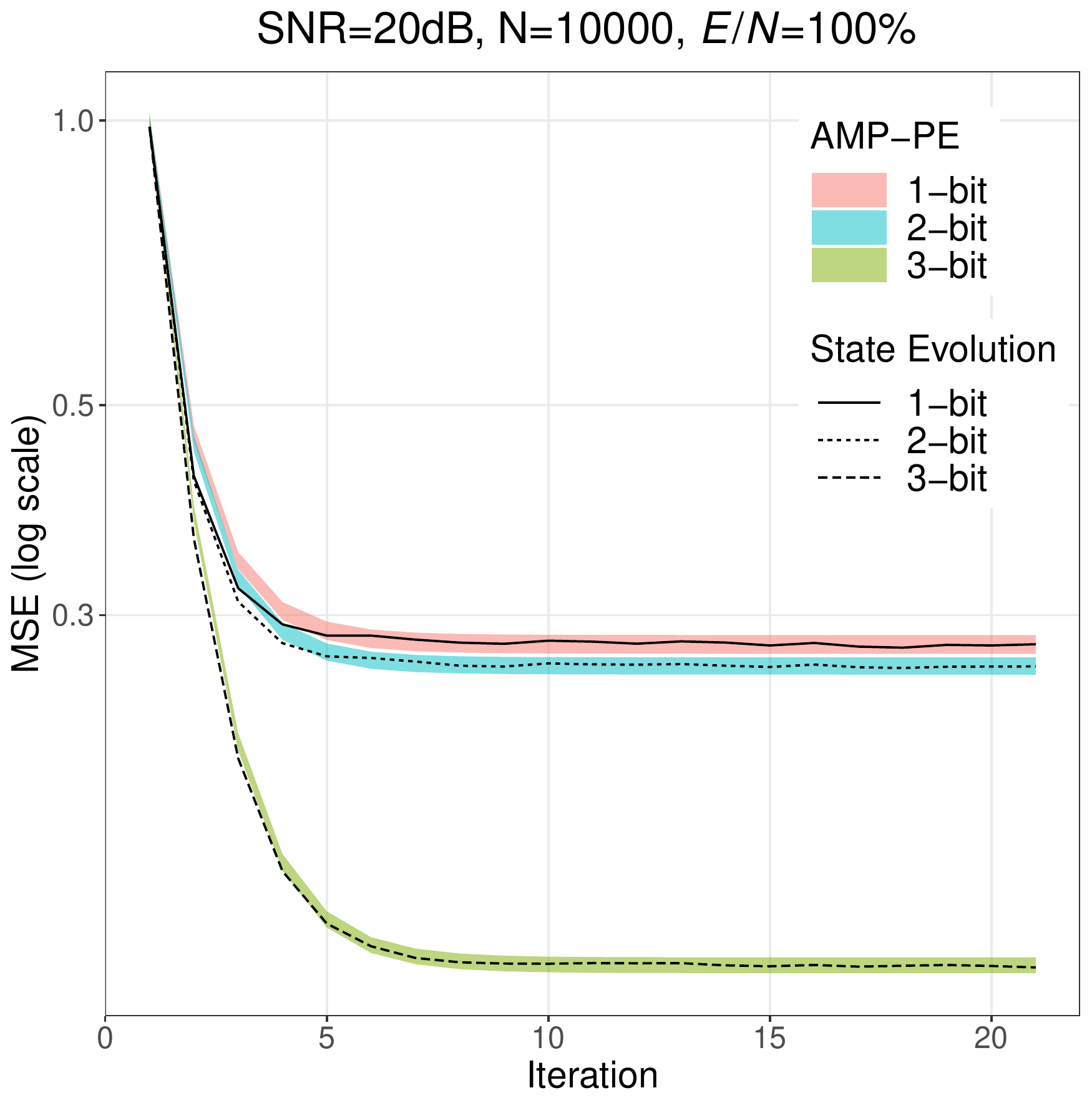}}
\subfigure{
\label{fig:se_100_50}
\includegraphics[width=0.3\textwidth]{figures/SE_s100_n10_10000_compare.pdf}}

\caption{Comparisons of the MSEs of AMP-PE estimations and the state evolution recursions for random Gaussian measurement matrices. The sampling ratio $\beta=\frac{M}{N}=2$, the sparsity level of the signal $\frac{S}{N}\in\{10\%,50\%,100\%\}$. The pre-quantization SNR varies from $30$dB, $20$dB to $10$dB.}

\label{fig:se_compare_vary}
\end{figure*}

With the signal length $N$ set to $10000$, we compare the MSEs of AMP-PE estimations and the state evolution recursions under different pre-QNT SNRs $\in\{30\textnormal{dB},20\textnormal{dB},10\textnormal{dB}\}$ and different sparsity levels $\frac{E}{N}\in\{10\%,50\%,100\%\}$. The results are shown in Fig. \ref{fig:se_compare_vary}. 

\newpage 

\begin{figure*}[htbp]
\centering
\subfigure[Estimated parameters]{
\label{fig:se_100_10_10000}
\includegraphics[height=0.35\textwidth]{figures/SE_s100_n10_10000_compare.pdf}}
\subfigure[True parameters]{
\label{fig:se_100_10_10000_oracle}
\includegraphics[height=0.35\textwidth]{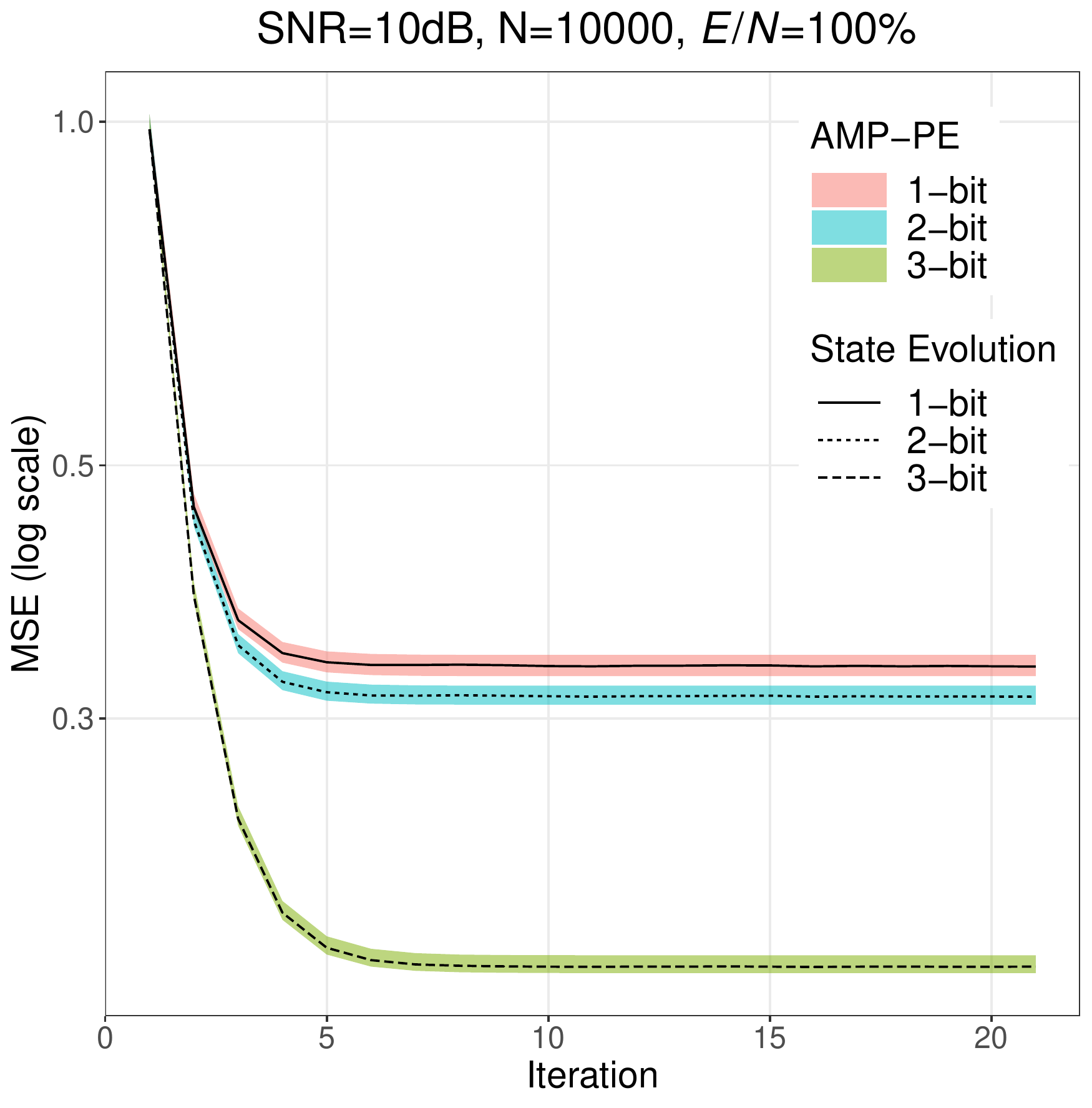}}
\caption{Comparison of the results obtained using the estimated parameters and the true parameters. The signal length $N=10000$, the sampling ratio $\beta=\frac{M}{N}=2$, the sparsity level of the signal $\frac{E}{N}=100\%$, pre-quantization SNR=10dB.}
\label{fig:se_compare_parameters}
\end{figure*}

When the pre-QNT SNR level is low or the sparsity level is high, the state evolution does not always match the MSEs of empirical experiments. This is caused by the nonconvexity of the parameter estimation problems. The computed maximizing parameters are only locally optimal; hence, they are slightly different between the AMP-PE algorithm and its state evolution. In Fig. \ref{fig:se_compare_parameters}, we can empirically verify it by comparing the results obtained using the estimated parameters with the results obtained using the true parameters. Fig. \ref{fig:se_100_10_10000} shows that state evolution does not always fall within the confidence region in the first few iterations. Fig. \ref{fig:se_100_10_10000_oracle} shows that the deviation could be corrected by using the true parameters.

\newpage
\section{1-bit and Multi-bit Compressive Sensing}

\subsection{Signal with the Bernoulli-Gaussian Mixture Prior}

\begin{figure*}[htbp]
\centering
\subfigure{
\includegraphics[width=0.3\textwidth]{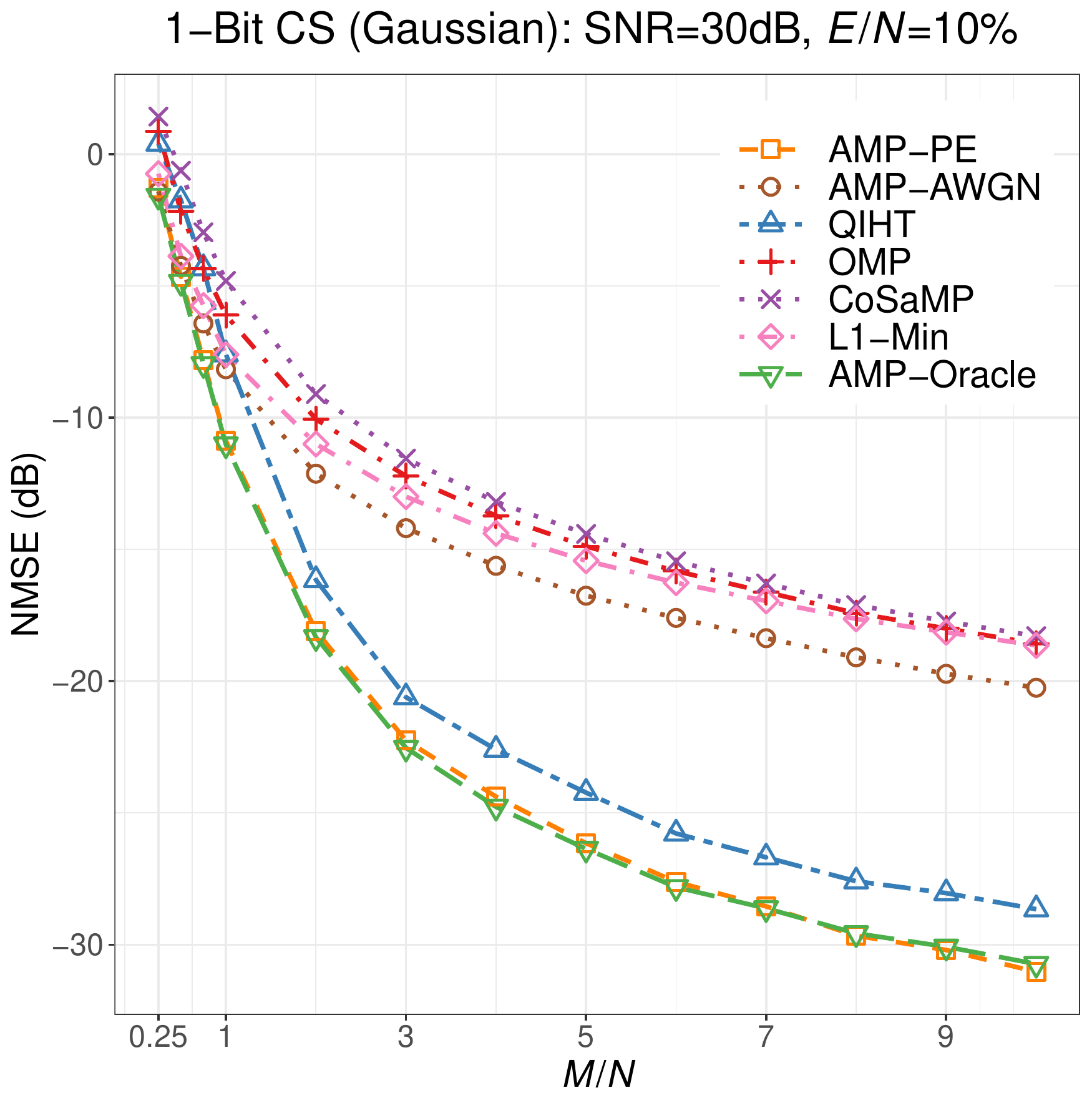}}
\subfigure{
\includegraphics[width=0.3\textwidth]{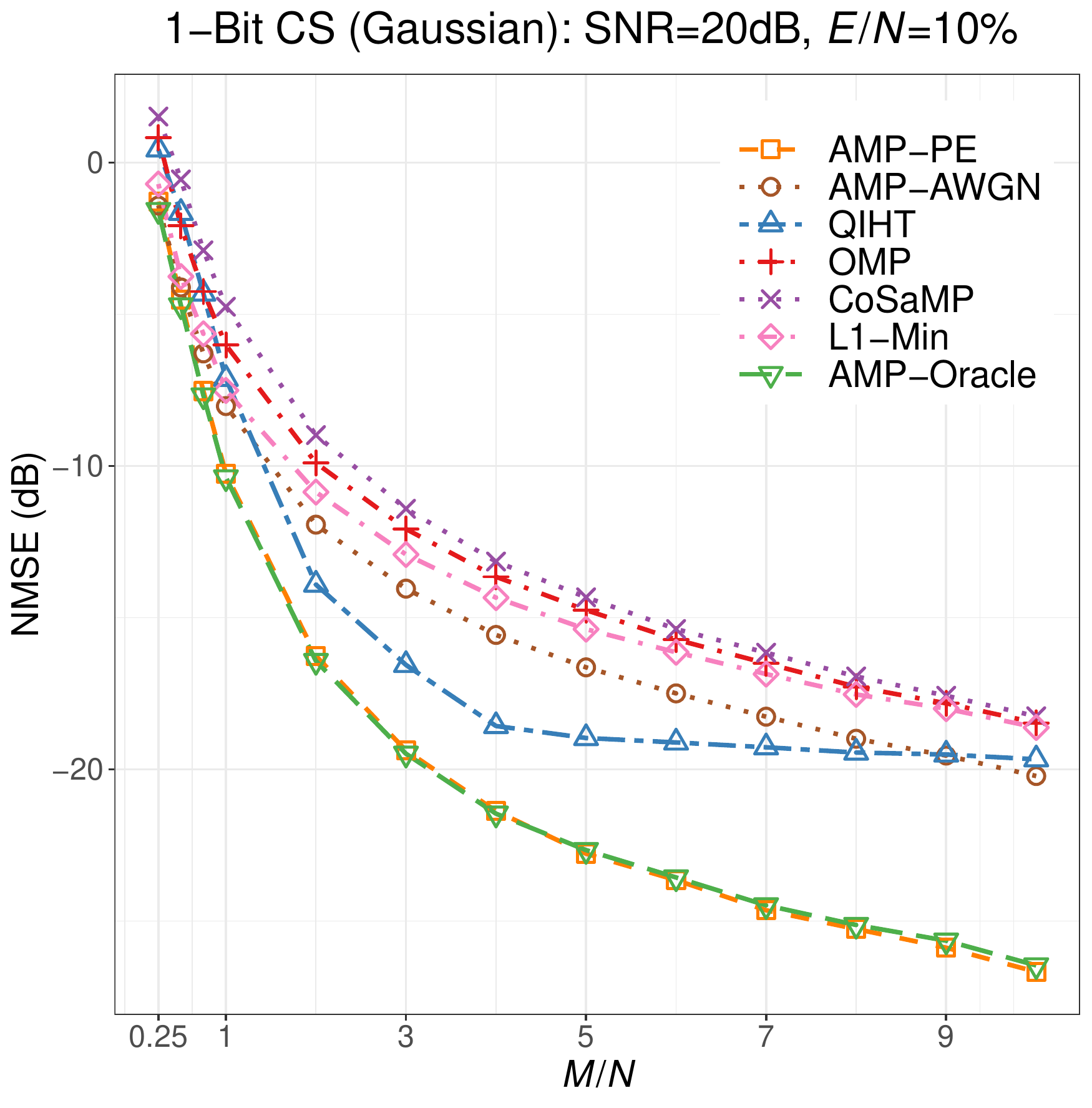}}
\subfigure{
\includegraphics[width=0.3\textwidth]{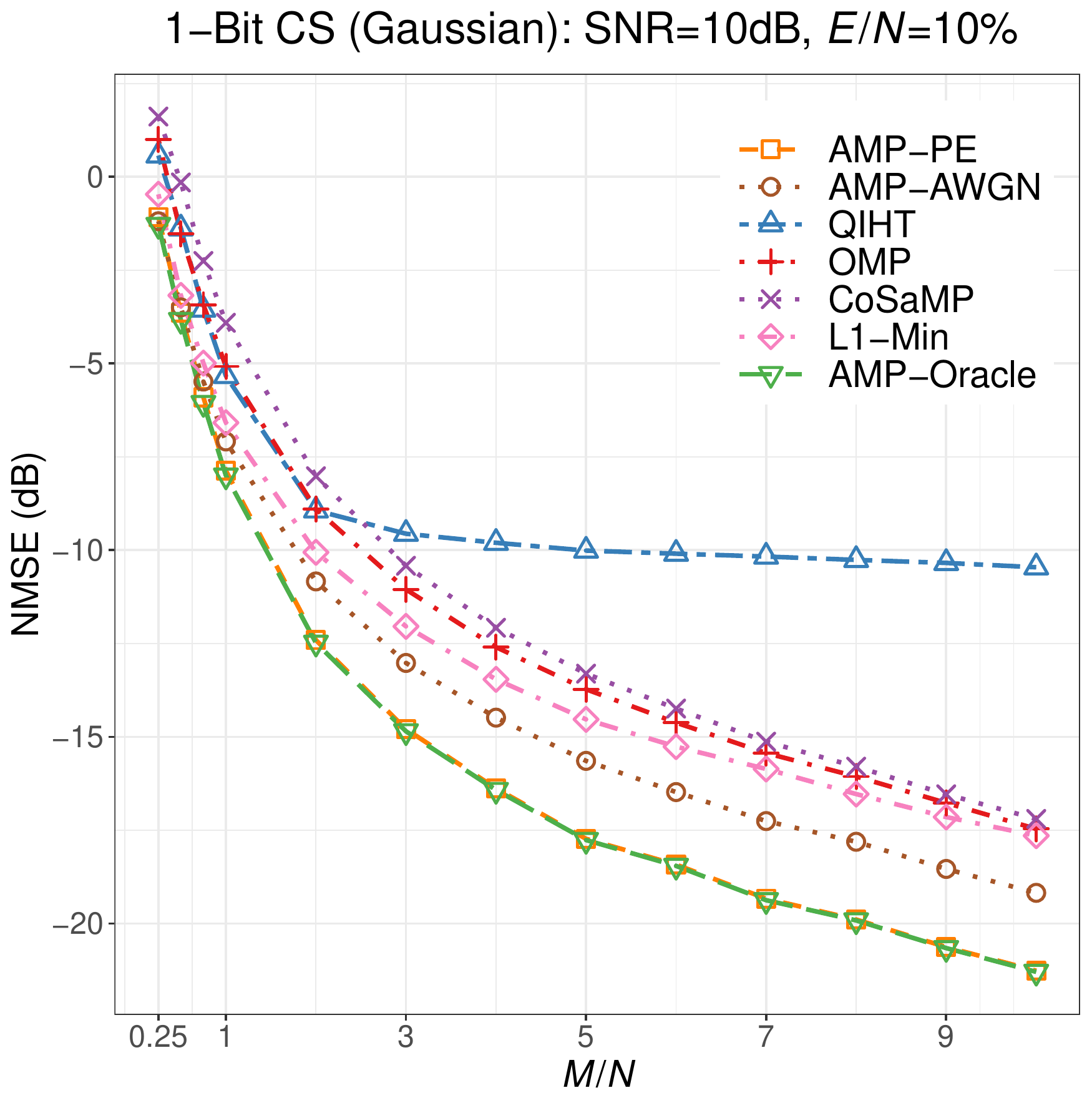}}\\

\subfigure{
\includegraphics[width=0.3\textwidth]{figures/1bit_s50_n30_compare_full_gaussian.pdf}}
\subfigure{
\includegraphics[width=0.3\textwidth]{figures/1bit_s50_n20_compare_full_gaussian.pdf}}
\subfigure{
\includegraphics[width=0.3\textwidth]{figures/1bit_s50_n10_compare_full_gaussian.pdf}}\\

\subfigure{
\includegraphics[width=0.3\textwidth]{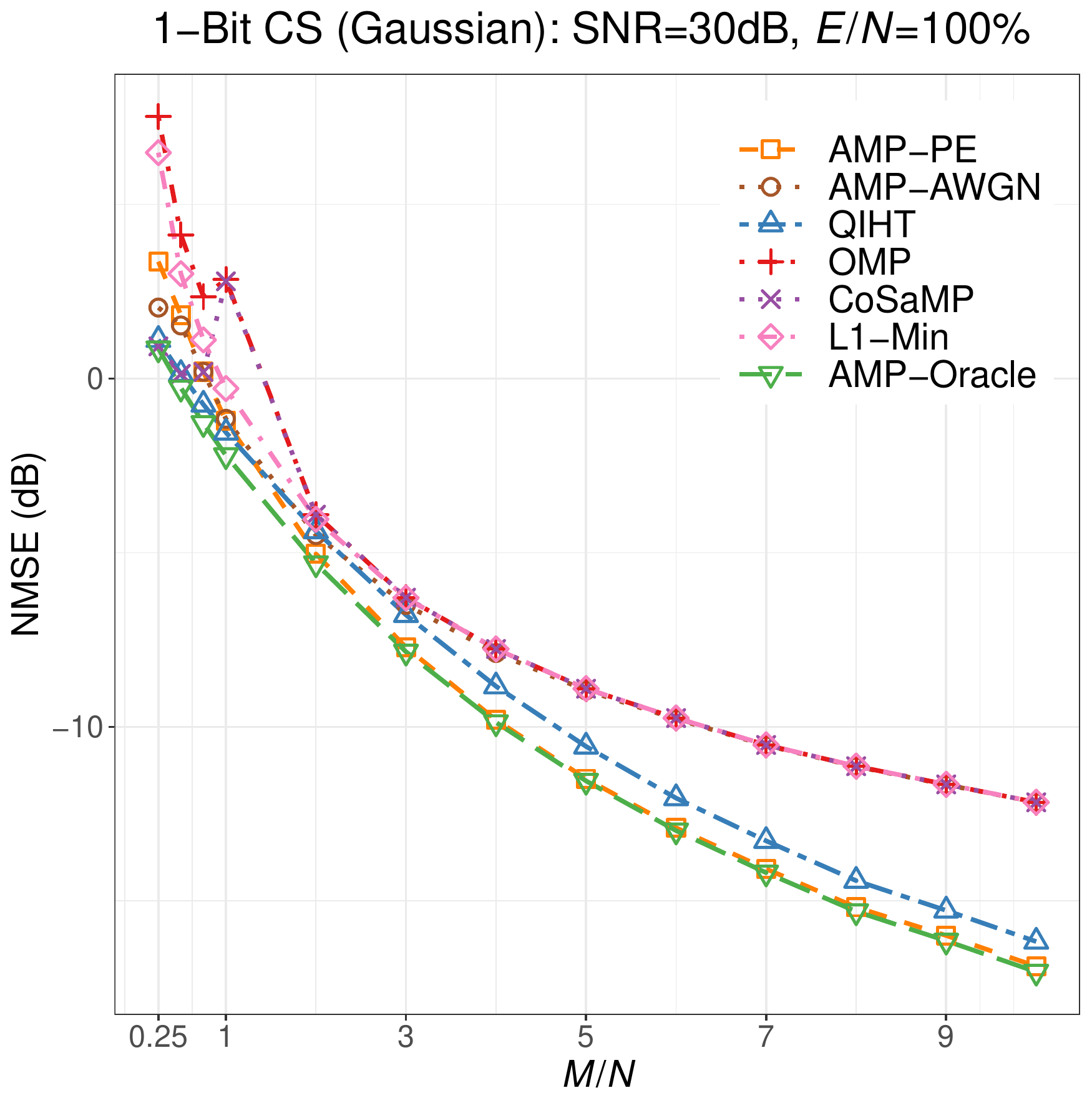}}
\subfigure{
\includegraphics[width=0.3\textwidth]{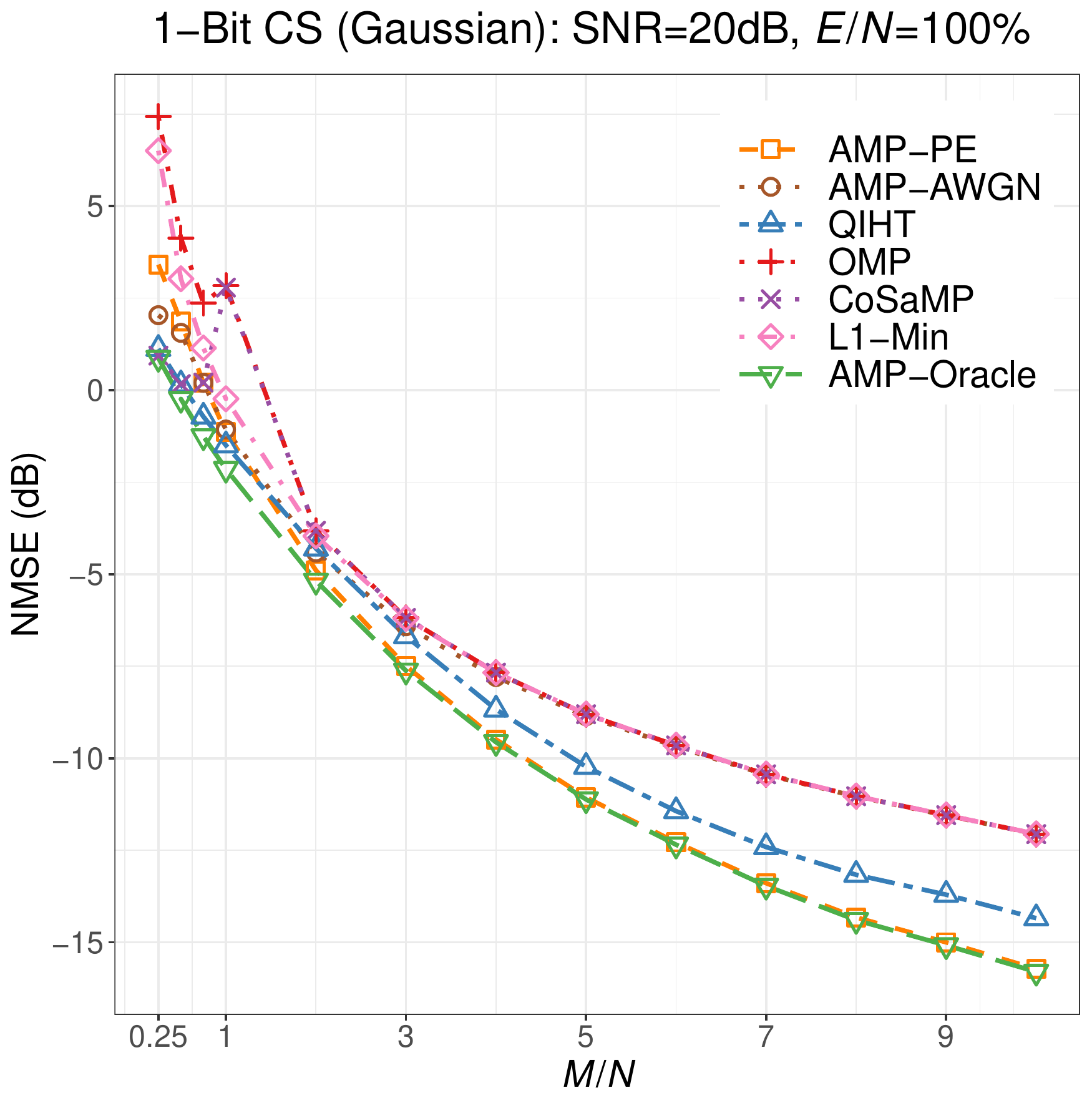}}
\subfigure{
\includegraphics[width=0.3\textwidth]{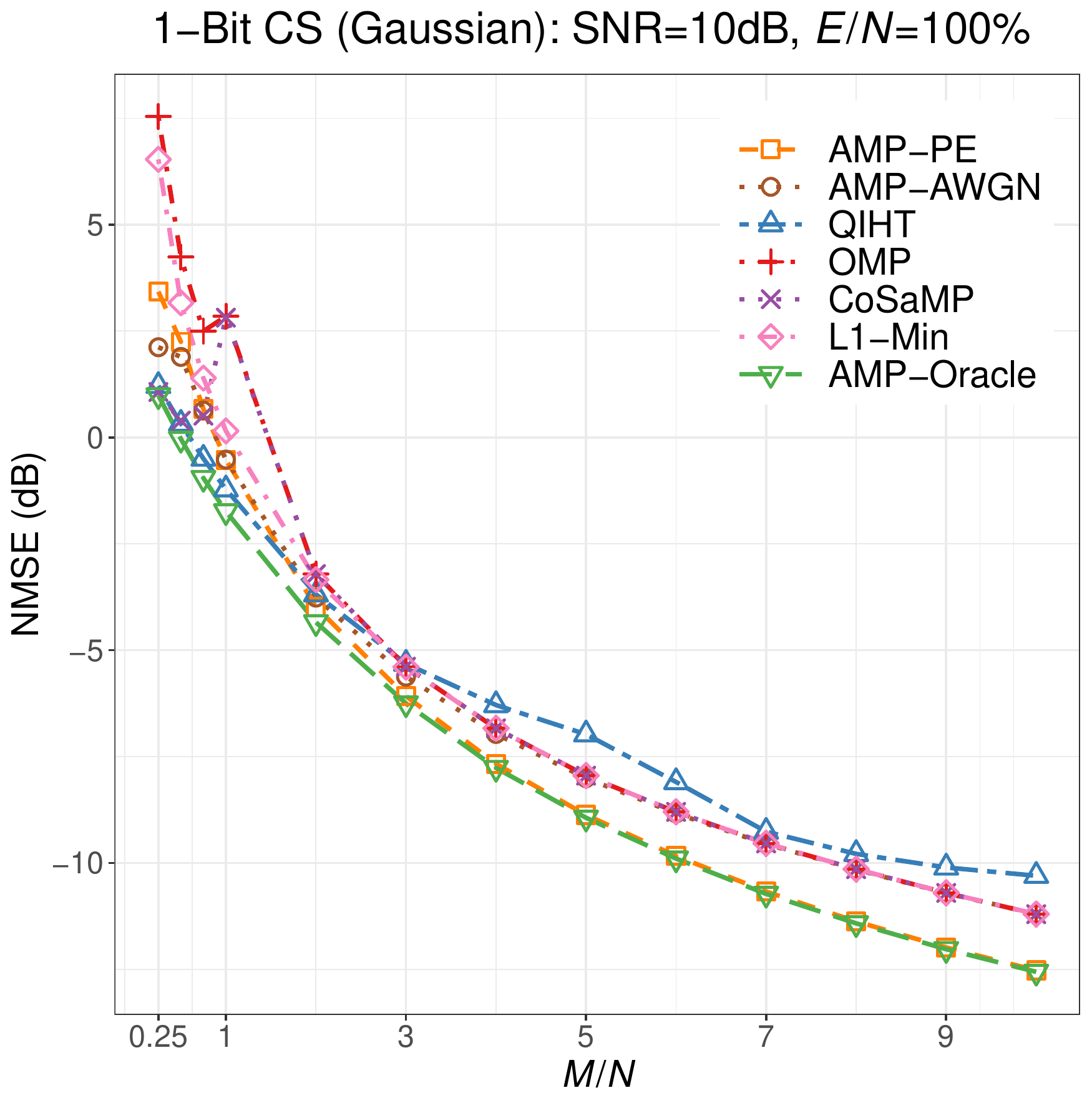}}

\caption{Comparison of different approaches in solving 1-bit CS. Nonzero entries of the signal follow the Gausian distribution. The sampling ratio $\frac{M}{N}\in\{0.25,\cdots,10\}$ and the sparsity level $\frac{E}{N}=\in\{10\%,50\%,100\%\}$. The pre-quantization SNR varies from $30$dB, $20$dB to $10$dB.}

\label{fig:1bit_experiments_gaussian}
\end{figure*}

\newpage
\begin{figure*}[htbp]
\centering
\subfigure{

\includegraphics[width=0.3\textwidth]{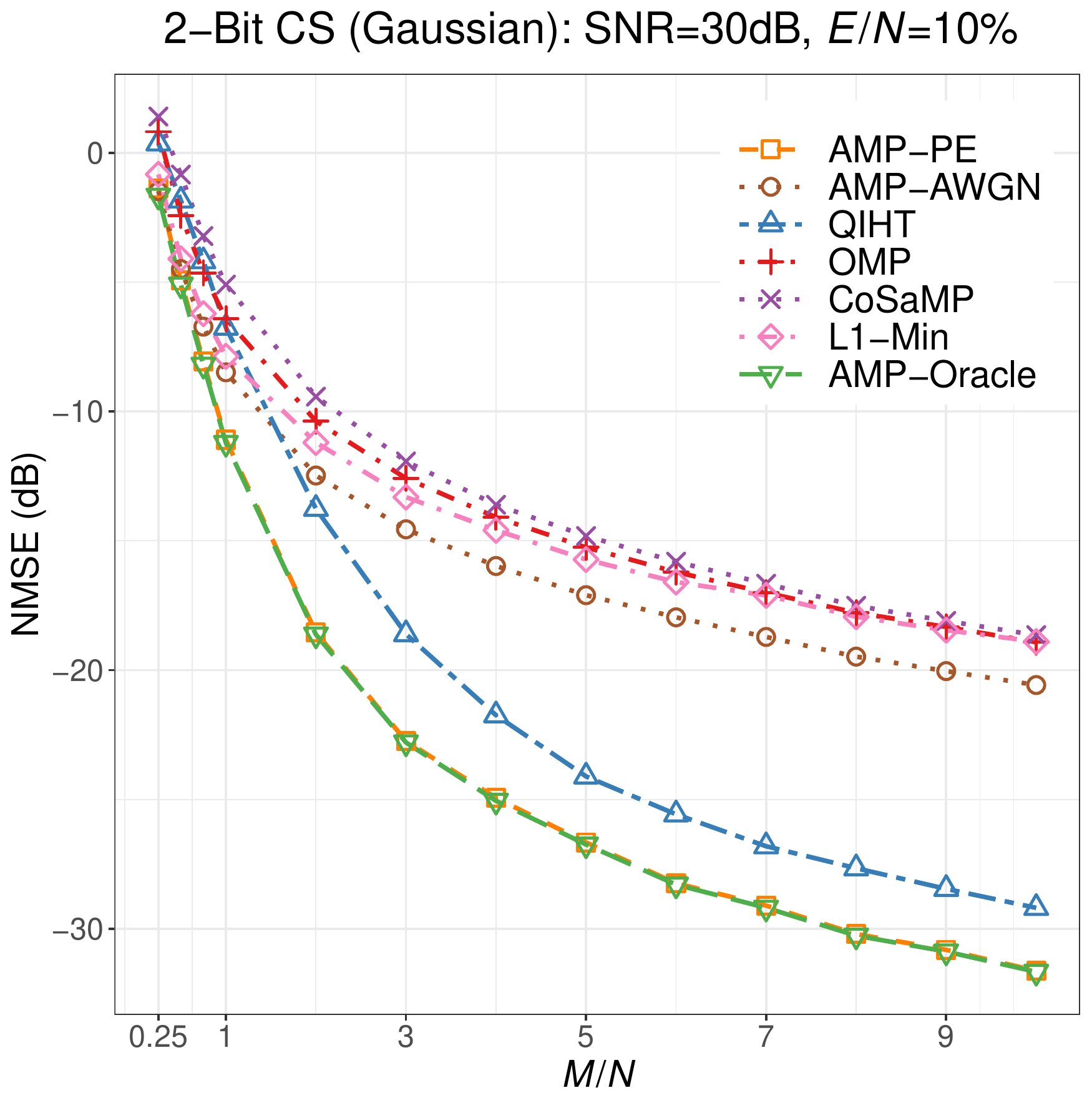}}
\subfigure{

\includegraphics[width=0.3\textwidth]{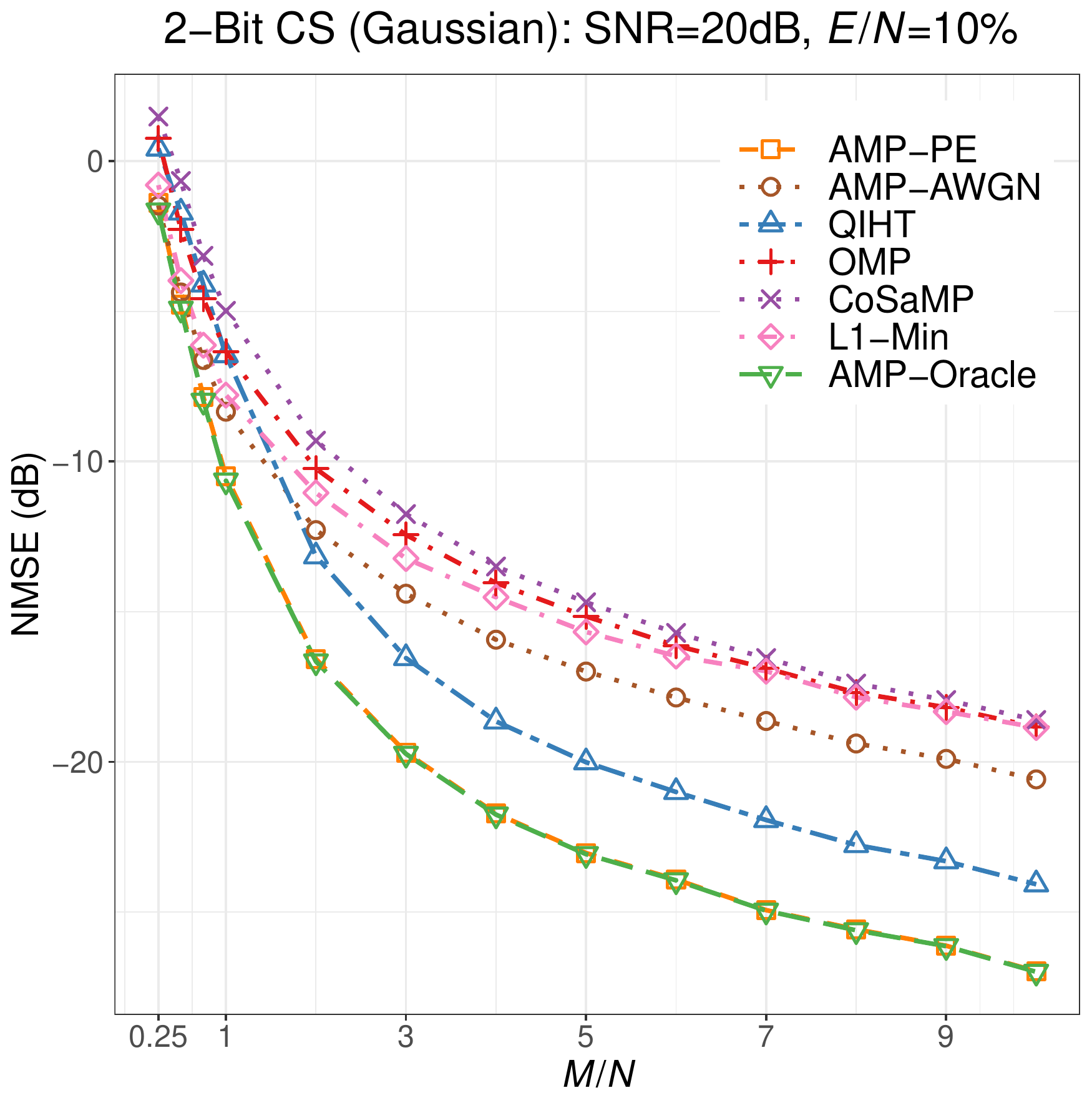}}
\subfigure{

\includegraphics[width=0.3\textwidth]{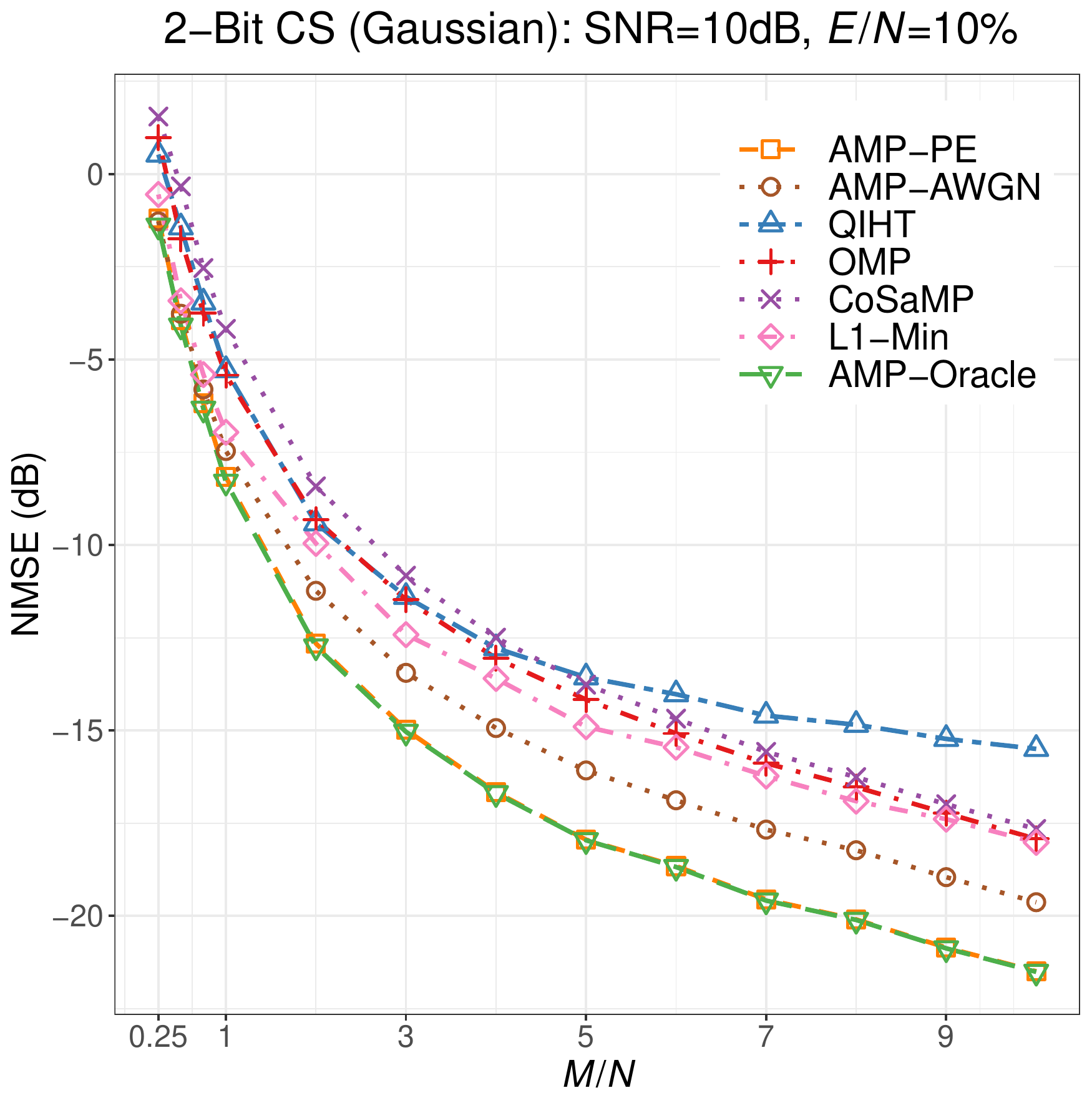}}\\

\subfigure{

\includegraphics[width=0.3\textwidth]{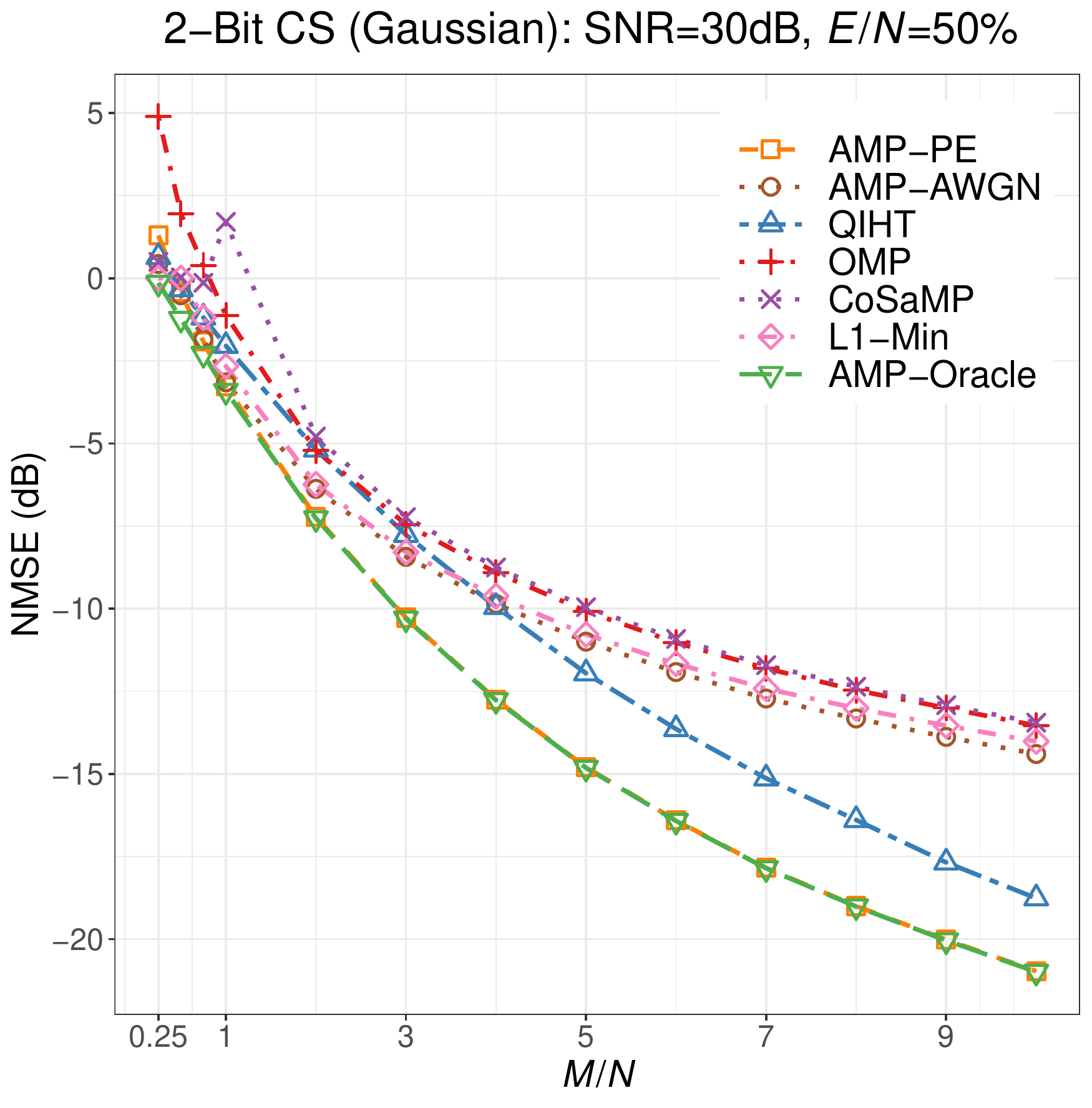}}
\subfigure{

\includegraphics[width=0.3\textwidth]{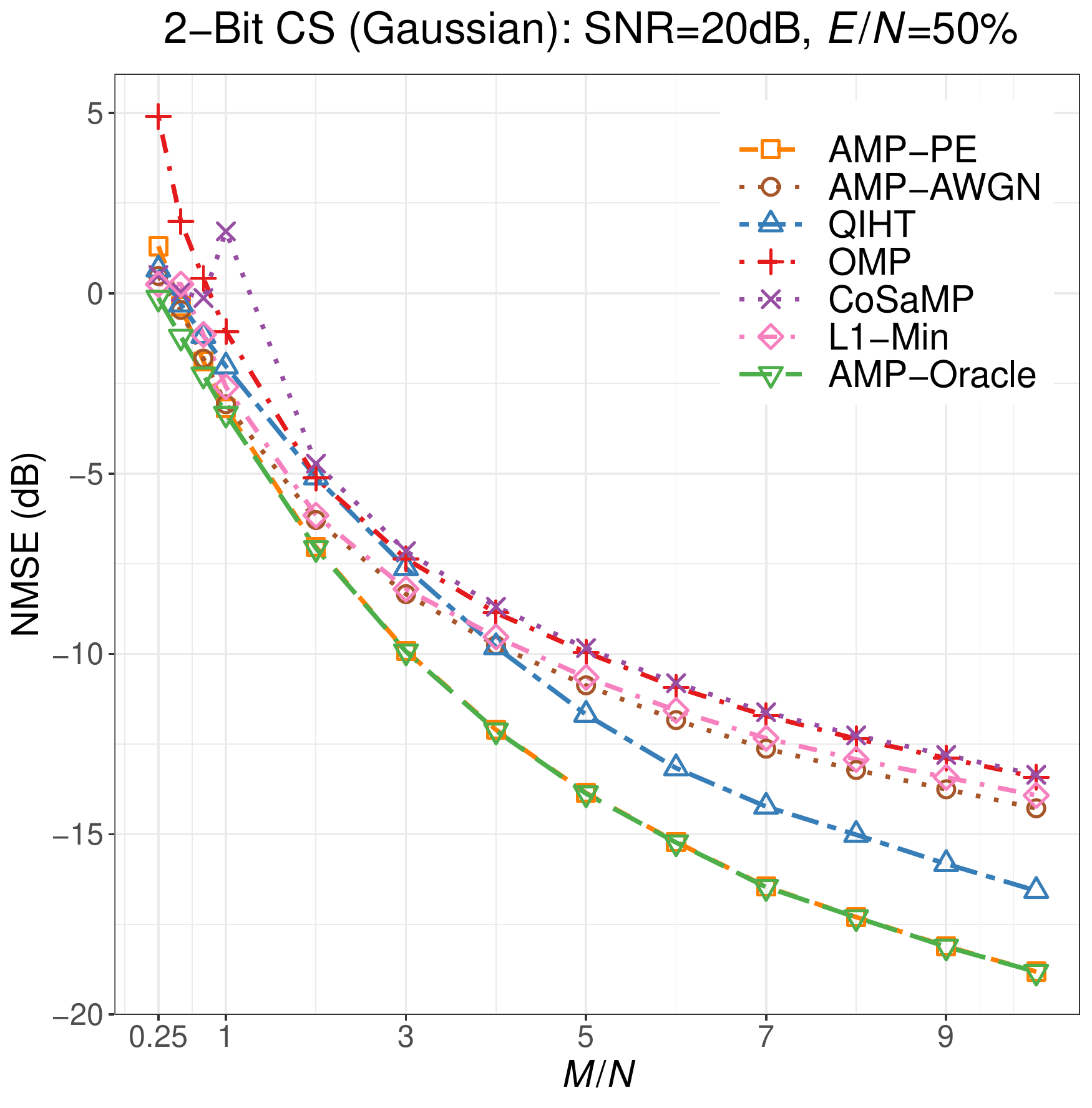}}
\subfigure{

\includegraphics[width=0.3\textwidth]{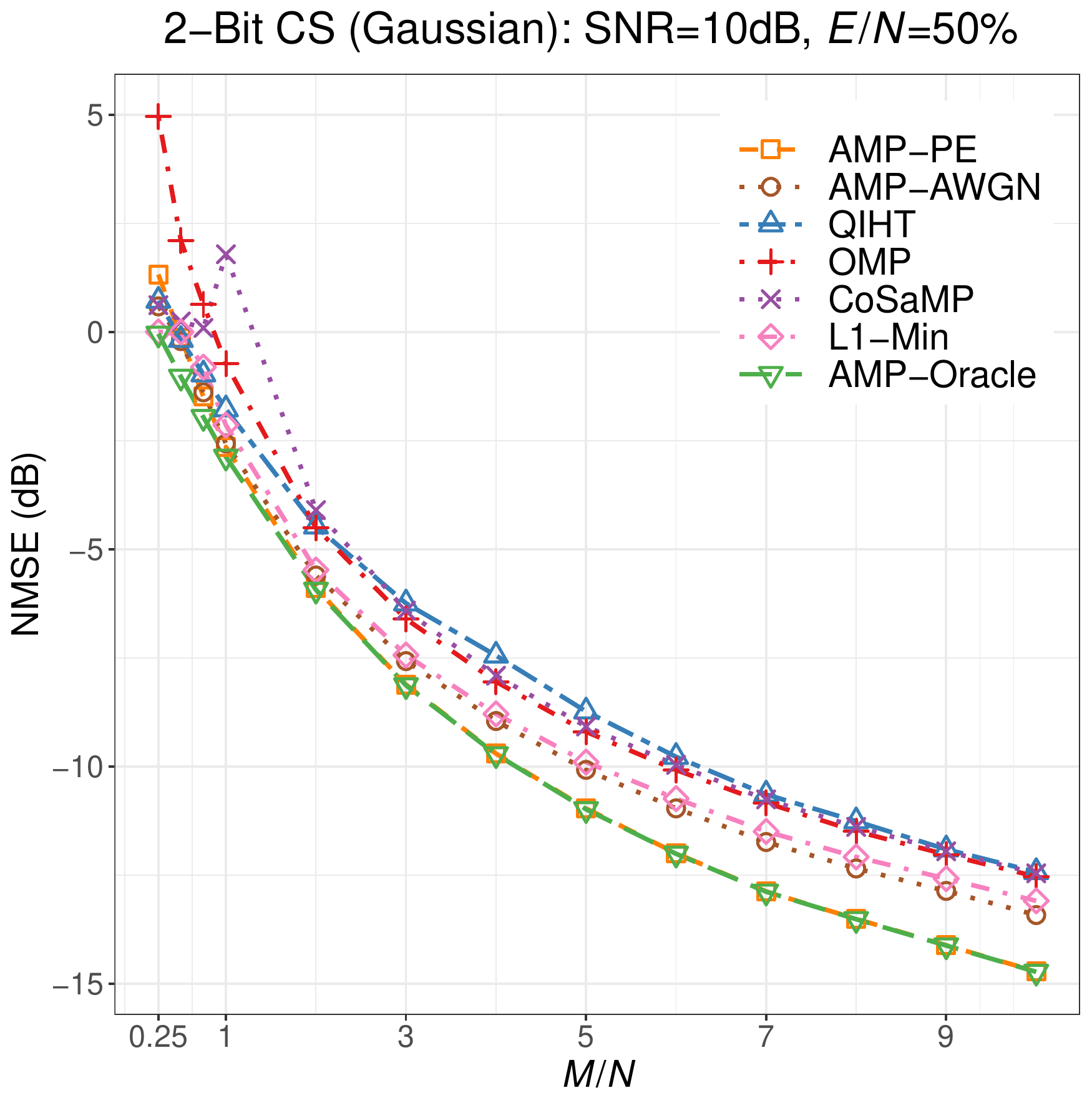}}\\

\subfigure{

\includegraphics[width=0.3\textwidth]{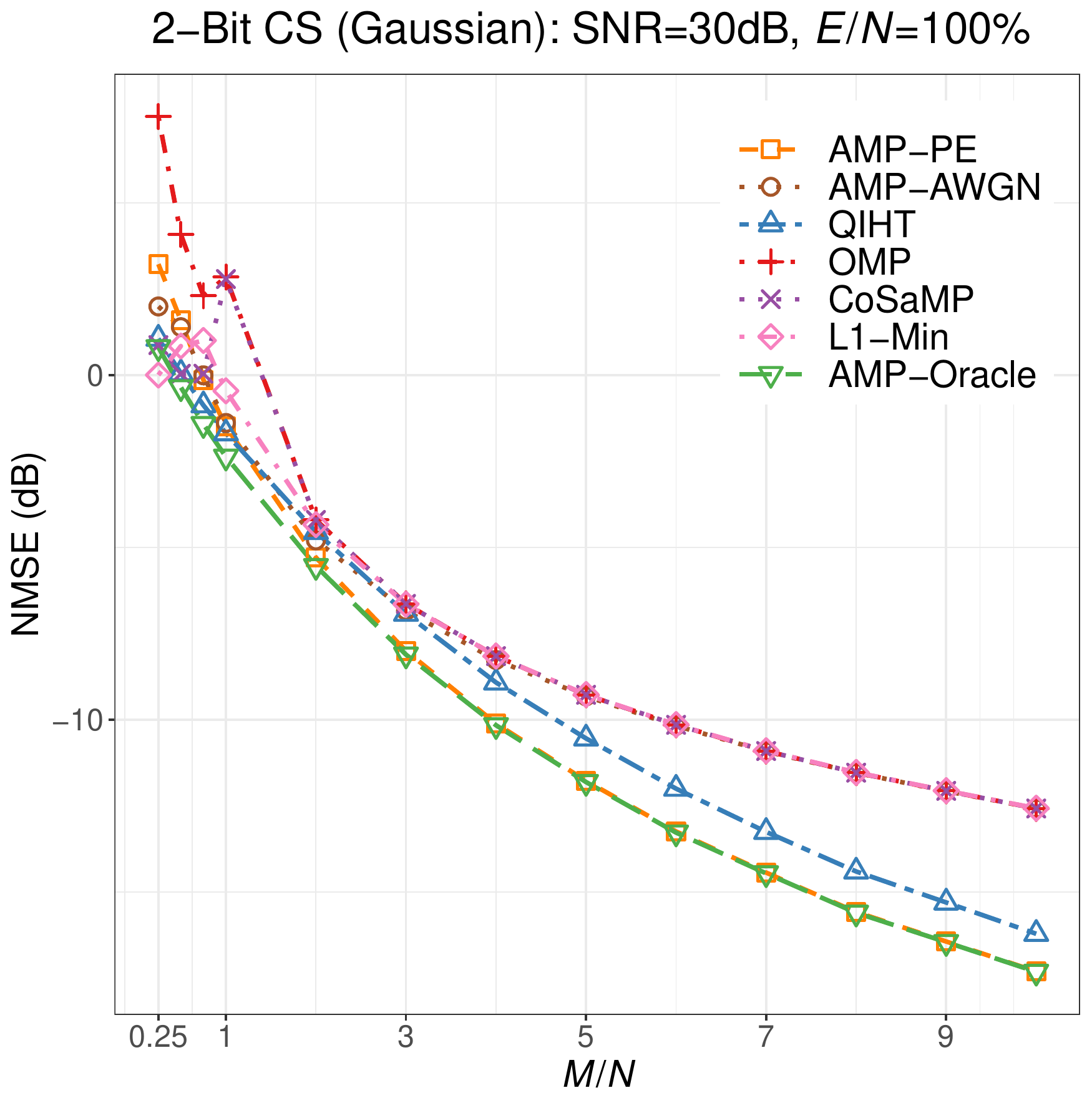}}
\subfigure{

\includegraphics[width=0.3\textwidth]{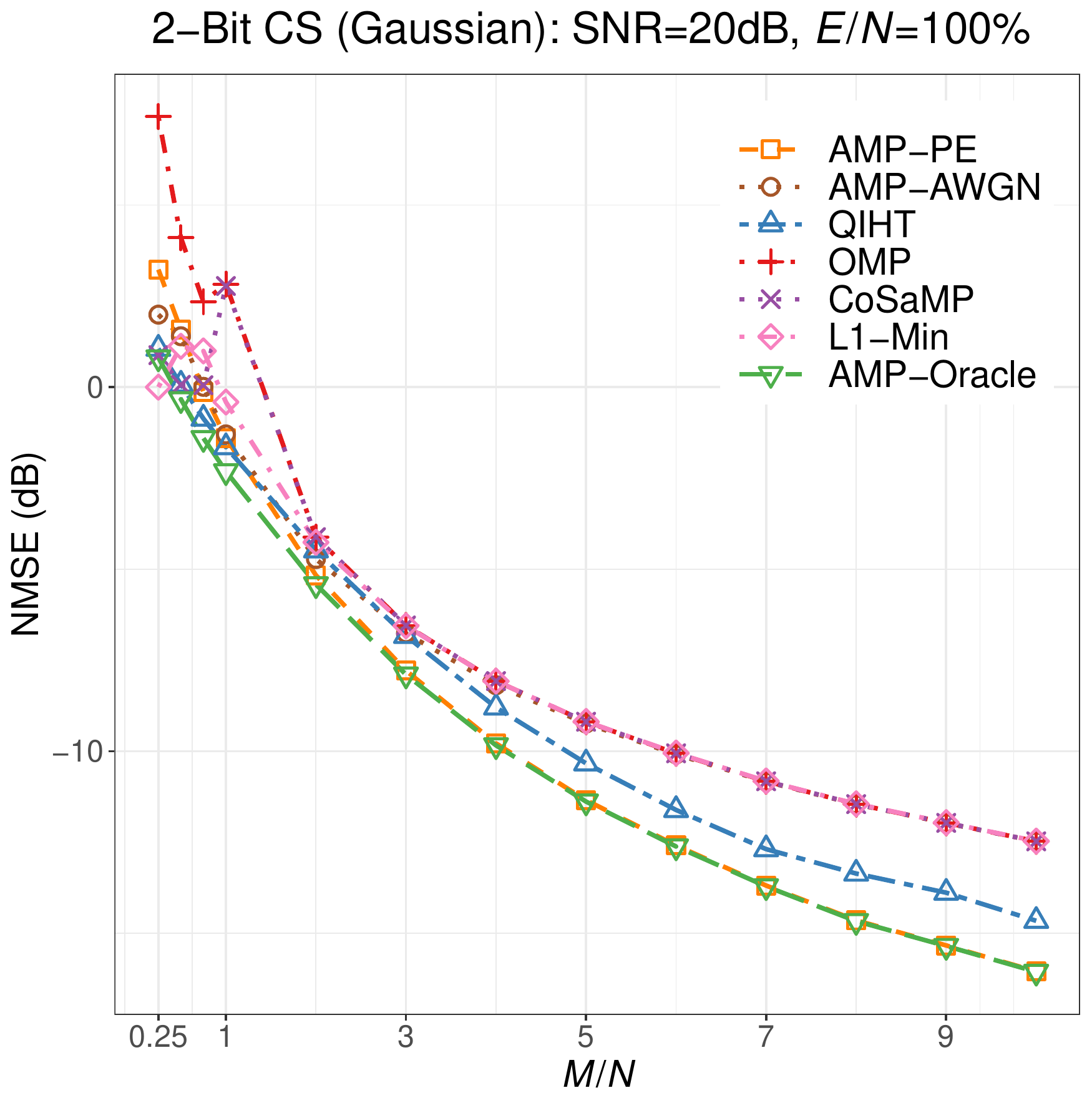}}
\subfigure{

\includegraphics[width=0.3\textwidth]{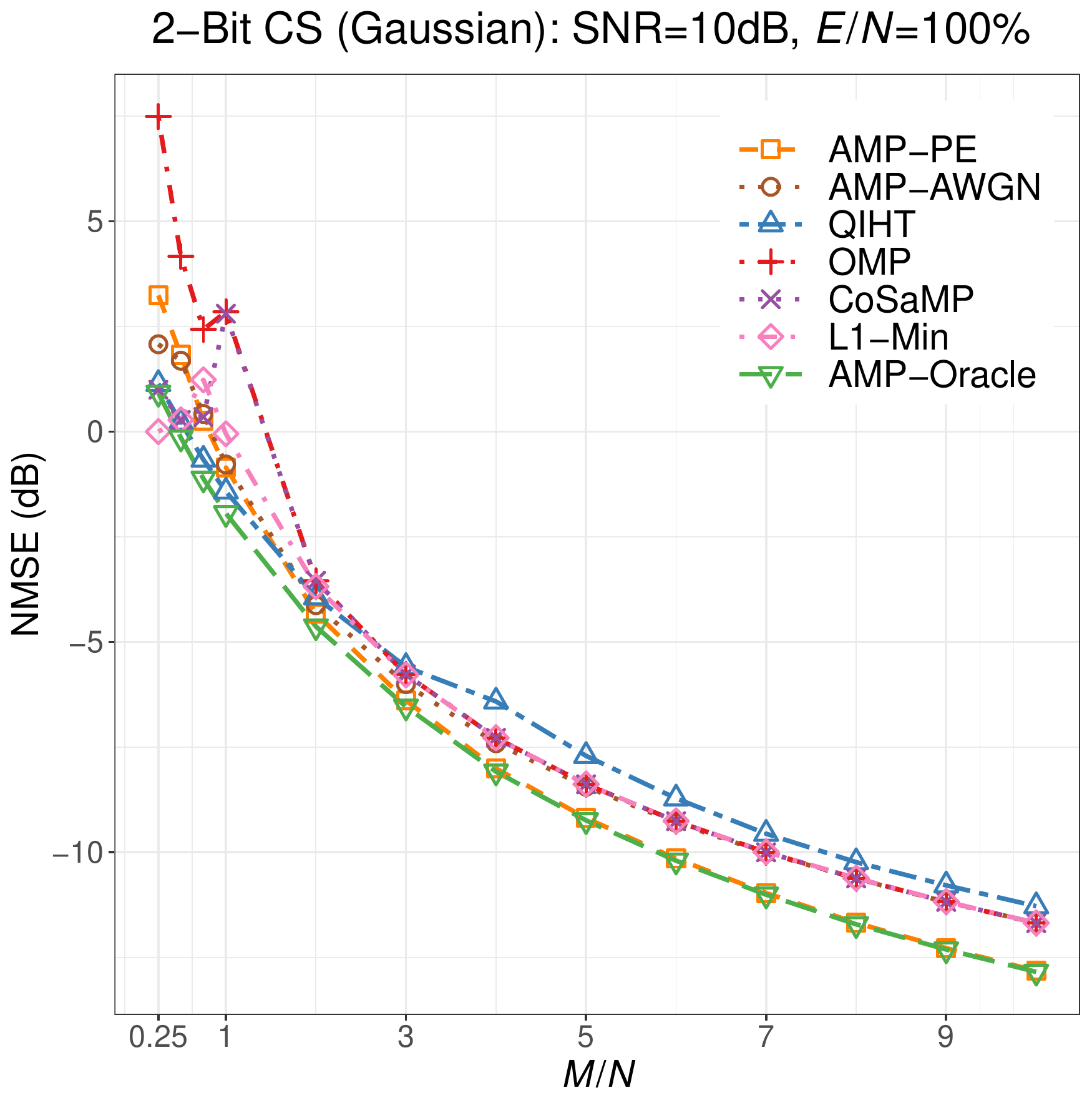}}

\caption{Comparison of different approaches in solving 2-bit CS. Nonzero entries of the signal follow the Gausian distribution. The sampling ratio $\frac{M}{N}\in\{0.25,\cdots,10\}$ and the sparsity level $\frac{E}{N}=\in\{10\%,50\%,100\%\}$. The pre-quantization SNR varies from $30$dB, $20$dB to $10$dB.}

\label{fig:2bit_experiments_gaussian}
\end{figure*}

\newpage
\begin{figure*}[htbp]
\centering
\subfigure{

\includegraphics[width=0.3\textwidth]{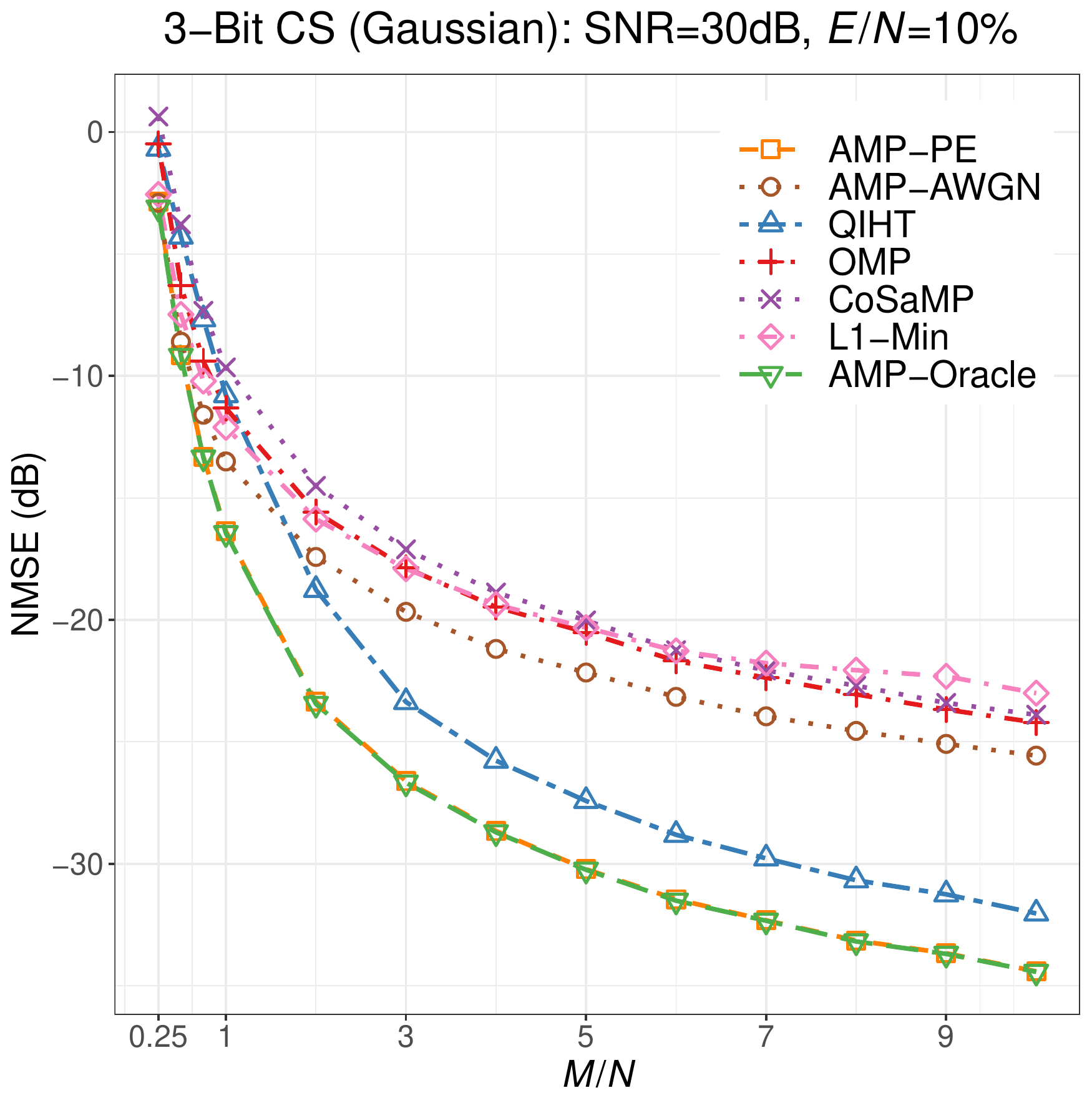}}
\subfigure{

\includegraphics[width=0.3\textwidth]{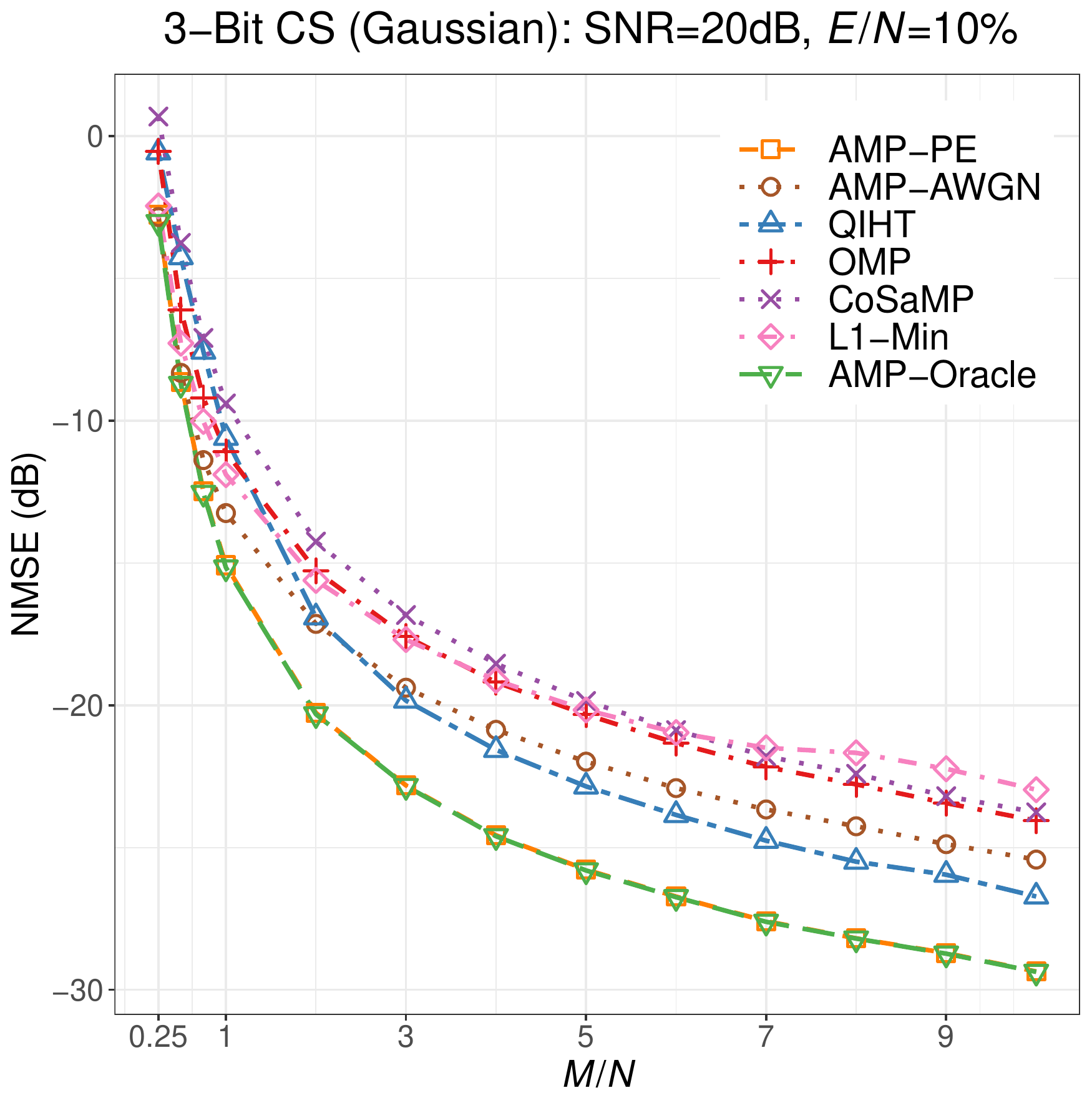}}
\subfigure{

\includegraphics[width=0.3\textwidth]{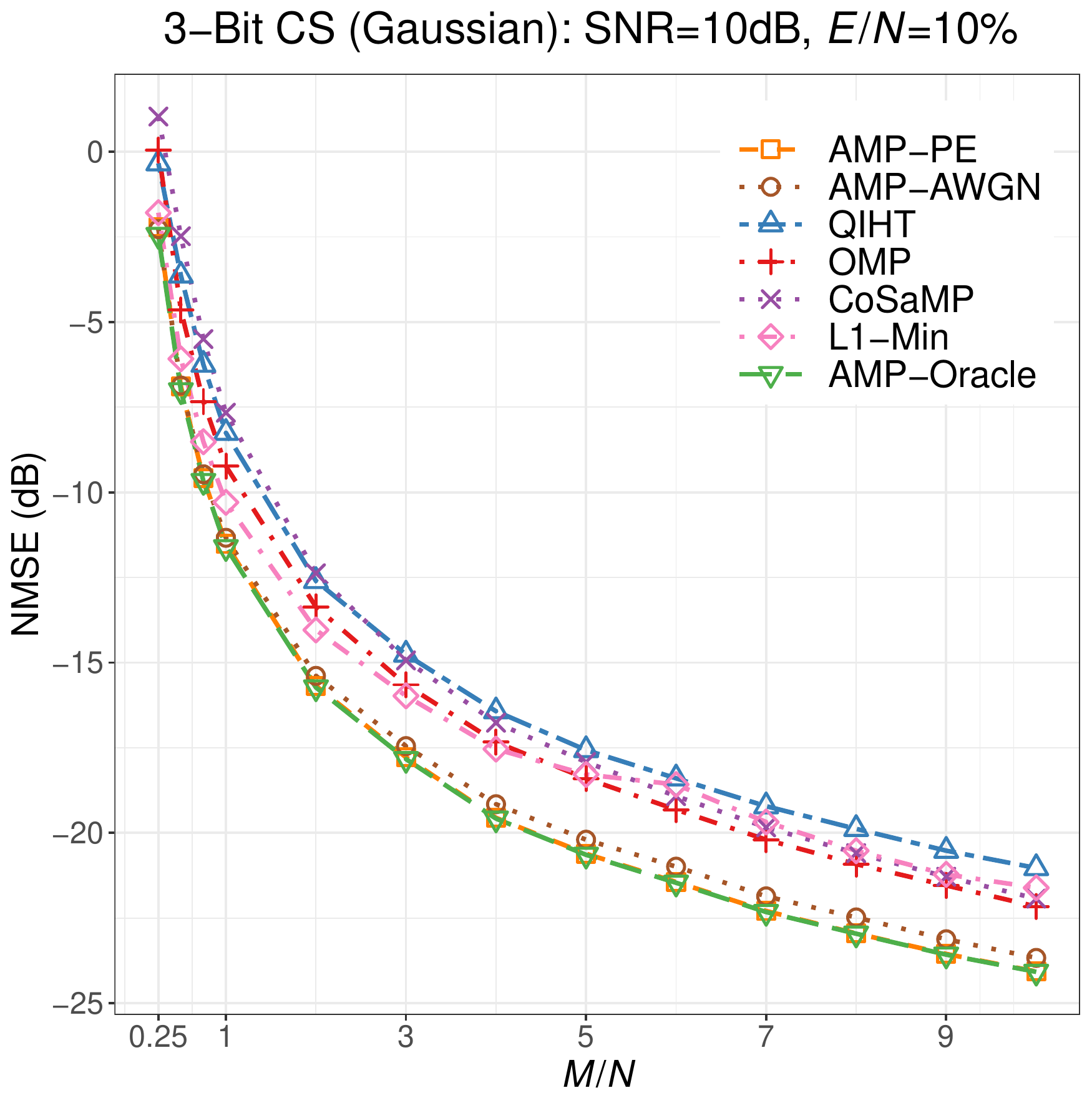}}\\

\subfigure{

\includegraphics[width=0.3\textwidth]{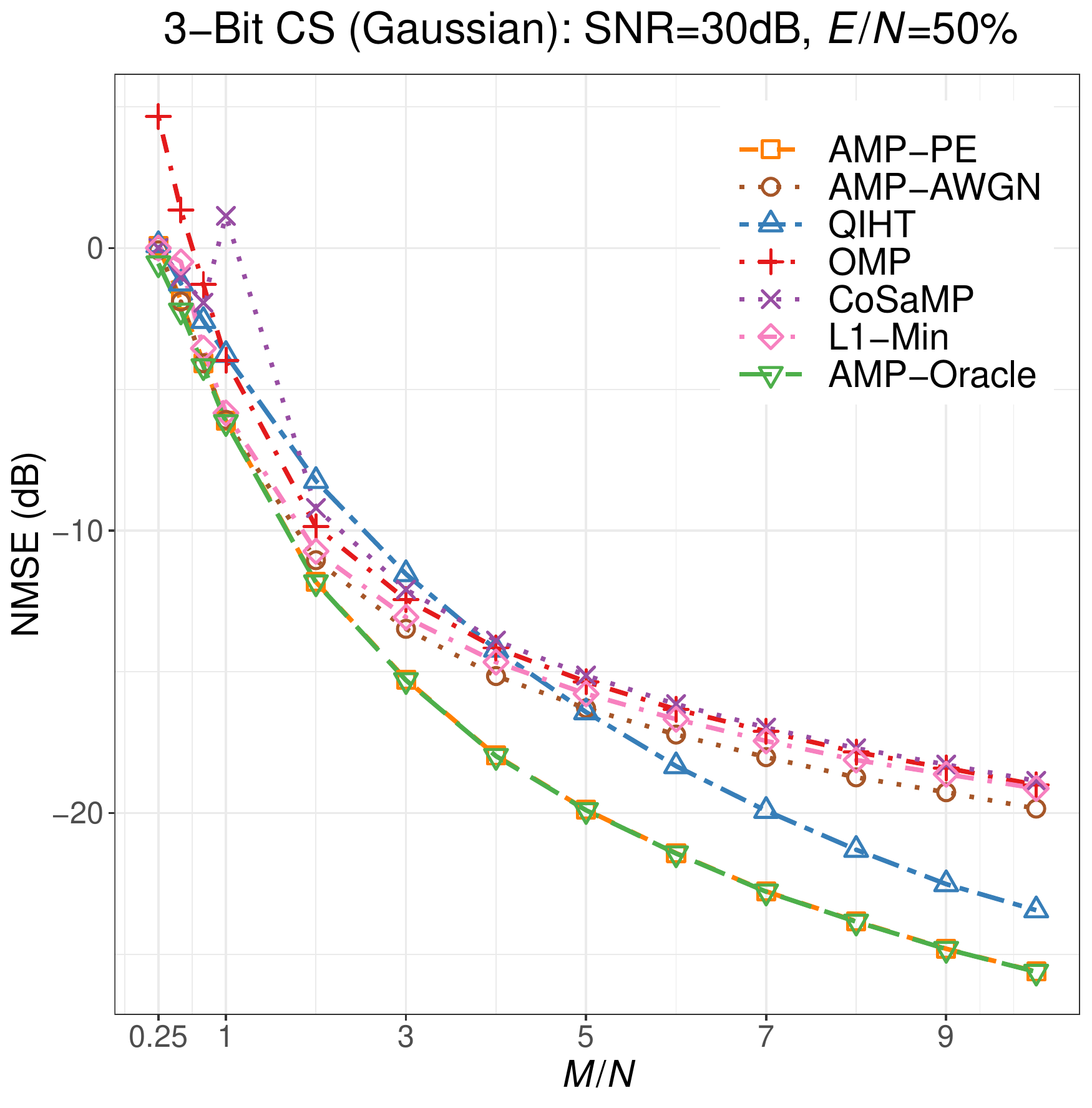}}
\subfigure{

\includegraphics[width=0.3\textwidth]{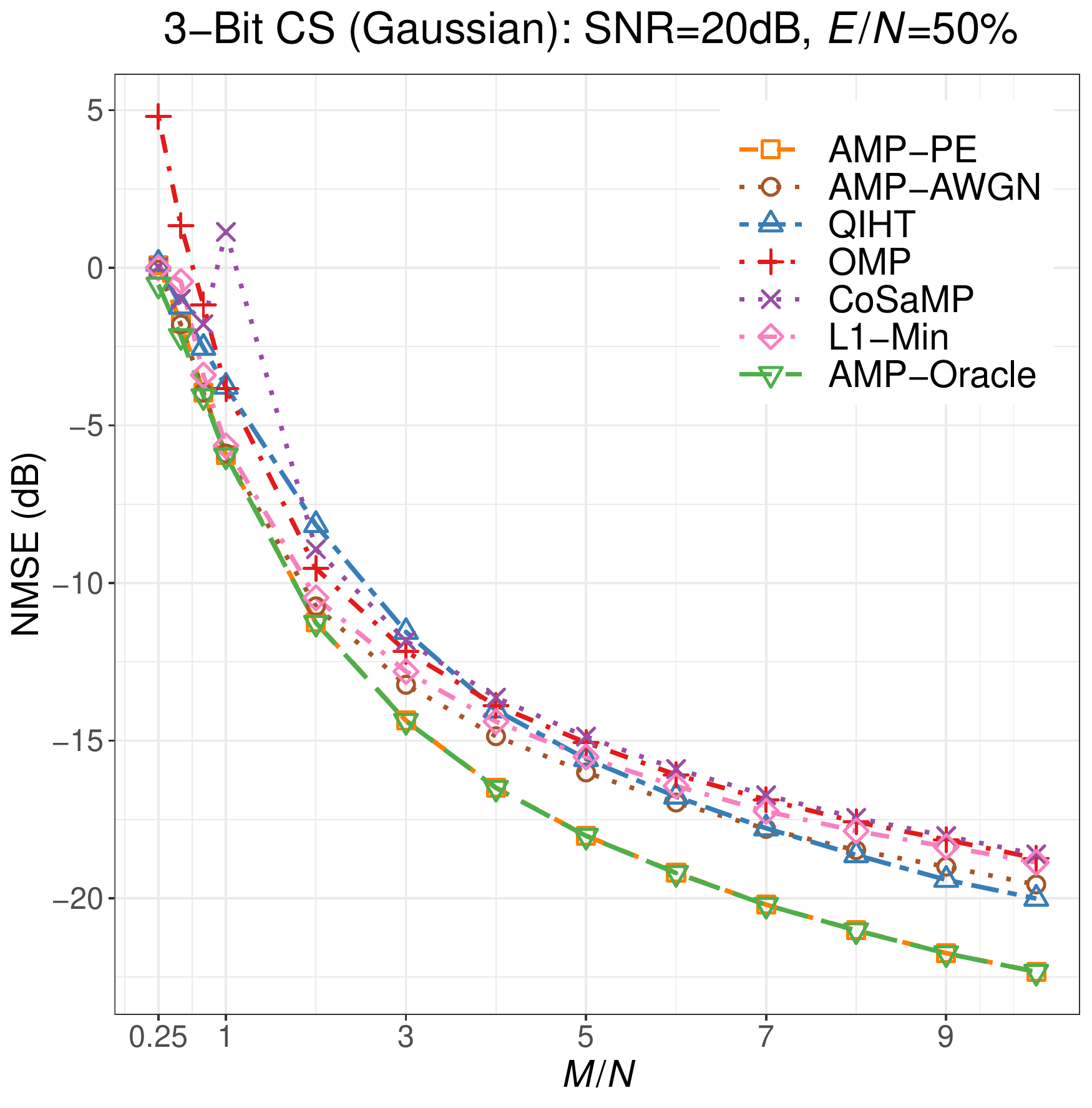}}
\subfigure{

\includegraphics[width=0.3\textwidth]{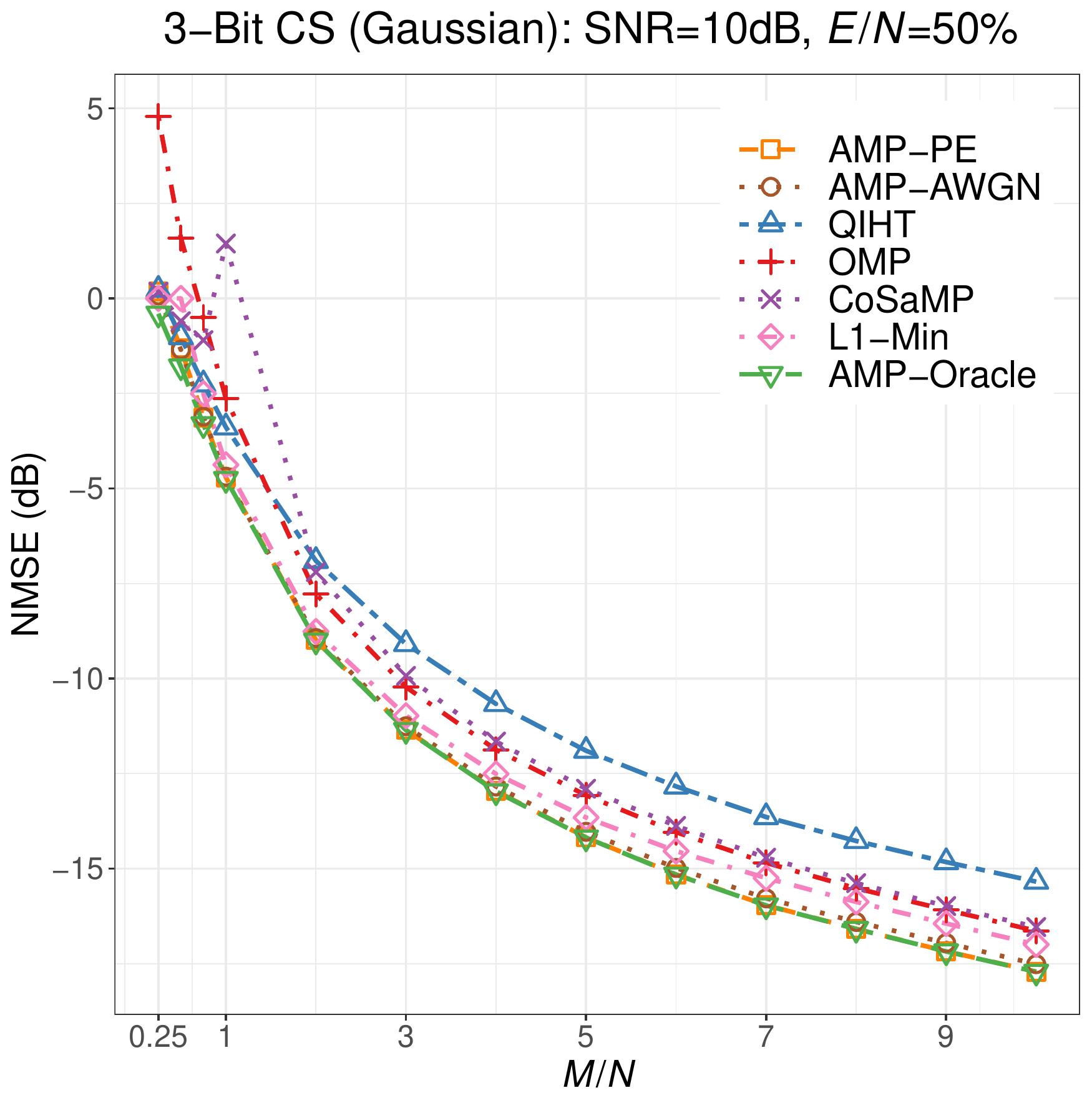}}\\

\subfigure{

\includegraphics[width=0.3\textwidth]{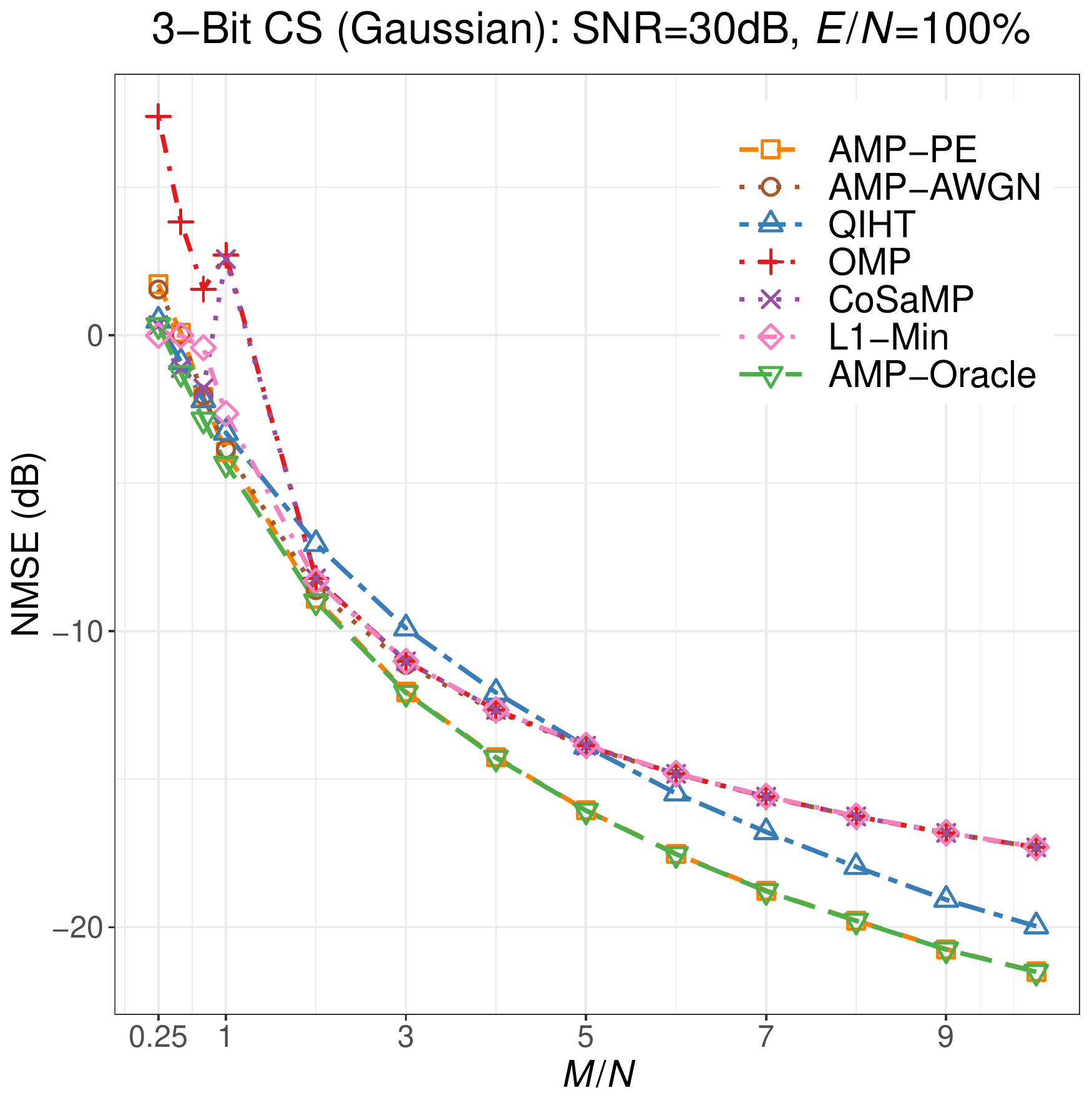}}
\subfigure{

\includegraphics[width=0.3\textwidth]{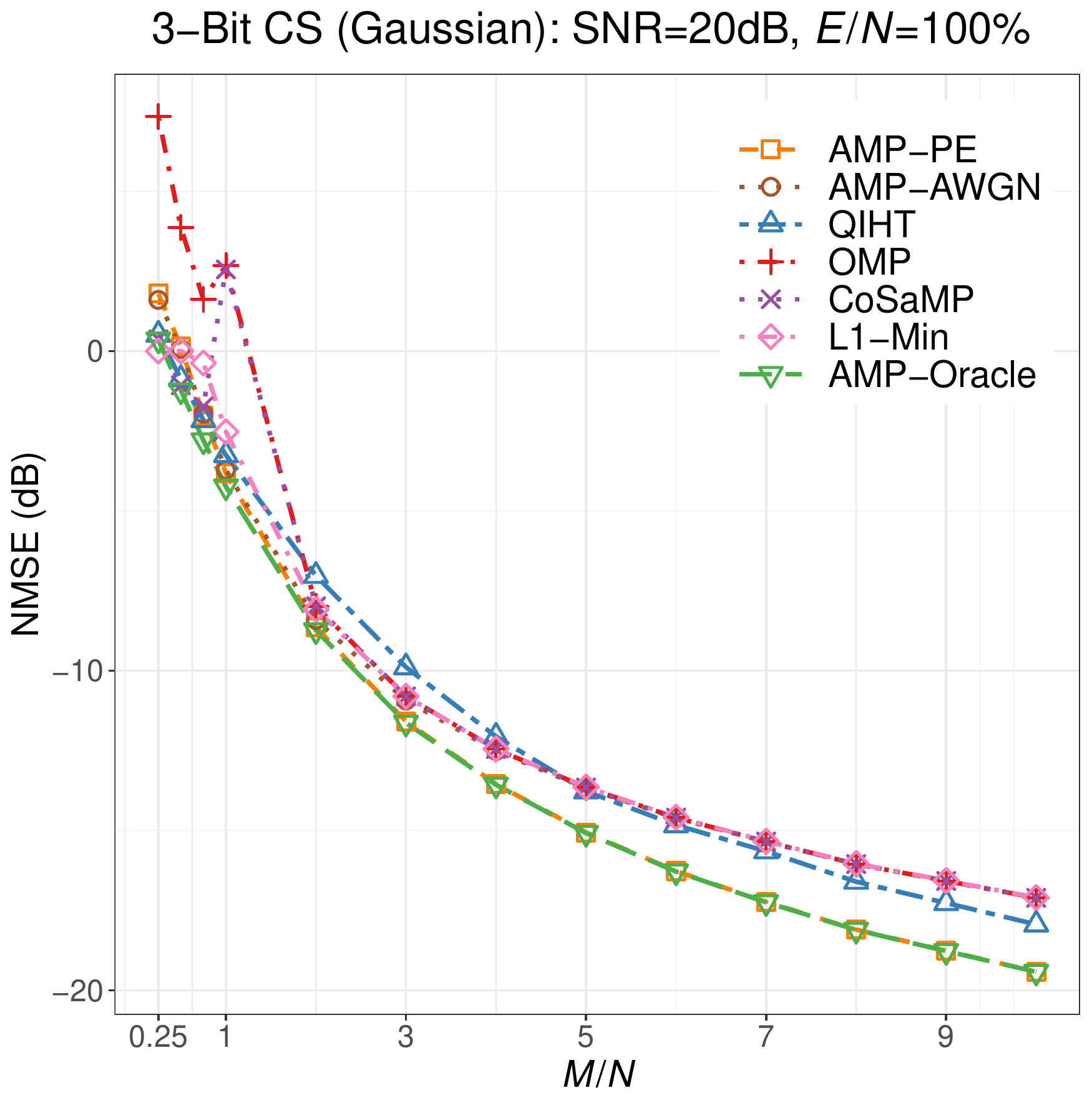}}
\subfigure{

\includegraphics[width=0.3\textwidth]{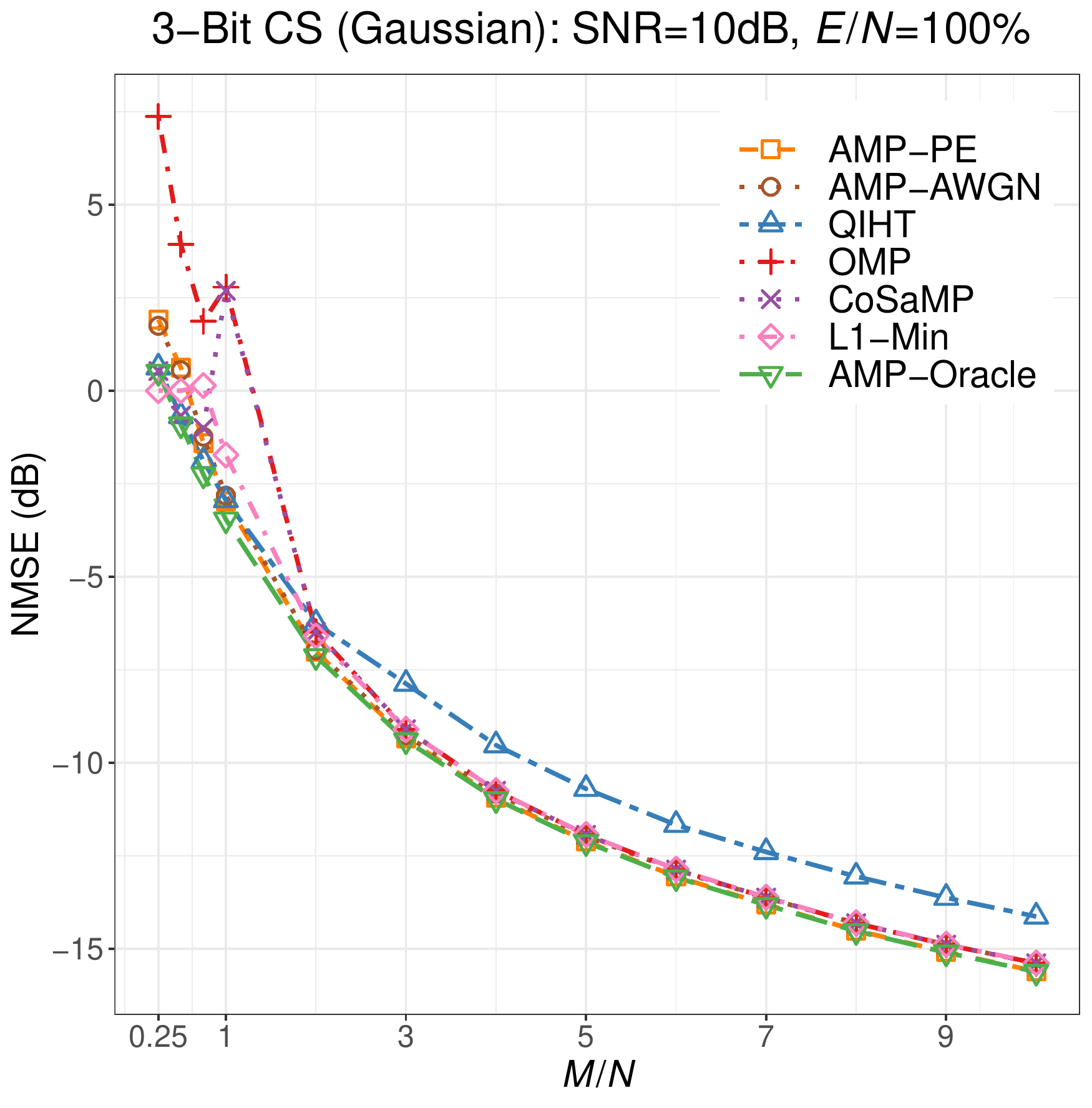}}

\caption{Comparison of different approaches in solving 3-bit CS. Nonzero entries of the signal follow the Gausian distribution. The sampling ratio $\frac{M}{N}\in\{0.25,\cdots,10\}$ and the sparsity level $\frac{E}{N}=\in\{10\%,50\%,100\%\}$. The pre-quantization SNR varies from $30$dB, $20$dB to $10$dB.}

\label{fig:3bit_experiments_gaussian}
\end{figure*}

\newpage
\subsection{Signal with the Bernoulli-Cauchy Mixture Prior}

\begin{figure*}[htbp]
\centering
\subfigure{

\includegraphics[width=0.3\textwidth]{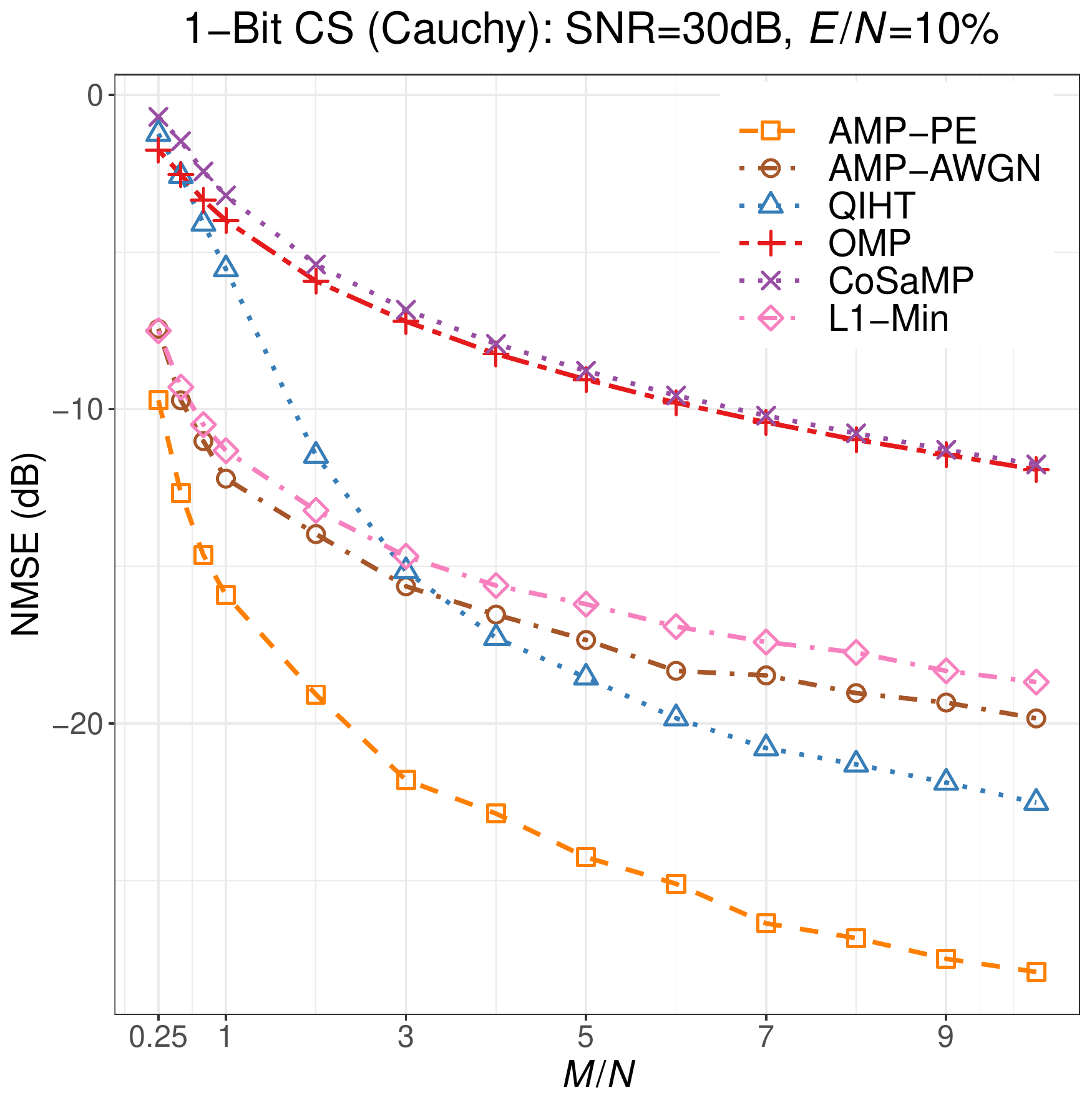}}
\subfigure{

\includegraphics[width=0.3\textwidth]{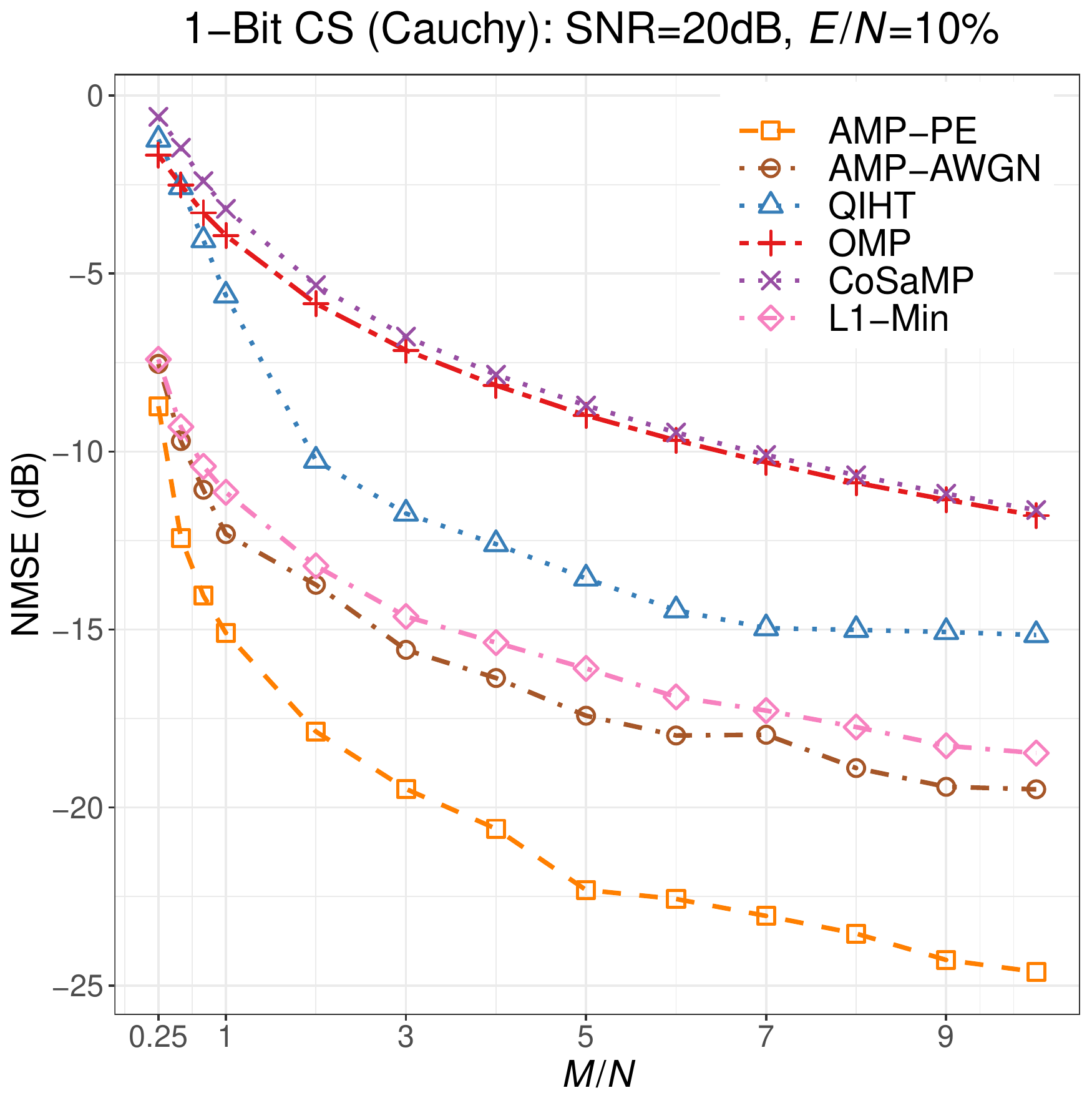}}
\subfigure{

\includegraphics[width=0.3\textwidth]{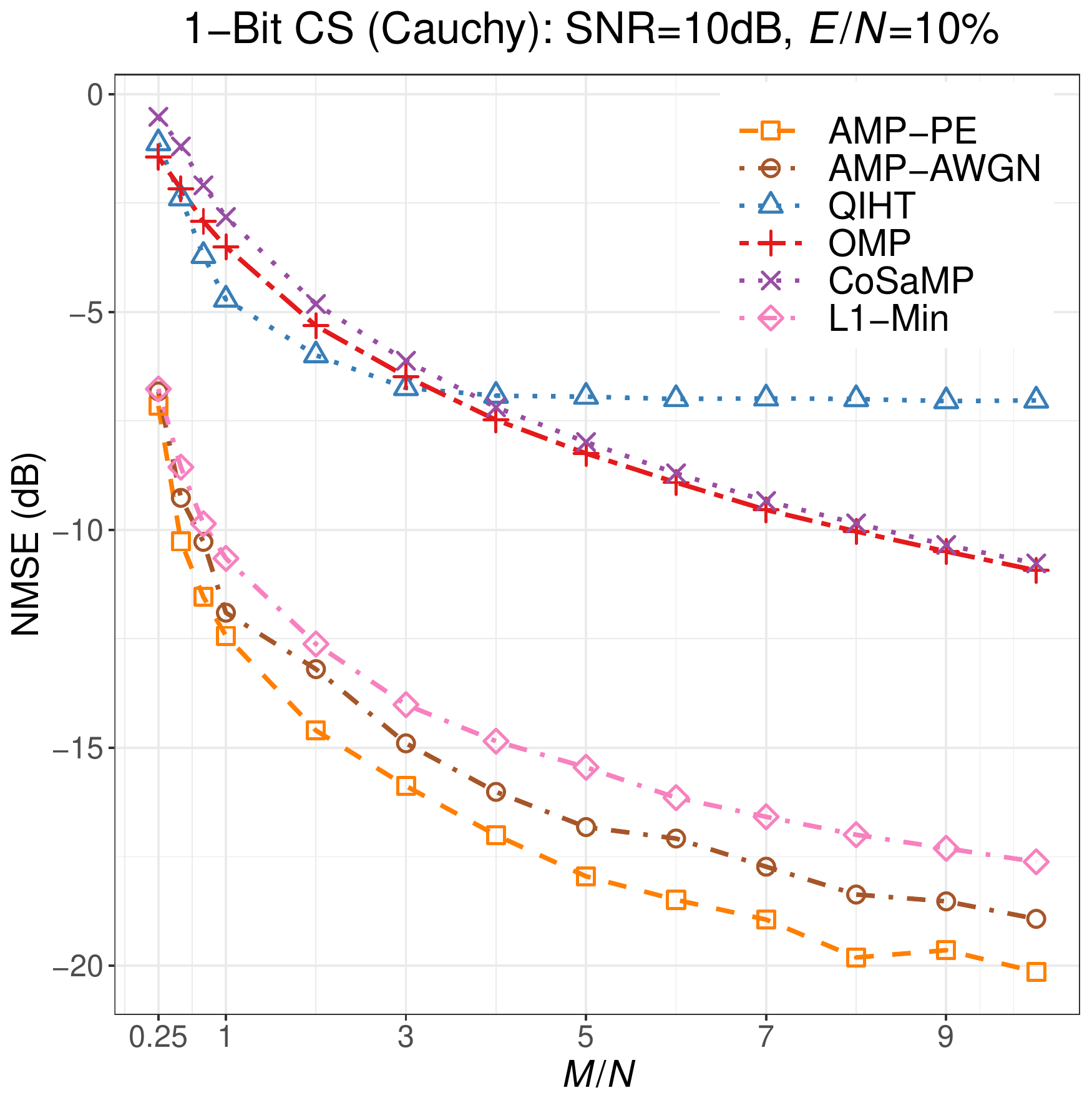}}\\

\subfigure{

\includegraphics[width=0.3\textwidth]{figures/1bit_s50_n30_compare_full_cauchy.pdf}}
\subfigure{

\includegraphics[width=0.3\textwidth]{figures/1bit_s50_n20_compare_full_cauchy.pdf}}
\subfigure{

\includegraphics[width=0.3\textwidth]{figures/1bit_s50_n10_compare_full_cauchy.pdf}}\\

\subfigure{

\includegraphics[width=0.3\textwidth]{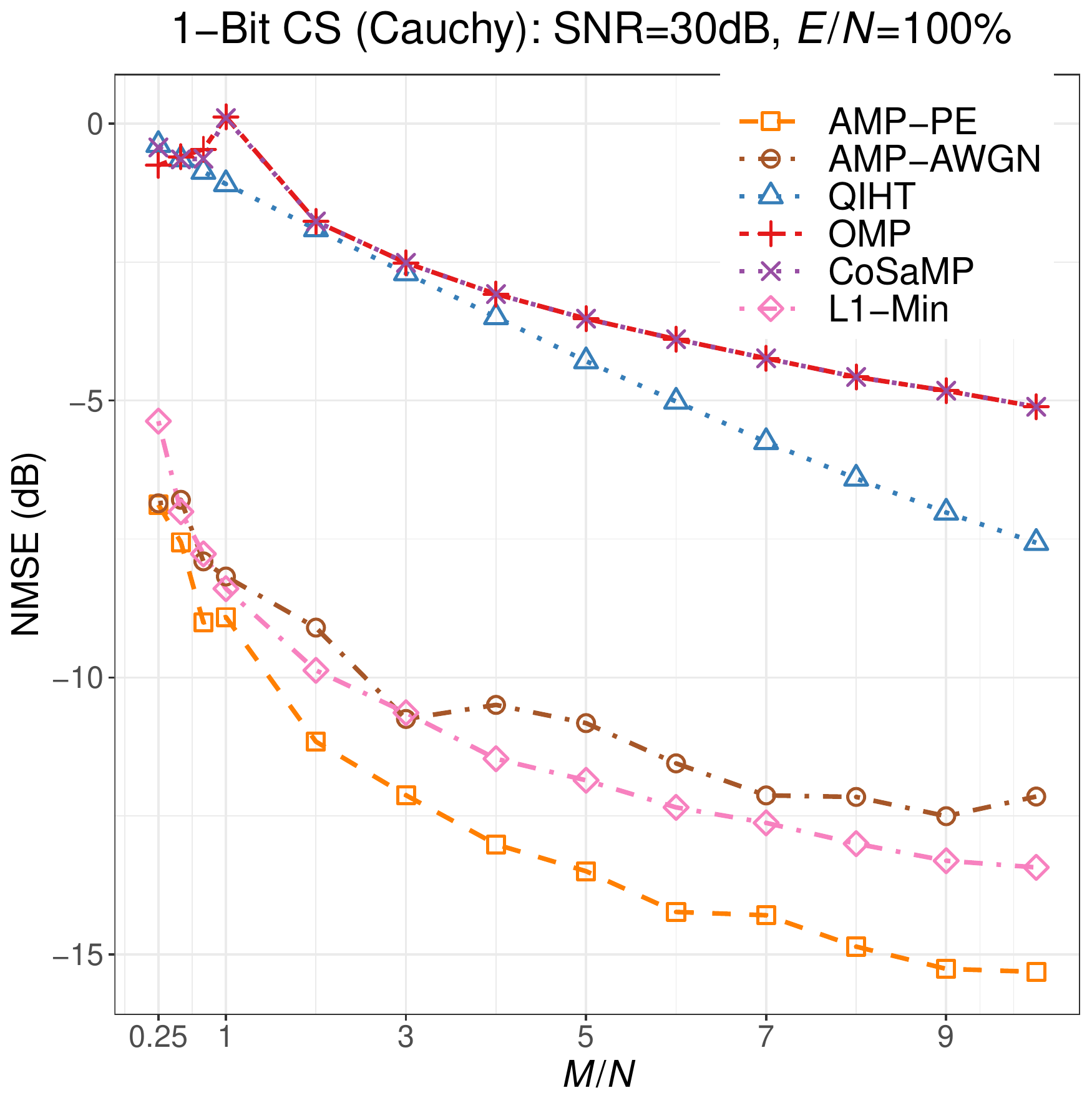}}
\subfigure{

\includegraphics[width=0.3\textwidth]{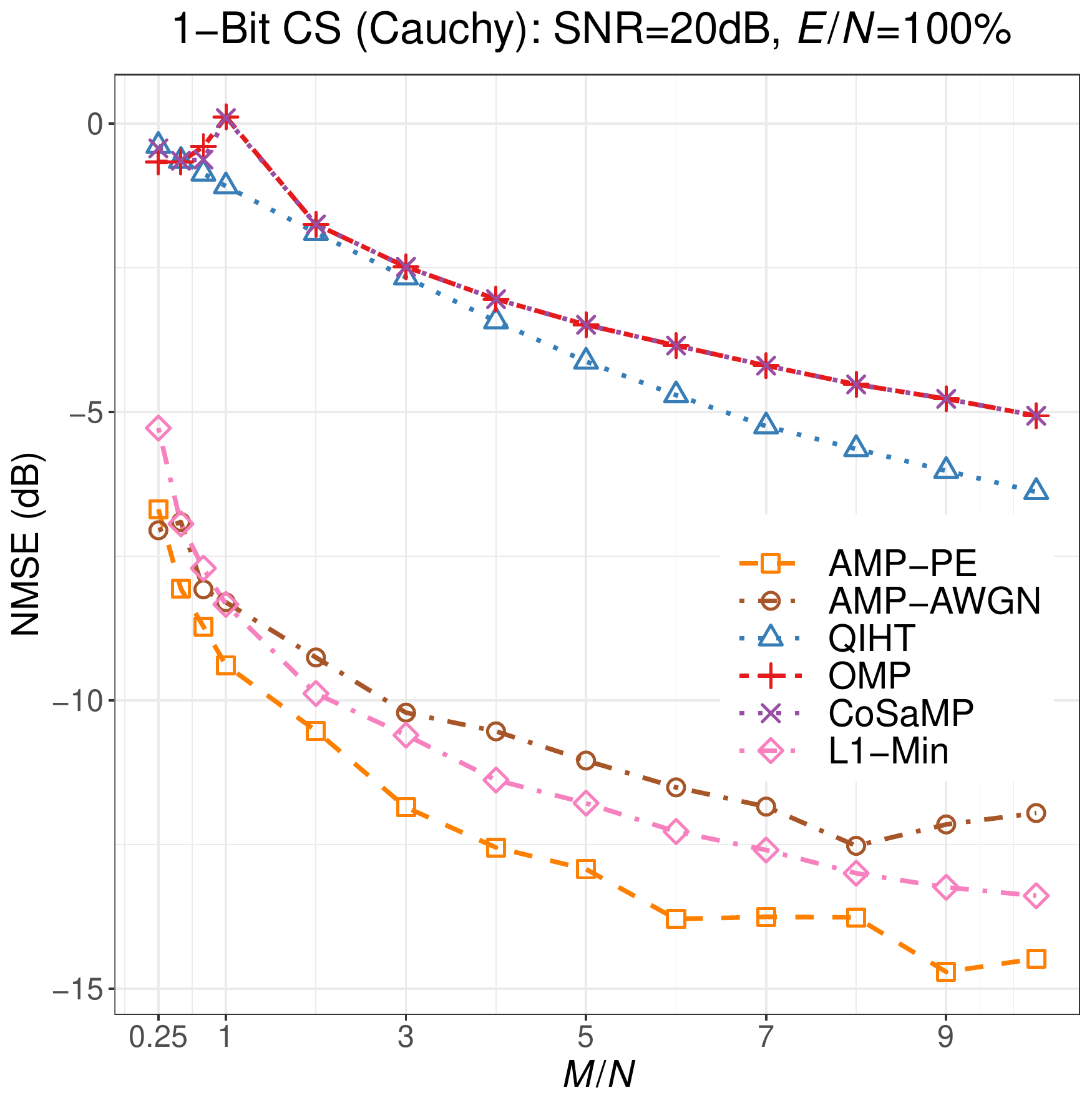}}
\subfigure{

\includegraphics[width=0.3\textwidth]{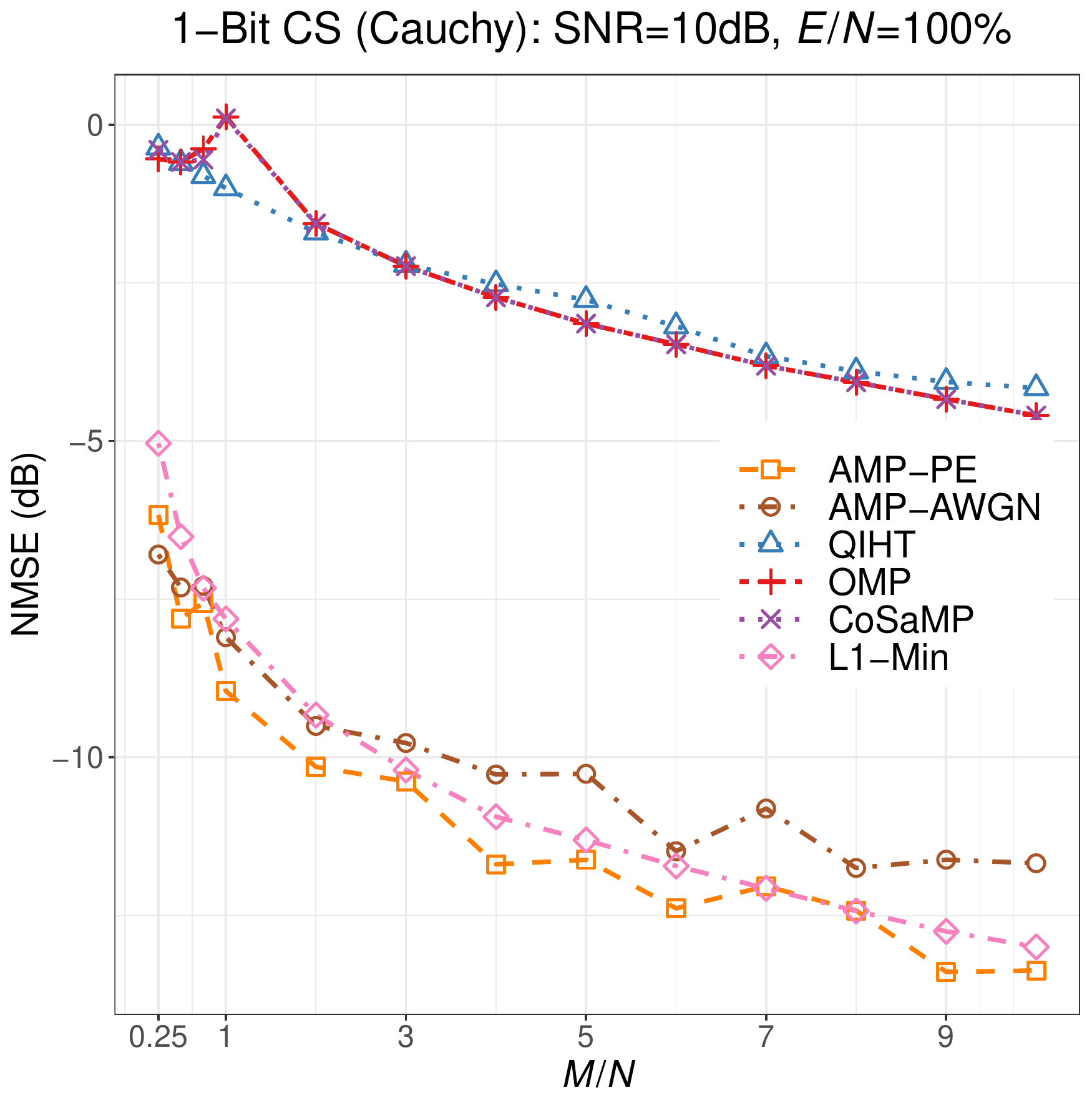}}

\caption{Comparison of different approaches in solving 1-bit CS. Nonzero entries of the signal follow the Cauchy distribution. The sampling ratio $\frac{M}{N}\in\{0.25,\cdots,10\}$ and the sparsity level $\frac{E}{N}=\in\{10\%,50\%,100\%\}$. The pre-quantization SNR varies from $30$dB, $20$dB to $10$dB.}

\label{fig:1bit_experiments_cauchy}
\end{figure*}

\newpage
\begin{figure*}[htbp]
\centering
\subfigure{

\includegraphics[width=0.3\textwidth]{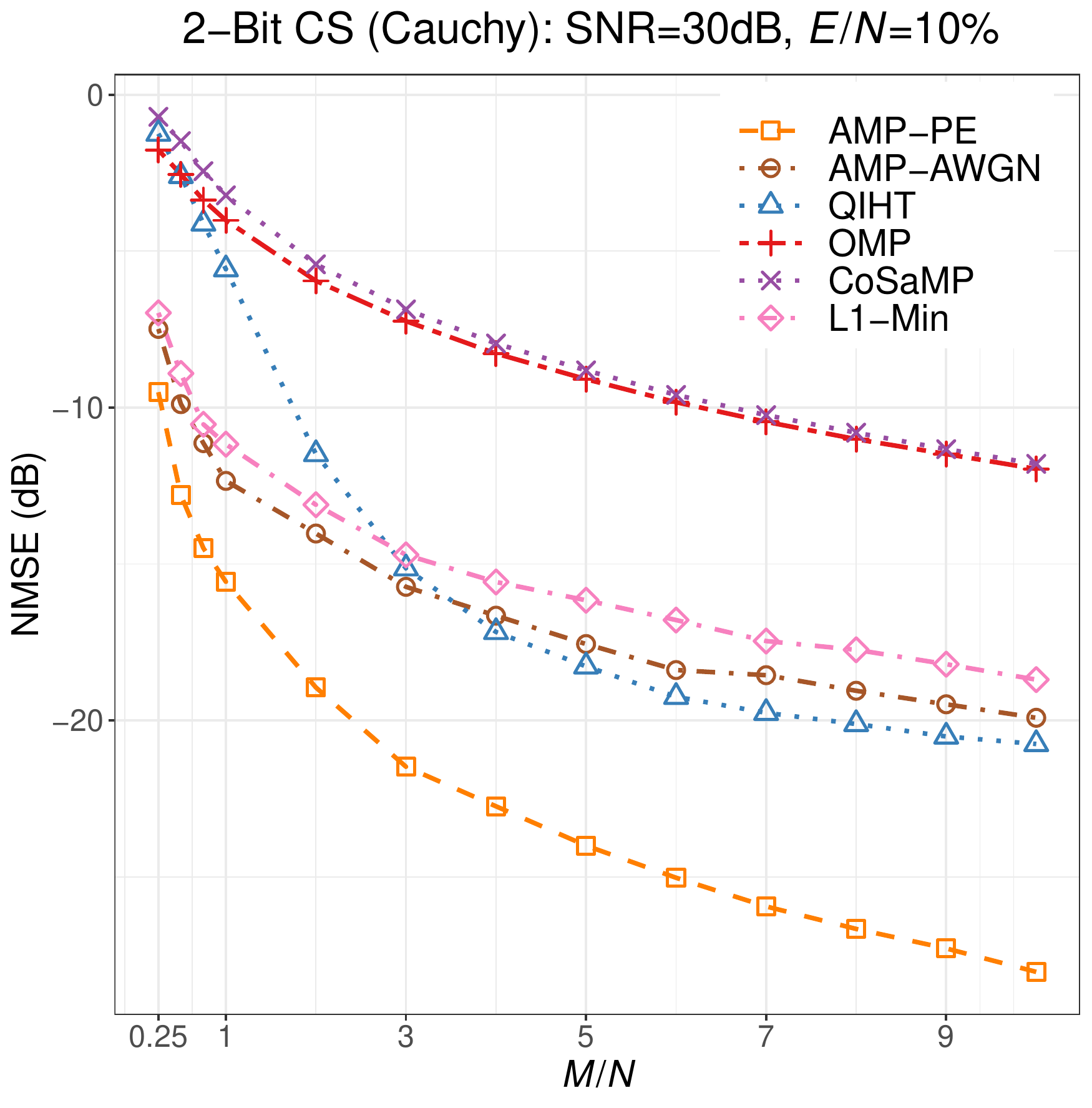}}
\subfigure{

\includegraphics[width=0.3\textwidth]{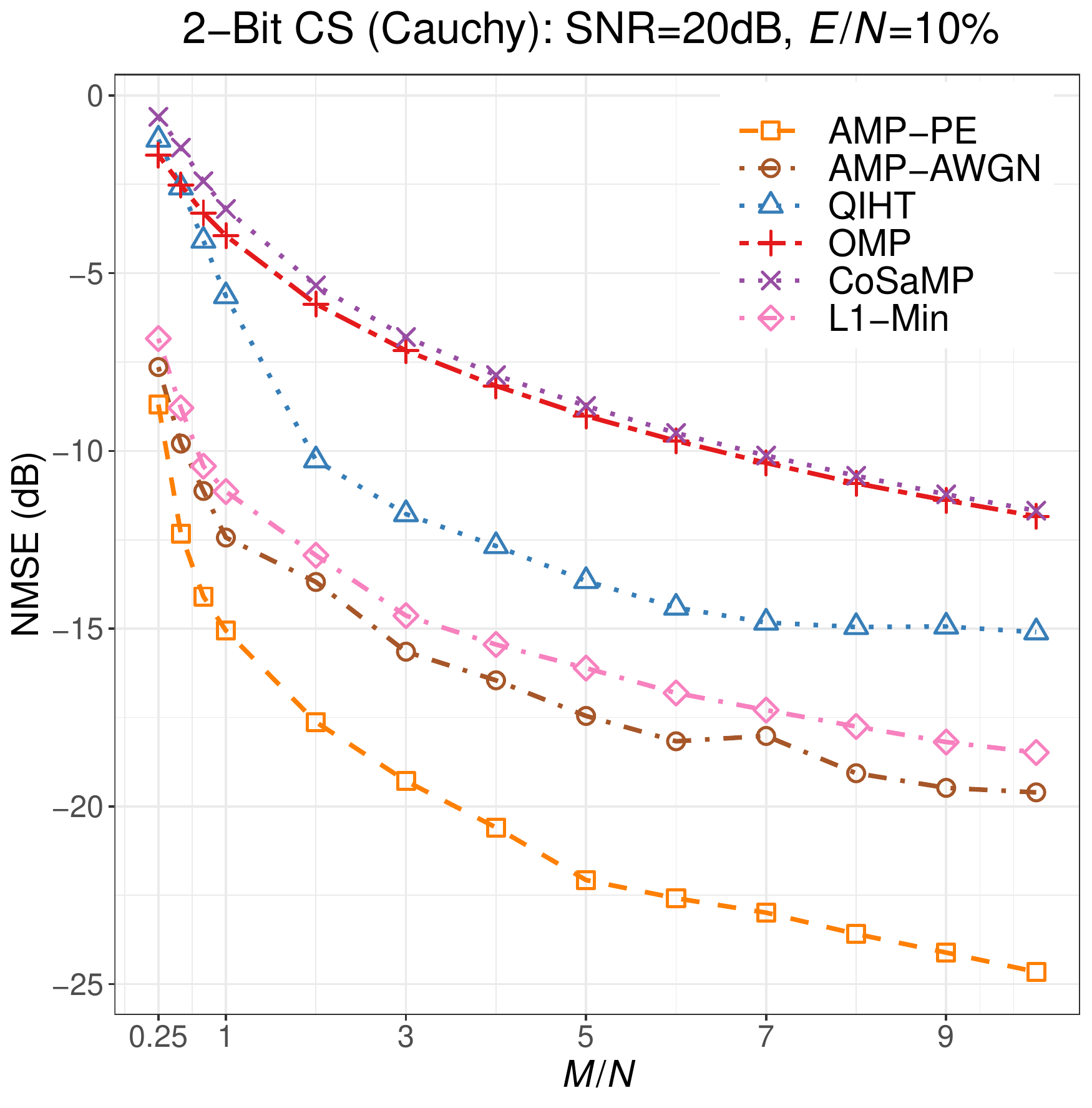}}
\subfigure{

\includegraphics[width=0.3\textwidth]{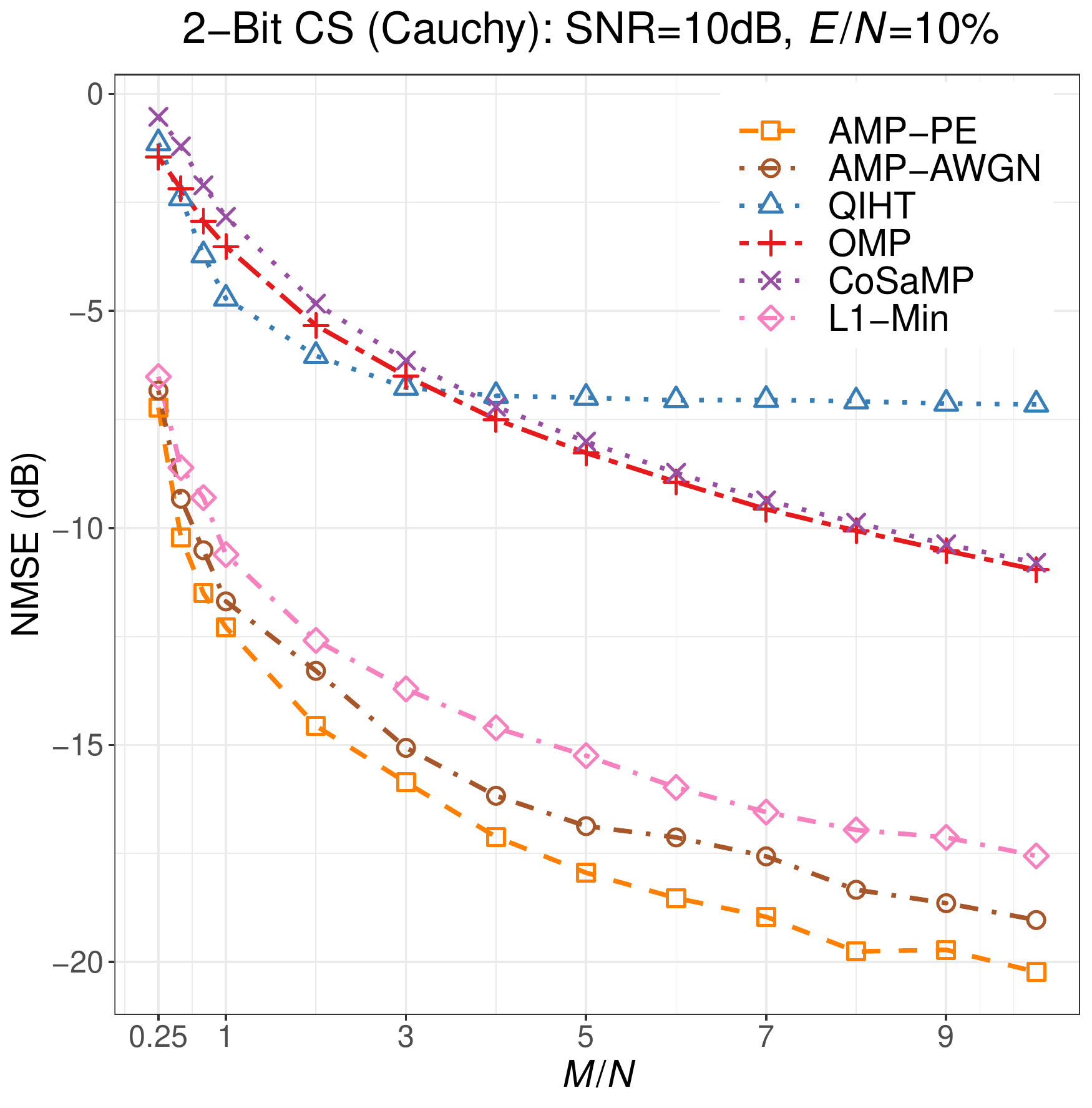}}\\

\subfigure{

\includegraphics[width=0.3\textwidth]{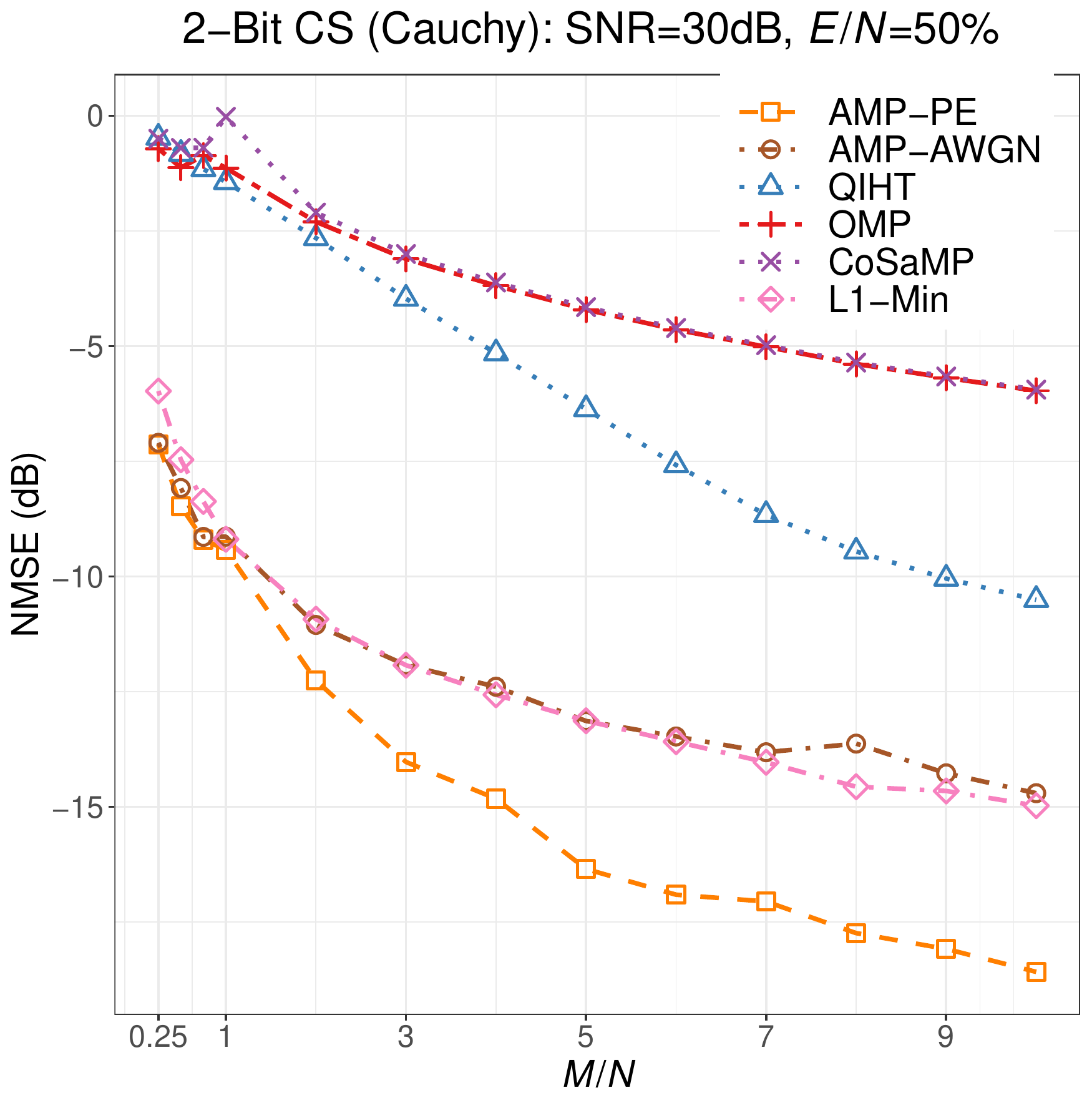}}
\subfigure{

\includegraphics[width=0.3\textwidth]{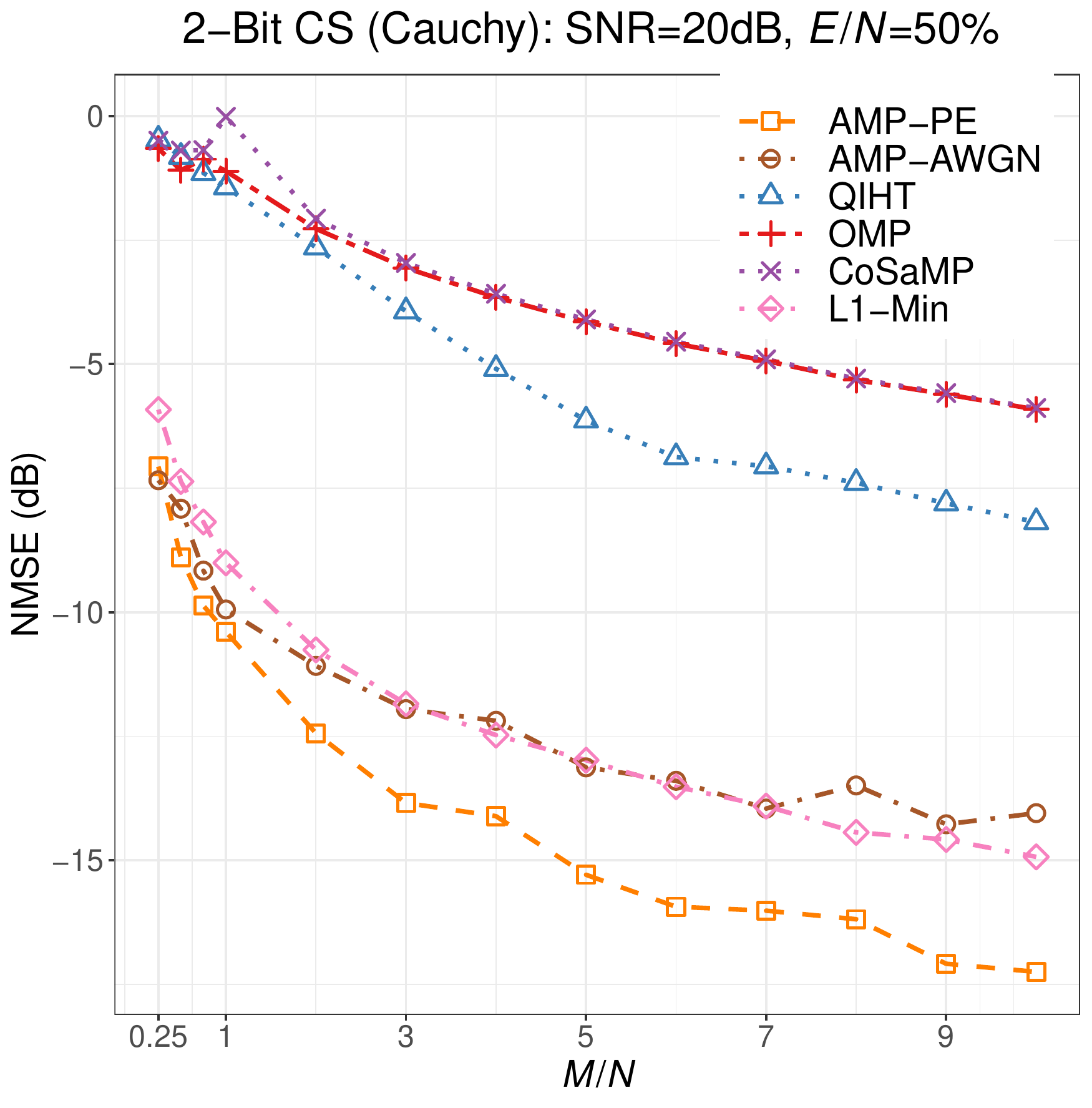}}
\subfigure{

\includegraphics[width=0.3\textwidth]{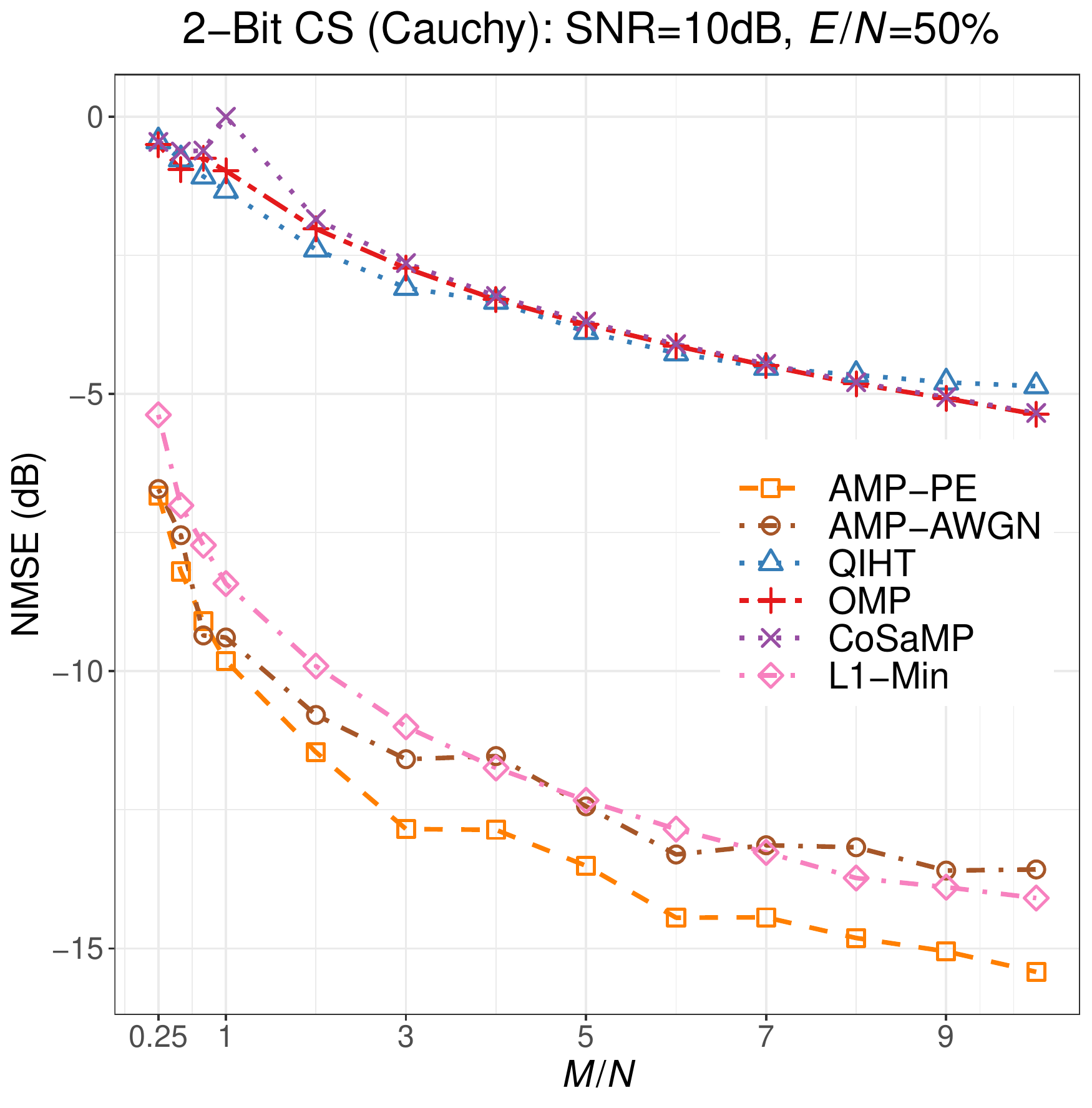}}\\

\subfigure{

\includegraphics[width=0.3\textwidth]{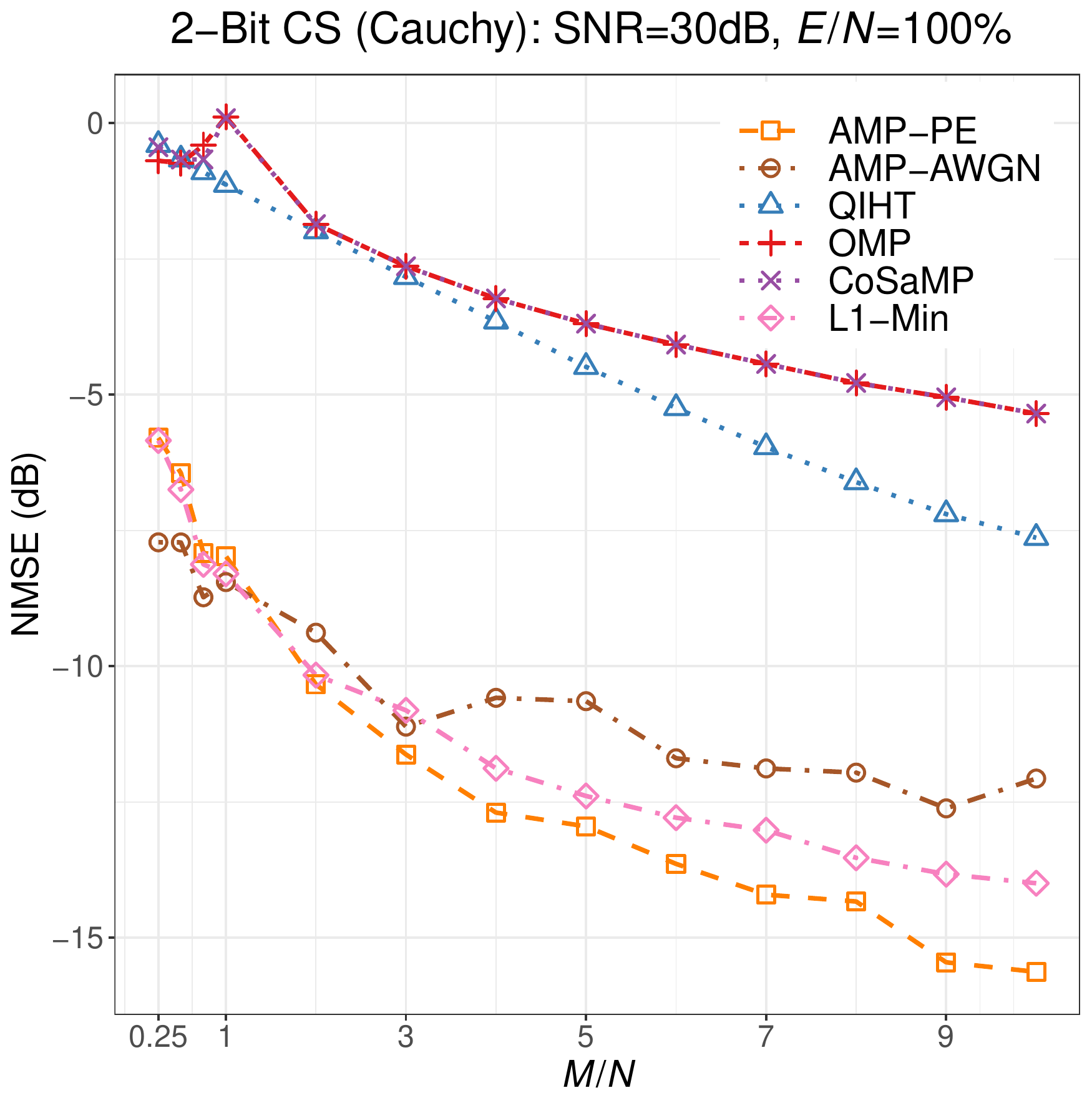}}
\subfigure{

\includegraphics[width=0.3\textwidth]{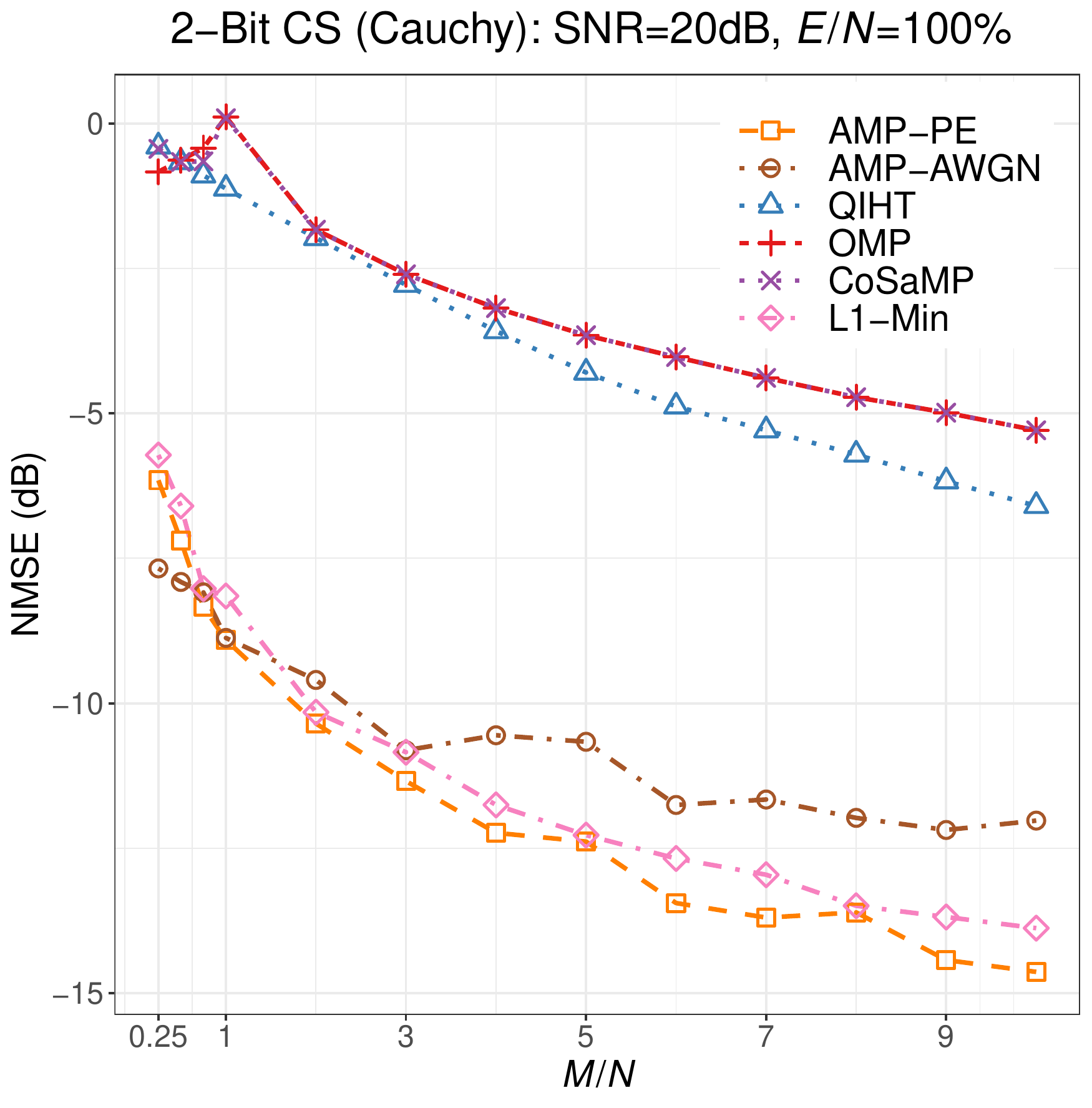}}
\subfigure{

\includegraphics[width=0.3\textwidth]{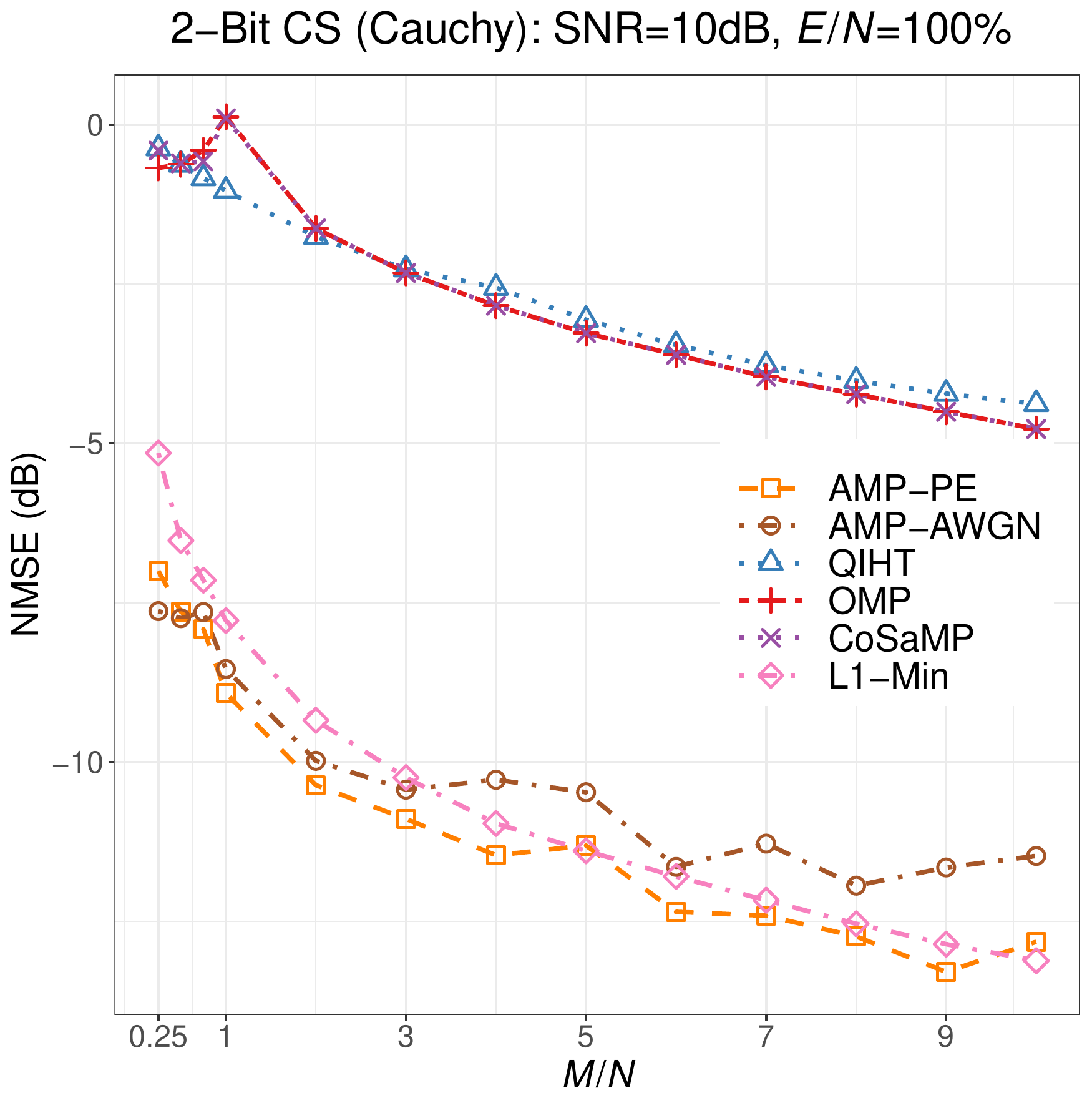}}

\caption{Comparison of different approaches in solving 2-bit CS. Nonzero entries of the signal follow the Cauchy distribution. The sampling ratio $\frac{M}{N}\in\{0.25,\cdots,10\}$ and the sparsity level $\frac{E}{N}=\in\{10\%,50\%,100\%\}$. The pre-quantization SNR varies from $30$dB, $20$dB to $10$dB.}

\label{fig:2bit_experiments_cauchy}
\end{figure*}

\newpage
\begin{figure*}[htbp]
\centering
\subfigure{

\includegraphics[width=0.3\textwidth]{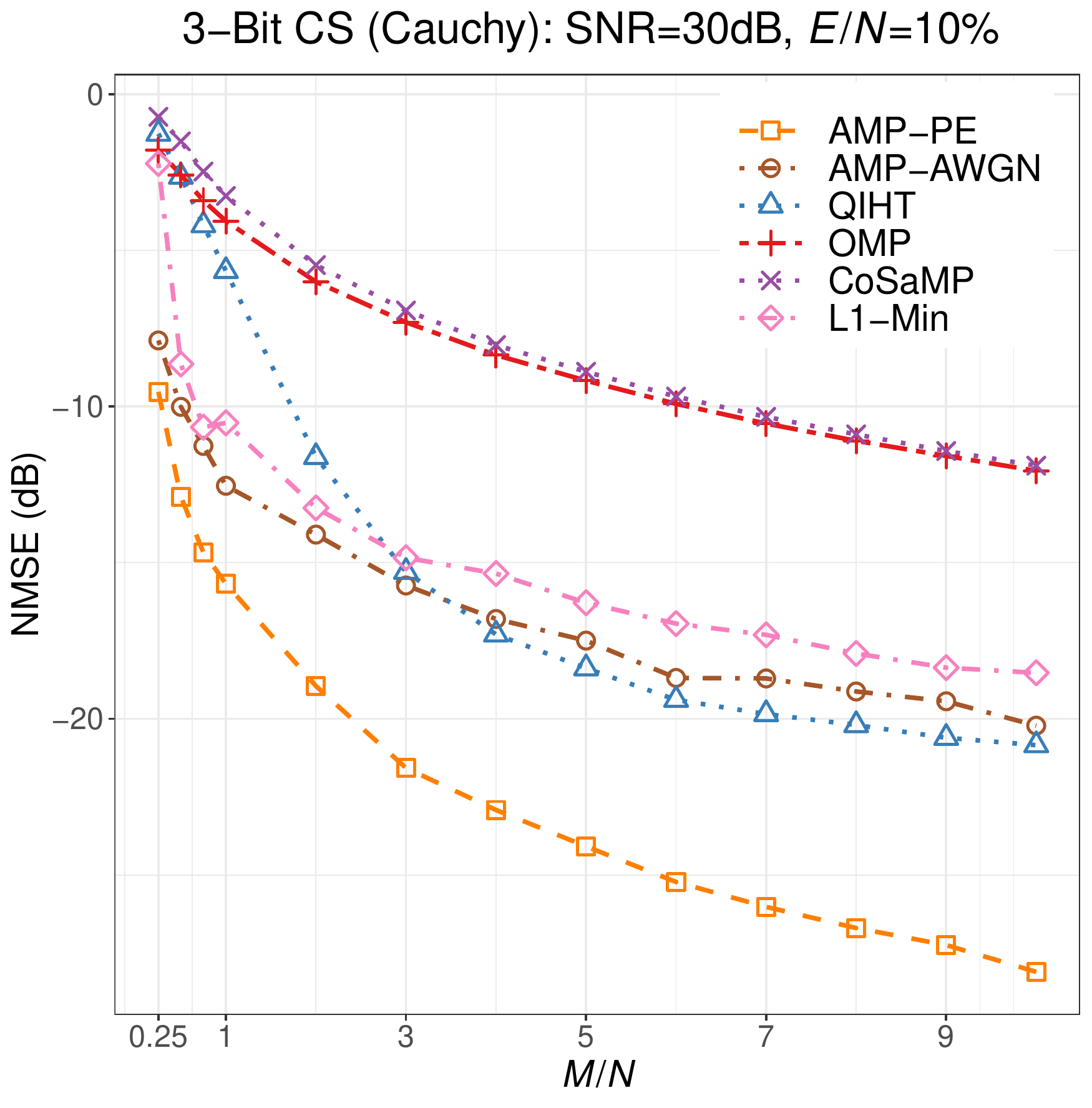}}
\subfigure{

\includegraphics[width=0.3\textwidth]{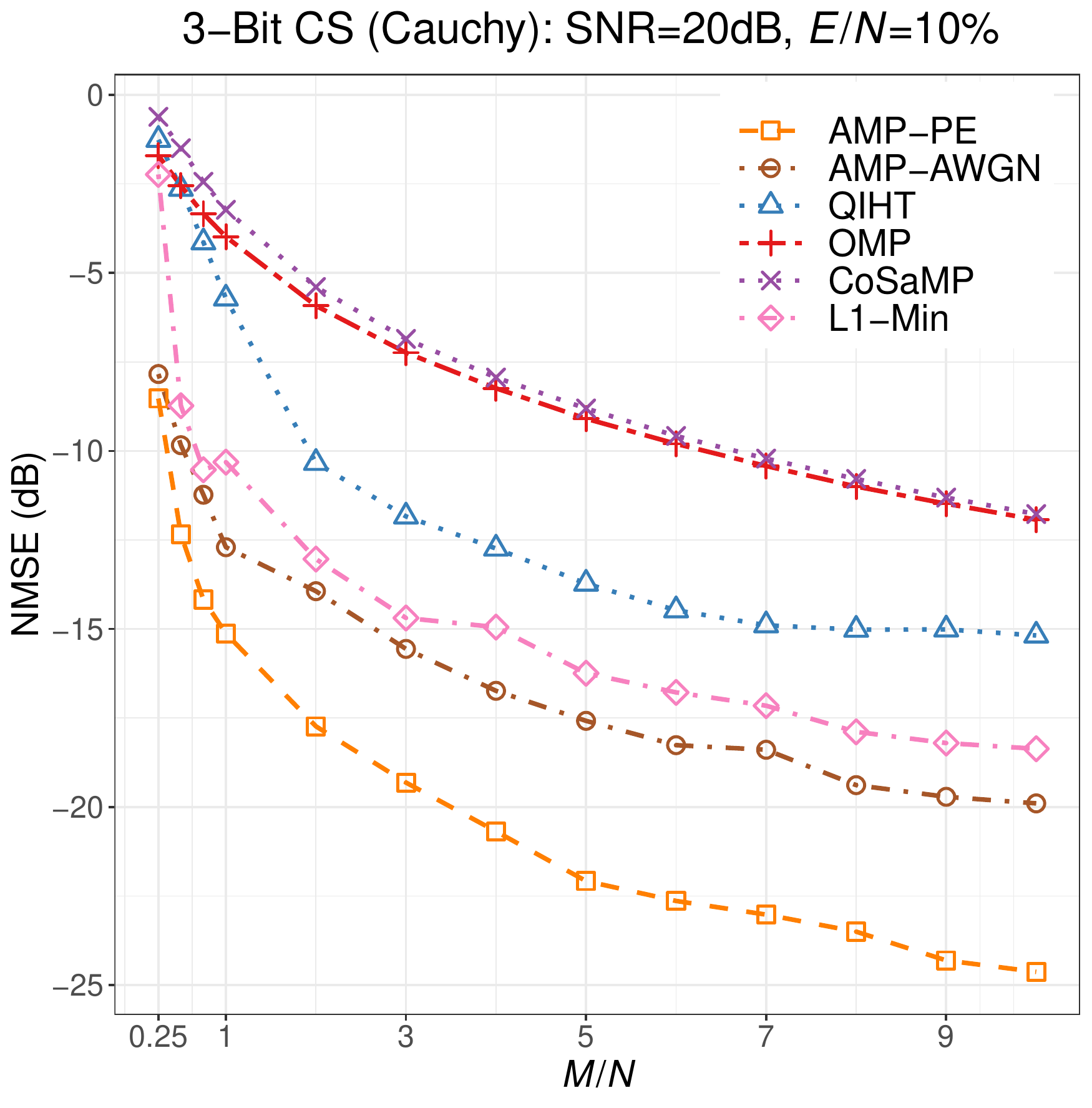}}
\subfigure{

\includegraphics[width=0.3\textwidth]{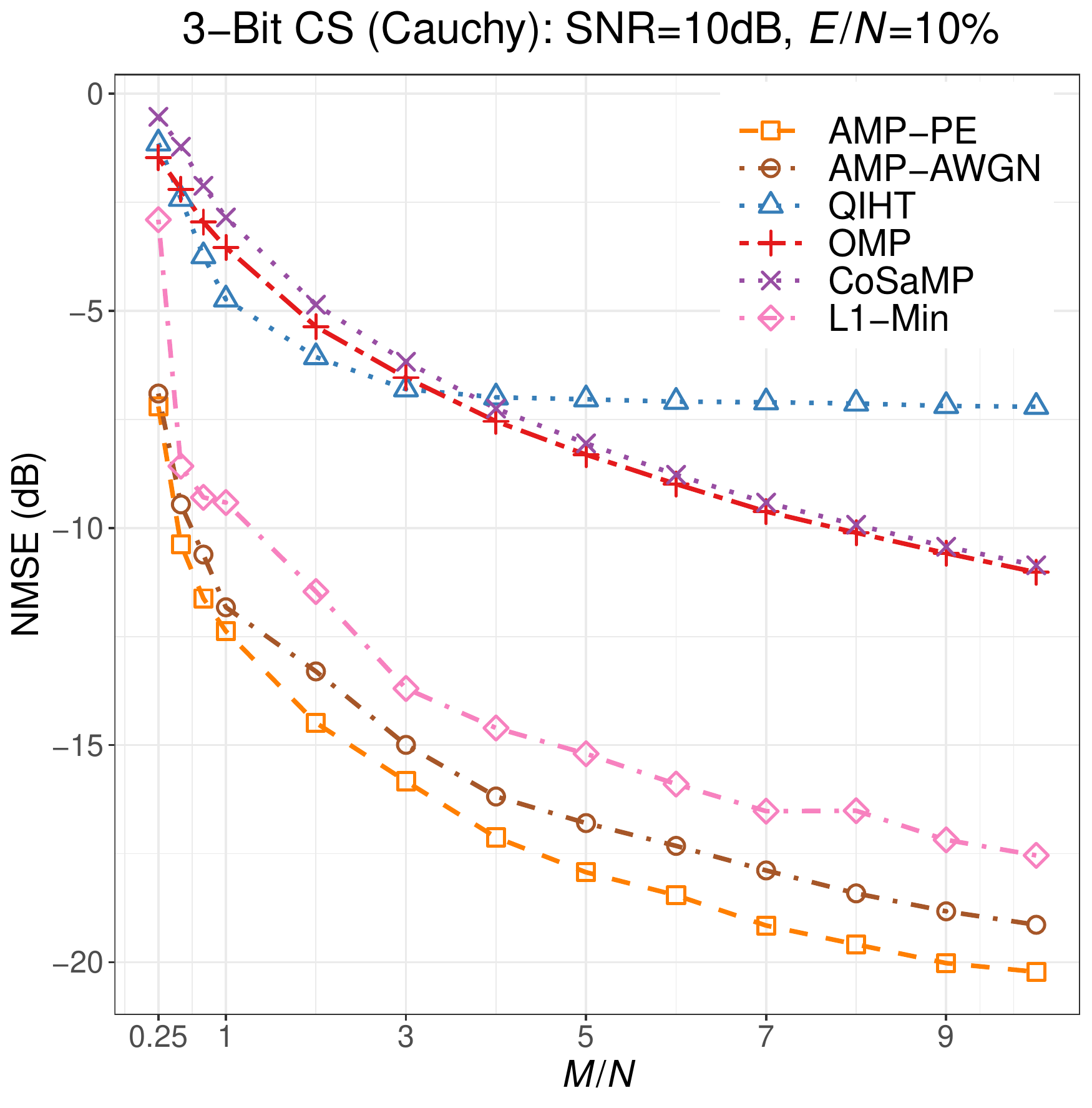}}\\

\subfigure{

\includegraphics[width=0.3\textwidth]{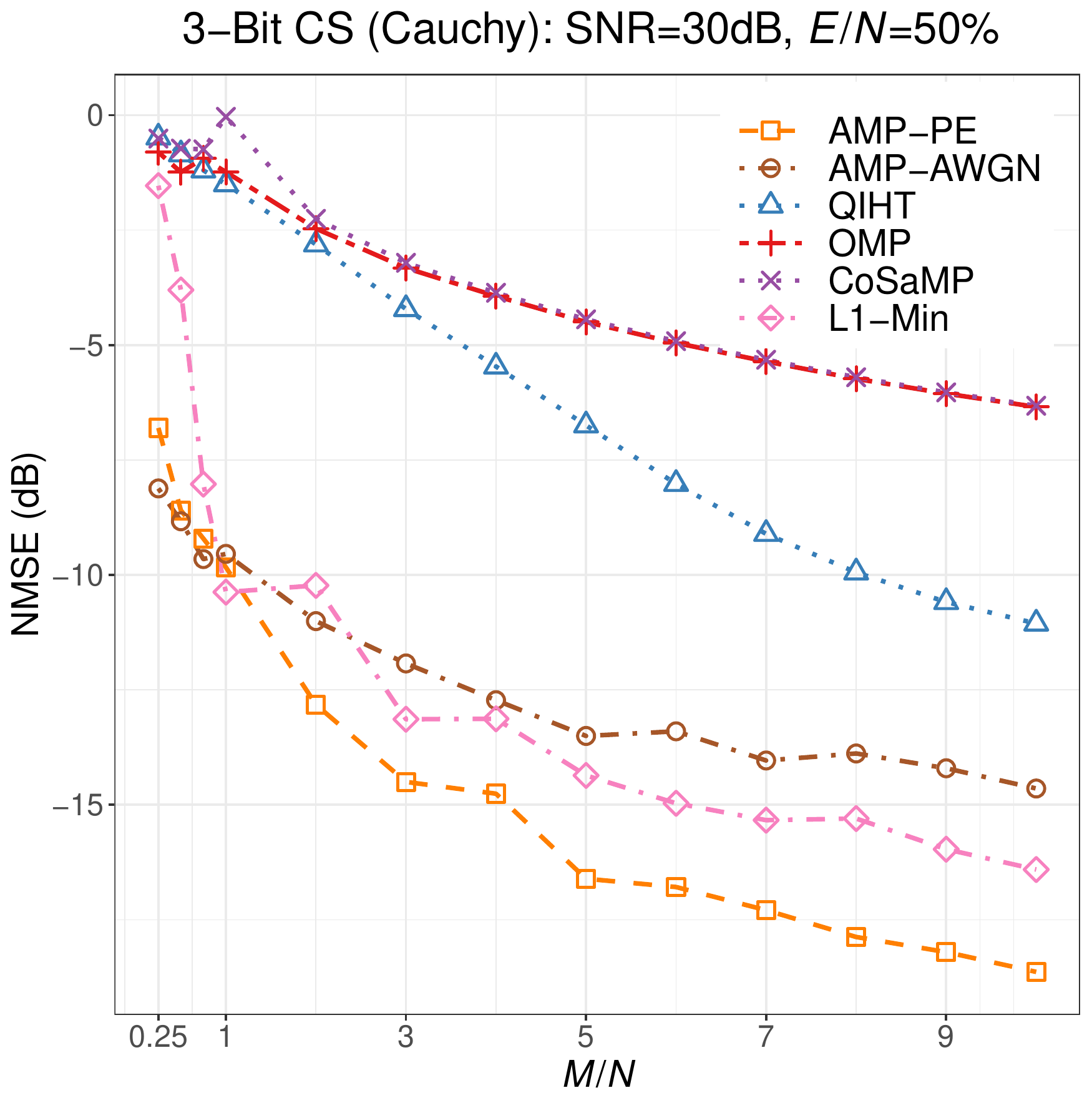}}
\subfigure{

\includegraphics[width=0.3\textwidth]{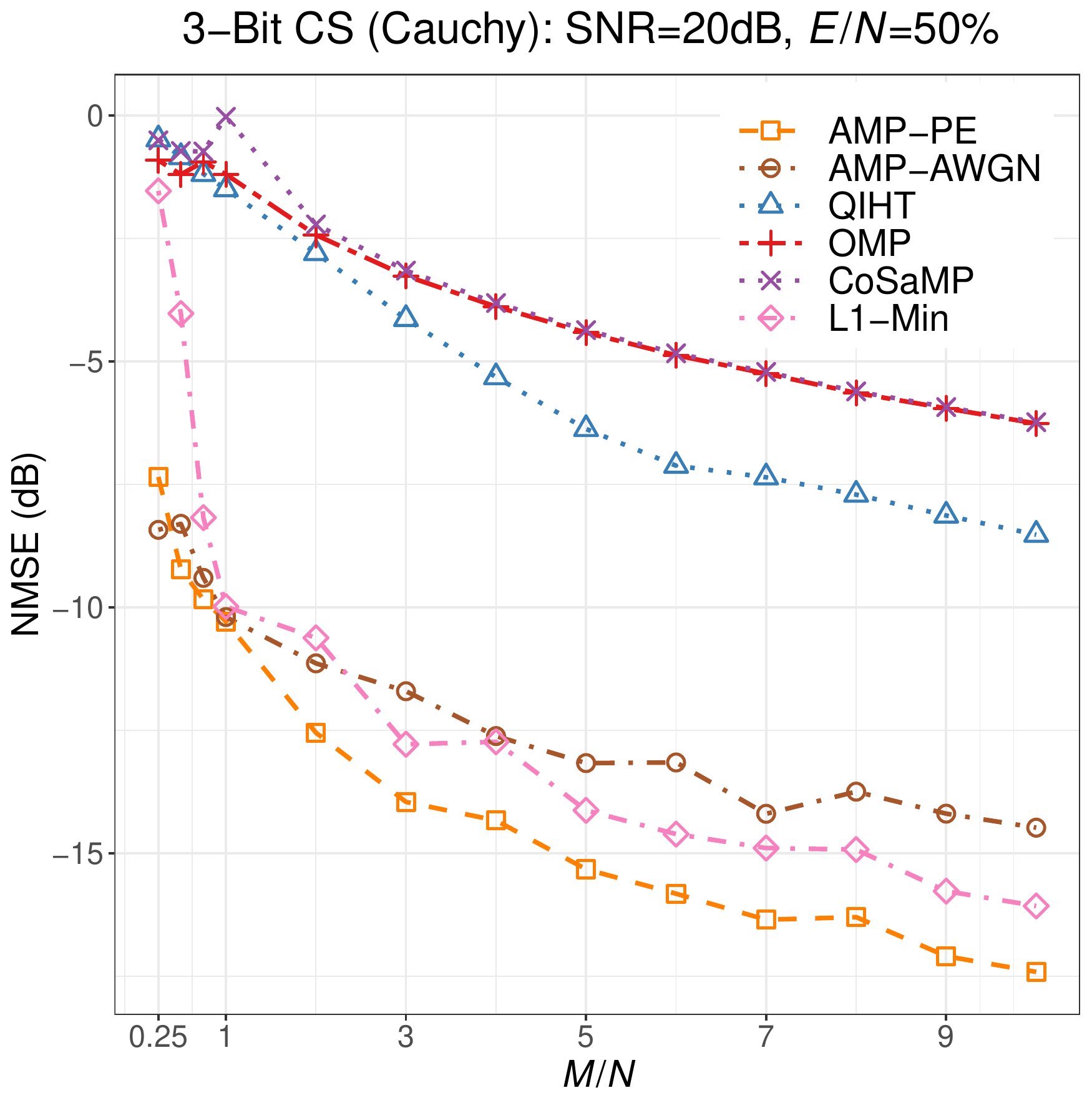}}
\subfigure{

\includegraphics[width=0.3\textwidth]{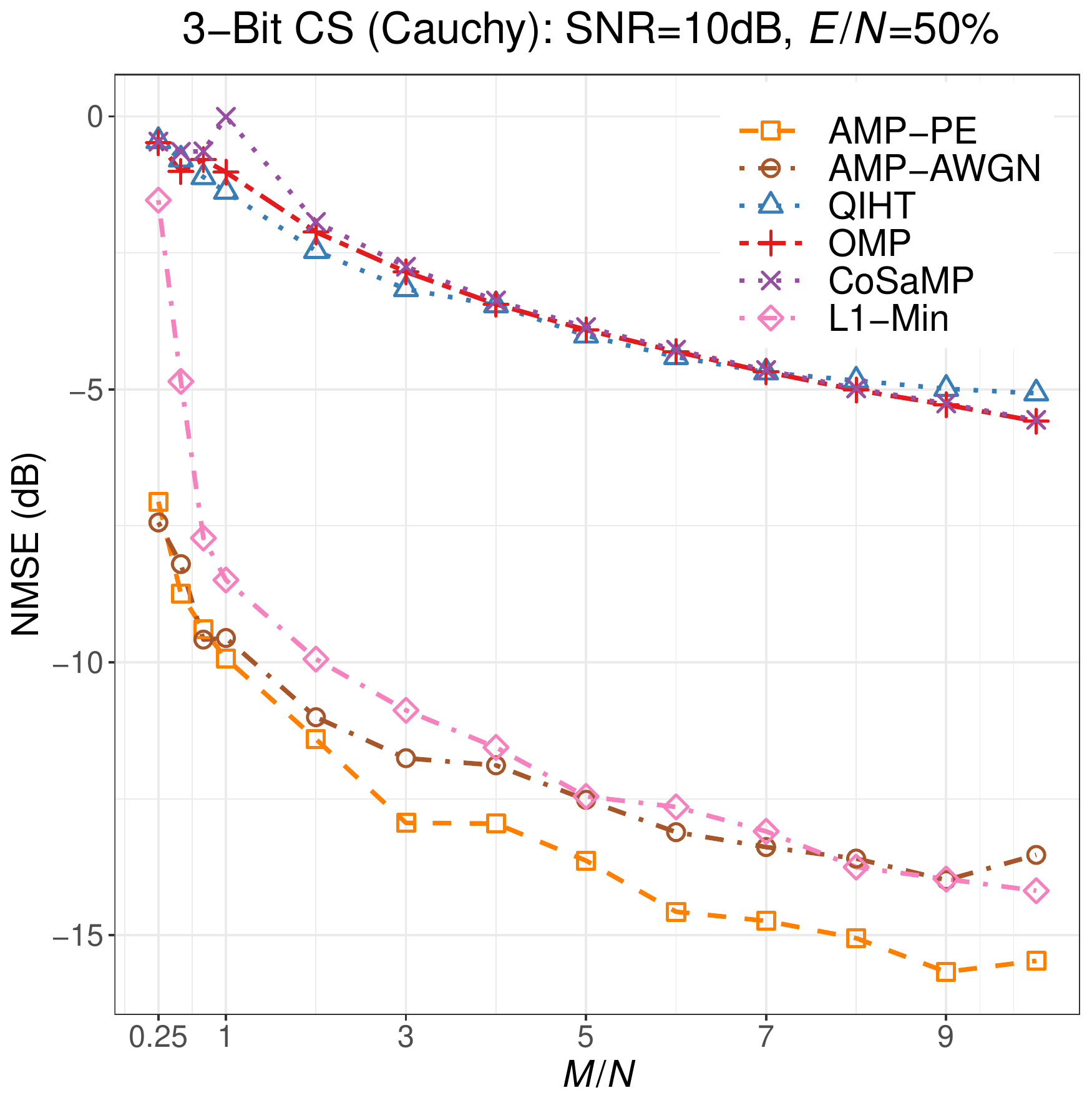}}\\

\subfigure{

\includegraphics[width=0.3\textwidth]{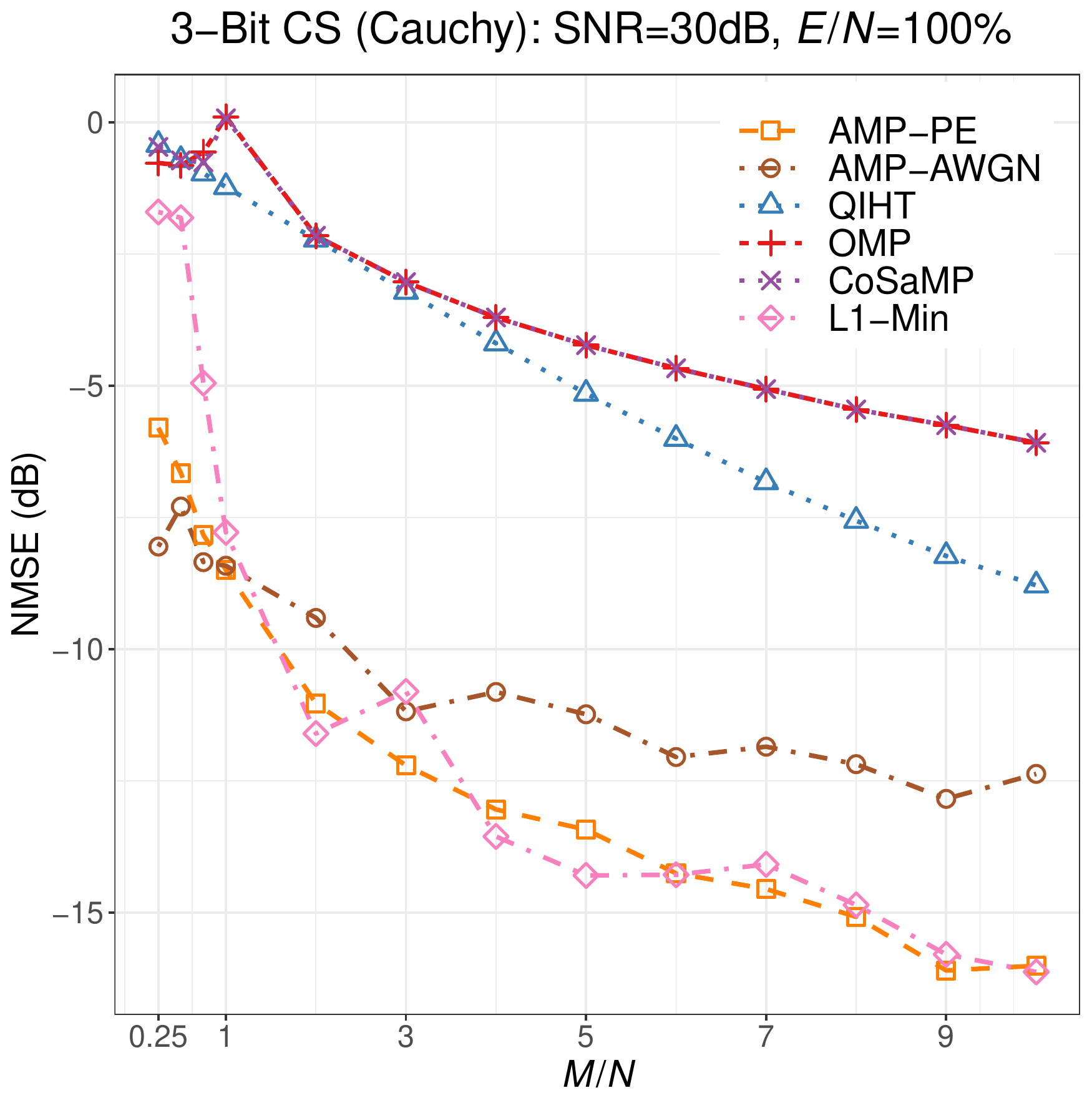}}
\subfigure{

\includegraphics[width=0.3\textwidth]{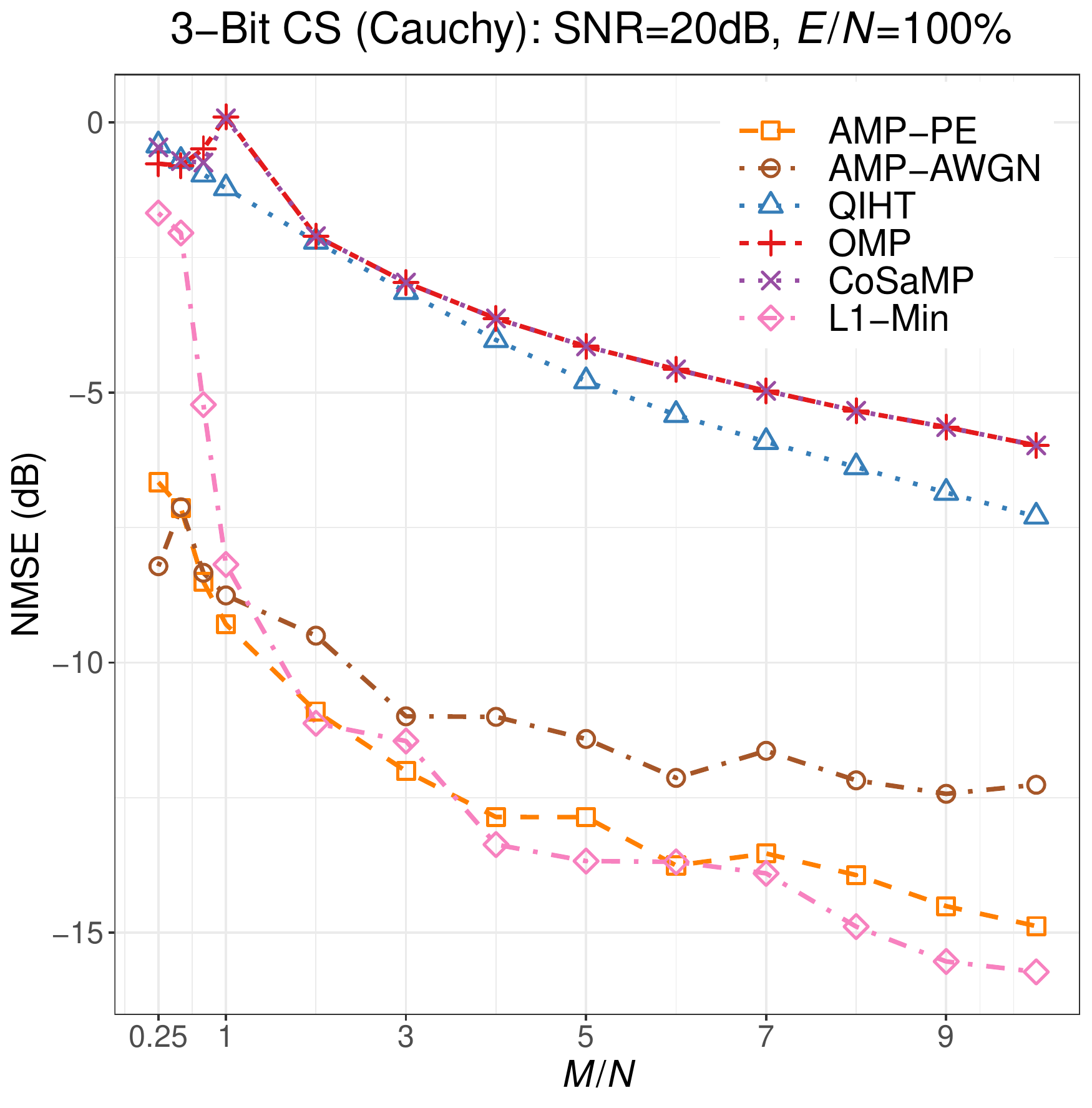}}
\subfigure{

\includegraphics[width=0.3\textwidth]{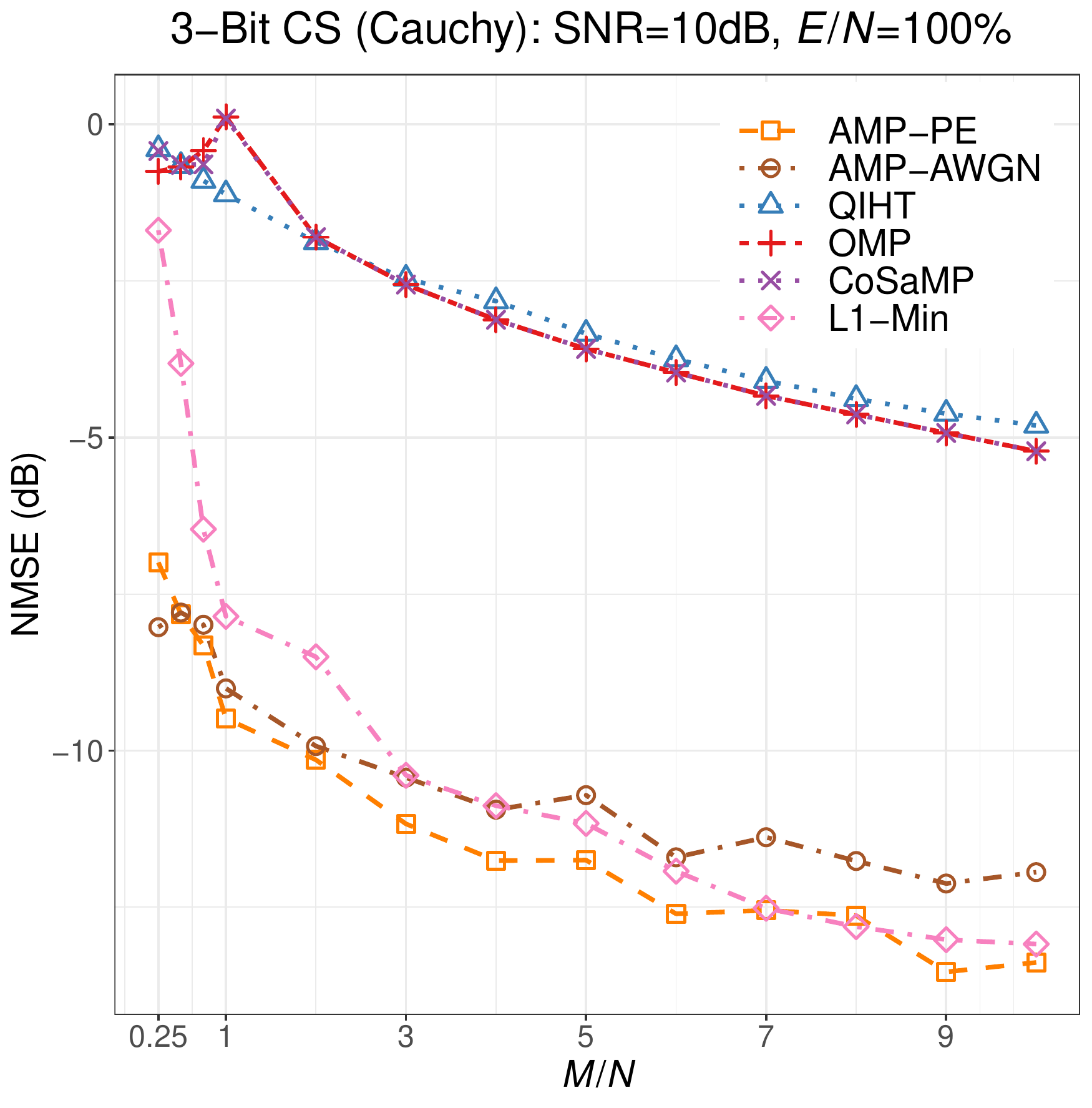}}

\caption{Comparison of different approaches in solving 3-bit CS. Nonzero entries of the signal follow the Cauchy distribution. The sampling ratio $\frac{M}{N}\in\{0.25,\cdots,10\}$ and the sparsity level $\frac{E}{N}=\in\{10\%,50\%,100\%\}$. The pre-quantization SNR varies from $30$dB, $20$dB to $10$dB.}

\label{fig:3bit_experiments_cauchy}
\end{figure*}

\newpage
\subsection{Signal with the Bernoulli-Laplace Mixture Prior}
\begin{figure*}[htbp]
\centering
\subfigure{

\includegraphics[width=0.3\textwidth]{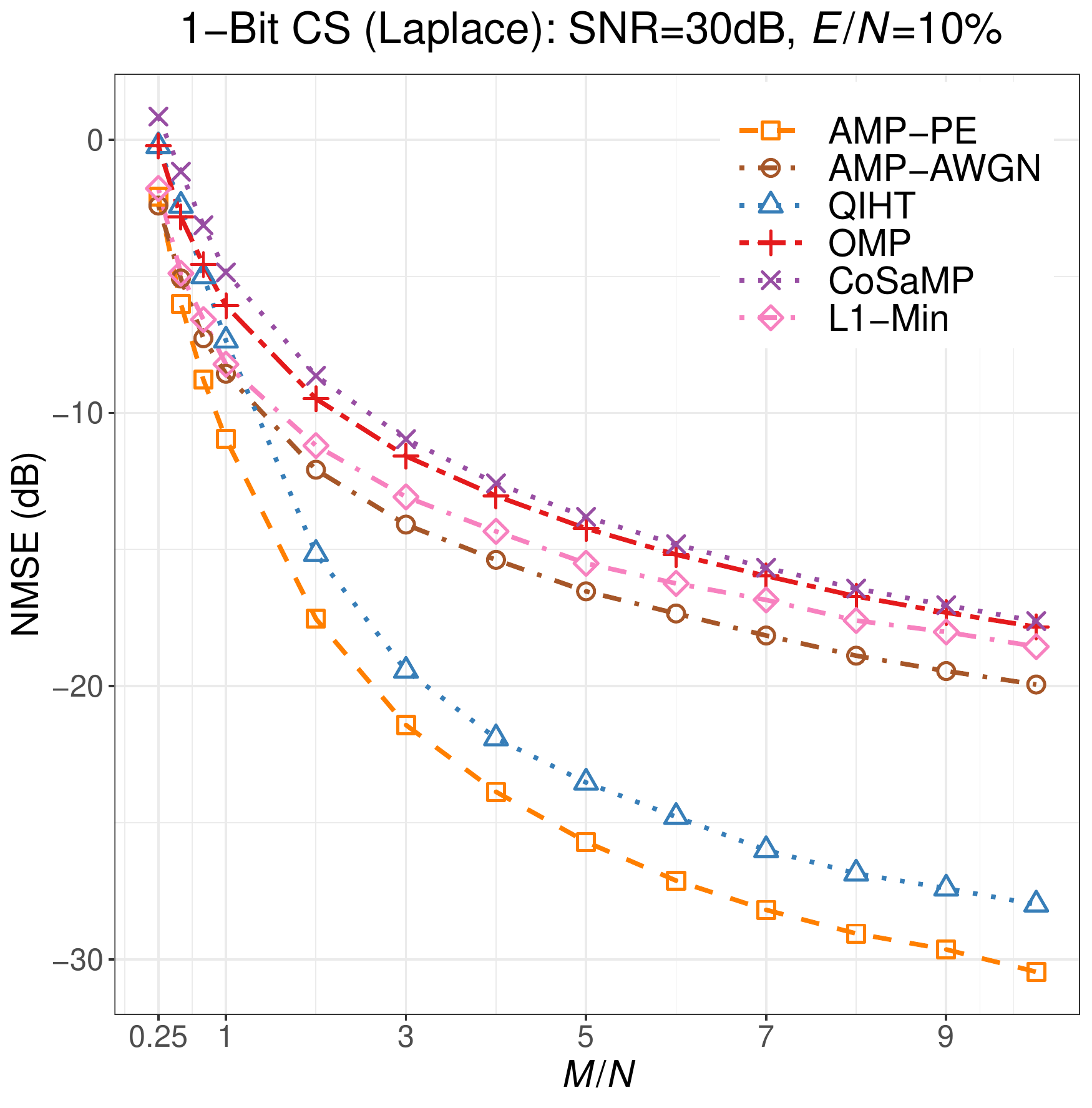}}
\subfigure{

\includegraphics[width=0.3\textwidth]{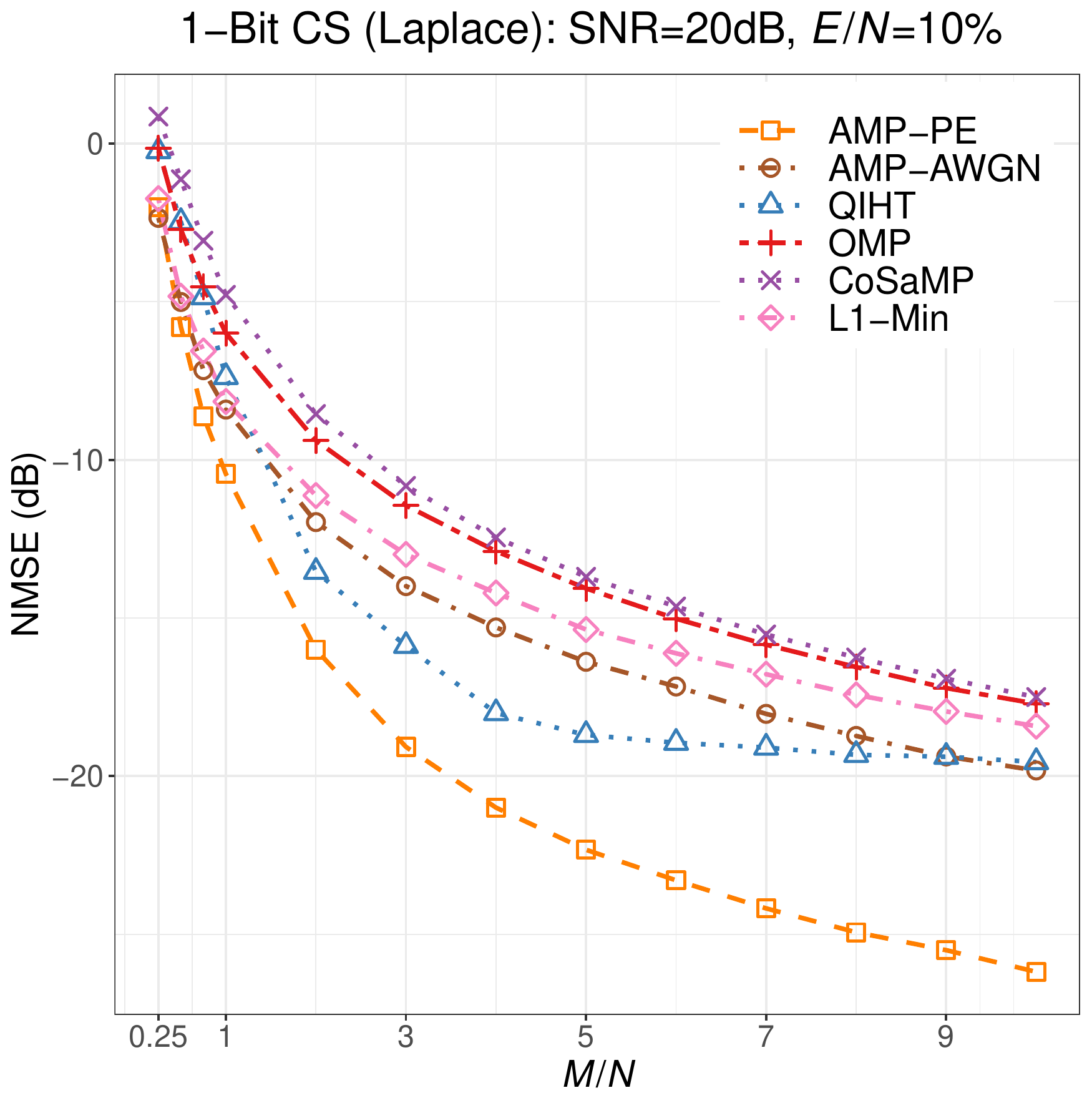}}
\subfigure{

\includegraphics[width=0.3\textwidth]{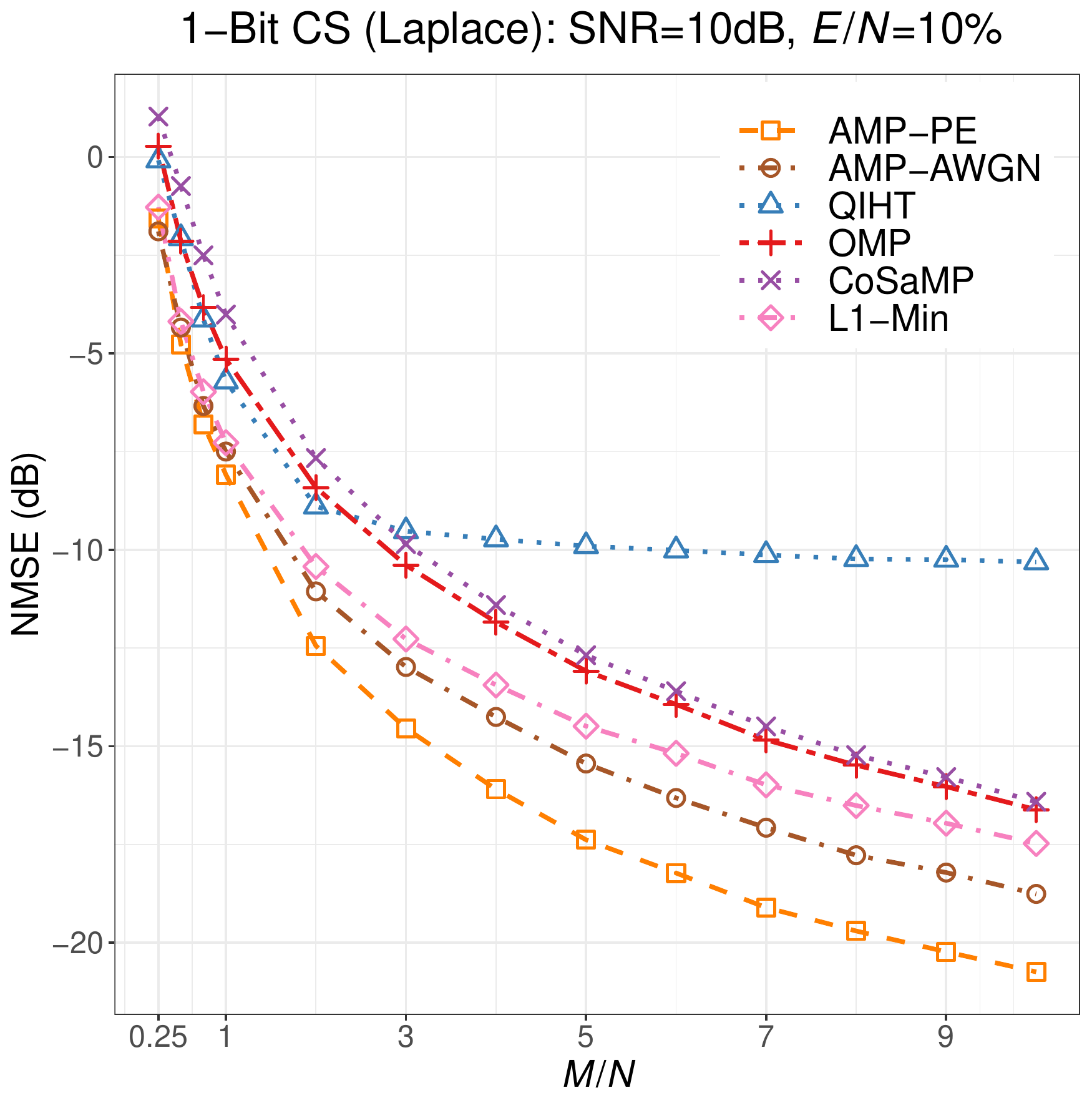}}\\

\subfigure{

\includegraphics[width=0.3\textwidth]{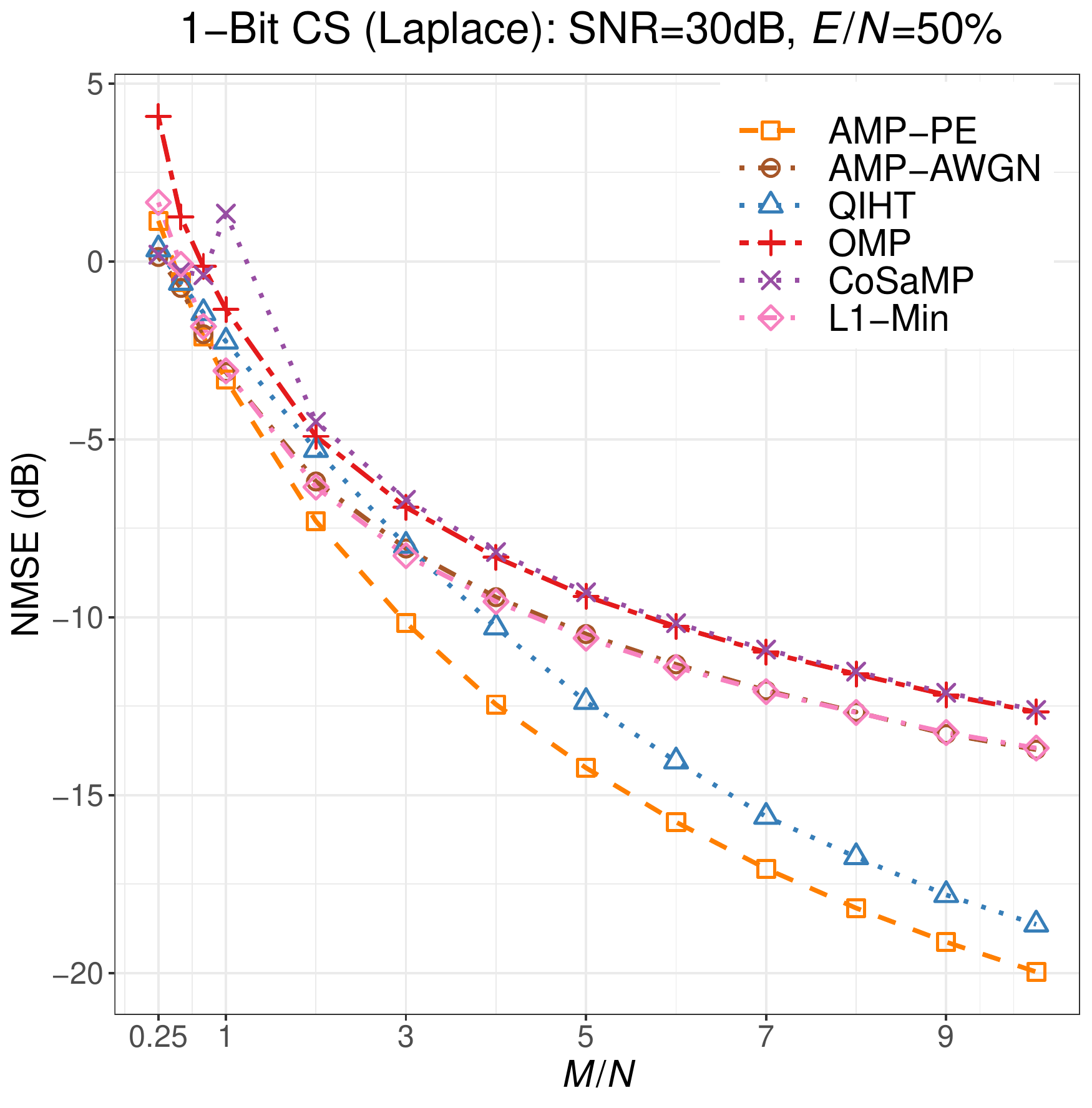}}
\subfigure{

\includegraphics[width=0.3\textwidth]{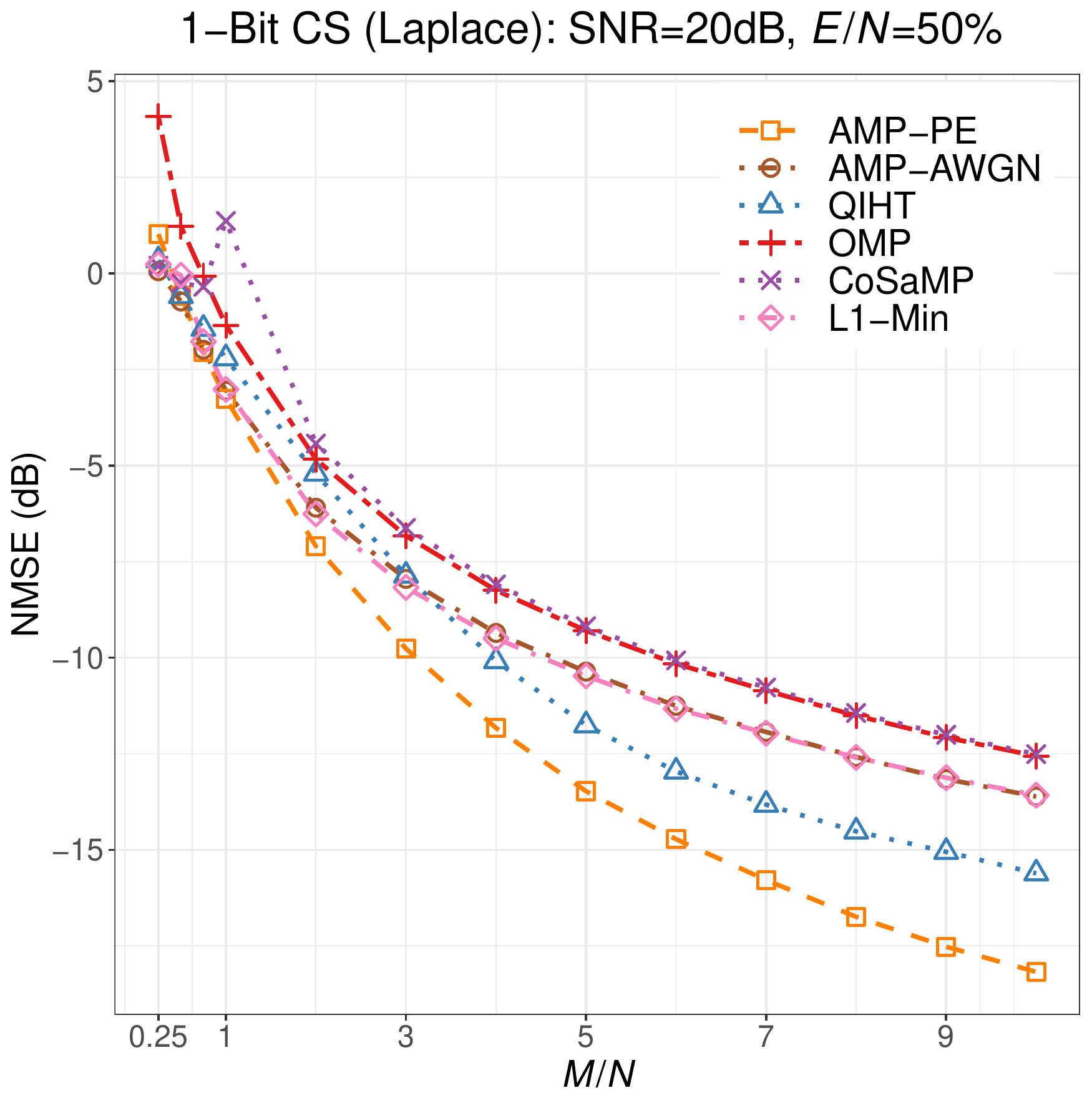}}
\subfigure{

\includegraphics[width=0.3\textwidth]{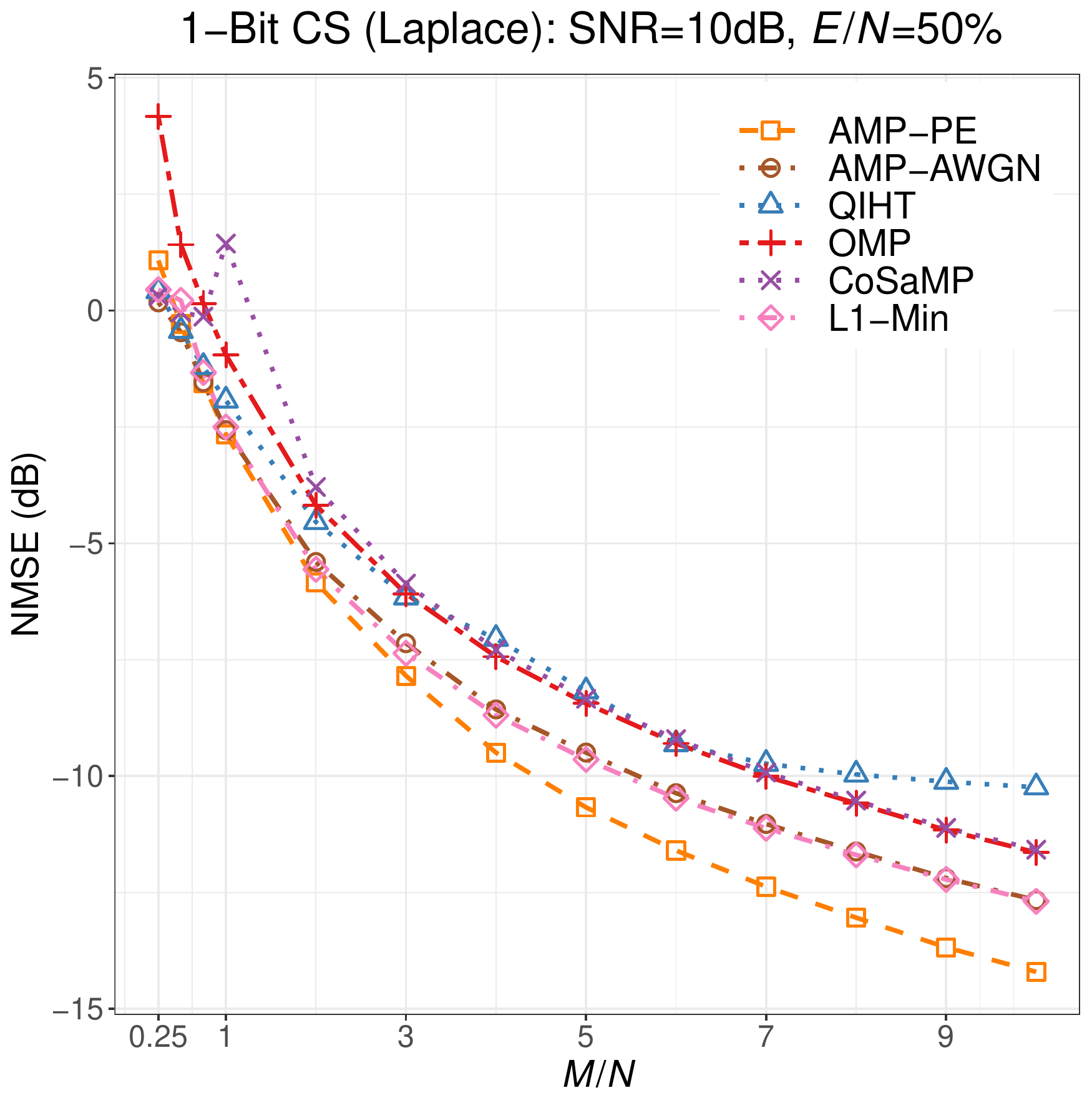}}\\

\subfigure{

\includegraphics[width=0.3\textwidth]{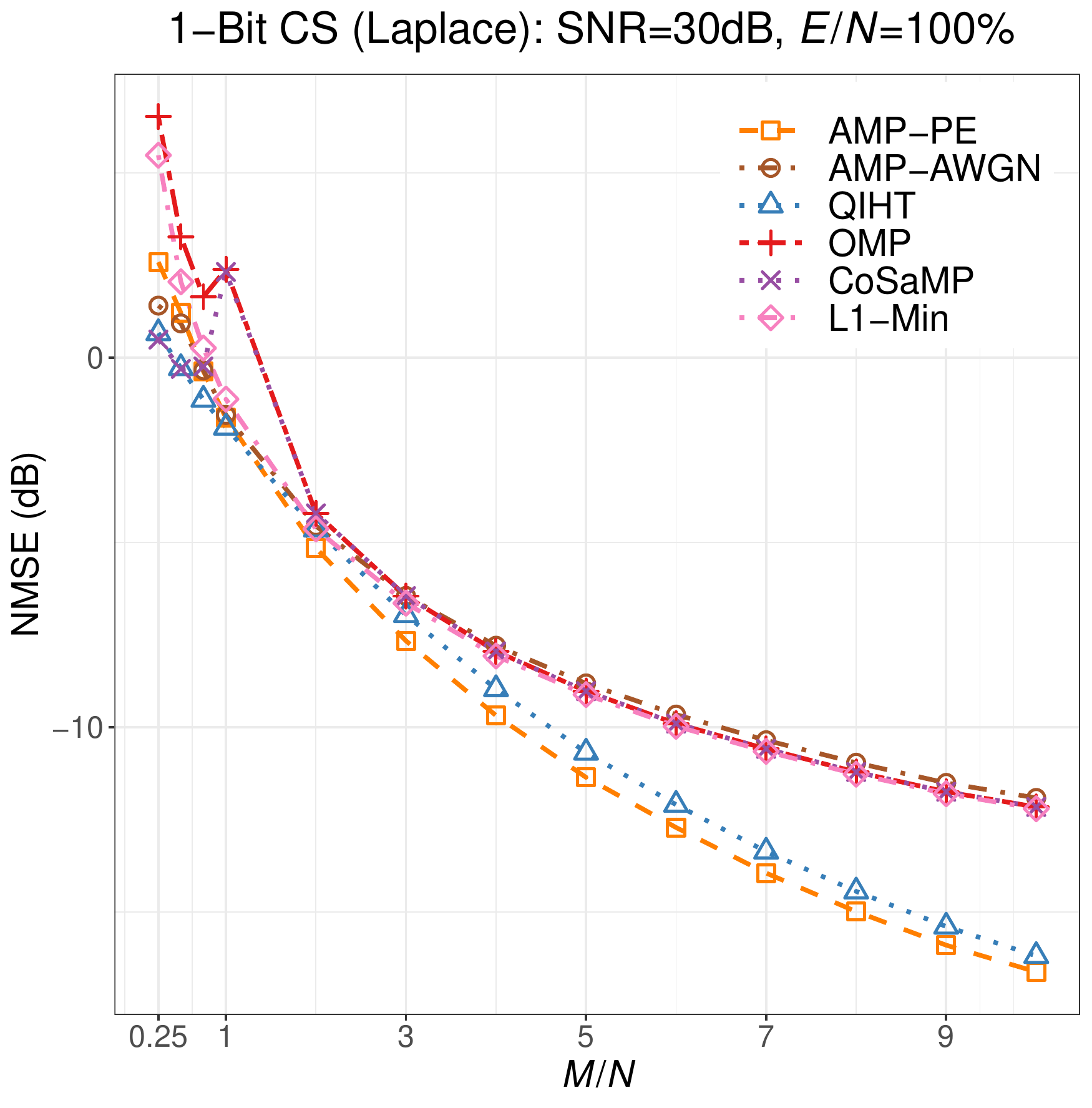}}
\subfigure{

\includegraphics[width=0.3\textwidth]{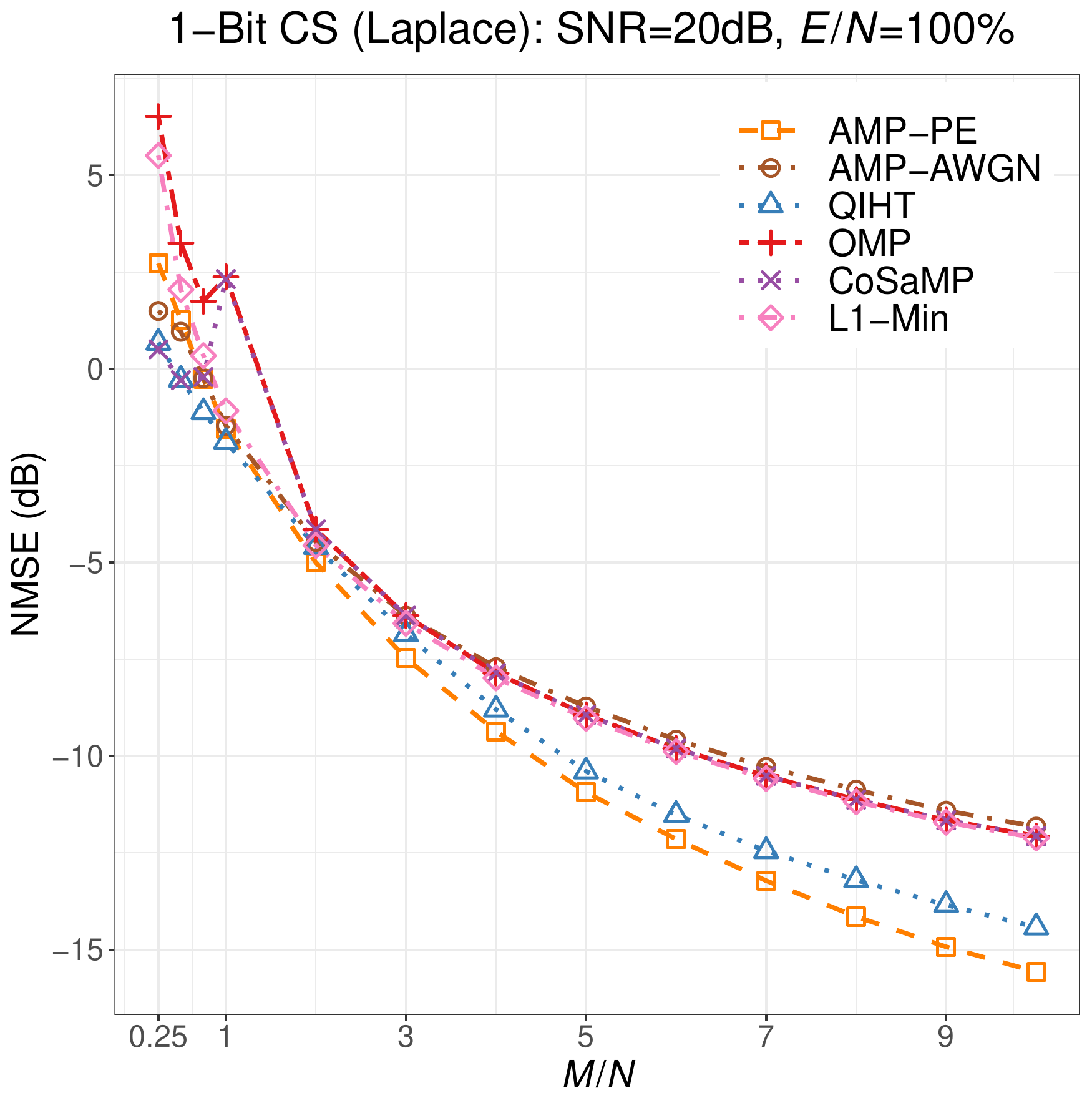}}
\subfigure{

\includegraphics[width=0.3\textwidth]{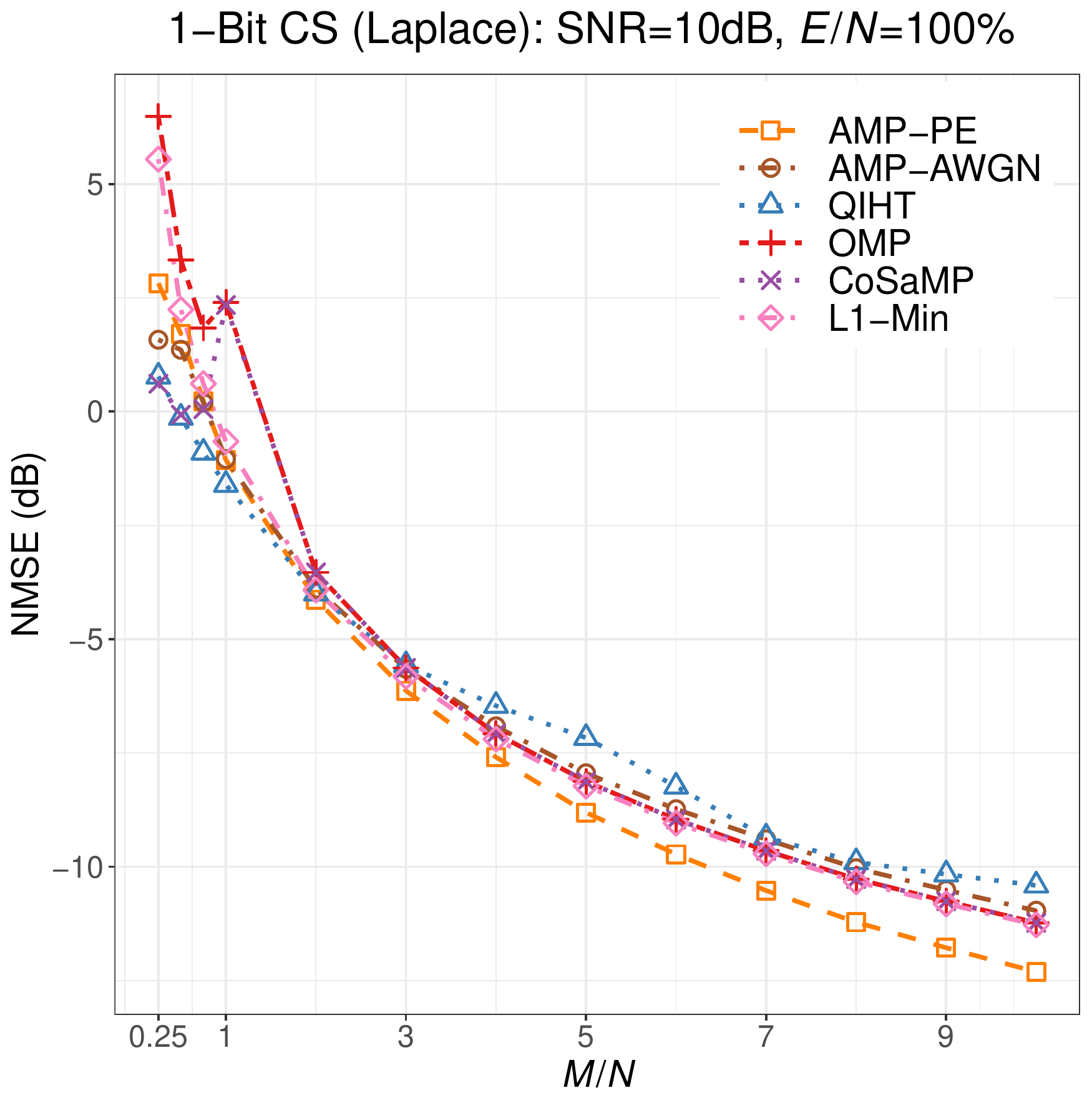}}

\caption{Comparison of different approaches in solving 1-bit CS. Nonzero entries of the signal follow the Laplace distribution. The sampling ratio $\frac{M}{N}\in\{0.25,\cdots,10\}$ and the sparsity level $\frac{E}{N}=\in\{10\%,50\%,100\%\}$. The pre-quantization SNR varies from $30$dB, $20$dB to $10$dB.}

\label{fig:1bit_experiments_laplace}
\end{figure*}

\begin{figure*}[htbp]
\centering
\subfigure{

\includegraphics[width=0.3\textwidth]{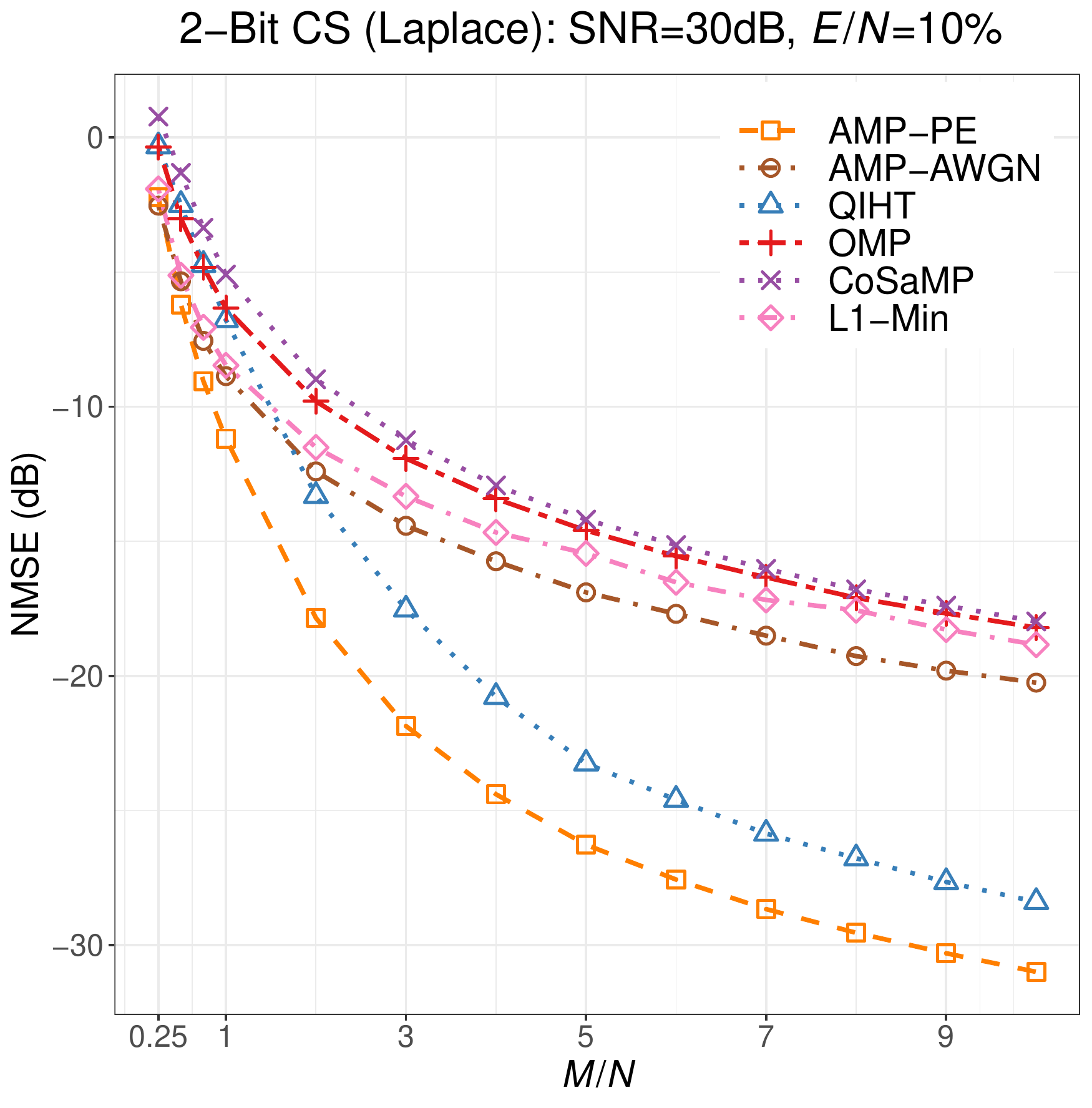}}
\subfigure{

\includegraphics[width=0.3\textwidth]{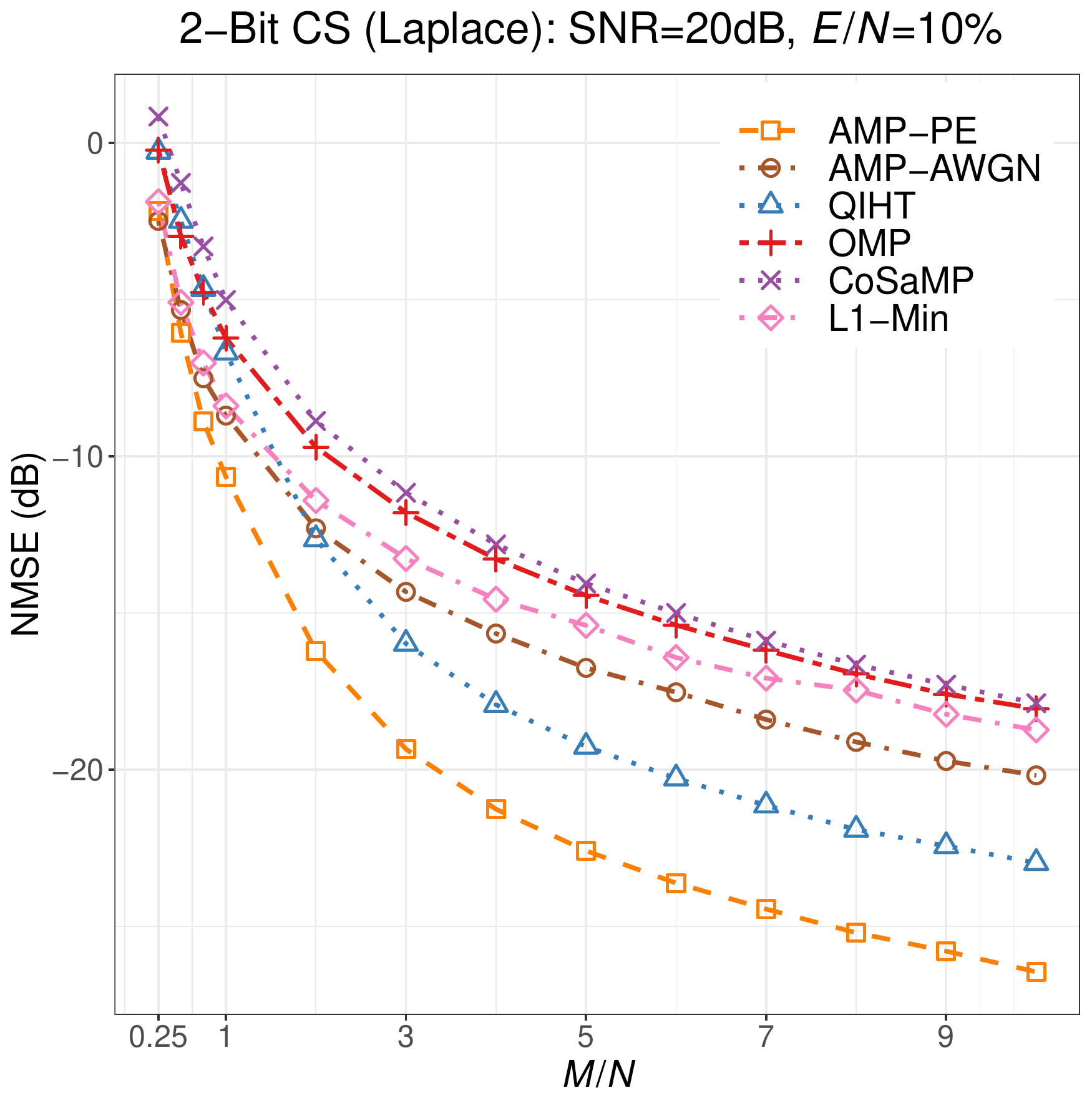}}
\subfigure{

\includegraphics[width=0.3\textwidth]{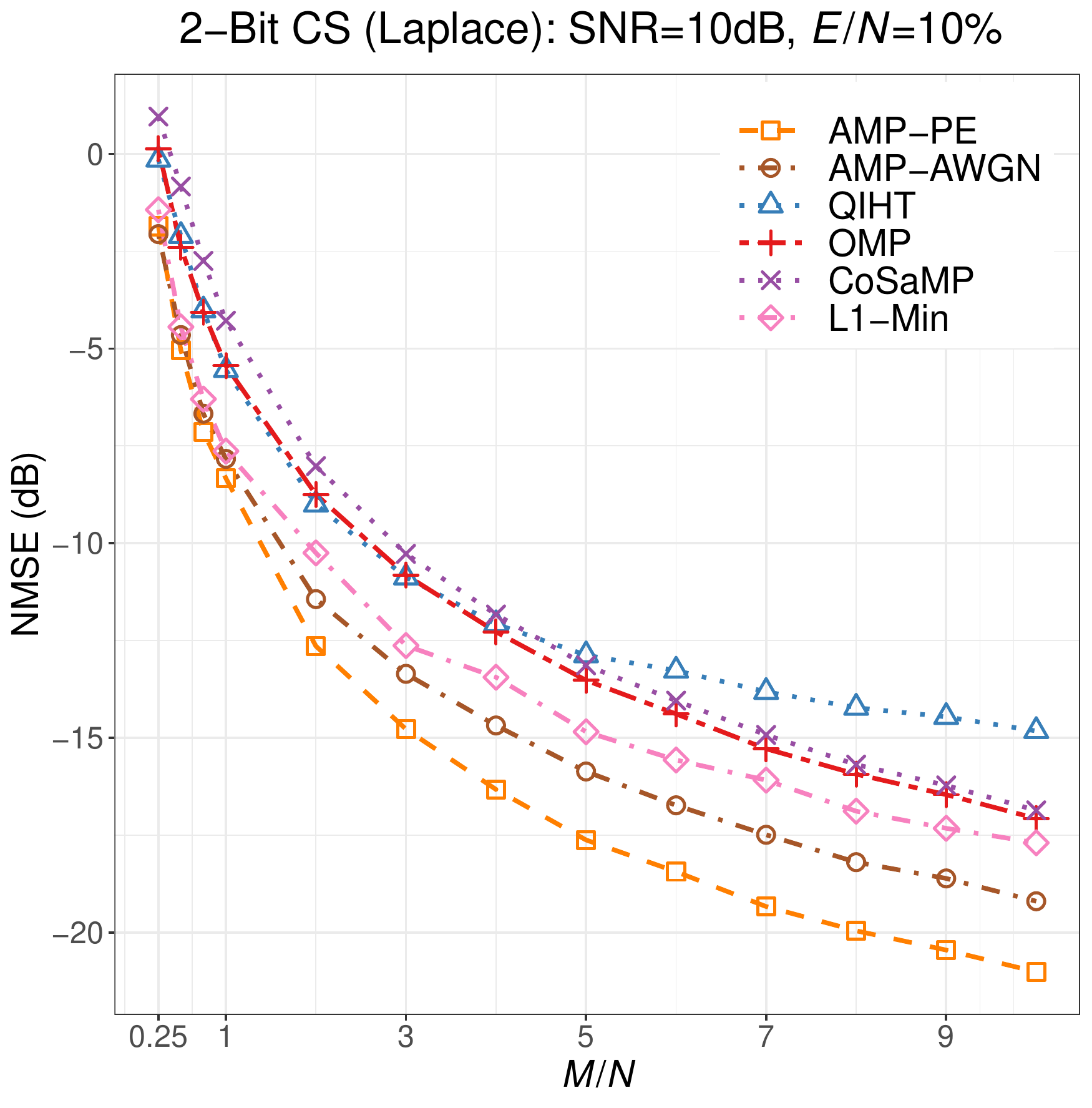}}\\

\subfigure{

\includegraphics[width=0.3\textwidth]{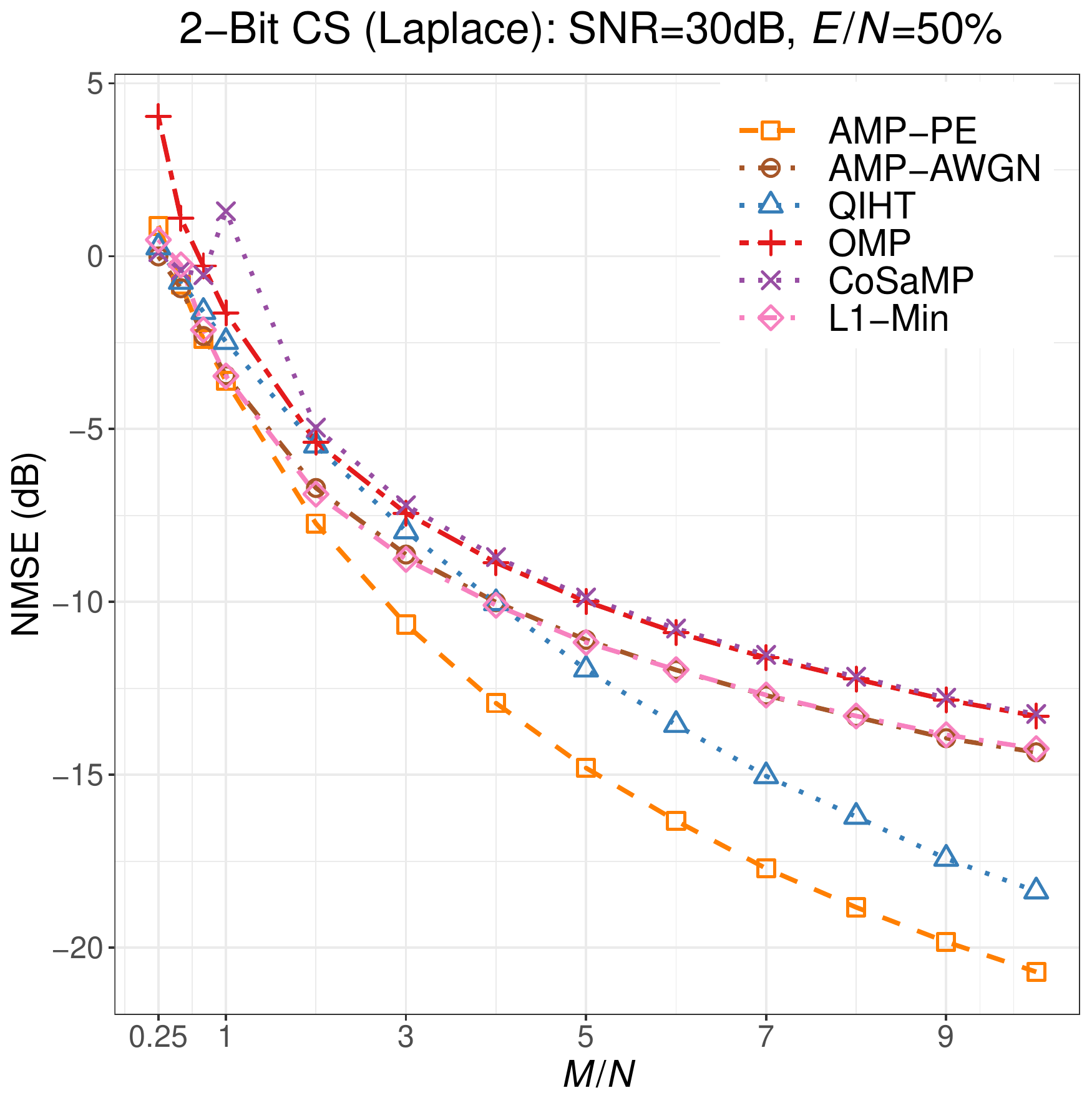}}
\subfigure{

\includegraphics[width=0.3\textwidth]{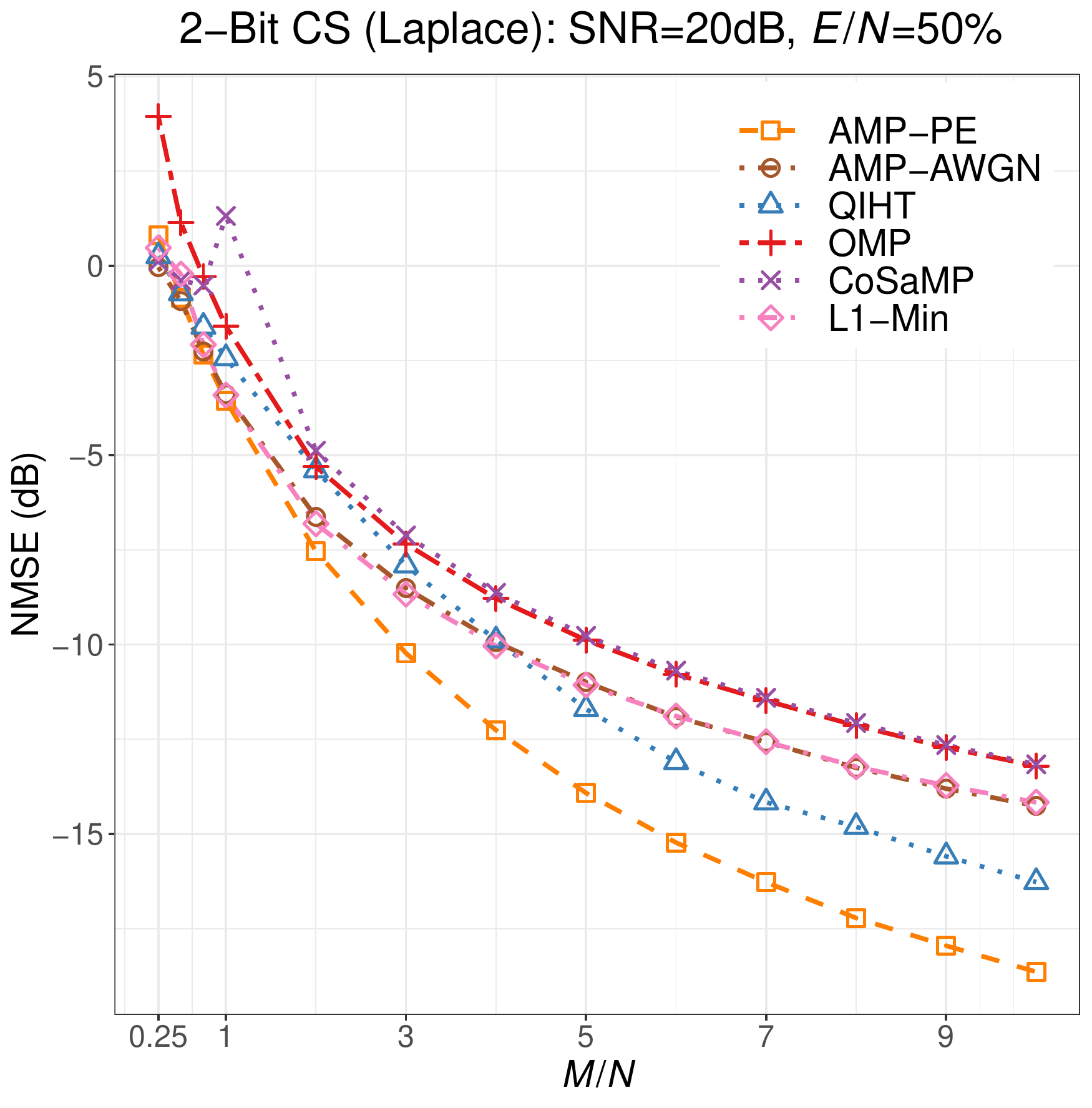}}
\subfigure{

\includegraphics[width=0.3\textwidth]{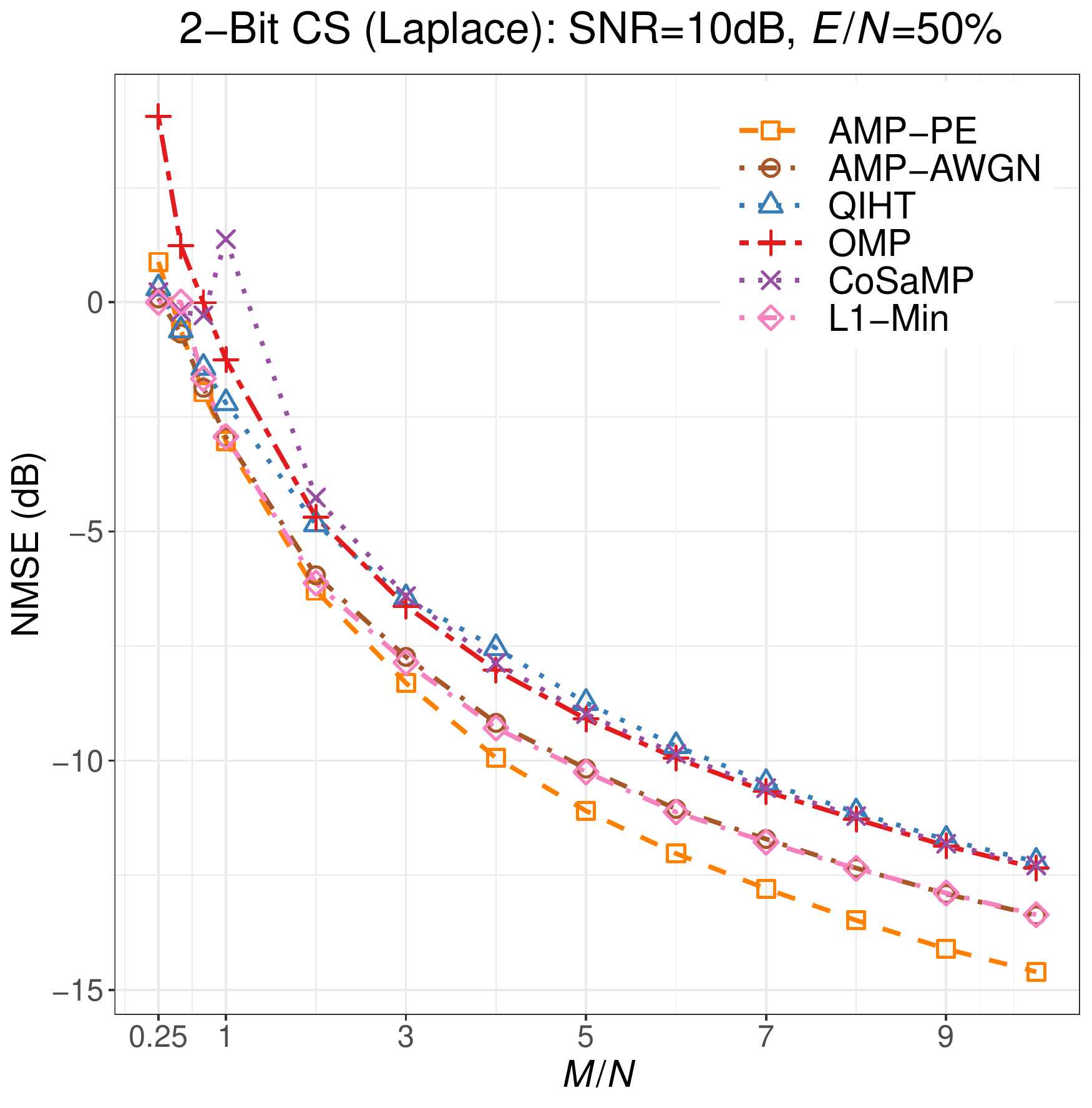}}\\

\subfigure{

\includegraphics[width=0.3\textwidth]{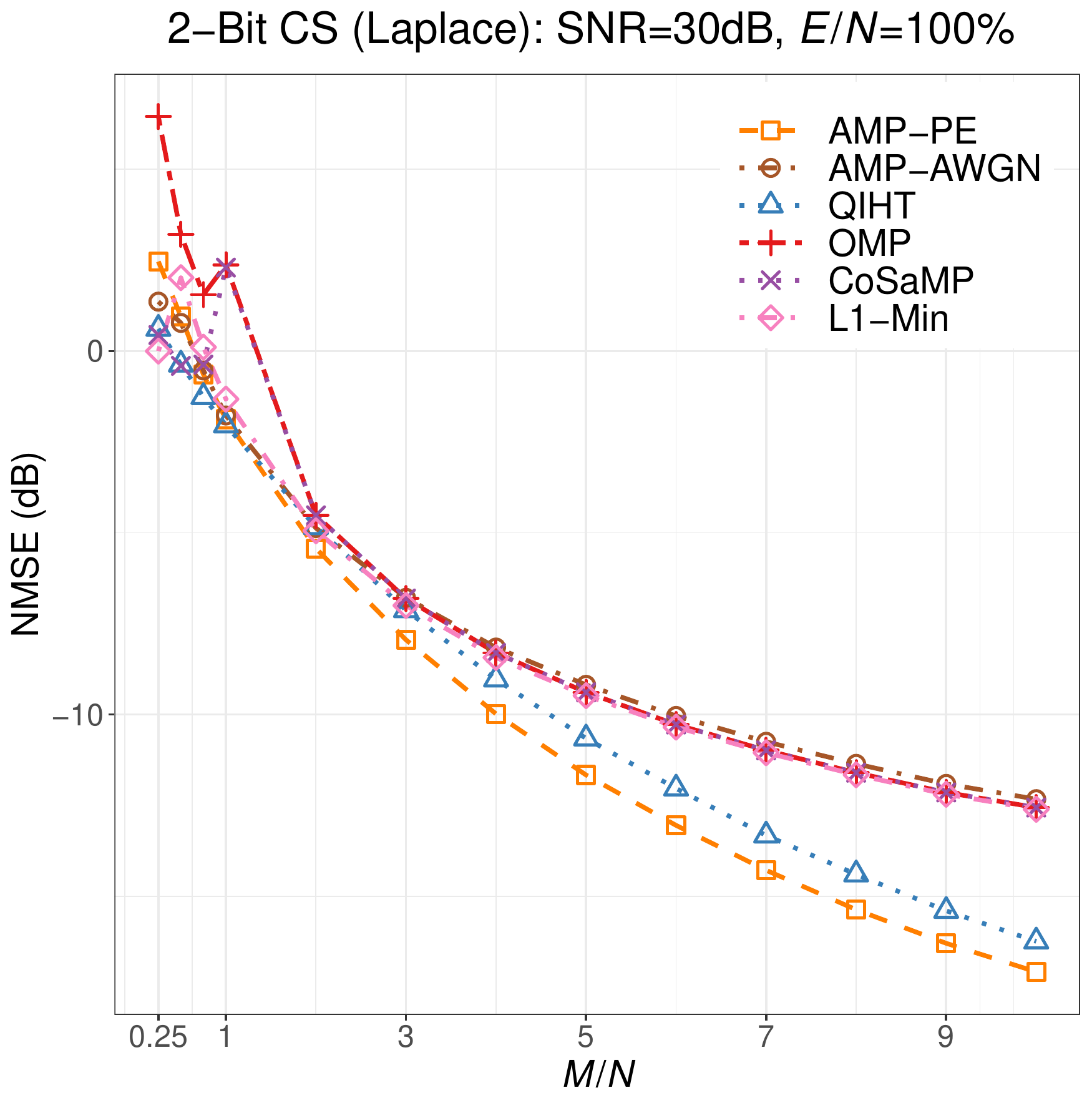}}
\subfigure{

\includegraphics[width=0.3\textwidth]{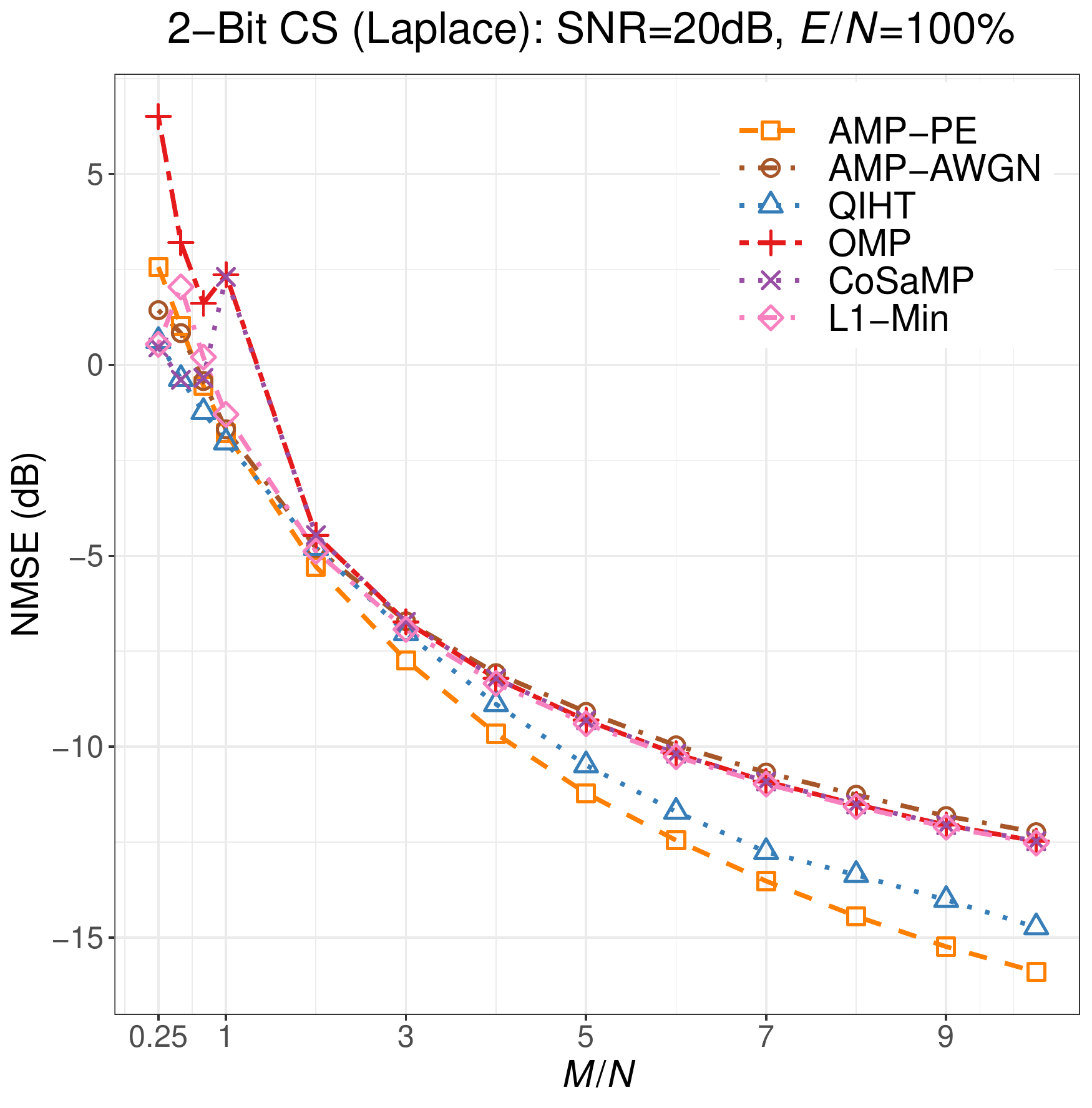}}
\subfigure{

\includegraphics[width=0.3\textwidth]{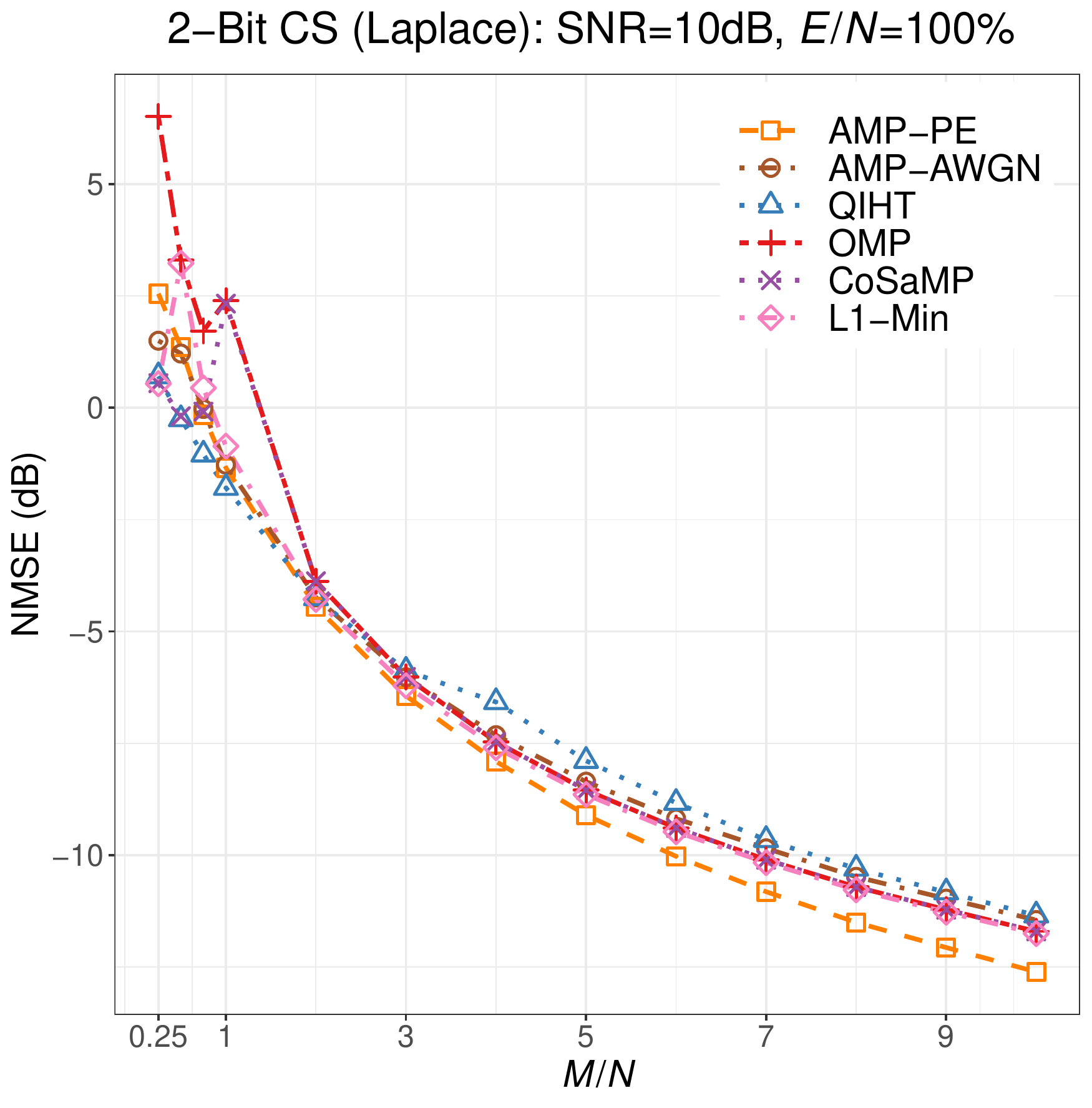}}

\caption{Comparison of different approaches in solving 2-bit CS. Nonzero entries of the signal follow the Laplace distribution. The sampling ratio $\frac{M}{N}\in\{0.25,\cdots,10\}$ and the sparsity level $\frac{E}{N}=\in\{10\%,50\%,100\%\}$. The pre-quantization SNR varies from $30$dB, $20$dB to $10$dB.}

\label{fig:2bit_experiments_laplace}
\end{figure*}

\begin{figure*}[htbp]
\centering
\subfigure{

\includegraphics[width=0.3\textwidth]{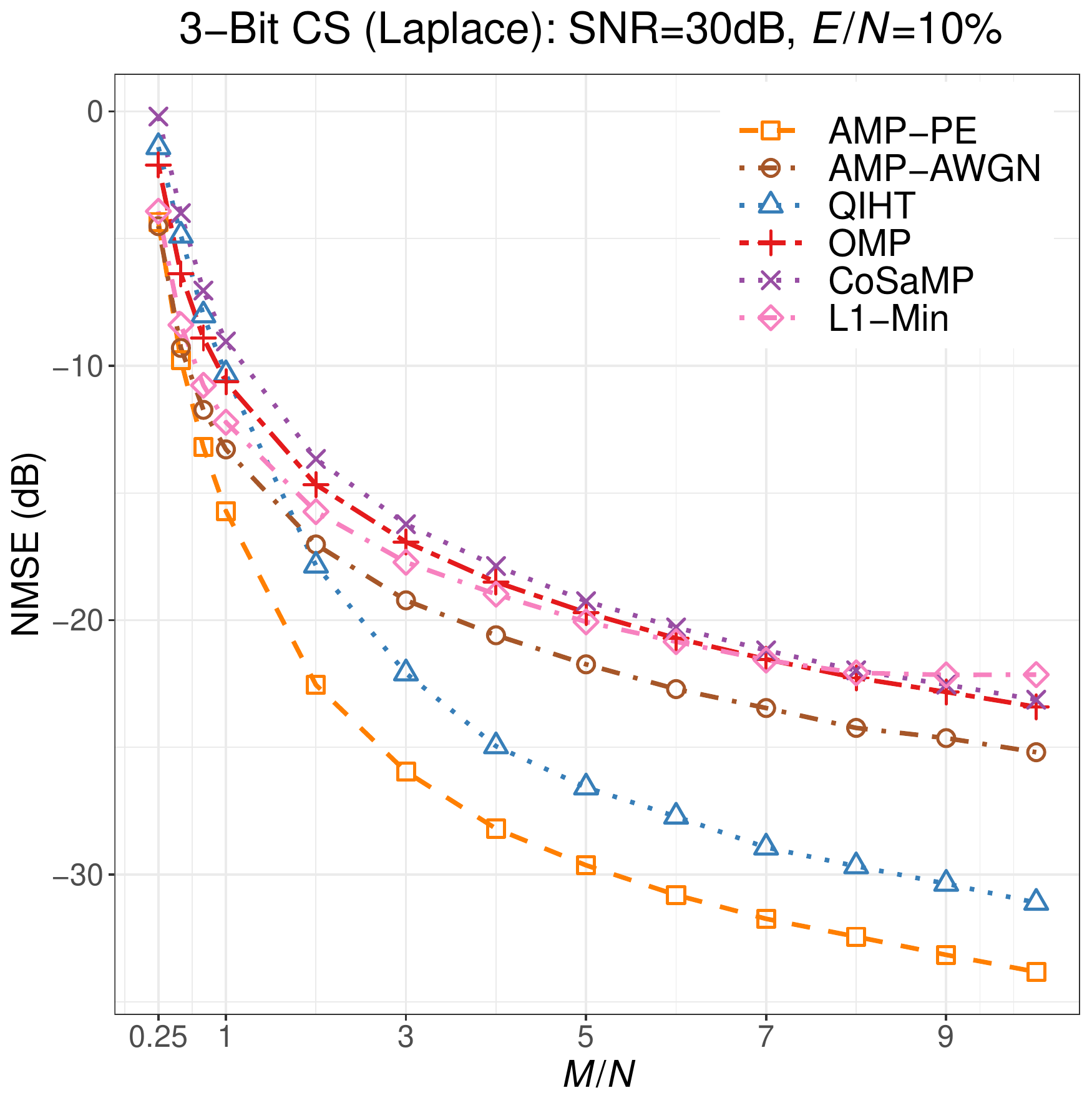}}
\subfigure{

\includegraphics[width=0.3\textwidth]{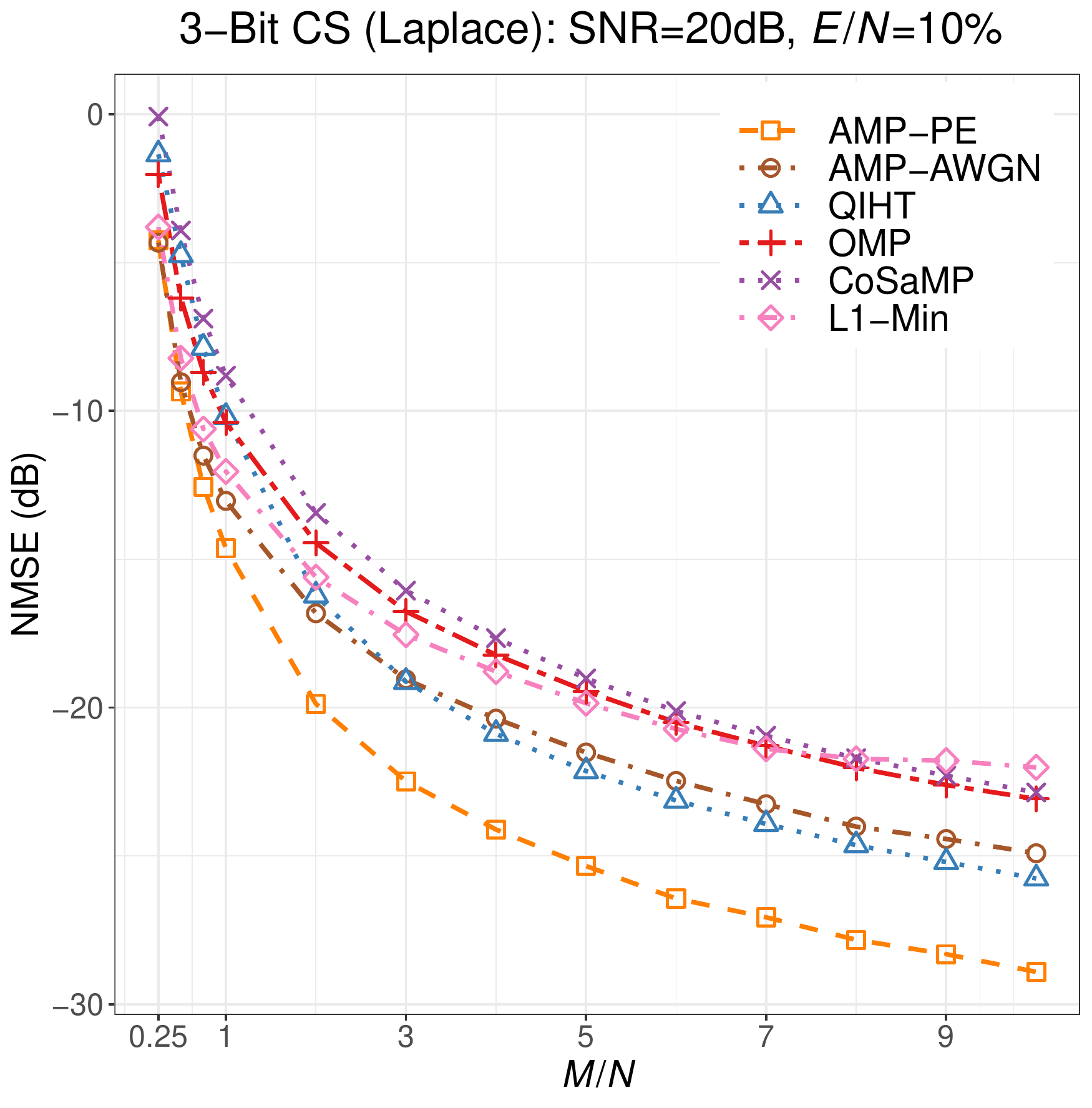}}
\subfigure{

\includegraphics[width=0.3\textwidth]{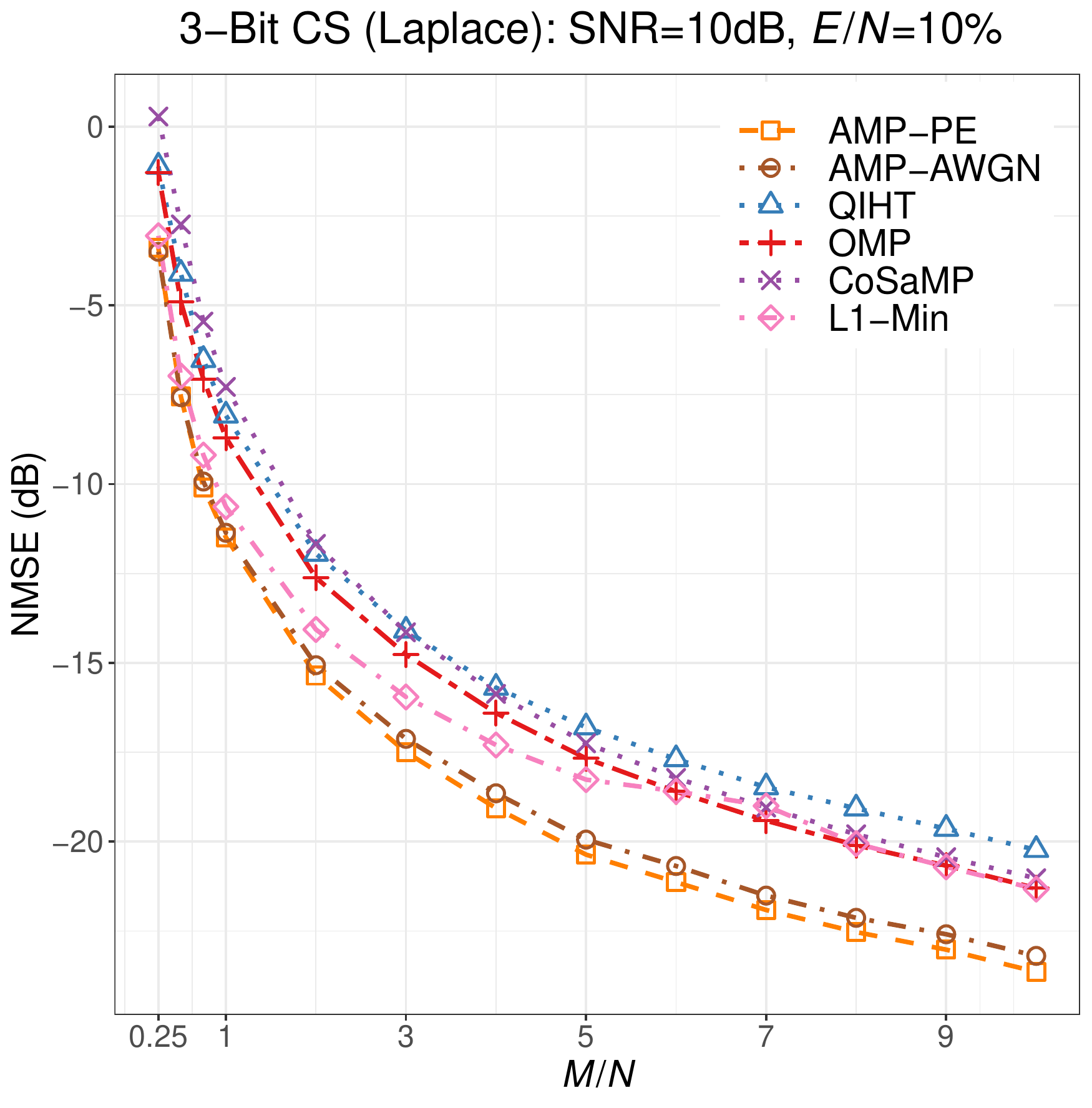}}\\

\subfigure{

\includegraphics[width=0.3\textwidth]{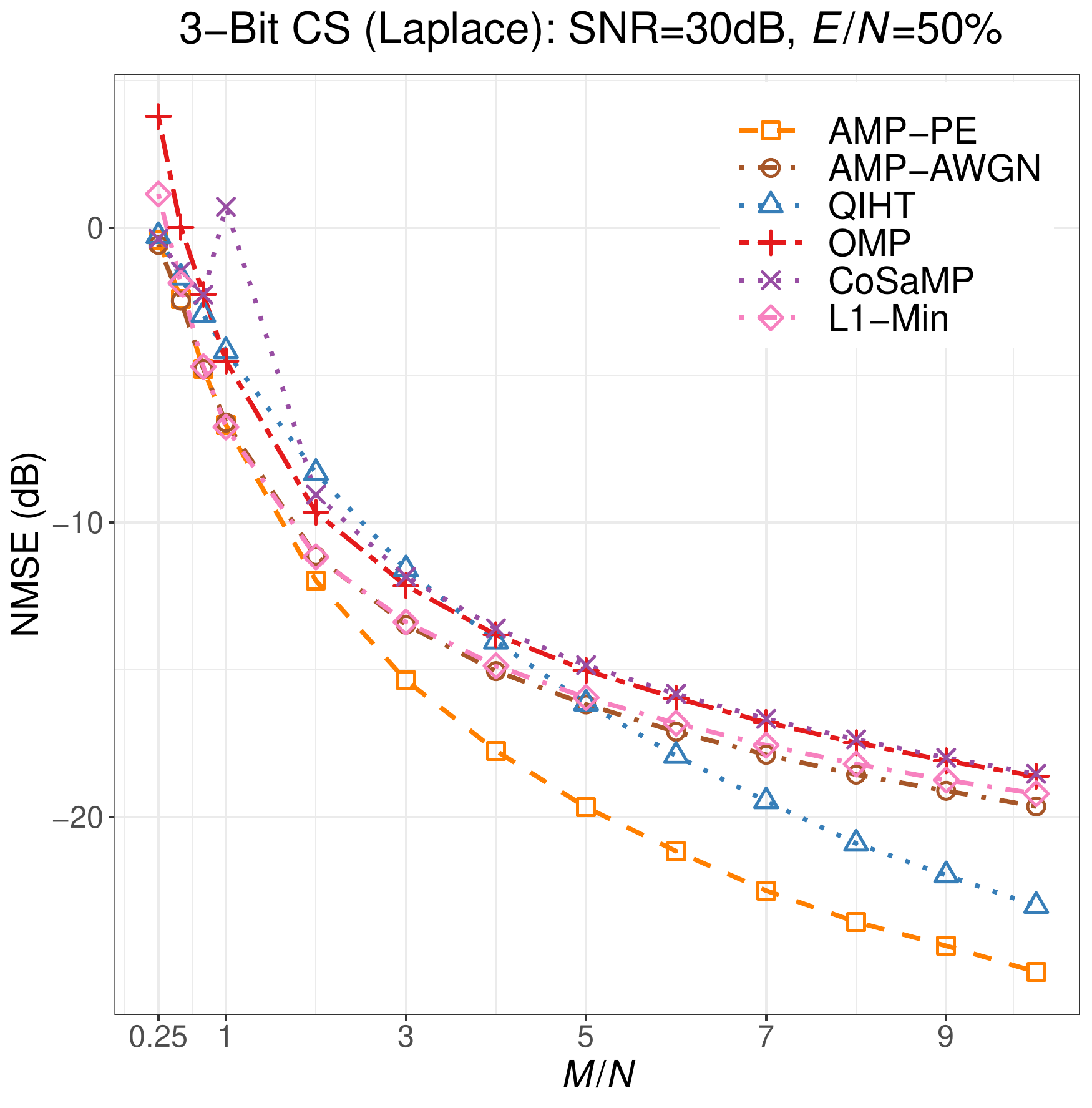}}
\subfigure{

\includegraphics[width=0.3\textwidth]{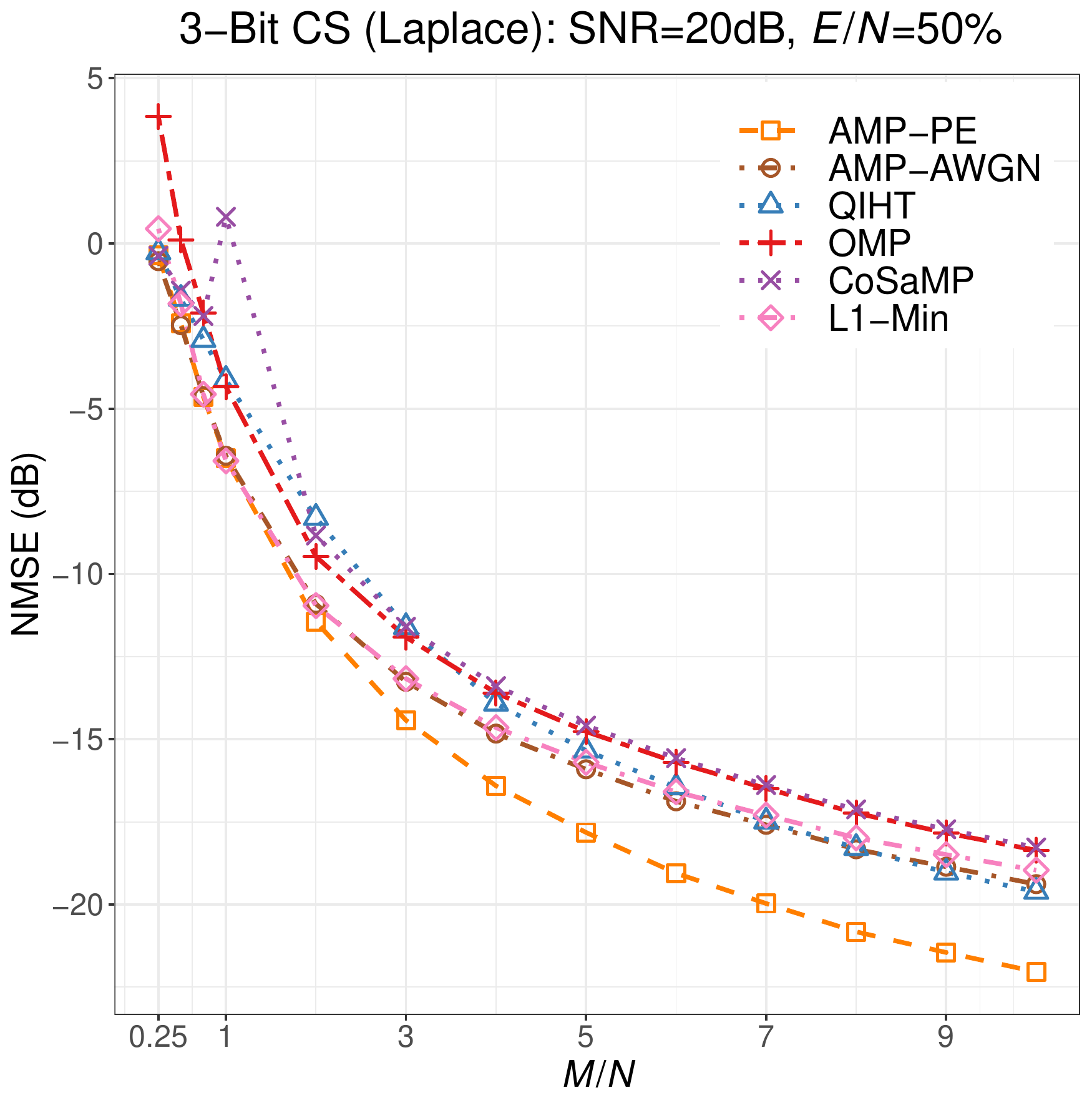}}
\subfigure{

\includegraphics[width=0.3\textwidth]{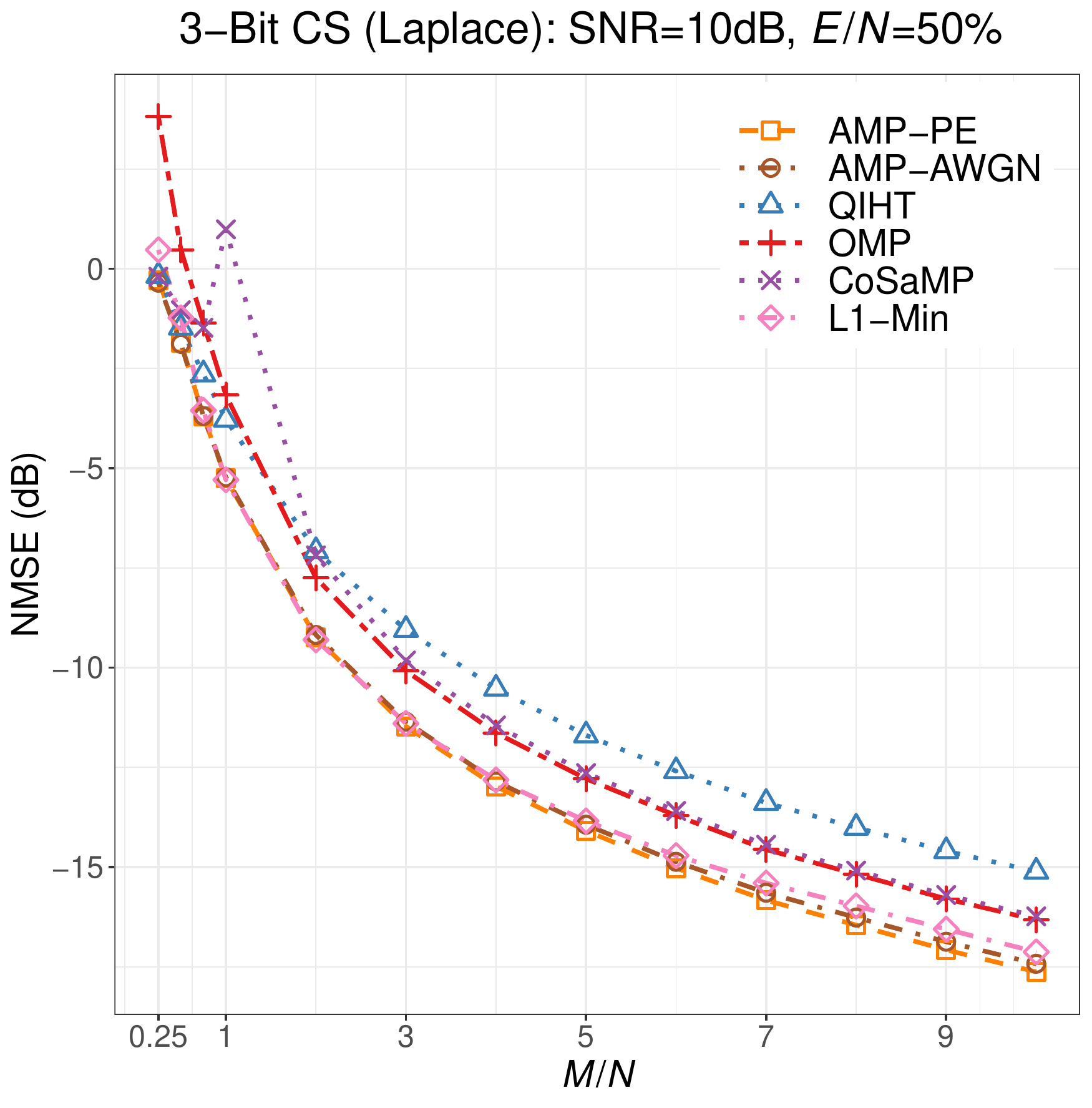}}\\

\subfigure{

\includegraphics[width=0.3\textwidth]{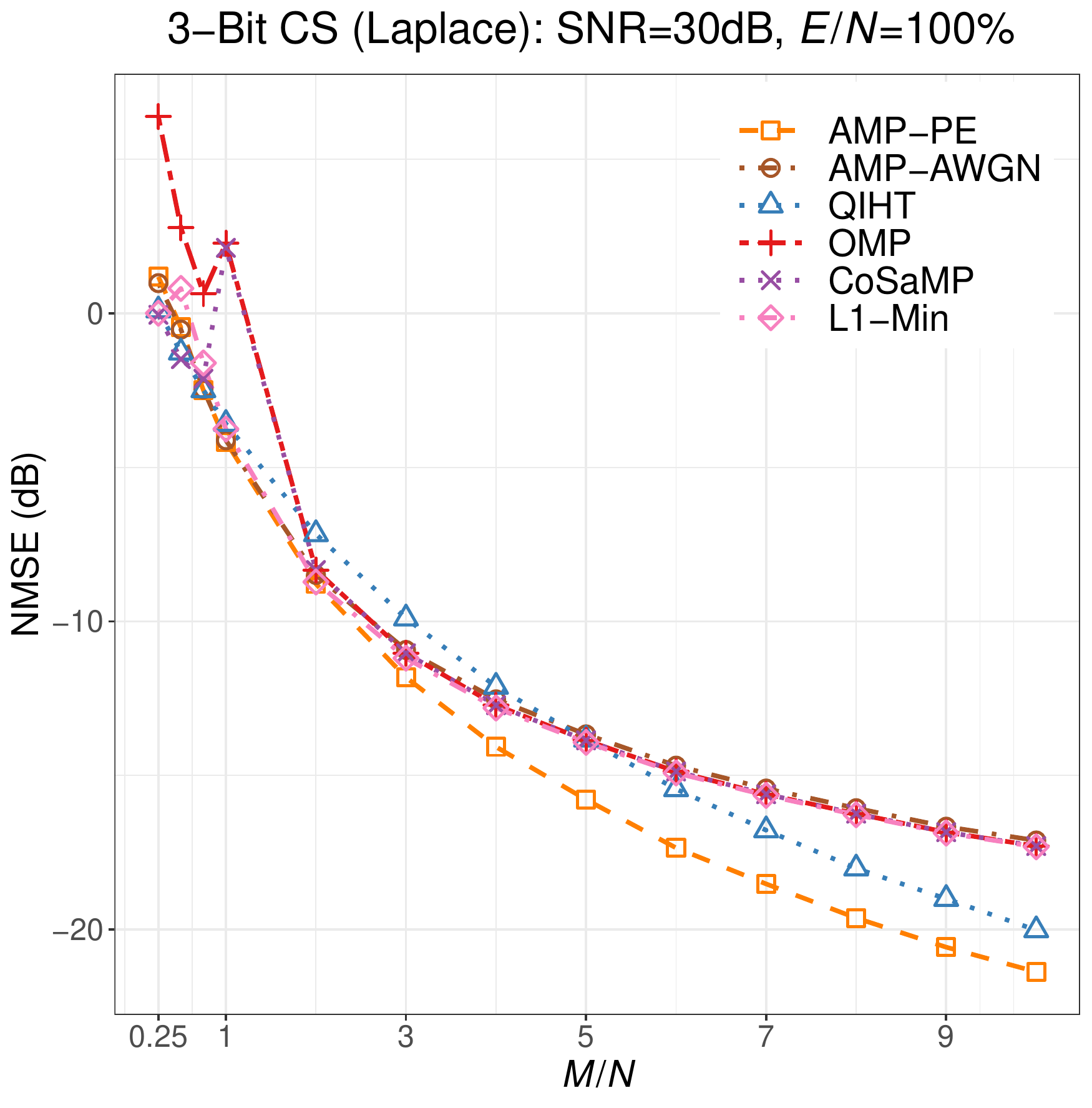}}
\subfigure{

\includegraphics[width=0.3\textwidth]{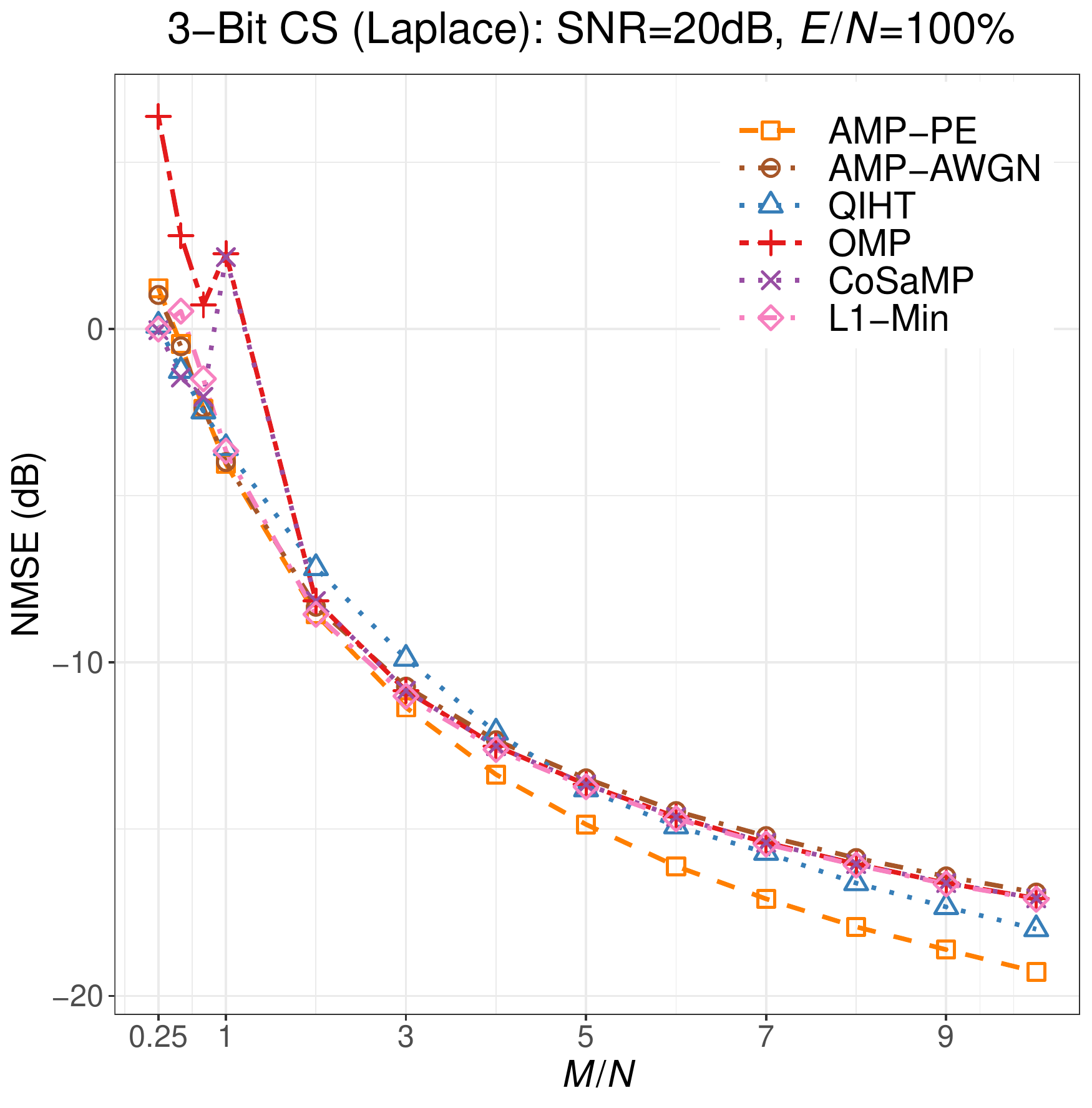}}
\subfigure{

\includegraphics[width=0.3\textwidth]{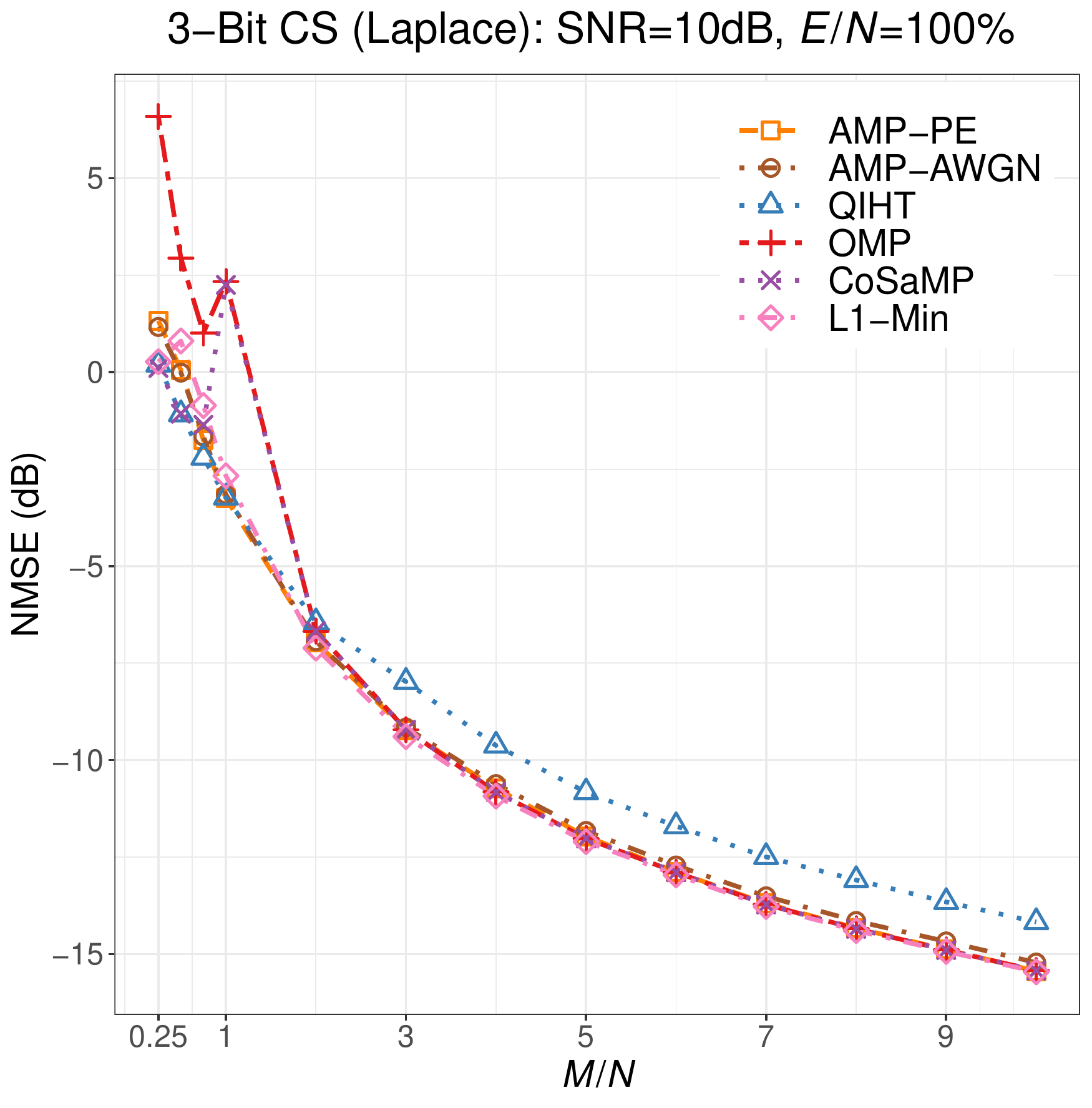}}

\caption{Comparison of different approaches in solving 3-bit CS. Nonzero entries of the signal follow the Laplace distribution. The sampling ratio $\frac{M}{N}\in\{0.25,\cdots,10\}$ and the sparsity level $\frac{E}{N}=\in\{10\%,50\%,100\%\}$. The pre-quantization SNR varies from $30$dB, $20$dB to $10$dB.}

\label{fig:3bit_experiments_laplace}
\end{figure*}

When nonzero entries are generated from the Laplace distribution, the results are shown in Fig. \ref{fig:1bit_experiments_laplace}-\ref{fig:3bit_experiments_laplace}. We can see that AMP-PE achieves leading performances across different noise and sparsity levels. When the pre-QNT SNR is 30dB, we can see that AMP-PE and QIHT perform much better than other sparse recovery methods as the sampling ratio $\frac{M}{N}$ increases, since they could work with quantized measurements directly. However, QIHT is not robust to noise. When the pre-QNT SNR is reduced to 10dB, we can see that QIHT performs worse than the other methods as the sampling ratio $\frac{M}{N}>7$.

\newpage

\newpage

\newpage
\section{Channel Estimation}

\begin{figure*}[htbp]
\centering
\subfigure{
\includegraphics[width=0.09\textwidth]{figures/random_qpsk_label.pdf}}
\subfigure{
\includegraphics[height=0.24\textwidth]{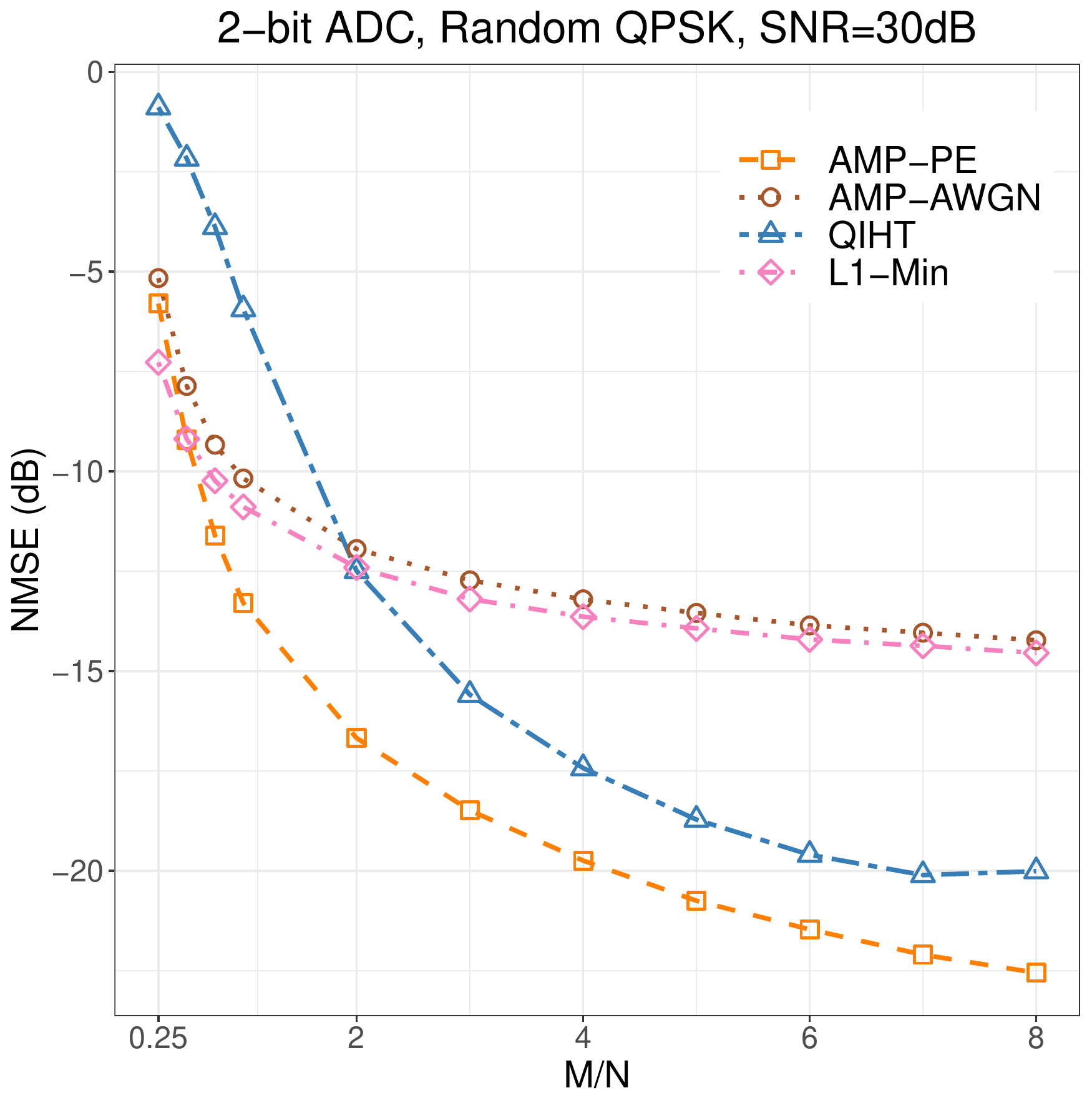}}
\subfigure{
\includegraphics[height=0.24\textwidth]{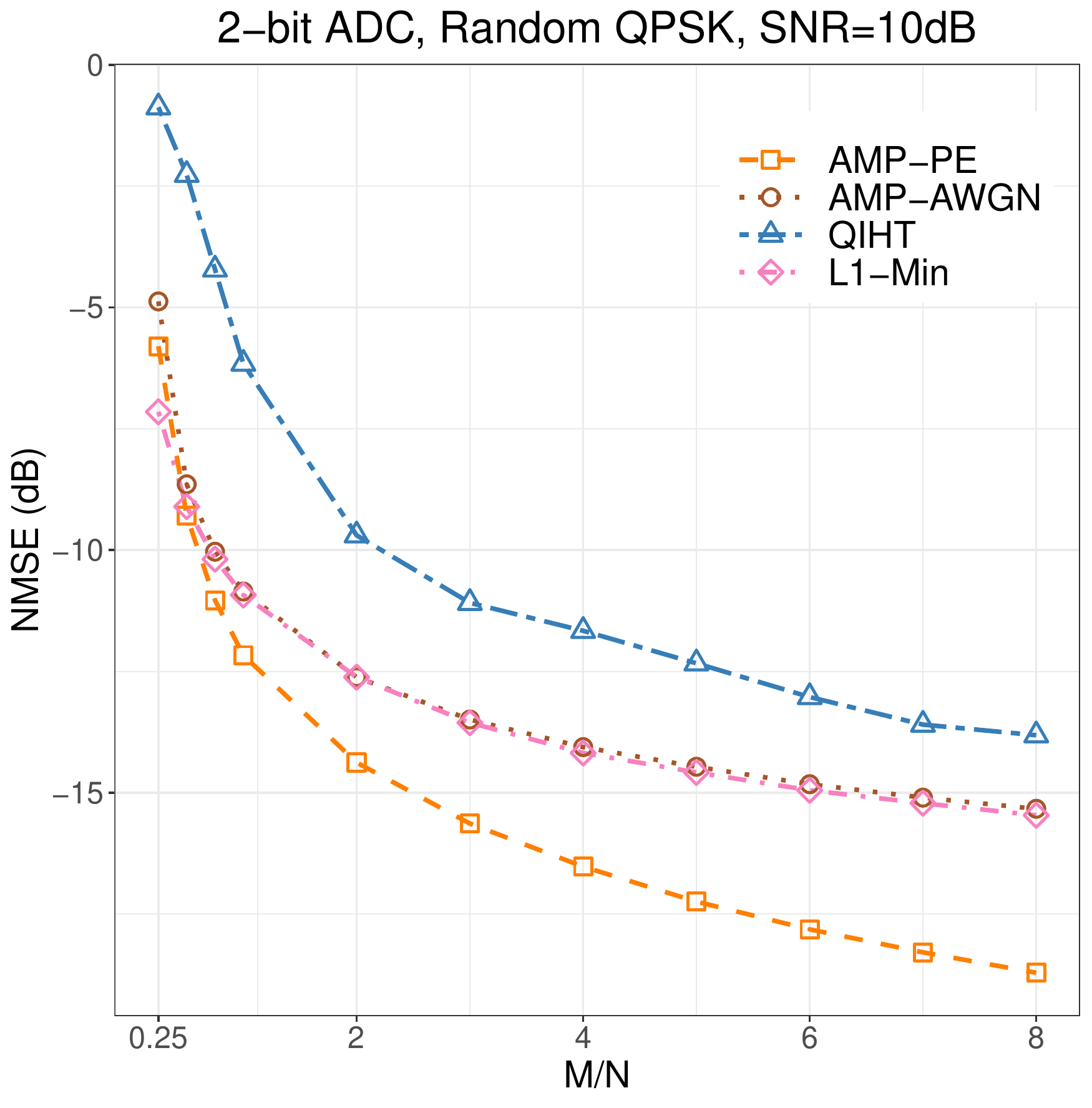}}
\subfigure{
\includegraphics[height=0.24\textwidth]{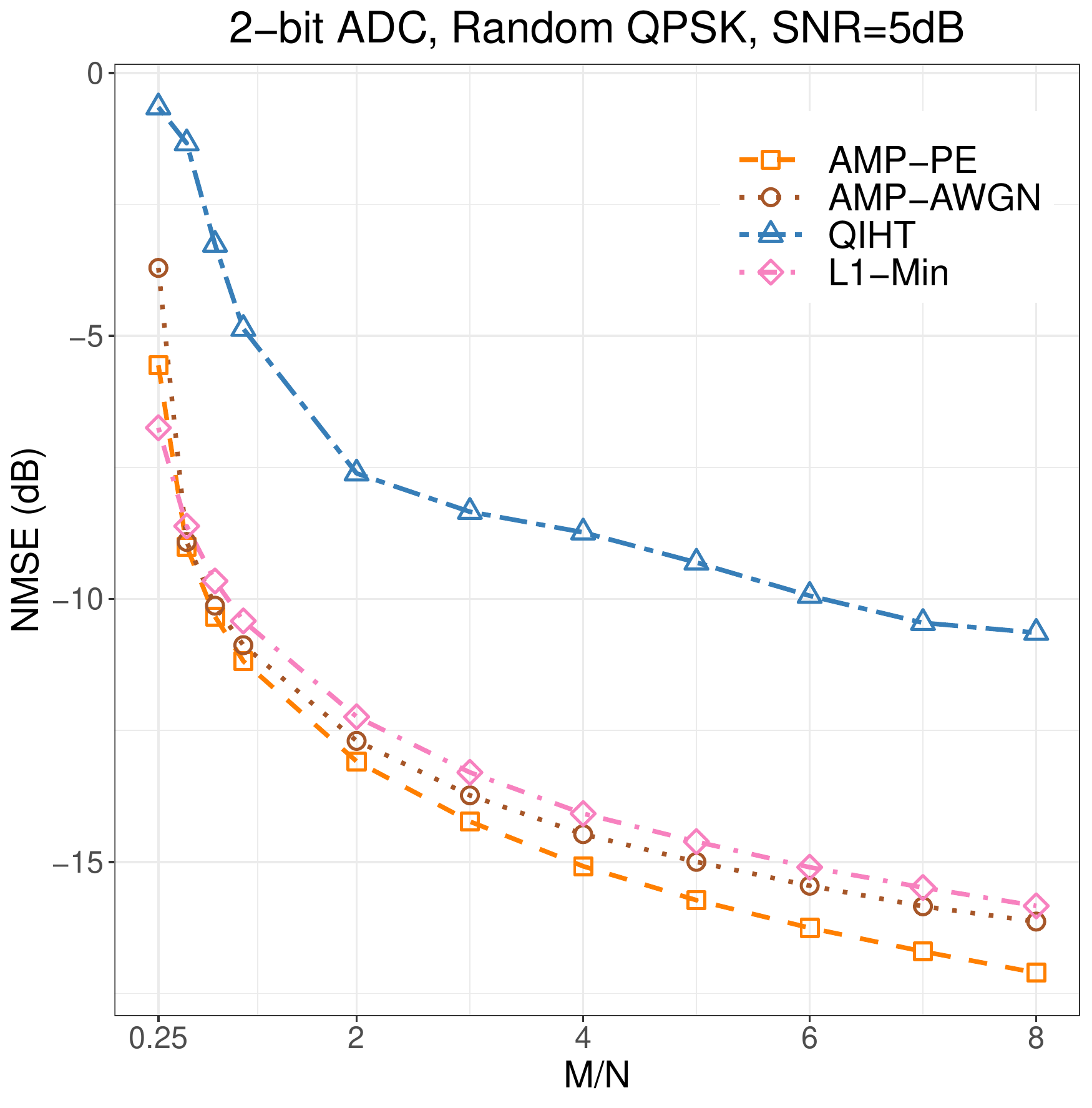}}\\
\subfigure{
\includegraphics[width=0.09\textwidth]{figures/random_gaussian_label.pdf}}
\subfigure{
\includegraphics[height=0.24\textwidth]{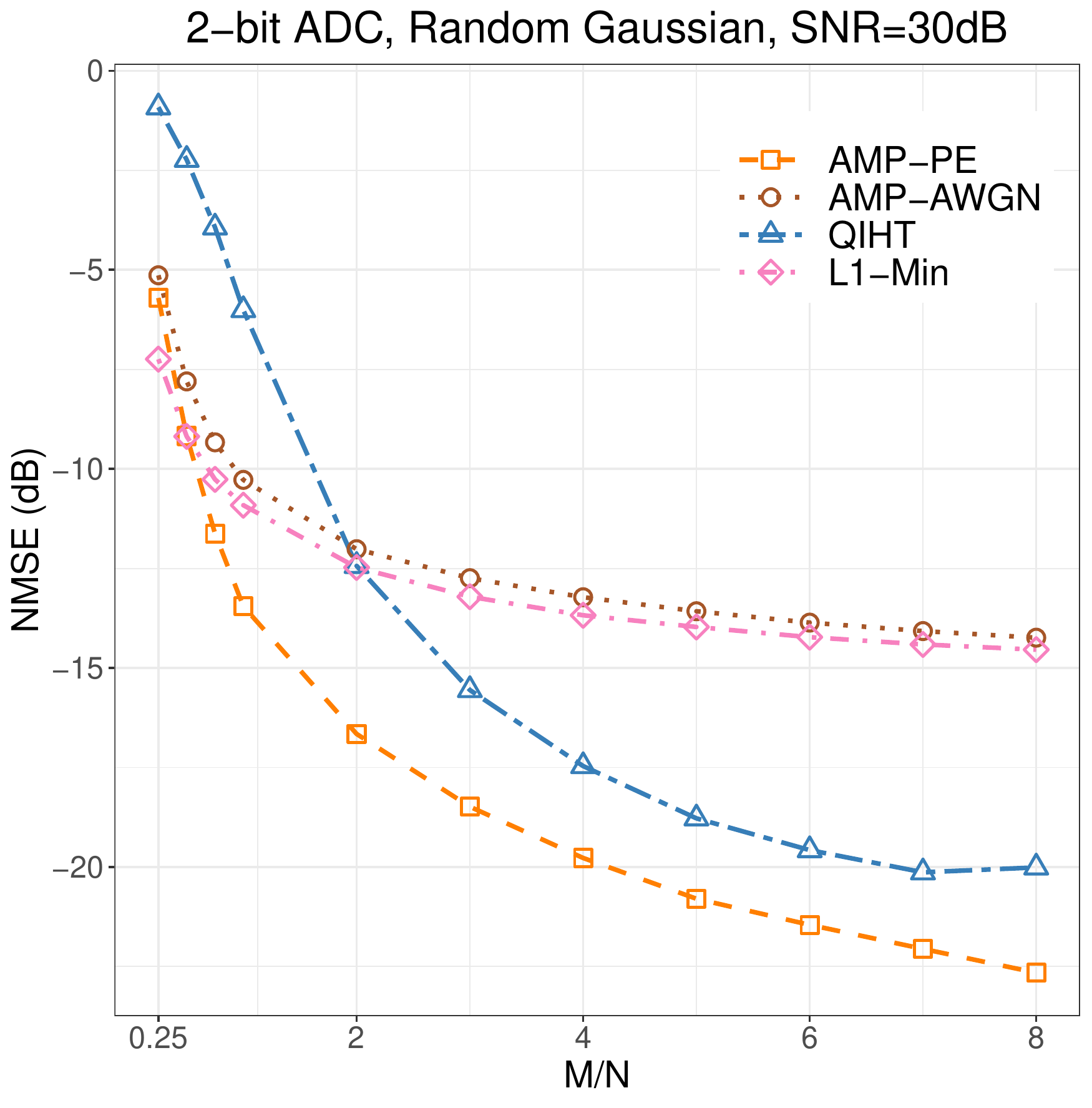}}
\subfigure{
\includegraphics[height=0.24\textwidth]{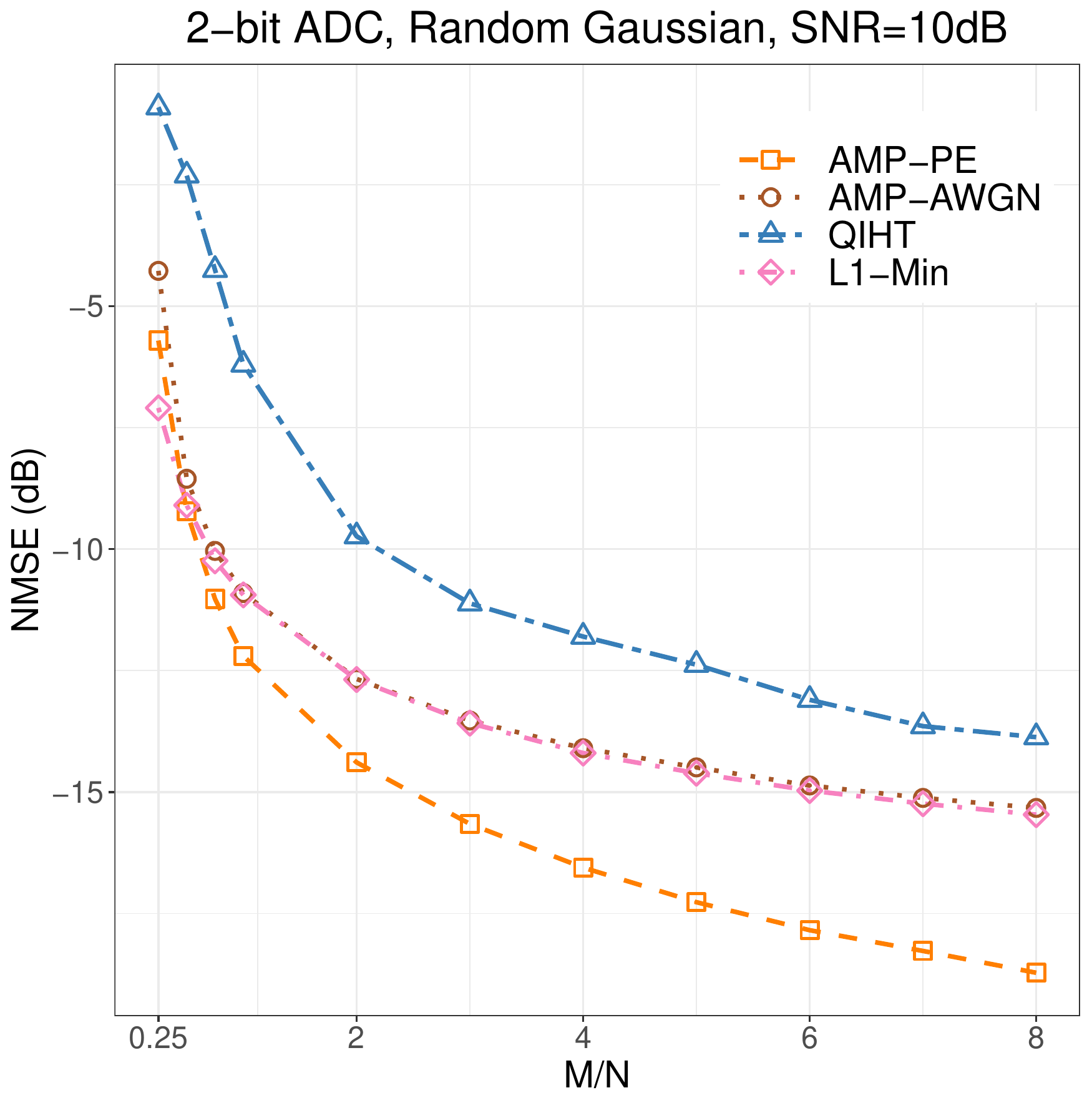}}
\subfigure{
\includegraphics[height=0.24\textwidth]{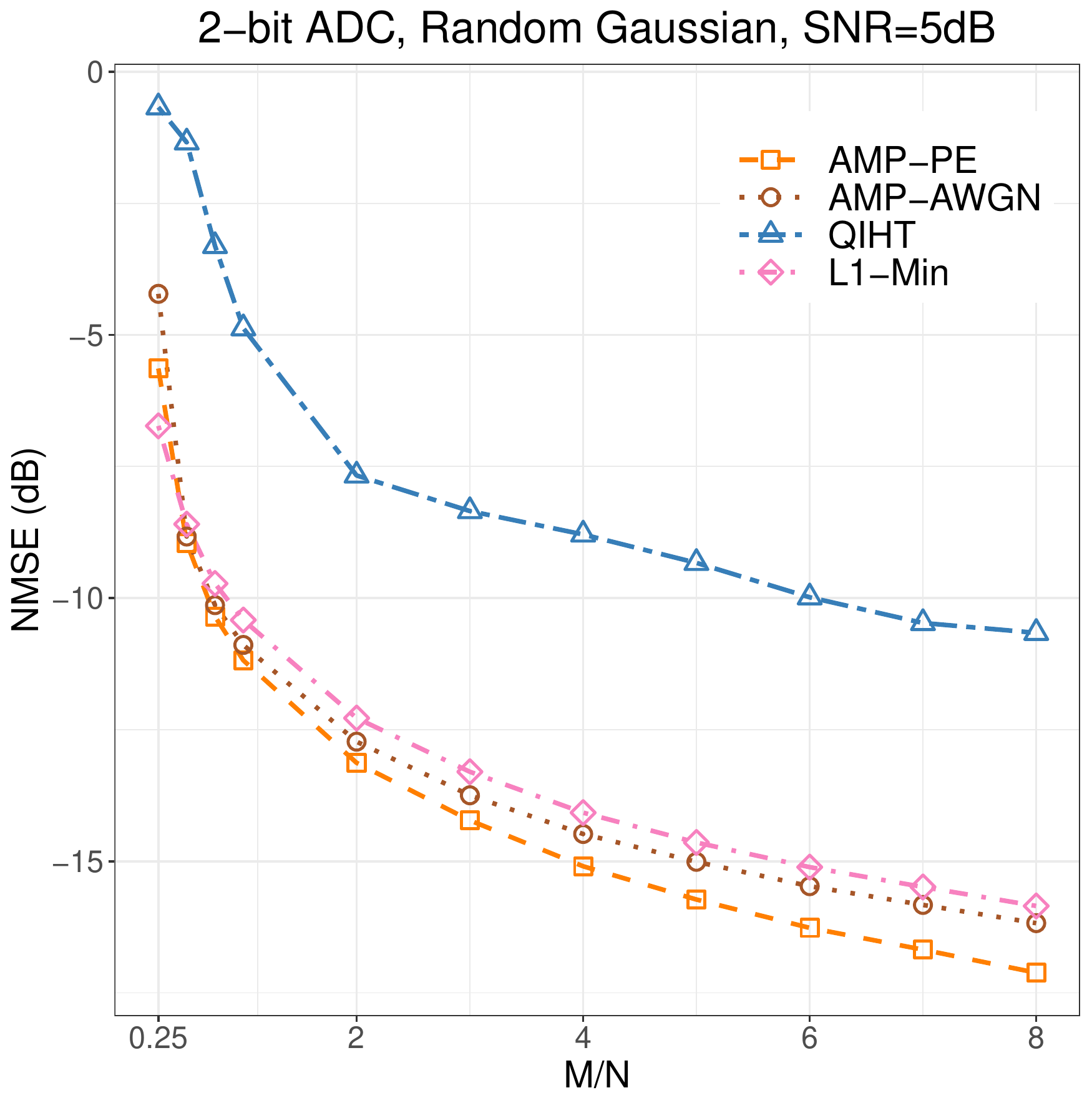}}

\caption{Comparison of different approaches in solving channel estimation from 2-bit measurements. The sampling ratio $\frac{M}{N}\in\{0.25,\cdots,8\}$. Two transmission sequences are used: random QPSK and random Gaussian. The pre-quantization SNR varies from $30$dB, $10$dB to $5$dB.}
\label{fig:2bit_ce_experiments}
\end{figure*}

\begin{figure*}[htbp]
\centering
\subfigure{
\includegraphics[width=0.09\textwidth]{figures/random_qpsk_label.pdf}}
\subfigure{
\includegraphics[height=0.24\textwidth]{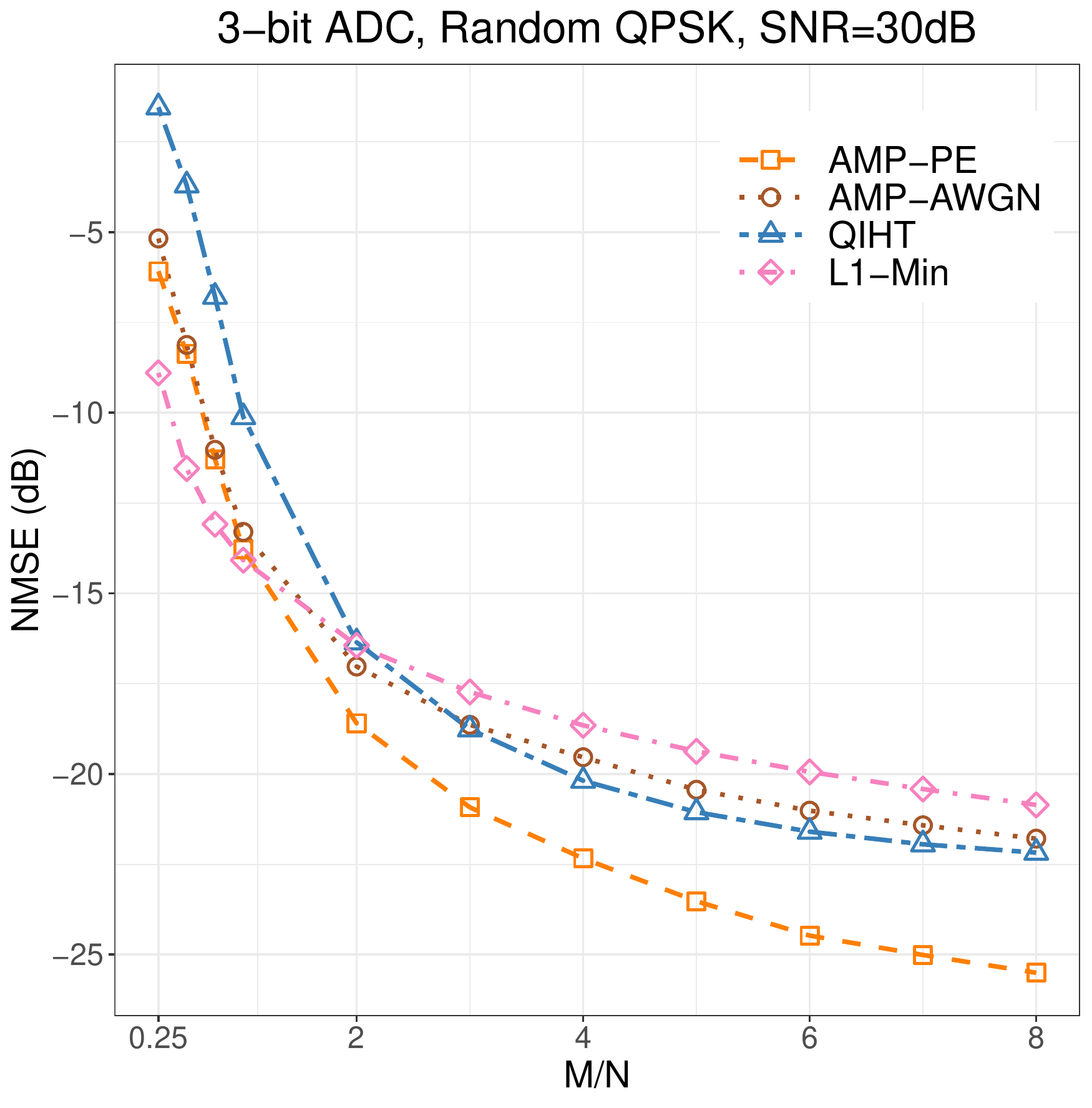}}
\subfigure{
\includegraphics[height=0.24\textwidth]{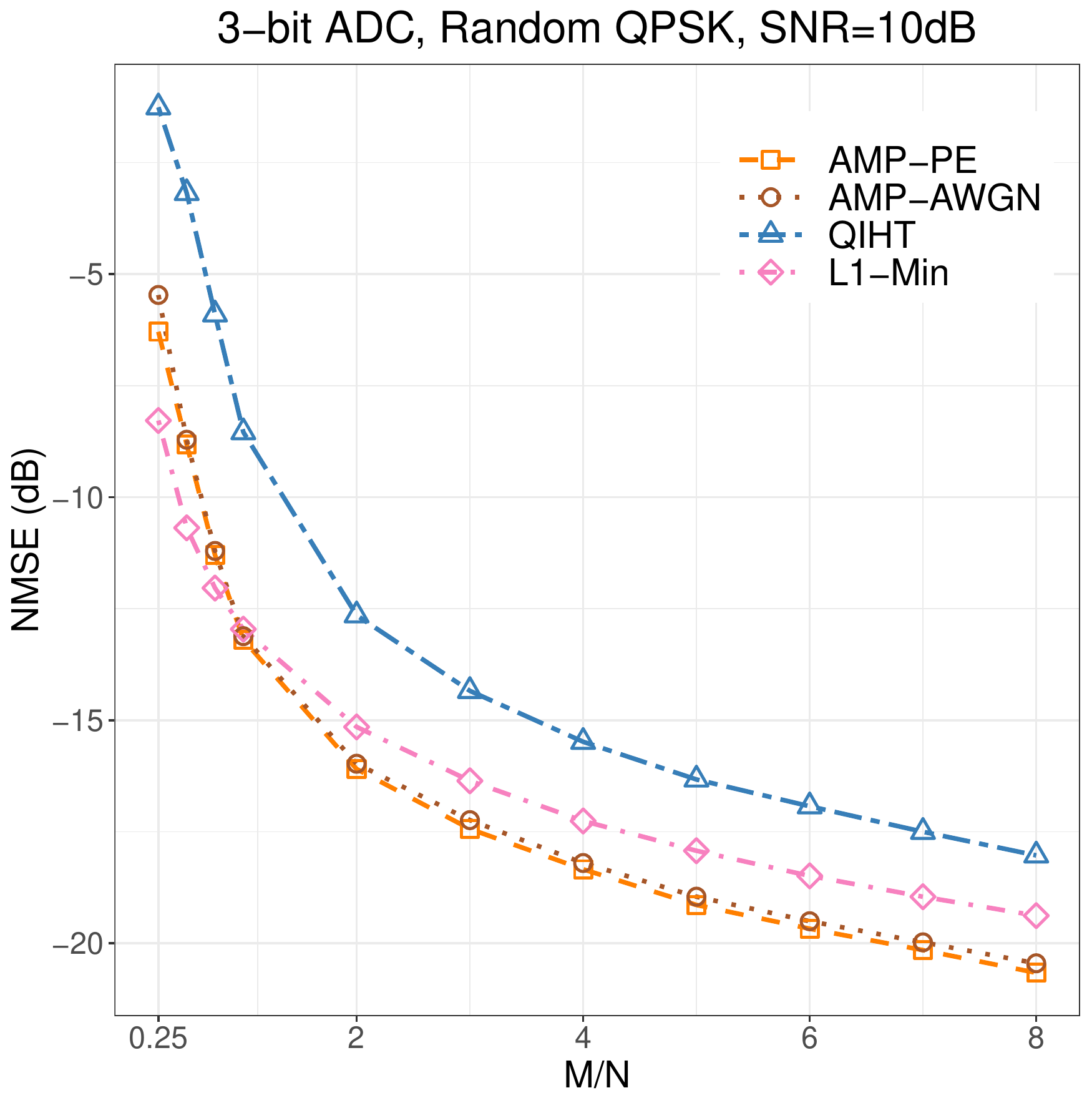}}
\subfigure{
\includegraphics[height=0.24\textwidth]{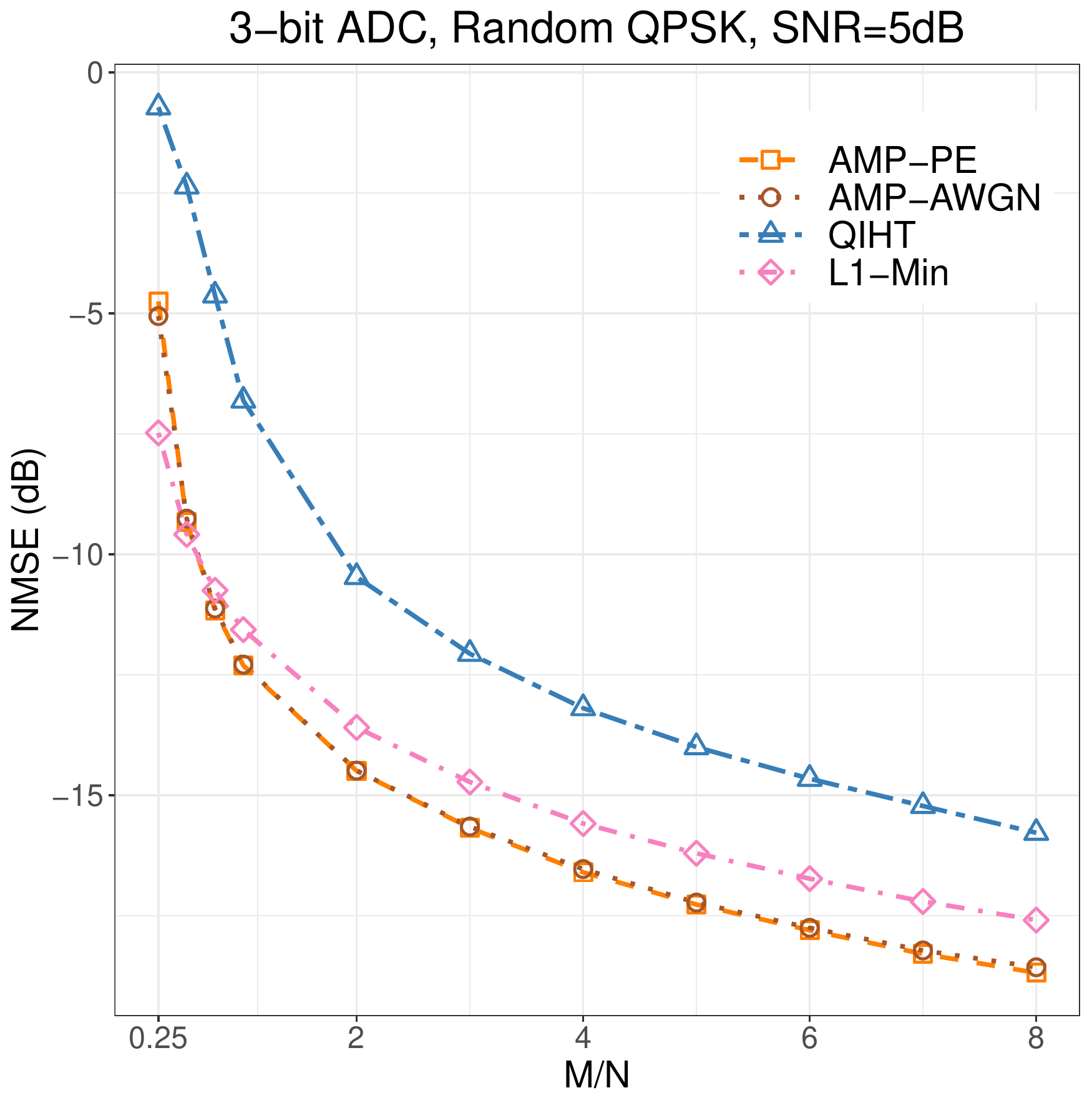}}\\
\subfigure{
\includegraphics[width=0.09\textwidth]{figures/random_gaussian_label.pdf}}
\subfigure{
\includegraphics[height=0.24\textwidth]{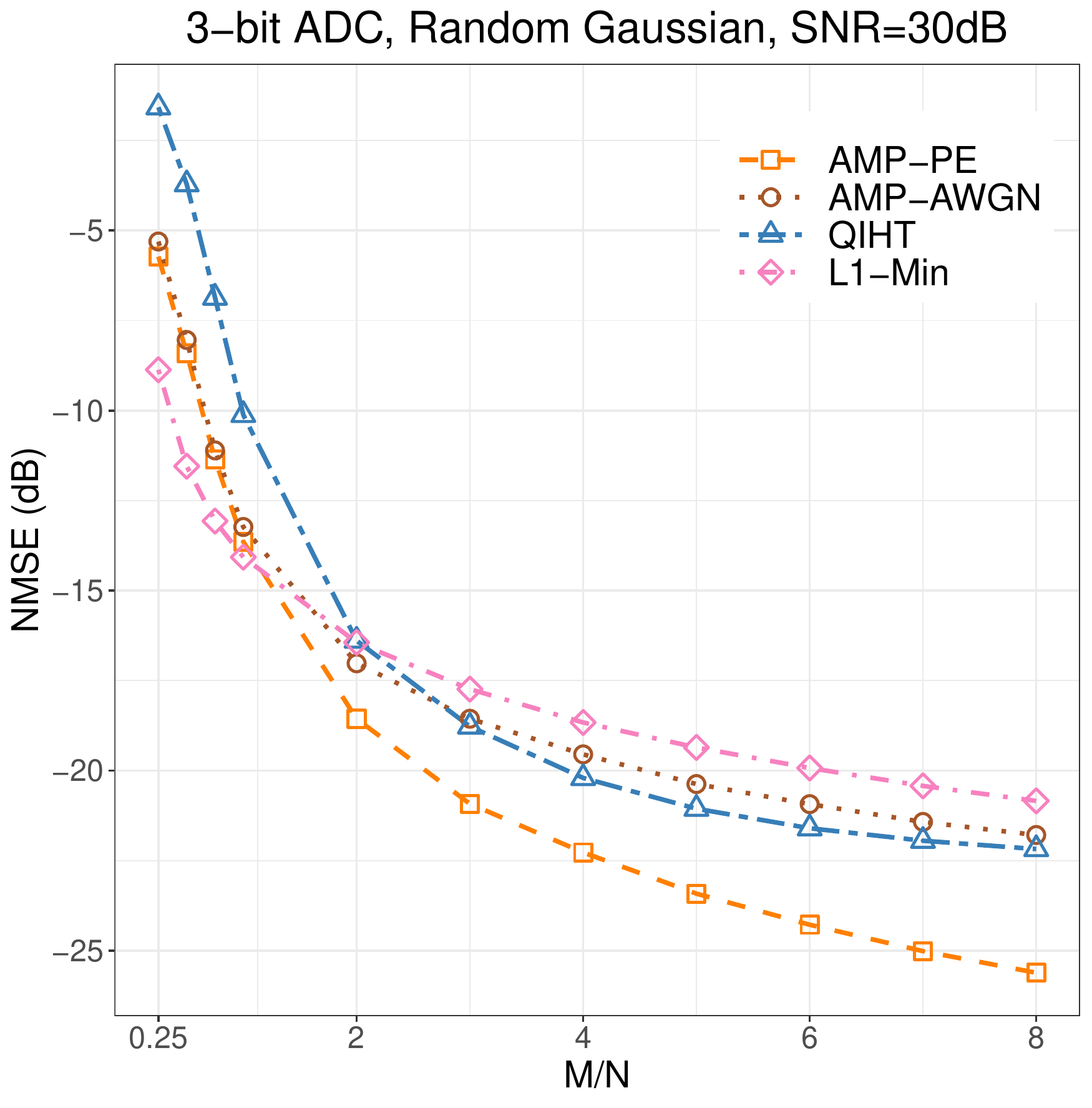}}
\subfigure{
\includegraphics[height=0.24\textwidth]{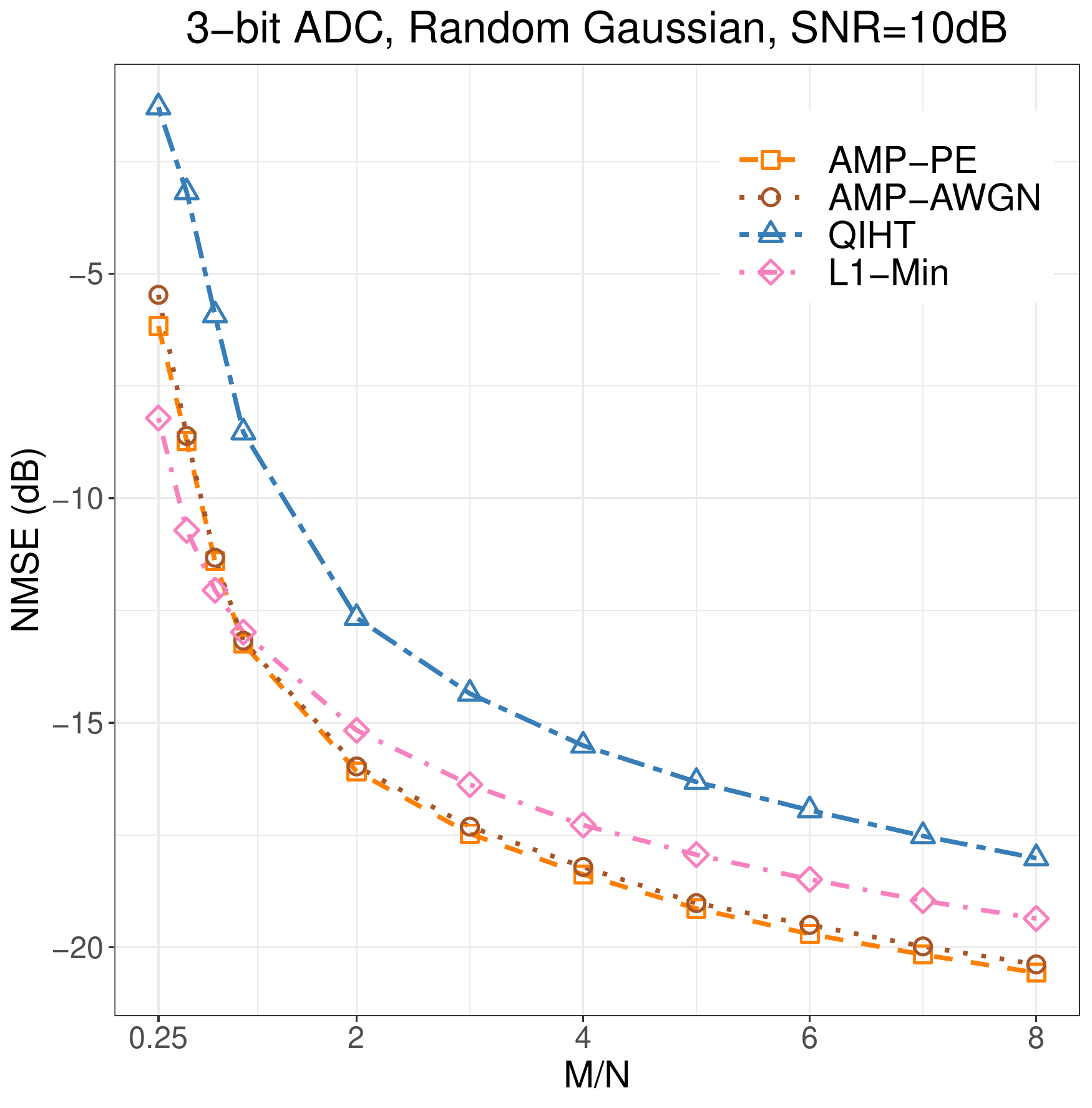}}
\subfigure{
\includegraphics[height=0.24\textwidth]{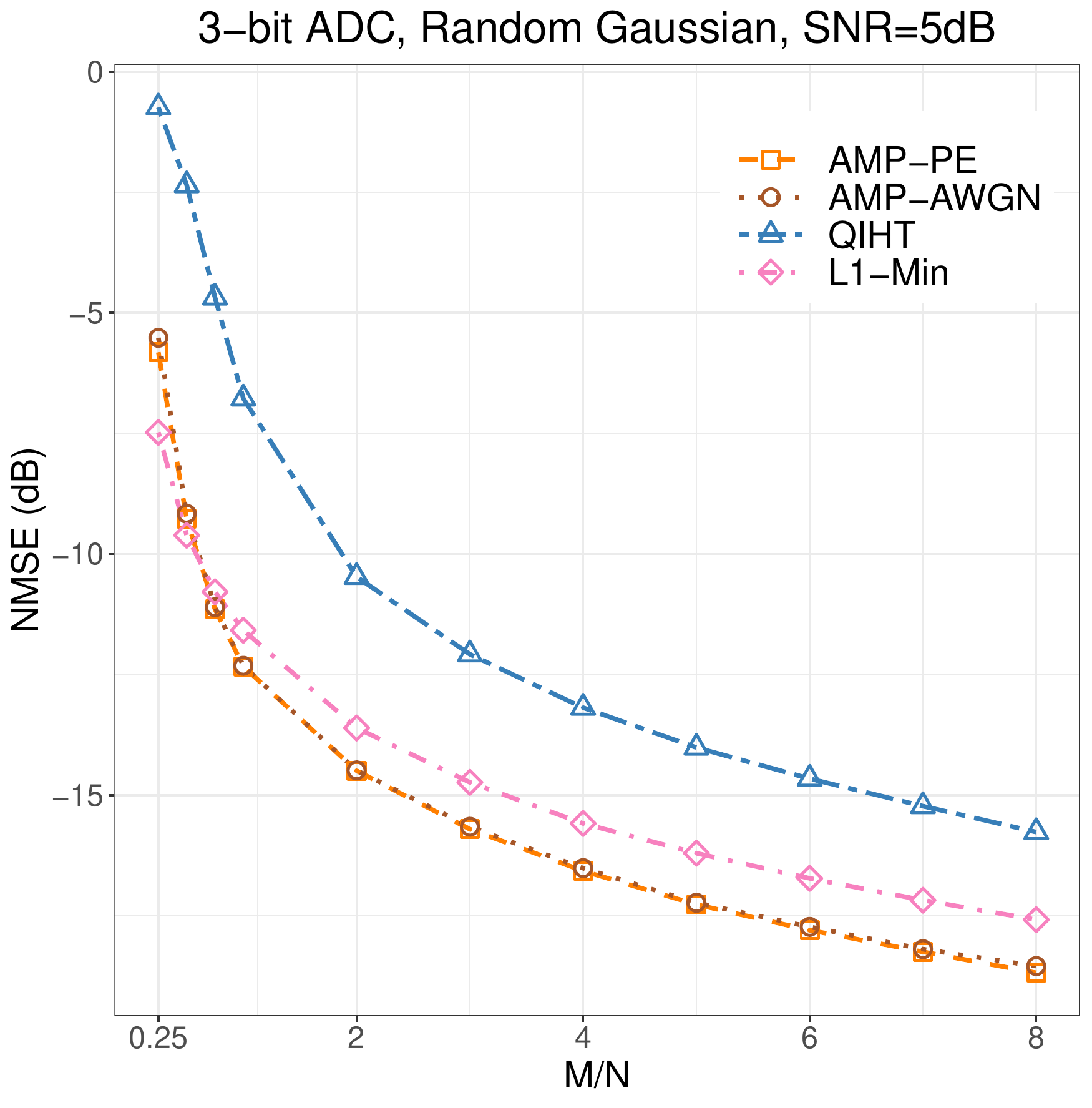}}

\caption{Comparison of different approaches in solving channel estimation from 3-bit measurements. The sampling ratio $\frac{M}{N}\in\{0.25,\cdots,8\}$. Two transmission sequences are used: random QPSK and random Gaussian. The pre-quantization SNR varies from $30$dB, $10$dB to $5$dB.}
\label{fig:3bit_ce_experiments}
\end{figure*}

The results from the channel estimation experiments with 2-bit and 3-bit measurements are shown in Fig. \ref{fig:2bit_ce_experiments}-\ref{fig:3bit_ce_experiments}.

\clearpage

\begin{figure*}[htbp]
\centering
\label{fig:compare_true_recovered_channel}
\subfigure{
\label{fig:true_channel}
\includegraphics[width=\textwidth]{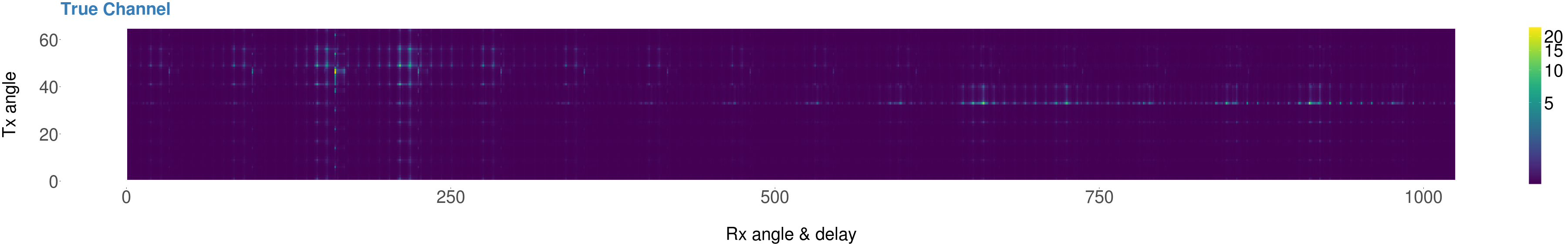}
}\\
\subfigure{
\label{fig:recovered_channel}
\includegraphics[width=\textwidth]{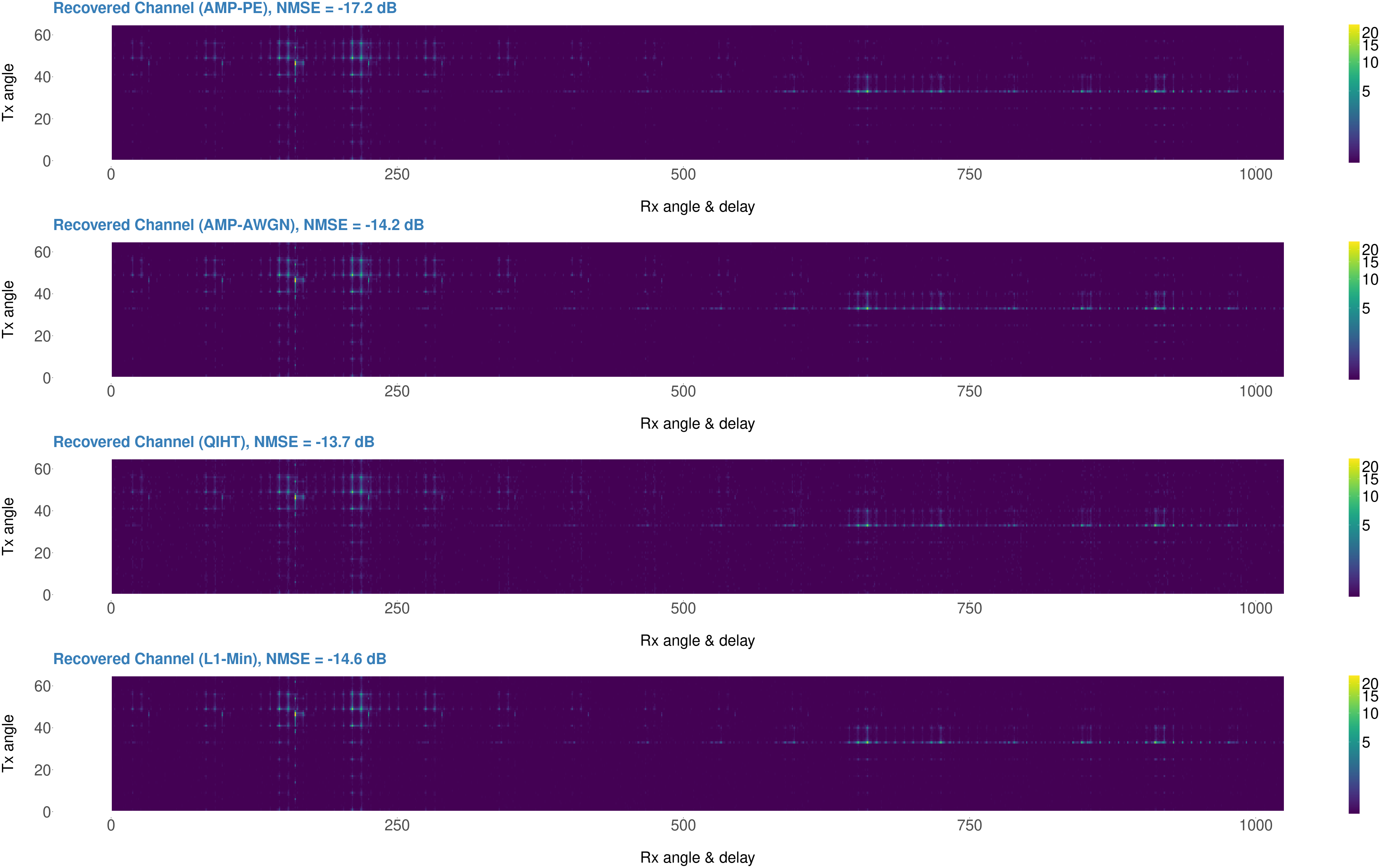}
}
\caption{The magnitudes (in log scale) of the true channel coefficients $|\vx|$ compared with those of the recovered channel coefficients $|\hat{\vx}|$ from 2-bit measurements produced with the random QPSK training sequence. The pre-QNT SNR=$10$dB and the sampling ratio $\frac{M}{N}=8$.}
\label{fig:compare_channel}
\end{figure*}

\begin{figure*}[htbp]
\centering
\label{fig:recovered_error}
\includegraphics[width=\textwidth]{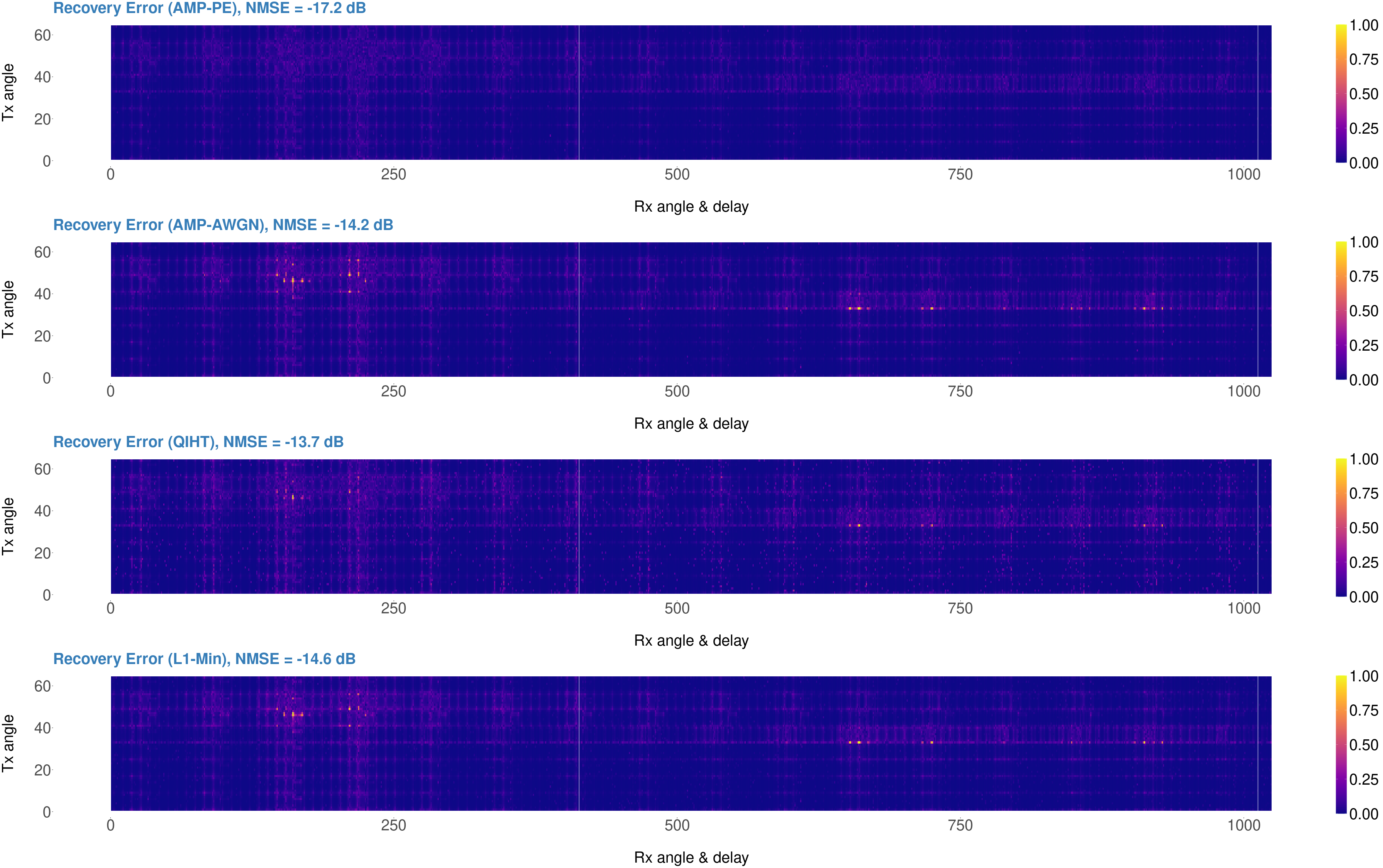}
\caption{The magnitudes of recovery errorr $|\vx-\hat{\vx}|$ between the true channel coefficients $\vx$ and the recovered channel coefficients $\hat{\vx}$ from 2-bit measurements produced with the random QPSK training sequence. The pre-QNT SNR=$10$dB and the sampling ratio $\frac{M}{N}=8$.}
\label{fig:compare_channel_error}
\end{figure*}

The recovered channels from 2-bit measurements when pre-QNT SNR=$10$dB and the sampling ratio $\frac{M}{N}=8$ are shown in Fig. \ref{fig:compare_channel} and Fig. \ref{fig:compare_channel_error}.

\end{document}